\documentclass[11pt]{Latex/Classes/GenThesis}

\title{Study of the Hall effect in two different strongly correlated fermion systems}

\author{GLADYS LE\'{O}N SUR\'{O}S}
\faculty{Sciences} \professor{T. Giamarchi} \departmentshort{DPMC}
\cityofbirth{Caracas} \country{Venezuela} \thesisnumber{0001}
\department{D\'{e}partement de Physique la Mati\`{e}re
condens\'{e}e} \university{Universit\'{e} de Gen\`{e}ve}
\yeargrad{2008} \degree{Philosophi\ae Doctor (PhD)}
\degreedate{30/6/2008}
\memberA{Christophe Berthod}{Codirecteur de
th\`{e}se}{Universit\'{e} de Gen\`{e}ve} \memberB{Dirk van der
Marel}{Professeur}{Universit\'{e} de Gen\`{e}ve} \memberC{Antoine
Georges}{Professeur}{\'{E}cole Polytechnique}

\approvaldate{12 Septembre 2008} \dean{Jean-Marc TRISCONE}

\usepackage[Sonny]{fncychap}        
\usepackage{epigraph}

\usepackage{color}

\begin{document}


\maketitle       
\fancyend

\maketitle
\fancyend
\frontmatter

\begin{dedication} 

Simplemente a mi familia

\end{dedication}


\fancyend

\setcounter{secnumdepth}{3} 
\setcounter{tocdepth}{3}    
\setcounter{page}{1}
\tableofcontents            

\pagestyle{plain}

Ce travail a été consacré a l'étude de l'effet Hall dans deux
systèmes différents de fermions fortement correlés: un système
constitué de liquides de Luttinger faiblement couplés et dans un
réseau triangulaire bidimensionnel. Dans le but d'avoir les outils
nécessaires pour attaquer ces problèmes, on a entrepris une
révision des propriétés principales des systèmes avec des
intéractions fortes, en nous concentrant dans les systèmes à deux
dimension et à basse dimensionnalité (1D et quasi 1D). Etant donné
que l'effet Hall est avant tout un phenonème de transport, on a
étudié les formalismes existants dans la littérature pour traiter
les propriétés de transport dans les systèmes fortement correlés,
et nous avons consacré tout un chapitre au traitement spécifique
de l'effet Hall.

Donc en un premier temps nous avons étudié le modèle des liquides
de Luttinger faiblement couplés. L'étude de l'effet Hall dans un
système quasi 1D a été partiellement motivée par les résultats
expérimentaux obtenus sur des conducteurs organiques quasi 1D. Ce
travail était basé sur l'application du formalisme de matrice
mémoire pour obtenir la constante de Hall ($\RH$) dans des chaines
a moitié remplis faiblement couplées, incluant la diffusion
umklapp. La géométrie du modèle a été choisie pour avoir le
courant circulant le long des chaines 1D. On a calculé la
dépendance en temperature et fréquence de $\RH$ en tenant compte
des intéractions particule-particule et particule-réseau
(processus umklapp). On a obtenu un coefficient de Hall $\RH$
constitué d'abbord par un terme d'éléctrons libres (valeur de
bande $\RH^0$) plus un terme de corection avec une dépendance en
temperature (fréquence) donné par une loi de puissance, due à la
présence de la diffusion umklapp. Ces lois des puissance sont des
signatures du comportement des liquides Luttinger, où les
exposants dépendent des paramètres d'interactions. Le coefficient
de Hall a aussi été calculé dans le système sans intéraction
particule-particule, condisuant à une dépendance logarithmique en
$T$ (où $\omega$), en accord avec la limite d'interaction nulle de
la loi de puissance. Etant donné que les conducteurs organiques
quasi 1D sont en même temps des systèmes à moitié remplis et au
quart remplis, nos résultats théoriques ne sont pas directement
applicables aux messures d'effect Hall existantes, mais ils nous
ont permis d'arriver a des conclusions intérésantes. En premier
lieu, la façon appropiée d'analyzer les messures d'effet Hall
realisées dans ces sytèmes quasi 1D (dans la même geométrie du
modèle), est de fitter les deviations de la valeur de bande

\fancyend


\begin{abstracts}        

Put your abstract or summary here, if your university requires it.

\end{abstracts}


\fancyend









\mainmatter


\selectlanguage{english}

\chapter{Introduction}

It is likely that Edwing Herbert Hall was not aware of the
enormous impact that his experimental results would have when he
measured the Hall voltage for the first time in 1879. Not only
because the Hall effect has allowed to determine the sign and
density of the charge carriers of many materials, but also because
it lead to a whole new field of research on the nature of the
particles responsible for the electrical current in any system.

Nowadays, the Hall resistivity has become a common tool when
studying the transport properties of any new material. However
real experiments have shown that this phenomenon cannot always be
interpreted in terms of the density of charge carriers and thus,
has to be interpreted in a different way. Compounds can have
strong interactions affecting the transport
properties. Such strongly interacting systems are now at the heart
of research in condensed matter physics and the understanding of
the effect of these interactions is crucial for the description of
these materials.

In this work we undertake the challenging task of studying the
Hall effect in two different systems taking into account
interactions. This requires the study of the existing methods for
treating strongly interacting systems and their application to the
Hall effect. We have chosen two model systems which are similar to
real compounds  where the measured Hall effect cannot be readily
explained. The work presented here is entirely theoretical but
some applications to real experiments are discussed. The plan of
the work is presented in what follows.

We begin this manuscript with a description of the standard
theoretical models used to describe two-dimensional and
low-dimensional fermions systems with strong correlations. In
particular, we review the main properties of the two-dimensional
square lattice and the 2D triangular lattice, and then those of
one-dimensional and quasi one-dimensional systems.

The second chapter is devoted to the transport phenomena in
systems with strong interactions. For this we explain the linear
response theory and the Kubo formulas. Such formulas are necessary
for the understanding of the memory function formalism developed
afterwards. We end this chapter with an application of the
previously mentioned theories in the calculation of transport
properties in low-dimensional systems.

The third chapter is completely dedicated to the Hall effect. We
begin with the explanation of the classical Hall effect. Next, we
obtain an expression for the Hall resistivity at infinite
frequency and finally, we use the memory formalism to include
interactions in the expression of the Hall resistivity.

Our first theoretical contribution is presented in the fourth
chapter. The Hall effect is investigated in a quasi
one-dimensional system made of weakly coupled 1D chains. For this
we begin with a review on quasi 1D organic conductors,
specifically in Bechgaard and Fabre salts, where various Hall
measurements have been accomplished. We explain the model and
methods used to calculate the Hall constant. We then present the
results and discuss their range of validity and application to
real experiments.

In the fifth chapter we present our second theoretical work. In
this case we consider the two-dimensional triangular lattice as
our model system. We start with the description of the sodium
cobalt oxide, a compound with a triangular lattice structure,
where the Hall effect have been investigated recently. As before,
we introduce the model and methods used to calculate the Hall
constant. We present the results obtained and discuss their range
of validity and application to the previously mentioned
experiments.

We close this manuscript with some general conclusions from the
two theoretical works presented and perspectives for future works.
Some appendices are presented at the end.

\chapter{Strongly correlated systems}\label{chap:strongly_correlated}





In condensed matter physics, understanding the effects of
interactions in real materials, has occupied theoretical and
experimental physicists for more than fifty years now. Depending
on the strength of interactions the convenient approach to be
implemented can be completely different. For example, if
interactions are very weak, like in metals, the free-electron
model works well to describe most of the physical properties. The
dimensionality also plays a crucial role in this choice. In
general, high-dimensional interacting systems are well described
by the Fermi Liquid theory, where electrons are dressed by the
density fluctuations around them, forming individual objects
called {\it{quasiparticles}}, as formulated by Landau and others
\cite{landau_fermiliquid_theory_microscopics,mahan_book,nozieres_book}.
The fraction of the electrons that remain in this quasiparticle
state is called the residue $Z$, and gives the amplitude of the
discontinuity in the occupation factor $n_{\vec{k}}$. The big
result of the Fermi liquid theory is that the properties of the
interacting system remain {\it{essentially}} the same of the free
fermionic particles.

In one dimension, on the contrary, the Fermi Liquid theory breaks
down and is the Luttinger Liquid theory that describes the
physics, even in systems with very strong interactions, as
explained in Sec.~\ref{sec:1D_case}. Furthermore, Fermi Liquid
theory is not always applicable in high-dimensions when one is
working with strong interactions. The common characteristic of
strongly correlated systems, where one finds materials as
different as heavy fermions, high Tc superconductors, cobalt
oxides and organic conductors, is a narrow band where fermions
interact strongly on a scale of the order of electronvolts at
short distances. In two-dimensional systems, for example, none of
the previously mentioned theories is applicable when strong
interactions are present. Then, basic models such as the Hubbard,
$t$-$J$, Anderson and Kondo lattice models had to be implemented
in an effort to understand the properties of these strongly
correlated layered materials.

As we are interested in the study of the Hall effect in a quasi
one-dimensional system and in a two-dimensional triangular
lattice, we will devote this chapter to the description of the
theoretical models used to describe these systems throughout this
work. First, we will study the main properties of the
two-dimensional Hubbard model, focusing on two geometries: the
rectangular and the triangular lattice. Then, for the case of
low-dimensional systems, we will make a short review on the
Luttinger liquid theory and its extension to describe quasi
one-dimensional systems.

\section{The Hubbard model}\label{sec:Hubbard_model}
\begin{figure}
\begin{center}
\includegraphics[width=8cm]{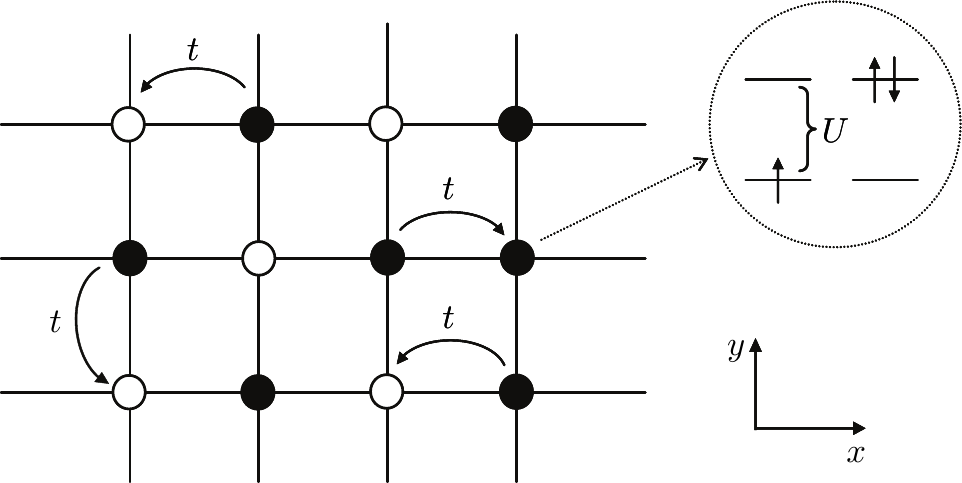}
\caption{\label{fig:Hubbard} The Hubbard model: electrons hops
from site to site with a hopping amplitude $t$. Full (empty)
circles denote sites with one (zero) electron. When a particle
hops into an occupied site, it has to pay a penalty $U$ (on-site
Coulomb interaction). This is shown in the image on the right.}
\end{center}
\end{figure}

The Hubbard model is one of the simplest models existing in
literature \cite{mahan_book} to describe the physics of
interacting fermions on a lattice. Hubbard (1963) proposed an
extension of the tight-binding model \cite{mahan_book}, where
electrons hop from site to site with a matrix element $t$, adding
a term that provides a penalty $U$ for any atomic site occupied by
more than one electron. This is depicted in
Fig.~\ref{fig:Hubbard}. In terms of fermionic operators, the
Hubbard Hamiltonian takes the following form \cite{mahan_book}
\begin{equation}\label{Hubbard}
\mathcal{H}=-t\sum_{\langle ij\rangle\sigma}c^{\dagger}_{i\sigma}
c^{\phantom{\dagger}}_{j\sigma} + U\sum_i
n_{i\uparrow}n_{i\downarrow}-\mu\sum_{i\sigma} n_{i\sigma},
\end{equation}
where $c^{\dagger}_{\alpha}\,(c_\alpha)$ is the creation
(annihilation) fermion operator,
$n_\alpha=c^{\dagger}_{\alpha}c_\alpha$ is the fermionic number
operator and $\langle ij\rangle$ are nearest-neighboring sites.
The first term in Eq.~(\ref{Hubbard}) represents the kinetic
energy of the electrons and the second term correspond to the
interaction between them. In addition, we include a chemical
potential $\mu$ in order to adjust the number of electrons in the
system.

There are extensions of the Hubbard model in which electron
hopping beyond nearest-neighbors and electron-electron interaction
at larger distances, are taken into account. We will not treat
these models here because they are beyond the scope of this work,
and we refer the reader to the literature for a complete
description of these cases \cite{mahan_book}. The Hubbard model
was solved exactly in one dimension, by Lieb and Wu (1968)
\cite{lieb_hubbard_exact}, as we will discuss it in the next
section. For dimensions greater than one, many attempts have been
undertaken to derive a phase diagram at zero temperature using
random phase approximation (RPA) \cite{mahan_book}, mean field
theory \cite{marder_book}, numerical methods as the dynamical mean
field theory (DMFT) \cite{georges_dmft}, etc; but there are still
many open questions and discrepancies between different
approaches.

When working with the Hubbard model, there are two important
energy scales to take into account: the on-site Coulomb
interaction $U$ and the bandwidth $W$. They determine the behavior
of the system by the value of the ratio $U/W$. If this ratio is
small, the free-electron model can be used to describe most of the
physical properties of the system. Simple metals like aluminium or
sodium have $U\ll W$. On the other hand if the ratio is large,
then $U\gg W$ and electron correlations are dominant. This case is
not fully understood but some of its properties are currently well
characterized. We list some of them below.

The electron density $n=N/N_s$ ($N$ being the number of electrons
and $N_s$ the number of lattice sites) varies from $0$ to $2$, due
to Pauli principle. When $n=2$ there are two electrons in each
site, they cannot move, and the system is an insulator. At half
filling, when there is exactly one electron per site ($n=1$), the
ground state is antiferromagnetic (AF), {\it{i.e.}}, neighboring
spins pointing in opposite directions. Furthermore, if $U\gg W$
and there is one particle per site, the system is in an insulating
state known as a {\it{Mott insulator}}
\cite{mott_historical_insulator,mott_metal_insulator}. Possible
phases of the Hubbard model are: the paramagnetic phase, the
ferromagnetic phase, the charge-density wave phase (CDW) and the
spin density wave phase (SPW) \cite{mahan_book}. Each of these
phases has particular properties which have been studied
extensively in the last decades. Nevertheless a unique phase
diagram for the Hubbard model for dimensions $d>1$, is still
lacking.

One important property of the Hubbard model at half-filling is the
{\it{particle-hole}} symmetry, which means that the Hamiltonian in
Eq.~(\ref{Hubbard}) is invariant under the electron-hole
transformation:
$c_{j\sigma}\rightarrow(-1)^{j}d^{\dagger}_{j\sigma}$, where
$d_{\alpha}$ ($d_{\alpha}^\dagger$) creates (annihilates) a hole.
This property is very peculiar to half-filled bi-partite lattices
(lattices that can be decomposed into a disjoint union of two
sublattices) and is not fulfilled, for example, on the triangular
lattice (see Sec.~\ref{sec:Hubbard_triangular}).


An special limit of the Hubbard model that will be studied in
Chapter.~\ref{chap:Hall_triangular} is called the {\it{atomic
limit}} and corresponds to a system where the hopping amplitude is
set to $t=0$. In this case the system consist on isolated sites
where particles cannot move from site to site. An exact solution
can be found for the atomic limit of the Hubbard model (each site
is considered individually), for any dimension (see
Appendix.~\ref{app_atomic}). Another exactly solvable limit is the
$U=0$ limit (tight-binding model), where the Hubbard Hamiltonian
can be also completely diagonalized. The properties of the Hubbard
model strongly depend on the type of the lattice and on the
dimensionality. That is why we will focus here only on two
geometries of the 2D Hubbard model: the square lattice and the
triangular lattice. But, before entering into these two particular
geometries, we will mention another variant of Hamiltonian
(\ref{Hubbard}) known as the $t$-$J$ model
\cite{hirsch_modele_tj}.

The $t$-$J$ model describes a system of interacting fermions on a
lattice where the on-site interaction $U$ is considered infinite,
in order to prohibit double occupancy, {\it{i.e.}}, sites with
$n_{i\uparrow}n_{i\downarrow}=1$. This constraint is implemented
using the Gutzwiller projection operator
$P_{G}=\Pi_{i}(1-n_{i\uparrow}n_{i\downarrow})$. Applying this
operator to Hamiltonian (\ref{Hubbard}) one obtains
$\mathcal{H}_{t-J}= P_{G}\mathcal{H}P_{G}\,+$``exchange term'',
where the exchange term is given by $J\sum_{\langle ij\rangle}
\vec{S}_i\cdot \vec{S}_j$, with $S_i$ the spin operator at site i.
The exchange term represents an effective antiferromagnetic
interaction between neighboring spins, which results from the
virtual hopping of electrons into and out of doubly-occupied
states, and has an amplitude $J=4t^2/U$. The $t$-$J$ model has no
exact solution either, but it is widely used in the study of
fermionic systems with very strong interactions. We will not study
the properties of the $t$-$J$ model here because we are interested
in the effect of interactions as a function of their strength, and
for this the Hubbard model is the more accurate model to describe
our two-dimensional systems. But it remains a good model to make
comparisons at very strong interactions, as will be done in
Chapter~\ref{chap:Hall_triangular}. We refer the reader to the
literature for a complete review on the $t$-$J$ model
\cite{matsumoto_tjmodel,Kotliar_Liu}.

\section{The 2D square lattice}

\begin{figure}
\begin{center}
\includegraphics[width=12cm]{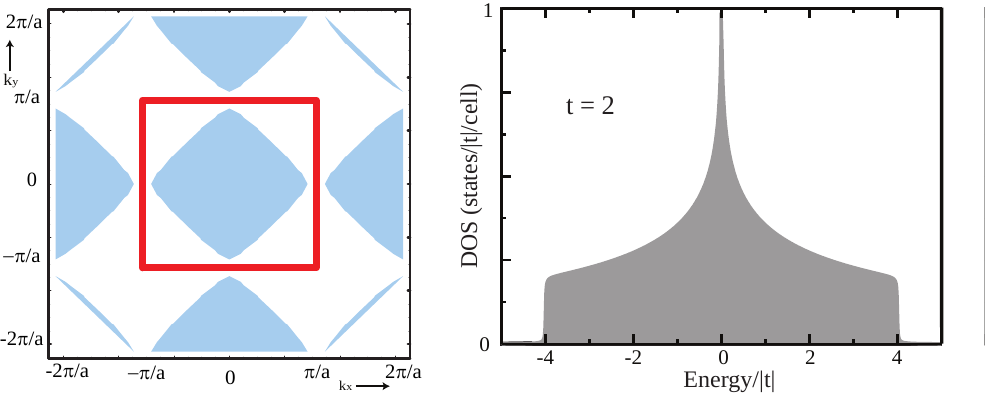}
\caption{\label{fig:Fermi_square} (Left) Fermi surface of the 2D
square lattice. The red square represent the first Brillouin zone
for the square lattice and the Fermi surface is the shaded area
inside this zone. (Right) Density of states of the square lattice.
The DOS exhibits a van Hove singularity at the center of the band.
The bandwidth of the system is equal to $8|t|$.}
\end{center}
\end{figure}

In this section we will study some general properties of the
square lattice and definitions that will be useful in the rest of
this work. Then, we will give the properties of the Hubbard model
in this particular geometry. The two-dimensional square lattice
has the structure shown in Fig.~\ref{fig:Hubbard} which is
invariant under translation (it has {\it{translational
symmetry}}). The lattice is defined by two primitive vectors
$\vec{a}_1=a(1,0)$ and $\vec{a}_2=a(0,1)$, where $a$ is the
intersite distance, called the {\it{lattice parameter}}. The real
space lattice, often referred as {\it{direct lattice}}, can also
be represented in momentum space. The latter is known as the
{\it{reciprocal lattice}} and is formed by the set of vectors
$\vec{K}$ satisfying $e^{i\vec{K}\cdot\vec{R}}=1$, with $\vec{R}$
a vector of the direct lattice \cite{ashcroft}. The two primitive
reciprocal vectors are $\vec{b}_1=\frac{2\pi}{a}(1,0)$ and
$\vec{b}_2=\frac{2\pi}{a}(0,1)$. We will see, in a minute, how any
vector on k-space is defined by $\vec{b}_1$ and $\vec{b}_2$. Due
to the periodicity of the lattice, it properties can be studied in
a reduced zone or primitive cell of the reciprocal space defined
by the {\it{Wigner-Seitz}} cell of the origin, which associates
with each lattice point all of space which is closer to it than
any other lattice point \cite{marder_book}. This is the
{\it{Brillouin zone}} (see Fig.~\ref{fig:Fermi_square}).

One important quantity that changes from one geometry to the
other, is the dispersion relation or momentum energy of the
particles. It also depends on the model used. Let us take the
previously studied Hubbard model to describe the 2D square
lattice. The dispersion relation is obtained by transforming the
Hamiltonian to momentum space, using the Fourier transform of
fermionic operators
\begin{equation}
c^{\dagger}_{i\sigma}=\frac{1}{\sqrt{\Omega}}\sum_{\vec{k}}e^{-i\vec{k}\cdot\vec{r}_i}c^{\dagger}_{\vec{k}\sigma},
\end{equation}
where $\Omega$ is the volume of the system. The inverse Fourier
transform is thus given by
$c^{\dagger}_{\vec{k}\sigma}=\left(1/\sqrt{\Omega}\right)\sum_{\vec{i}}e^{i\vec{k}\cdot\vec{r}_i}c^{\dagger}_{i\sigma}$.
We will also need the Fourier transform of the number operator,
which is
\begin{equation}
n_{i\sigma}=\frac{1}{\sqrt{\Omega}}\sum_{\vec{k}}e^{i\vec{k}\cdot\vec{r}_i}n_{\vec{k}\sigma}.
\end{equation}
Then, Fourier transforming Hamiltonian (\ref{Hubbard}) we obtain \cite{mahan_book}
\begin{equation}\label{Hubbard_kspace}
\mathcal{H}=\sum_{\vec{k}\sigma}\varepsilon_{\vec{k}}
c^{\dagger}_{\vec{k}\sigma} c^{\phantom{\dagger}}_{\vec{k}\sigma}
+ U\sum_{\vec{k}}
n_{\vec{k}\uparrow}n_{\vec{-k}\downarrow}-\mu\sum_{\vec{k}\sigma}c^{\dagger}_{\vec{k}\sigma}
c^{\phantom{\dagger}}_{\vec{k}\sigma}.
\end{equation}
The coefficient of the first term in Eq.~(\ref{Hubbard_kspace})
(the kinetic energy term) gives the dispersion relation of the 2D
square lattice. Thus, the dispersion relation is
\begin{equation}\label{square_dispersion}
\varepsilon_{\vec{k}}= -2t\left[\cos(k_xa)+\cos(k_ya)\right],
\end{equation}
where $a$ is the lattice parameter ($|r_i-r_j|=a$, with $i,j$
nearest-neighboring sites). In Eq.~(\ref{square_dispersion}) we
supposed an isotropic square lattice. The components $k_x$ and
$k_y$ of any vector in the reciprocal lattice are obtained from
the primitive vectors $\vec{b}_1$ and $\vec{b}_2$ trough the
relation: $k_x=n\vec{b}_1/2\pi$ and $k_y=m\vec{b}_2/2\pi$, where
$m$ and $n$ are integers. The dispersion relation determines the
bandwidth ($W$) of the system, which is given by the range of
energies between the maximum and minimum of
$\varepsilon_{\vec{k}}$. As can be seen from
Eq.~(\ref{square_dispersion}), the bandwidth of the 2D square
lattice is equal to $W=8|t|$. The dispersion relation is often
written with reference to the chemical potential, combining the
1st and 3rd terms on the right hand side of
Eq.~(\ref{Hubbard_kspace}):
$\xi_{\vec{k}}=-2t\left[\cos(k_xa)+\cos(k_ya)\right]-\mu$.
The ground state of the non-interacting system ($U=0$ in
Hamiltonian (\ref{Hubbard_kspace})) is constructed by occupying
all the energy-levels with two electrons with opposite spins,
until reaching the total number of particles $N$. The highest
energy level is called the {\it{Fermi}} level and has an energy
$\varepsilon_{\vec{k_{\text{F}}}}$, with $k_{\text{F}}$ the Fermi
momentum. This energy level allows to define the "Fermi surface",
which is an important quantity when studying real materials
(nowadays measured experimentally with different methods, such as
angle resolved photoemission spectroscopy). The Fermi surface is
defined by the collection of points in the reciprocal space with
energy $\varepsilon_{\vec{k_{\text{F}}}}=\mu$, {\it{i.e}},
$\xi_{\vec{k}}=0$. Fig.~\ref{fig:Fermi_square} shows the Fermi
surface and first Brillouin zone of the 2D square lattice
described with the Hubbard model. Following the {\it{Luttinger
Theorem}}, interactions do not change the volume of the Fermi
surface, only its shape. Another important quantity is the Density
of states (DOS) of the system, which describes the number of
states at each energy level that are available to be occupied. It
is important to known the shape of the DOS when one is working
with the Hall effect, due to the relation between the Hall
resistivity and the electronic density, as we will see in
Chapter~\ref{chap:Hall_effect}. Fig.~\ref{fig:Fermi_square} shows
the DOS of the non-interacting square lattice, which exhibits a
van Hove singularity at the center of the band.

The particle-hole symmetry mentioned in the previous section is
fulfilled in the square lattice, because it is a bipartite
lattice. The particle-hole symmetry at half-filling is evident
from the shape of the DOS, which is symmetric with respect to the
center of the band. Although we will not work on the square
lattice for the Hubbard model, it is important to know this
geometry because it is the simpler model, it has been extensively
studied and thus, it always works as a point of reference. Let us
now move to the triangular lattice, which is the geometry that
will be used with the Hubbard model in
Chapter.~\ref{chap:Hall_triangular}

\section{The 2D triangular lattice}\label{sec:Hubbard_triangular}

\begin{figure}
\begin{center}
\includegraphics[height=4cm,width=8cm]{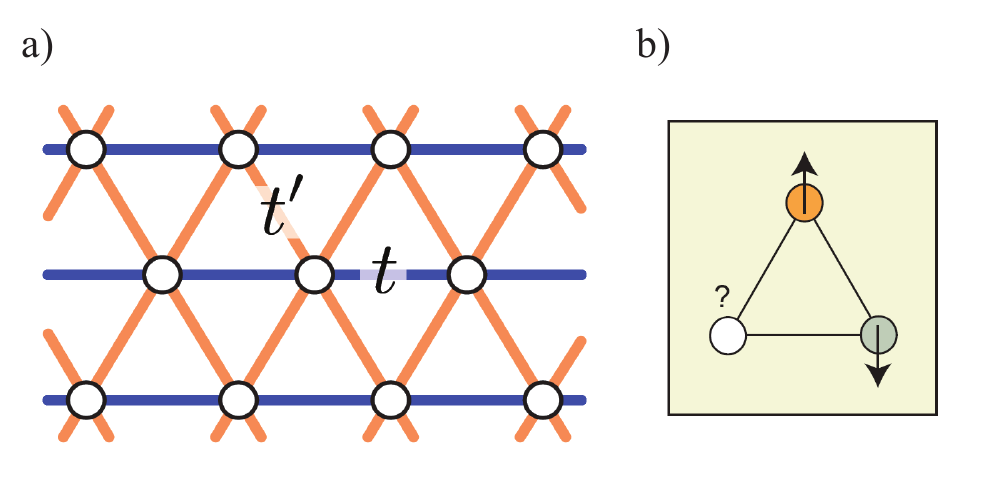}
\caption{\label{fig:triangular_frustation} a) Two-dimensional
triangular lattice structure with hopping amplitudes $t$ and $t'$.
b) Geometrical frustration on a triangle. Should the empty site be
spin-up or spin-down?. From Ref.~\cite{Ong_Cava}}
\end{center}
\end{figure}

\begin{figure}
\begin{center}
\includegraphics[width=13cm]{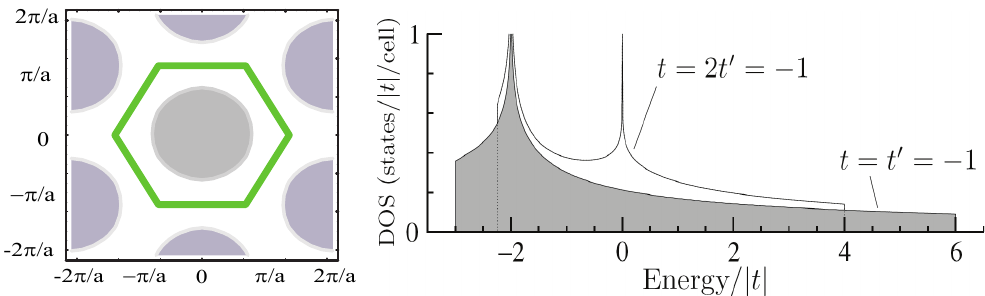}
\caption{\label{fig:Fermi_triangular}.(Left)  Fermi surface of the
2D triangular lattice. The green hexagon represent the first
Brillouin zone for the triangular lattice and the Fermi surface is
the shaded area inside this zone. (Right) Density of states of the
triangular lattice. The bandwidth of the system is $W=9|t|$. The
density of states (DOS) exhibits two van Hove singularities which
are degenerate when $t=t'$.}
\end{center}
\end{figure}

The structure of the two-dimensional triangular lattice is shown
in Fig.~\ref{fig:triangular_frustation}a. In the triangular
lattice, hopping $t$ occurs along the longitudinal direction and
hopping $t'$ along a direction defined by an angle of $60$ degrees
from the longitudinal direction, as can be seen in
Fig.~\ref{fig:triangular_frustation}a. The lattice is called
isotropic when the hopping amplitudes satisfy $t=t'$. The
primitive vectors defining the lattice in real space are
$\vec{a}_1=a(1,0)$ and $\vec{a}_2=a(1,\sqrt{3})/2$, with $a$ the
lattice parameter. And for the reciprocal space vectors we have
$\vec{b}_1=\frac{2\pi}{a}(1,-\frac{1}{\sqrt{3}})$ and
$\vec{b}_2=\frac{2\pi}{a}(0,\frac{2}{\sqrt{3}})$.
As before, we obtain the dispersion relation from the Fourier
transformed Hamiltonian (\ref{Hubbard_kspace}) for the 2D
triangular lattice. Thus, $\varepsilon_{\vec{k}}$ is this geometry
is given by
\begin{equation}
\varepsilon_{\vec{k}} = -2\left[t\cos(k_xa)+2t'\cos(k_xa/2)\cos(k_ya\sqrt{3}/2)\right].
\end{equation}
Thus, the bandwidth is $W=9|t|$. With this dispersion relation we
obtain the Fermi surface shown in Fig.~\ref{fig:Fermi_triangular},
as well as the Brillouin zone which has an hexagonal character.
The corresponding density of states (DOS) exhibits two van Hove
singularities which are degenerate when $t=t'$ (see
Fig.~\ref{fig:Fermi_triangular}). Unlike in the square lattice,
the DOS has no particle-hole symmetry, irrespective of the value
of $t$ and $t'$. Thus, the particle-hole symmetry is not fulfilled
because the lattice is no a bipartite one.

Two important properties arise from the topology of
the triangular lattice. The first one is that it has the smallest
number of steps necessary to make a loop, namely three. This will
be reflected in the study of the Hall effect on the triangular
lattice in Chapter~\ref{chap:Hall_triangular}. The second one is
called the electronic {\it{frustration}} and it appears because
the geometrical arrangement in the triangular lattice frustrates
the alternation of the up and down spins that minimizes the
Coulomb repulsion. As shown in
Fig.~\ref{fig:triangular_frustation}, two of the three electrons
in each triangle share the same spin orientation \cite{Ong_Cava}.
The effects of frustration in transport properties are not yet
fully understood. We will not treat frustration here because it is
out of the scope of this work. We refer the reader to the
literature \cite{Ong_Cava} for an explanation on this phenomena.
The ground state of the triangular lattice is expected to be
antiferromagnetic with the spins oriented forming 120 degrees
between them.

At this point we have seen numerous properties of the
two-dimensional Hubbard model in two different geometries. In
particular, the properties of the triangular lattice will be used
in Chapter.~\ref{chap:Hall_triangular}. In the following, we will
study other types of systems called low dimensional systems where
the physics vary substantially from the one studied previously.

\section{Low dimensional systems}

Until now we have seen the theoretical models, that will be used
in the next chapters, to describe strongly correlated systems in
two dimensions. The next step, is to investigate the rest of the
systems that interest us for this work, the one- and quasi
one-dimensional cases, in order to study their most important
properties. In particular those related to transport phenomena. We
will begin this section by studying the main features of electrons
moving in one dimension, and the Luttinger liquid theory that
allows us to describe the physics of these systems. Then, we will
focus on a special 1D system called the one-dimensional Mott
insulator. It is essential to understand the physics of the latter
if one is interested in electrons moving on a lattice, as will be
the case in Chapter~\ref{chap:Hall_quasi1d}. Finally, we will
investigate the main properties of quasi one-dimensional systems,
and the way to extend the Luttinger liquid theory with the aim of
describing their physics.

\subsection{One-dimensional case}\label{sec:1D_case}
\begin{figure}
\begin{center}
\includegraphics{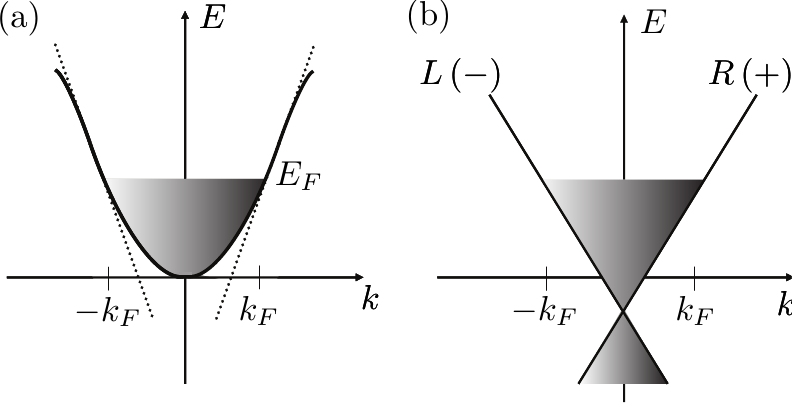}
\caption{\label{fig:linearization} (a) Single-particle dispersion
in a one-dimensional system. The Fermi surface consists of two
points. (b) Linear spectrum of the Luttinger model. The particles
are now separated in right (R$+$) and left (L$-$) going fermions.
The spectrum is extended to $\pm\infty$, giving rise to an
infinite number of negative states (\textquoteleft
unphysical\textquoteright  states). This requires the introduction
of a cutoff on the momentum to make the model well defined.}
\end{center}
\end{figure}
In order to understand the physics and, in particular, the
transport properties of strongly correlated systems in one
dimension, we must first realize that fermions in 1D have quite
different behaviors than fermions in higher dimensions. In one
dimension, an individual electron cannot move without pushing its
neighbors, thus any individual excitation has to become a
collective one. The main consequence of these differences is that,
instead of the Fermi liquid theory, the proper theoretical
description for one-dimensional systems is the {\it{Luttinger
Liquid}} theory. In the following we will focus on the basic ideas
necessary to describe transport properties in one-dimensional
systems, and we refer the reader to the many existing reviews and
books in the literature
\cite{haldane_bosonisation,vondelft_bosonization_review,giamarchi_book_1d}
for a detailed description of all the properties of 1D systems.
\begin{figure}
\begin{center}
\includegraphics[width=9cm]{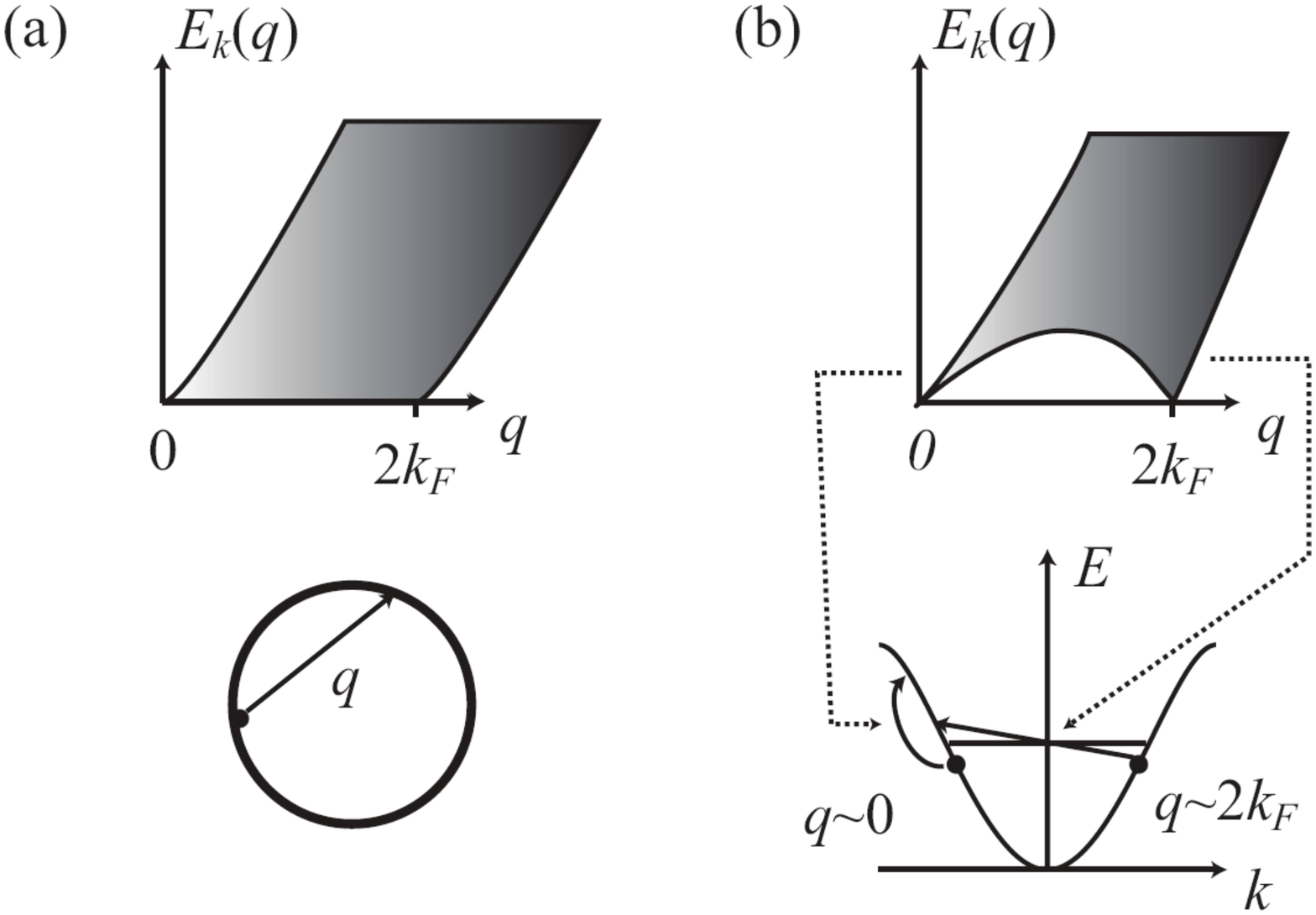}
\caption{\label{fig:spectrum} Particle-hole spectrum for
high-dimensional systems (a) and one-dimensional ones (b). In 1D
systems, particle-hole excitations have well defined energy and
momentum for small values of $q$ (low energy modes at $q\sim0$ and
$q\sim2k_F$), which is not the case in high-dimensional systems
(From Ref.~\cite{giamarchi_review_chemrev}).}
\end{center}
\end{figure}

Fig.~\ref{fig:linearization}(a) shows the single-particle
dispersion for a one-dimensional system, where the Fermi surface
consists of only two discrete points. Because the states
contributing to transport are the low-energy ones, {\it{i.e.}}
excitations close to the Fermi level, we may replace the original
model by one with a purely linear spectrum (see
Fig.~\ref{fig:linearization}(b)) extended to $\pm\infty$. This
linearization was first proposed by Luttinger
\cite{luttinger_model} and requires the introduction of two
species of fermions, right (R) and left (L) going fermions with a
dispersion relation given by
\begin{equation}\label{1d_dispersion}
    \varepsilon_{\pm}(k)= v_{\text{F}}(\pm k- k_{\text{F}}),
\end{equation}
where $v_{\text{F}}$ is the Fermi velocity, $k_{\text{F}}$ is the
Fermi momentum and $+$ $(-)$ refers to right (left) going fermions.

\begin{figure}
\begin{center}
\includegraphics[height=5cm,width=6cm]{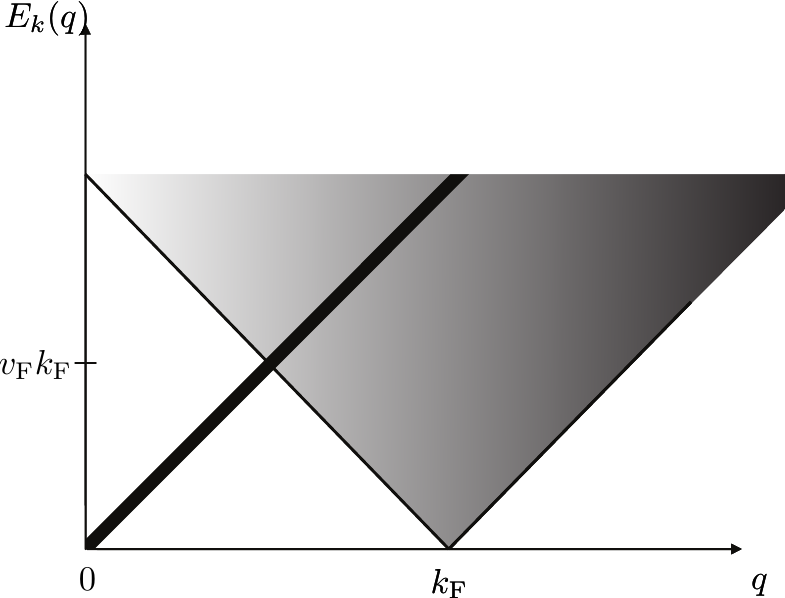}
\caption{\label{fig:part_hole} Particle-hole spectrum in the
Luttinger model. The states on the left of the black line are
\textquoteleft unphysical\textquoteright states coming from the
extension to $\pm\infty$ of the single-particle spectrum (see
Fig.~\ref{fig:linearization}). That is why a cutoff must be
introduced in momentum to make the model well defined}
\end{center}
\end{figure}

Another property of one-dimensional systems is that low-energy
particle-hole excitations, where an electron with momentum $k$ is
removed from below the Fermi surface and created above with
momentum $k+q$, do not exist for $0<q<k_\text{F}$. This arises
because the Fermi surface consist of only two points and thus the
only places where the particle-hole energy can reach zero are
$q=0$ and $q=2k_\text{F}$. This is depicted in
Fig.~\ref{fig:spectrum}. Thus, these low-energy excitations have a
well defined momentum $q$, a well-defined energy $E(q)\sim q$
(independent of $k$) and an energy dispersion $\delta E(q)$ which
goes to zero much faster than the average energy. These properties
make them well defined \textquoteleft particles\textquoteright
with a lifetime that increases when one goes closer to the Fermi
level ($E=0$) \cite{giamarchi_book_1d}. In the linearized model,
these excitations have an energy (for right going fermions)
\begin{equation}
E_{R,k}(q)= v_{\text{F}}(k+q)-v_{\text{F}}k=v_{\text{F}}q,
\end{equation}
which is indeed independent of $k$. They have a well defined
momentum $q$ and energy $E(q)=v_{\text{F}}q$. The particle-hole
spectrum for the Luttinger model is shown in
Fig.~\ref{fig:part_hole}.
These excitations are bosonic in nature (product of two fermion
operators) and because they are well defined, they can be used as
a basis to represent the Hamiltonian and all fermionic operators.
Indeed, the density fluctuations which are a superposition of
particle-hole excitations, are used as such basis.

To further understand the previous arguments, let us take a system
of fermions moving in 1D and write down the linearized kinetic
energy term in momentum space
\begin{equation}\label{kinetic_momentum}
\mathcal{H}_{\text{kin}}=\sum_{k,\sigma,r=R,L}v_{\text{F}}(\pm k- k_{\text{F}})c^{\dagger}_{rk\sigma}c^{\phantom{\dagger}}_{rk\sigma},
\end{equation}
where the $+$ ($-$) sign corresponds to $r=R$ ($r=L$), right (left) going
fermions. To transform this term to direct space, the single
particle fermion operator $\psi_\sigma(x)$ must be rewritten in
terms of right and left going fermions, taking into account only
the parts acting close to the Fermi surface
\begin{equation}\label{k_representation}
\psi_\sigma(x)=\frac{1}{\Omega}\sum_{k}e^{ikx}c_{k\sigma}\simeq\frac{1}{\Omega}\sum_{k\sim+k_F}e^{ikx}c_{k\sigma}
+\frac{1}{\Omega}\sum_{k\sim-k_F}e^{ikx}c_{k\sigma}=\psi_{\sigma
R}(x)+\psi_{\sigma L}(x),
\end{equation}
where $\Omega$ is the volume of the system. Thus, in terms of
$\psi_{R}$ and $\psi_{L}$, the kinetic energy term in direct space
is given by
\begin{equation}\label{kinetic_direct}
\mathcal{H}_{\text{kin}}=\int
dx\sum_{\sigma=\uparrow,\downarrow}\left[v_{\text{F}}
    \psi_{\sigma}^{\dagger}(x)\tau_3(-i\partial_x)
        \psi_{\sigma}^{\phantom{\dagger}}(x)\right],
\end{equation}
where $\psi^{\dagger}=(\psi^{\dagger}_R\;\psi^{\dagger}_L)$ is the
two-component vector composed of right- and left-moving fermions
and $\tau_3$ is the Pauli matrix
$\left(\begin{smallmatrix}1&0\\0&-1\end{smallmatrix}\right)$.
Eq.~(\ref{kinetic_direct}) is easily Fourier transformed into
Eq.~(\ref{kinetic_momentum}) using representations
(\ref{k_representation}). It can be proved, by simple commutation
relations \cite{giamarchi_book_1d}, that the kinetic energy term
(\ref{kinetic_momentum}) is well represented in a basis generated
by boson operators $b^\dagger_p$ and $b_p$:
$\mathcal{H}_{\text{kin}}\simeq
\sum_{p\neq0}v_{\text{F}}|p|b^\dagger b^{\phantom{\dagger}}$,
showing how this term can be {\it{quadratic}} in such a basis.
Similarly, in a system of fermions with spins, the interaction
term has typically the following form
\begin{equation}\label{H_int}
\mathcal{H}_{\text{int}}=\sum_{\sigma\sigma'}\int dx \,
dx'V(x-x')\rho_{\sigma}(x)\rho_{\sigma'}(x'),
\end{equation}
where $\rho_{\sigma}(x)$ is the density operator defined as
$\rho_{\sigma}(x)=\psi^{\dagger}_\sigma(x)\psi^{\phantom{\dagger}}_\sigma(x)$.
This operator is made of a product of two fermions operators and
thus it can also be represented in the boson basis $\rho(q)\sim
b_q$ or $\rho(q)\sim b^\dagger_q$ (or some linear combination).
The interaction term thus remains quadratic in the boson basis and
can be easily diagonalized, as shown below. In terms of right and
left going fermions the density operator becomes
\begin{equation}
\rho_\sigma(x)=\psi^{\dagger}_{\sigma L}(x)\psi^{\phantom{\dagger}}_{\sigma L}(x)
+ \psi^{\dagger}_{\sigma R}(x)\psi^{\phantom{\dagger}}_{\sigma R}(x)
+ \psi^{\dagger}_{\sigma L}(x)\psi^{\phantom{\dagger}}_{\sigma R}(x)
+ \psi^{\dagger}_{\sigma R}(x)\psi^{\phantom{\dagger}}_{\sigma L}(x).
\end{equation}
It must be kept in mind that the relevant processes in the
interaction are those close to the Fermi surface ($q\sim0$ and
$q\sim 2k_{\text{F}}$). Because the Fermi surface in 1D consist
only on two points, the interaction term can be decomposed in
three different types. The first process called $g_4$ couples
fermions on the same side of the Fermi surface. The second type is
called $g_2$ and couples fermions from one side with fermions on
the other side of the Fermi surface. But each species stays on the
same side after the interaction (forward scattering). And finally
we have the $g_1$ process which corresponds to a scattering of
$2k_{\text{F}}$, where fermions exchange sides (backscattering)
\cite{giamarchi_book_1d}. The processes between fermions with
parallel spins are denoted $g_{\|}$ and with opposite spins,
$g_{\perp}$. We suppose our system is made of fermions with spin
rotational symmetry (for simplicity), moving in one dimension.
Adding the interactions terms to the Hamiltonian we obtain the
following expression
\begin{eqnarray}\label{H_1d_continium}
&&\mathcal{H}_{\text{1D}}=\int
dx\sum_{\sigma=\uparrow,\downarrow}\left[v_{\text{F}}
    \psi_{\sigma}^{\dagger}(x)\tau_3(-i\partial_x)
        \psi_{\sigma}^{\phantom{\dagger}}(x)
    +g_{2\|}\,\rho_{\sigma R}(x)\rho_{\sigma L}(x)
    \right.\\ \nonumber
    &&+g_{2\perp}\,\rho_{\sigma R}(x)\rho_{-\sigma L}(x)+\sum_{r=R,L}\left.
    \left(\frac{g_{4\|}}{2}\,\rho_{\sigma r}(x)\rho_{\sigma r}(x)
    +\frac{g_{4\perp}}{2}\,\rho_{\sigma r}(x)\rho_{-\sigma r}(x)\right)\right].
\end{eqnarray}
The first term is the kinetic energy (\ref{kinetic_direct}) and
the other terms in Eq.~(\ref{H_1d_continium}) refer to forward
scattering. In $\mathcal{H}_{\text{1D}}$ we have omitted the
backscattering terms ($g_1$ processes) which are, for spin
rotationally invariant systems, marginally irrelevant
\cite{giamarchi_book_1d}. We therefore take
$g_{1\perp}=g_{1\parallel}=0$. If only backscattering terms are
considered, it must be kept in mind that logarithmic corrections
can be introduced by the forward scattering terms. This will be
the case in Chapter~\ref{chap:Hall_quasi1d}, in our study of the
Hall effect in a quasi 1D chain.

The representation of fermionic operators in term of bosonic
fields, is an important tool of the Luttinger liquid theory to
calculate correlations functions, as explained at the end of this
section. This technique is known as {\it{bosonization}}
\cite{vondelft_bosonization_review,giamarchi_book_1d}. We will not
present the bosonization method in details here, but only the
important results needed for the understanding of the transport
properties in 1D. We refer the reader to the literature for a
detailed description of this technique
\cite{vondelft_bosonization_review,giamarchi_book_1d}. Hamiltonian
(\ref{H_1d_continium}) will be used in
Chapter~\ref{chap:Hall_quasi1d}, both in its fermionic and bosonic
representations (see Eq.~(\ref{K_u}))

Let us see first how the fermionic fields
$\psi_{\sigma}=\psi_{\sigma,R}+\psi_{\sigma,L}$ are represented in
terms of bosonic fields denoted $\theta_{\nu}$ and $\phi_{\nu}$
\footnote{The boson fields are defined as \[
\phi(x),\theta(x)=\pm\left(N_R\pm N_L\right)\frac{\pi
x}{L}\mp\frac{i\pi}{L}\sum_{p\ne0}\frac{1}{p}e^{-a|p|/2-ipx}(\rho^{\dagger}_R(p)\pm\rho^{\dagger}_L(p)).\]
The $N_r$ terms disappear in the thermodynamic limit $L\to\infty$
\cite{giamarchi_book_1d}.} (fields written in the boson basis $b$
and $b^\dagger$), where $\nu=\rho(\sigma)$ denotes the charge
(spin) degrees of freedom. The charge and spins degrees of freedom
must be separated in order to diagonalize Hamiltonian
(\ref{H_1d_continium}). The fermionic field yields
\begin{equation}\label{fermionic}
    \psi_{\sigma,r}(x)=\frac{e^{irk_{\text{F}}x}}{\sqrt{2\pi a}}
    e^{-\frac{i}{\sqrt{2}}\left\{r\phi_{\rho}(x)-\theta_{\rho}(x)
    +\sigma\left[r\phi_{\sigma}(x)-\theta_{\sigma}(x)\right]\right\}}
\end{equation}
with $r=+1(-1)$ for right (left) moving fermions, and $a$ a cutoff
which precise value is irrelevant for the low-energy, long-wave
length properties of the model. The limit $a\to0$ should be taken
in principle. All the formulas given here are found after taking
the thermodynamic limit $L\to\infty$, with $L$ the size of the
system \cite{vondelft_bosonization_review,giamarchi_book_1d}. The
fields $(\phi_{\rho},\theta_{\rho})$ and
$(\phi_{\sigma},\theta_{\sigma})$ obey the following commutation
relations
\begin{equation}\label{commutation_relations}
 \left[\phi_{\nu}(x),\frac{1}{\pi}\nabla\theta_{\nu}(x')\right]=i\delta(x-x').
\end{equation}
The bosonic fields $\phi$ and $\theta$ can be represented in terms
of the density operators $\rho_R$ and $\rho_L$
\cite{giamarchi_book_1d}, giving rise to the following relations
\begin{eqnarray}\nonumber
\nabla\phi_{\uparrow}(x)=-\pi\left[\rho_{R\uparrow}(x)+\rho_{L\uparrow}(x)\right]\\
\nabla\theta_{\uparrow}(x)=-\pi\left[\rho_{R\uparrow}(x)-\rho_{L\uparrow}(x)\right],
\end{eqnarray}
and the same is valid for the other spin ($\downarrow$). The boson
fields for the total charge and spin degrees of freedom are
related to the formers as
\begin{eqnarray}\nonumber
\phi_{\rho}(x)=\frac{1}{\sqrt{2}}\left[\phi_{\uparrow}(x)+\phi_{\downarrow}(x)\right]\\
\phi_{\sigma}(x)=\frac{1}{\sqrt{2}}\left[\phi_{\uparrow}(x)-\phi_{\downarrow}(x)\right].
\end{eqnarray}
and the $\theta$ fields obey the same relations. From this
expressions we can already see that interactions of the form
(\ref{H_int}), introduce term such as $(\nabla\phi_{\nu}(x))^2$
and $(\nabla\theta_{\nu}(x))^2$.

Using representation (\ref{fermionic}), the Hamiltonian
$\mathcal{H}_{\text{1D}}$ can be bosonized obtaining a quadratic
Hamiltonian of the form
\begin{eqnarray}\label{H_1D}
        \mathcal{H}_{\text{1D}}&=&\mathcal{H}_{\rho}+\mathcal{H}_{\sigma} \\ \label{K_u}
        \mathcal{H}_{\nu}&=&\int\frac{dx}{2\pi}\,\left\{u_{\nu}K_{\nu}
        \left[\nabla\theta_{\nu}(x)\right]^2+\frac{u_{\nu}}{K_{\nu}}
        \left[\nabla\phi_{\nu}(x)\right]^2\right\}\qquad,
    \end{eqnarray}
where $\nu=\rho(\sigma)$ denotes the charge (spin) degrees of
freedom, $u_{\nu}$ is a velocity, $K_{\nu}$ a dimensionless
parameter depending on the interactions, and $\theta_{\nu}$ and
$\phi_{\nu}$ are the \textquotedblleft new\textquotedblright
bosonic fields. The parameters $u$ and $K$ are given by the
interaction parameters
\begin{eqnarray}\nonumber
 u_\nu&=& v_F\left[(1+y_{4\nu}/2)^2-(y_\nu/2)^2 \right]^{1/2}\\
 K_\nu&=& \left[\frac{1+y_{4\nu}/2+y_\nu/2}{1+y_{4\nu}/2-y_\nu/2}\right]^{1/2}\\ \nonumber
 g_\nu&=& g_{1\|}-g_{2\|}\mp g_{2\perp}\\ \nonumber
 g_{4\nu}&=& g_{4\|}\pm g_{4\perp}\\ \nonumber
 y_\nu&=& g_{\nu}/(\pi v_{\text{F}}),
\end{eqnarray}
where the upper sign refers to $\rho$ and the lower one to
$\sigma$. In the non-interacting case $u_\rho=u_\sigma=v_{F}$ and
$K_\rho=K_\sigma=1$. For systems with spin rotation symmetry we
have $g_{1\|}=g_{1\perp}$ and thus $K_\sigma=1$. $K_\rho=1$ in the
absence of interactions and $K_\rho<1$ for repulsive interactions.

The representation of a fermionic Hamiltonian in a quadratic form
is one of the big advantages of the Luttinger liquid theory. It
takes into account all the effects of fermionic interactions
(momentum conserving interactions) and put them in a simple
quadratic Hamiltonian of the form of Eq.~(\ref{K_u}). Although we
will not treat the case of $g_{1\perp}$ scattering here, it is
important to know that this term cannot be written in the same
quadratic form of Hamiltonian (\ref{H_1D}) because it gives a
cosine term (called {\it{sine-Gordon}} Hamiltonian) and must be
treated with {\it{renormalization group}} equations
\cite{anderson_kondo_1}. The splitting of Hamiltonian (\ref{H_1D})
into a charge part $\mathcal{H}_{\rho}$ and a spin part
$\mathcal{H}_{\sigma}$ reveals the complete separation of charge
and spin degrees of freedom in 1D systems, which is an important
property of Luttinger liquids. It forbids single particle
excitations (free electron) carrying spin and charge together.

There are various properties of Luttinger liquids that we will not
treat here since it would require more than an entire chapter, but
there is one special property which is very important to know when
one is working with one-dimensional systems: the power-law decay
of correlation functions. In addition to the particle-hole
excitations mentioned at the beginning of this section there are
also particle-particle excitations. These are collective
excitations that represent charge-density and spin-density
fluctuations respectively, with $q\sim0$ and $q\sim 2k_F$
components (see Fig.~\ref{fig:spectrum}). When calculating
charge-density or spin-density correlations functions in one
dimension, the $q\sim0$ part gives a free fermion decay of the
correlation and the $q\sim 2k_F$ (or multiples $q\sim
2nk_{\text{F}}$) behaves as a {\it{non-universal}} power law, with
an exponent depending on interactions. This arises because the
charge and spin densities fluctuate in space and time. That is why
ordered states do not exist in one-dimensional systems
\cite{giamarchi_book_1d}. Let us take for example the
density-density correlation function of a system described by
Hamiltonian (\ref{K_u}),
\begin{eqnarray}\nonumber
\langle\rho(x,\tau)\rho(0)\rangle=\rho_0^2+\frac{K_{\rho}}{\pi^2}\frac{y_\alpha^2-x^2}{(x^2+y_\alpha^2)^2}+
\rho_0^2A_2\cos(2\pi\rho_0x)\left(\frac{a}{r}\right)^{K_\rho+K_\sigma}\\
\label{den-den}
+\rho_0^2A_4\cos(4\pi\rho_0x)\left(\frac{a}{r}\right)^{4K_\rho}+\cdots
\end{eqnarray}
Where $y_\alpha=u\tau+a\text{Sign}(\tau)$ and $\tau$ is an
imaginary time related to $t$ by the Wick rotation
\cite{mahan_book}: $\tau=it+\epsilon \text{Sign}(t)$. $\rho_0$ is
the average density of particles. In Chapter~\ref{chap:transport}
we will give the technical details necessary to compute a
correlation function of this type. At this point we are interested
in the physical meaning of Eq.~(\ref{den-den}). The $q\sim0$ part
of the correlation decays as $1/x^2$. This is a Fermi liquid like
decay and only the amplitude is renormalized by the interactions.
The $2k_{\text{F}}$ and $4k_{\text{F}}$ parts, however, decay as a
{\it{non-universal}} power law, with exponents depending on
interactions. In this specific correlation, the $4k_{\text{F}}$
component does not depend on the spin part, thus if $K_\sigma=1$
and $K_\rho<1/3$, it becomes the dominant component. A system with
this type of correlations is referred as a Luttinger liquid.
Another important quantity showing a power-law decay is the
retarded single-particle Green's function, which will be used
repeated times in the next chapters. It is defined as
\cite{mahan_book}
\begin{equation}
G^{\text{ret}}_{r,\sigma}(x,t)=-i\theta(t)\langle [\psi_{r\sigma}^{\phantom{\dagger}}(x,t),\psi_{r\sigma}^{\dagger}(0,0)]_{+} \rangle,
\end{equation}
where $[]_{+}$ is the anticommutator, $\theta(t)$ is the
Heaviside-function, $r=+1$ ($-1$) for right (left) movers, and
$\sigma$ denotes the spin. The retarded Green's function is
usually obtained form the Green's function in imaginary time
\cite{mahan_book}. The relation between these correlation
functions will be explained in Sec.~\ref{sec:Kubo}. In a Luttinger
liquid, the retarded Green's function is given by
\begin{eqnarray}\label{Green_1D}
G^{\text{ret}}_{r,\sigma}(x,t)&=& -i\frac{\theta(t)}{2\pi}e^{irk_{\text{F}}}\lim_{\epsilon\to0}
\left\{\frac{a+i(v_{\text{F}}t-rx)} {\epsilon+i(v_{\text{F}}t-rx)}\times \right.\\ \nonumber
&&\left.\prod_{\nu=\rho,\sigma}\frac{1}{\sqrt{a+i(u_{\nu}t-rx)}}\left(\frac{a^2}{(a+iu_{\nu}t)^2+x^2} \right)^{\gamma_{\nu}}+\left(\begin{matrix} x\to&-x\\t\to&-t\end{matrix}\right)\right\},
\end{eqnarray}
with $a$ the momentum cutoff. The exponent is
\begin{equation}
\gamma_{\nu}=(K_{\nu}+K_{\nu}^{-1}-2)/8>0
\end{equation}
For a spin rotation invariant system, $K_{\sigma}=1$ and
$\gamma_{\sigma}=0$. It is not our intention to explain how to
obtain Eq.~(\ref{Green_1D}) because is out of the scope of this
work. We refer the reader to Ref.~\cite{giamarchi_book_1d} for a
detail explanation on the calculation of the 1D Green's function.
We are interested in the asymptotic behavior of
Eq~(\ref{Green_1D}), {\it{i.e.}}, at $x\to\infty$
\begin{equation}\label{Green_largex}
G^{\text{ret}}_{r\sigma}(x\to\infty)\sim
\left(\frac{a}{x}\right)^{2\gamma_{\nu}}.
\end{equation}
The behavior of correlation functions at large distances will be
used constantly in the next chapters, to detect power-law
behaviors in the calculations using scaling analysis. Because most
of the functions in one dimension are power laws, most result can
be obtain by simple scaling analysis. The single-particle Green's
function in imaginary time $G_{r\sigma}(x,\tau)$ follows the same
asymptotic power-law behavior of Eq.~(\ref{Green_largex}). From
$G_{r\sigma}(x,\tau)$ we can obtain the occupation factor $n(k)$,
which is simply the Fourier transform of the equal time Green's
function \cite{mahan_book}
\begin{equation}
n_{r\sigma}(k)=\int dx e^{-ikx}G_{r\sigma}(x,\tau=0^{-}).
\end{equation}
The occupation factor is thus the Fourier transform of a power-law
and, for a spin rotational invariant system $K_{\sigma}=1$, is
given by \cite{giamarchi_book_1d}
\begin{equation}\label{occupation_factor}
n(k)\sim |k-k_{\text{F}}|^{1/4[K_{\rho}+K_{\rho}^{-1}]-1/2}.
\end{equation}
Instead of the discontinuity at $k_{\text{F}}$ characteristic of a Fermi liquid,
in one-dimension one finds a power-law singularity. This means
that the quasi-particle residue is $Z=0$, which is an
evidence that single-particle excitation cannot exist in 1D.

At this point we can summarize the features of Luttinger liquids
that we have discussed until now. Firstly, we saw that in one
dimension the dispersion relation can be linearized keeping the
same low-energy properties of the system. This requires the
definition of right- and left-going fermions. We learned that
particle-hole excitations are well defined in 1D and thus can be
used as a basis to represent fermionic operators, thought a
transformation known as bosonization. And we saw that a
fundamental property of Luttinger liquid is the power-law decay of
correlation functions. Now we will discussed the case of fermions
moving on a lattice. For this we dedicate the next section to the
one-dimensional Mott insulator and all the relevant properties for
the study of transport physics in it.

\subsection{1D Mott insulator}\label{sec:Mott_1D}

Let us begin by considering a system made of fermions moving on a
one-dimensional lattice. This system is described with the Hubbard
model studied in Sec.~\ref{sec:Hubbard_model}. We will apply the
bosonization transformation to the Hubbard Hamiltonian of
Eq.~(\ref{Hubbard}) in one dimension, where the sum over $j$
disappears. The result is again a quadratic Hamiltonian of the
form given in Eq.~(\ref{H_1D}), but now parameters $u$ and $K$
are:
\begin{eqnarray}\nonumber
u_{\rho}K_{\rho}&=&u_{\sigma}K_{\sigma}=v_F\\ \label{K_u_Hubbard}
u_{\rho}/K_{\rho}&=& v_F\left(1+\frac{U}{\pi v_F} \right)\\
\nonumber u_{\sigma}/K_{\sigma}&=& v_F\left(1-\frac{U}{\pi v_F}
\right),
\end{eqnarray}
functions of the Coulomb interaction $U$.

\begin{figure}
\begin{center}
\includegraphics[width=7cm]{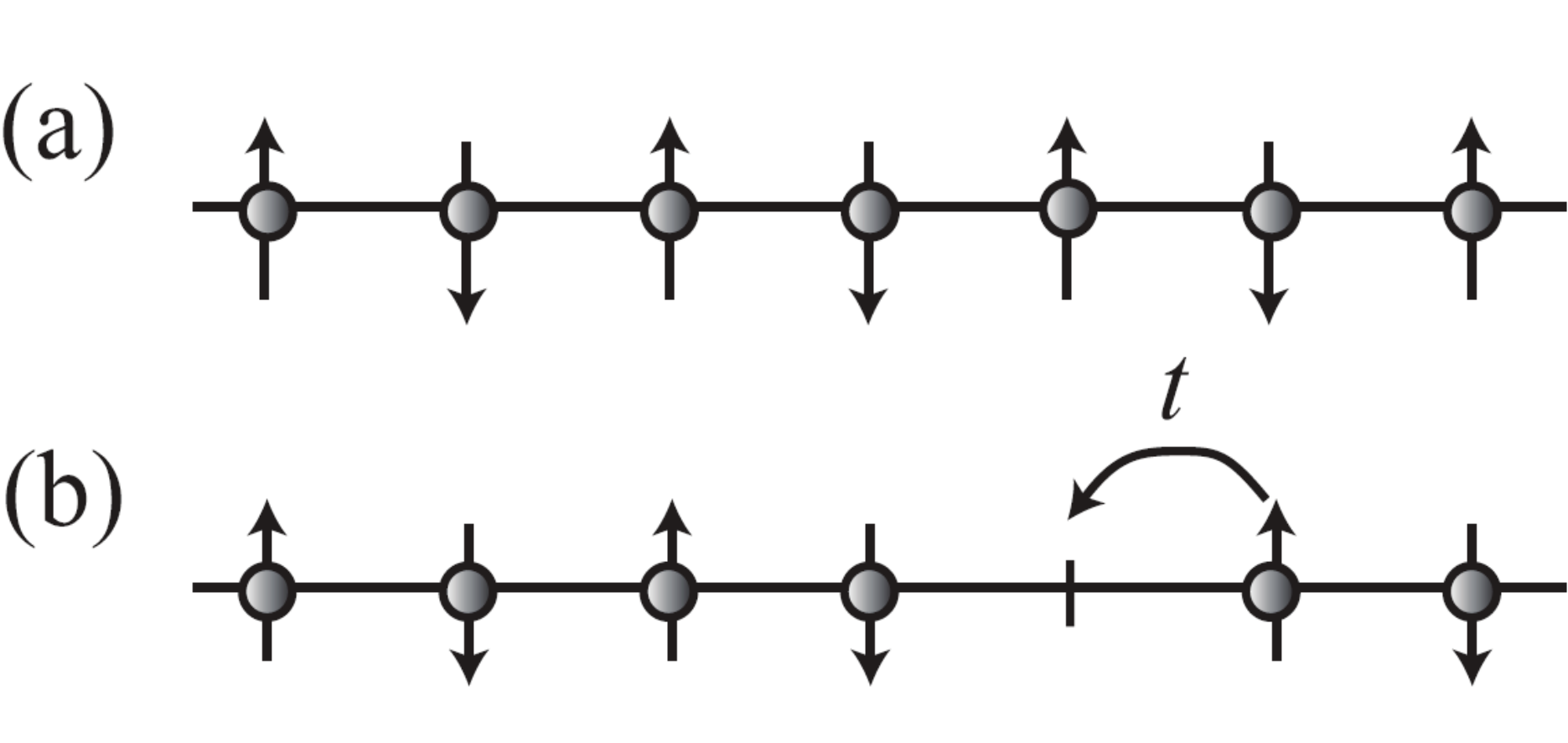}
\caption{\label{fig:mott_1d}  (a) One dimensional Mott insulator:
at half filling (one particle per site) due to the repulsion
between particles, the system prefers to localize one particle on
each lattice site. In this case, antiferromagnetism order is
favored. (b) If the system is doped with holes or electrons, the
particles (electrons or holes) can move and we thus have a
metallic state (From Ref.~\cite{giamarchi_review_chemrev}).}
\end{center}
\end{figure}

\begin{figure}
\begin{center}
\includegraphics[width=6cm]{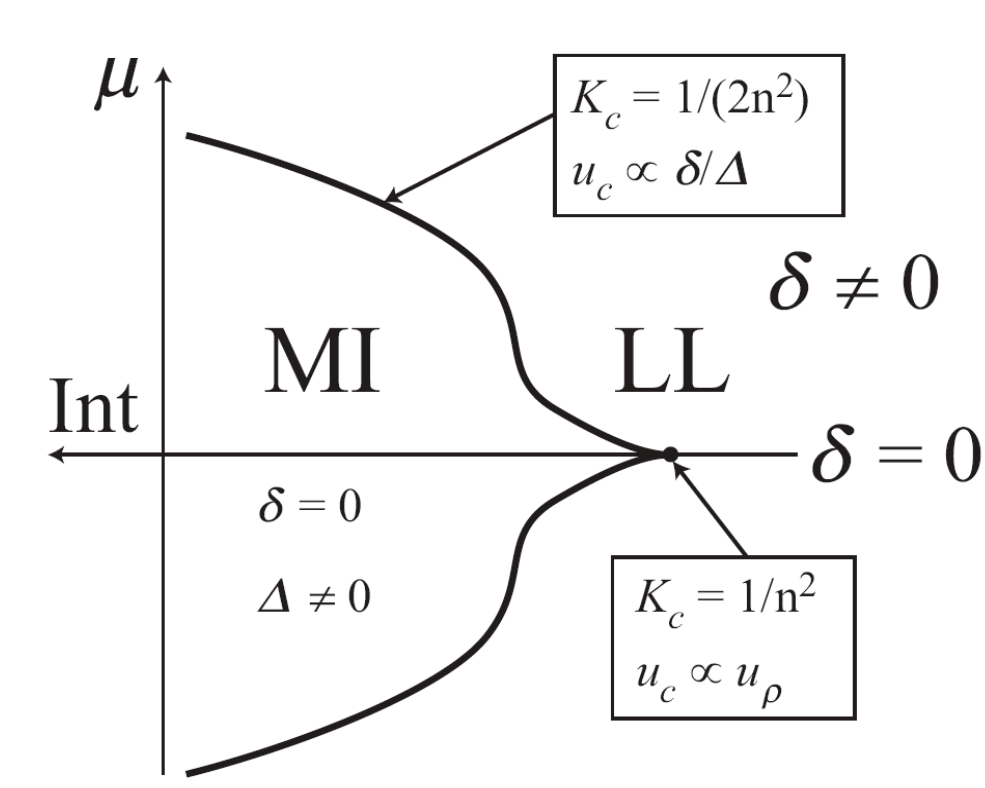}
\caption{\label{fig:mott_transition} Phase diagram of the
one-dimensional Mott insulator. $n$ denotes the order of the
commensurability: $n=1$ for $1/2$-filled and $n=2$ for
$1/4$-filled systems. Int denotes the repulsive interaction
between particles, $\mu$ is the chemical potential, $\delta$ is
the doping and $\Delta$ is the Mott gap. MI and LL correspond to
the Mott insulator and Luttinger liquid (metallic) phases,
respectively. The critical exponent $K_c$ and velocity $u_c$
depend on whether it is a Mott-U or Mott-$\delta$ transition (From
Ref.~\cite{giamarchi_book_1d}).}
\end{center}
\end{figure}

If there is one particle per site (half-filled system) and the
repulsion $U$ is larger than the kinetic energy $t$, the particles
are localized on the lattice sites to minimize repulsion and the
system is called a Mott insulator
\cite{mott_historical_insulator}. This is depicted in
Fig.~\ref{fig:mott_1d}. In the Mott insulating state a gap
$\Delta$ appears in the charge excitation spectrum, but the spin
properties are totally unaffected. To attain a metallic state, the
system must be weakly doped away from half-filled case in order to
have particles propagating. For other commensurate fillings, like
quarter-filling (one particle every two sites), it is necessary to
have a nearest neighbor repulsion ($V$) to form a Mott insulator
\cite{giamarchi_book_1d}.  The transition between the insulator
and metallic state is known as Mott transition
\cite{mott_metal_insulator} and is depicted in
Fig.~\ref{fig:mott_transition}. It can occur changing the value of
the repulsive interaction, in this case is called a Mott-U
transition; or changing the doping $\delta$ and is called
Mott-$\delta$ transition (see Fig.~\ref{fig:mott_transition}).
Here we will not treat the case of doping and thus we refer to the
Mott-$U$ transition as just the Mott transition. The properties of
Luttinger liquid discussed in the previous section remain
essentially the same for the 1D Hubbard model, knowing that
interaction $K_\nu$ and velocity $u_\nu$ parameters, depend on $U$
as shown in Eq.~(\ref{K_u_Hubbard}).

\begin{figure}
\begin{center}
\includegraphics[width=9cm]{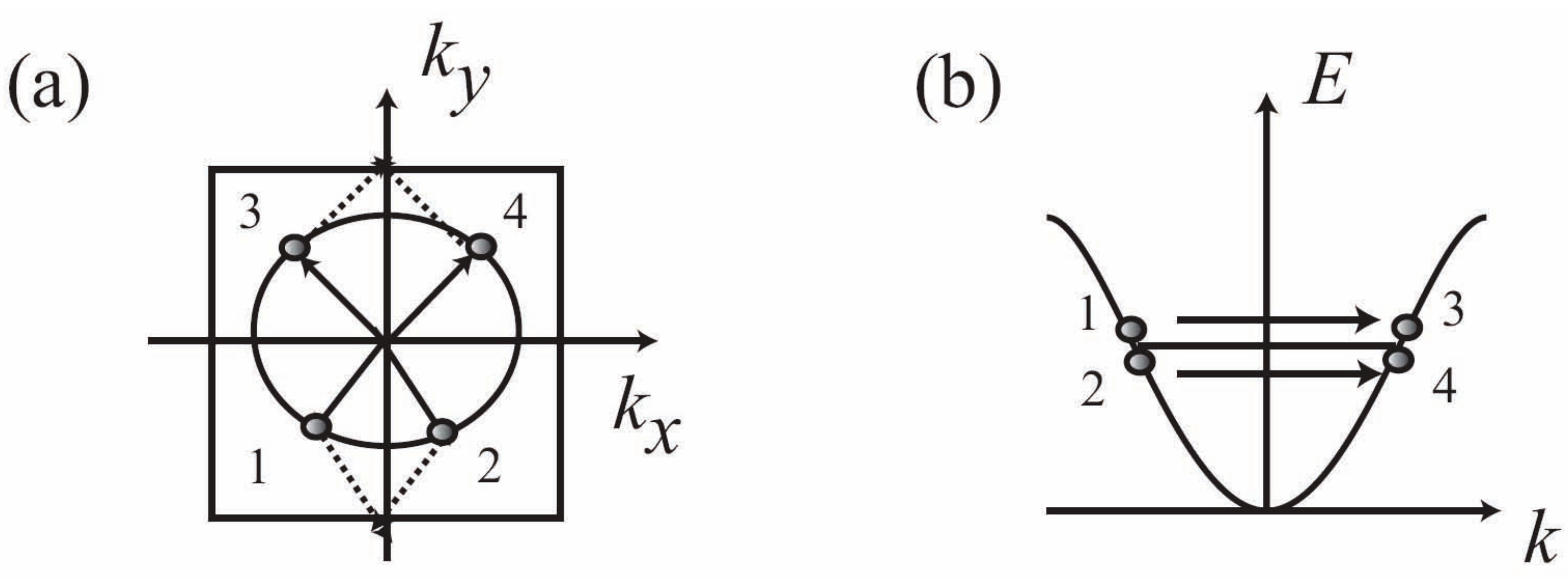}
\caption{\label{fig:umklapp} (a) Umklapp processes in
high-dimensional systems appear regardless of the filling
(provided $|k_{\text{F}}|$ is large enough). (b) In one dimension
an umklapp process appears when two particles are scattered from
one side to the other on the Fermi surface  $4k_F=2\pi/a$, that
is, for half filling. There are also umklapp processes for other
commensurate fillings (From
Ref.~\cite{giamarchi_review_chemrev}).}
\end{center}
\end{figure}

To study transport properties in commensurate systems (as will be
done in Chapter ~\ref{chap:transport}), we must take into account
the effect of the underlaying lattice. Due to the presence of the
lattice, the wavevector is defined modulo a vector of the
reciprocal lattice, {\it{i.e.}}, $2\pi/a$ in one dimension with
$a$ the lattice spacing. As a consequence, new interaction
processes will appear in which the total momentum is not
conserved, such that $k_1+k_2+k_3+k_4=Q$, where $Q$ is a vector of
the reciprocal lattice (in momentum conserving processes, such as
(\ref{H_int}), $k_1+k_2+k_3+k_4=0$). In other words, the particles
exchange momentum with the lattice. These processes are known as
{\it{umklapp}} processes and they are the only ones responsible
for dissipation in 1D systems \cite{giamarchi_book_1d}. The
umklapp processes appear only at commensurate fillings. For
example in a half-filled system we have $4k_F=2\pi/a$ and it
corresponds to a process where two electrons jump from one side to
the other of the Fermi surface, transferring a momentum $4k_F$ to
the lattice \cite{giamarchi_book_1d,giamarchi_review_chemrev}.
This is illustrated in Fig.~\ref{fig:umklapp}.
The umklapp term modifies only the charge part of Hamiltonian
(~\ref{H_1D}), by adding a term of the form
\begin{equation}\label{general_umklapp}
\mathcal{H}_{1/2n}=\text{g}_{1/2n}\int dx\cos(n\sqrt{8}\phi_\rho(x)),
\end{equation}
where $n$ is the order of the commensurability ($n=1$ for
half-filling and $n=2$ for quarter-filling), $\text{g}_{1/2n}$ is
the coupling constant, and the commensurability $n$ is related to
the reciprocal lattice vectors by $2pk_F=2\pi q/a$, where $p=2n$
and $q$ are integers. In the case of one particle per site ($n=1$,
half-filled band), we rewrite Eq.~(\ref{general_umklapp}) as
\cite{giamarchi_book_1d}
\begin{equation}\label{umklapp_half}
\mathcal{H}=\frac{2\text{g}_{3}}{(2\pi a)^2}\int
dx\cos(\sqrt{8}\phi_\rho(x)).
\end{equation}
We have replaced the amplitude $\text{g}_{1/2}$ by the factor
$2\text{g}_{3}/(2\pi a)^2$, where the coefficient $\text{g}_{3}$
is of the order of the Coulomb interaction $U$ and $a$ is the
lattice spacing. The umklapp process of Eq.~({\ref{umklapp_half}})
will be the only term, producing dissipation, considered in our
study of the Hall effect in a quasi one-dimensional system.

After reviewing the necessary tools to understand the transport
properties of fermions moving on a one-dimensional lattice, we
will describe an extension of the pure 1D case which is the lowest
dimensional situation where the Hall effect has a sense: quasi
one-dimensional systems. This will be the type of strongly
correlated low-dimensional system treated on
Chapter~\ref{chap:Hall_quasi1d}.

\subsection{Quasi one-dimensional case}\label{sec:quasi_1d_systems}
In the group of low-dimensional systems we find, beside the purely
one-dimensional ones, systems with a dimensionality between one
and two or one and three. These systems are known as {\it{quasi
one-dimensional}} and they consist of one-dimensional chains
coupled by an {\it{interchain coupling}} term of the form
\begin{equation}\label{hopping_term}
\mathcal{H}_{\perp}=-\int dx \sum_{\langle\mu,\nu\rangle\sigma}
t_{\perp,\mu,\nu}
\left[\psi_{\mu\sigma}^{\dagger}\psi_{\nu\sigma}^{\phantom{\dagger}}
+ h.c. \right],
\end{equation}
with $\langle\mu,\nu\rangle$ a pair of chains and
$t_{\perp,\mu,\nu}$, the hopping integral between these two
chains. The simplest example of a quasi 1D system is composed of
two one-dimensional chains coupled to form a fermionic ladder
\cite{giamarchi_book_1d}. Here we are interested in systems made
of a large or infinite number of coupled chains. In
Chapter~\ref{chap:Hall_quasi1d}, we will use
Eq.~(\ref{hopping_term}) together with Hamiltonian
(\ref{H_1d_continium}) to describe our system of weakly coupled
one-dimensional chains.

In addition to the coupling term (\ref{hopping_term}), there exist
hopping processes of second order in $t_\perp$, where two
particles jump between chains. These terms have a well defined
classical limit (unlike the coupling term in
Eq.~(\ref{hopping_term})) and for a sufficiently large number of
chains they can be treated using a mean field approximation
\cite{schulz_coupled_spinchains}. Their most important effect is
to drive the system into an ordered state
\cite{giamarchi_review_chemrev}. The ratio $t_\perp/t_\|$, where
$t_\|$ is the intrachain hopping, determines the effective
dimensionality of the system. When the system is nearly isotropic,
$t_\perp\sim t_\|$, we deal with a high-dimensional situation
where even small interchain interactions must be taken into
account because they lead to a Fermi liquid state or another
correlated state. In the opposite limit $t_\perp\ll t_\|$, the
system is highly anisotropic and the chains are in a well defined
Luttinger liquid regime. Thus, the processes of second order in
$t_{\perp}$ are less likely to occurs and can be neglected. This
is the limit that interest us for the discussion of transport in
quasi one-dimensional systems. We will therefore focus only on the
interchain hopping term of Eq.~(\ref{hopping_term}), and we refer
the reader to the literature for an extensive discussion of the
effects of interchain interaction terms
\cite{giamarchi_book_1d,giamarchi_review_chemrev}. Therefore, in
the rest of this manuscript the word \textquotedblleft
interaction\textquotedblright will only refer to intrachain
interaction.
\begin{figure}
\begin{center}
\includegraphics[width=8cm]{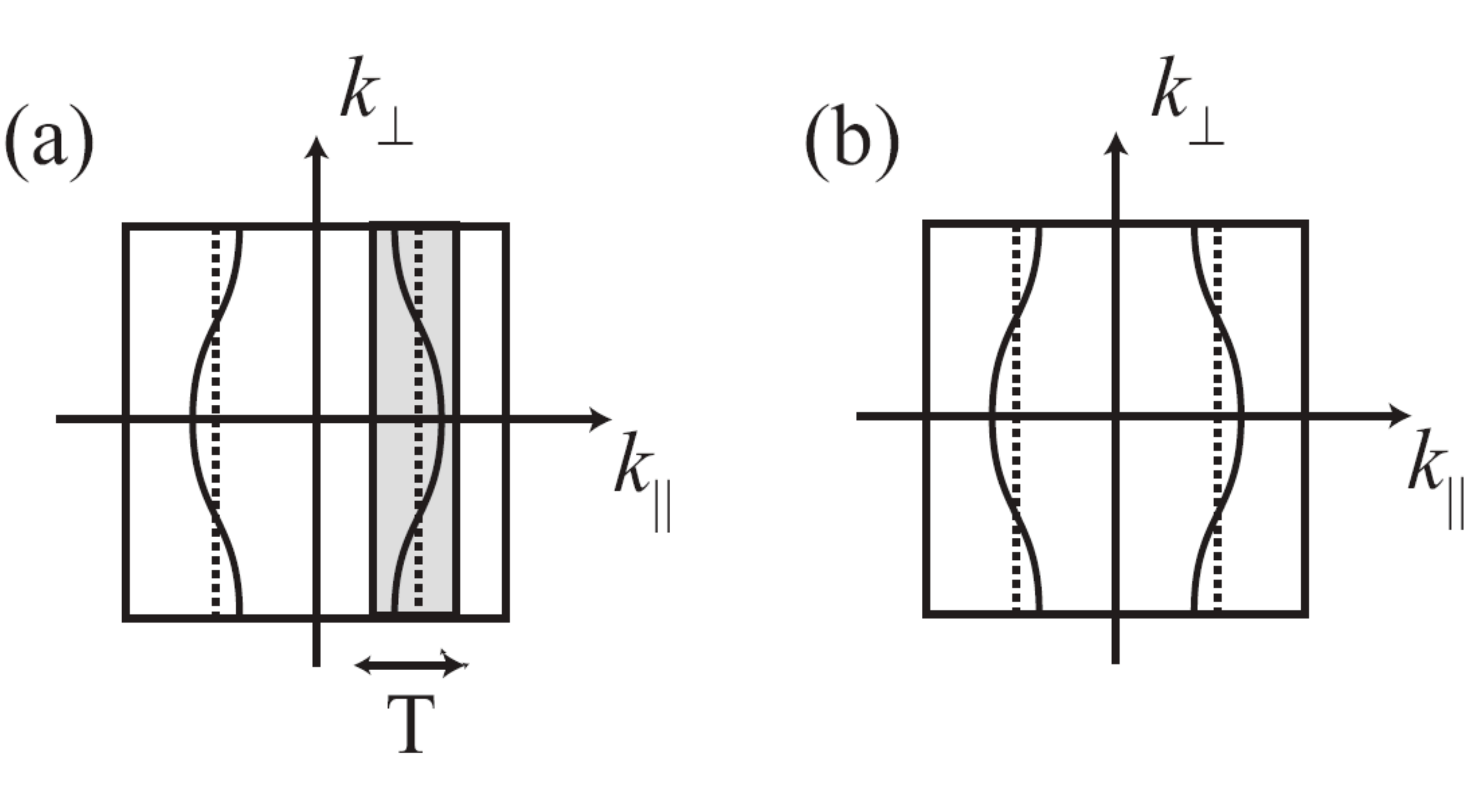}
\caption{\label{fig:Fermi_surface_1d} Open Fermi surface of a
quasi one-dimensional system. (a) If the temperature (or any other
energy scale) is larger than the warping of the Fermi surface
produced by the interchain hopping, the system cannot feel it and
thus behaves as a one-dimensional one. (b) For temperature/energy
smaller than the warping, the system feel the two- or
three-dimensional nature of the dispersion and thus behaves as a
high-dimensional system (From
Ref.~\cite{giamarchi_review_chemrev}). }
\end{center}
\end{figure}

One important effect of the coupling term (\ref{hopping_term}) is
to induce a dimensional crossover, as a function of decreasing
temperature (or another energy scale), from a one-dimensional
regime to a high-dimensional one ($2$D or $3$D). To understand
this dimensional crossover we consider the dispersion relation of
the system in the absence of interactions:
\begin{equation}\label{open_Fermi}
\varepsilon (k_\|,k_{\perp})=
-2t_{\|}\cos(k_{\|}a)-2t_{\perp}\cos(k_{\perp}b)
\end{equation}
where $b$ is the distance between chains. In the limit $t_\perp\ll
t_\|$, the Fermi surface given by Eq.~(\ref{open_Fermi}) is an
open surface as the one shown in Fig.~\ref{fig:Fermi_surface_1d}.
If the temperature (or energy) is larger than the
warping of the Fermi surface due to interchain hopping, the system
cannot be sensitive to it (see Fig.~\ref{fig:Fermi_surface_1d})
and thus feels a flat Fermi surface corresponding to a
one-dimensional regime. In contrast, when the temperature (energy)
is much smaller than the warping, the system behaves as a two- or
three-dimensional one. This dimensional crossover occurs, in the
non-interacting case, at an energy scale of the order of the
interchain hopping $t_\perp$. Interactions renormalize this energy
scale to (unities are taken in order to have $k_B=1$)
\begin{equation}
E^*=T_{x1}\sim W\left(\frac{t_\perp}{W} \right)^{1/(1-\zeta)},
\end{equation}
where $\zeta=\sum_\nu \gamma_\nu$ and
$\gamma_\nu=(K_\nu+K_\nu^{-1}-2)/8>0$ \cite{giamarchi_book_1d}.
For a spin rotation invariant system $K_\sigma=1$ and
$\gamma_\sigma=0$. Furthermore, in the non-interacting case
$\zeta=0$ and we recover $T_{x1}\sim t_\perp$. For interacting
systems we have $\zeta>0$ and thus the energy scale at which the
dimensional crossover takes place is reduced, making the system
effectively more one-dimensional. This can be understood in the
following manner: the interchain coupling involves single-particle
hopping processes which are unstable excitations in a Luttinger
liquid, then in order to have an electron jumping from one chain
to the other a collective excitation must brake and then
recombine in the new chain. This makes the single-particle hopping
very difficult. As mentioned previously, there exist hopping
processes of second order in $t_\perp$ which are more favorable,
where two particles jump between chains. These processes lead to
an ordered state at a temperature $T_{x2}$ which depends on the
precise coupling (spin-spin, Josephson term or density-density)
\cite{giamarchi_book_1d}. In the limit $t_\perp\ll t_\|$, the
dimensional crossover always occurs first ($T_{x1}>T_{x2}$)
because two-particle processes are of the order $t_\perp^2\ll 1$
and thus are less likely to occur.

\begin{figure}
\begin{center}
\includegraphics[width=7cm]{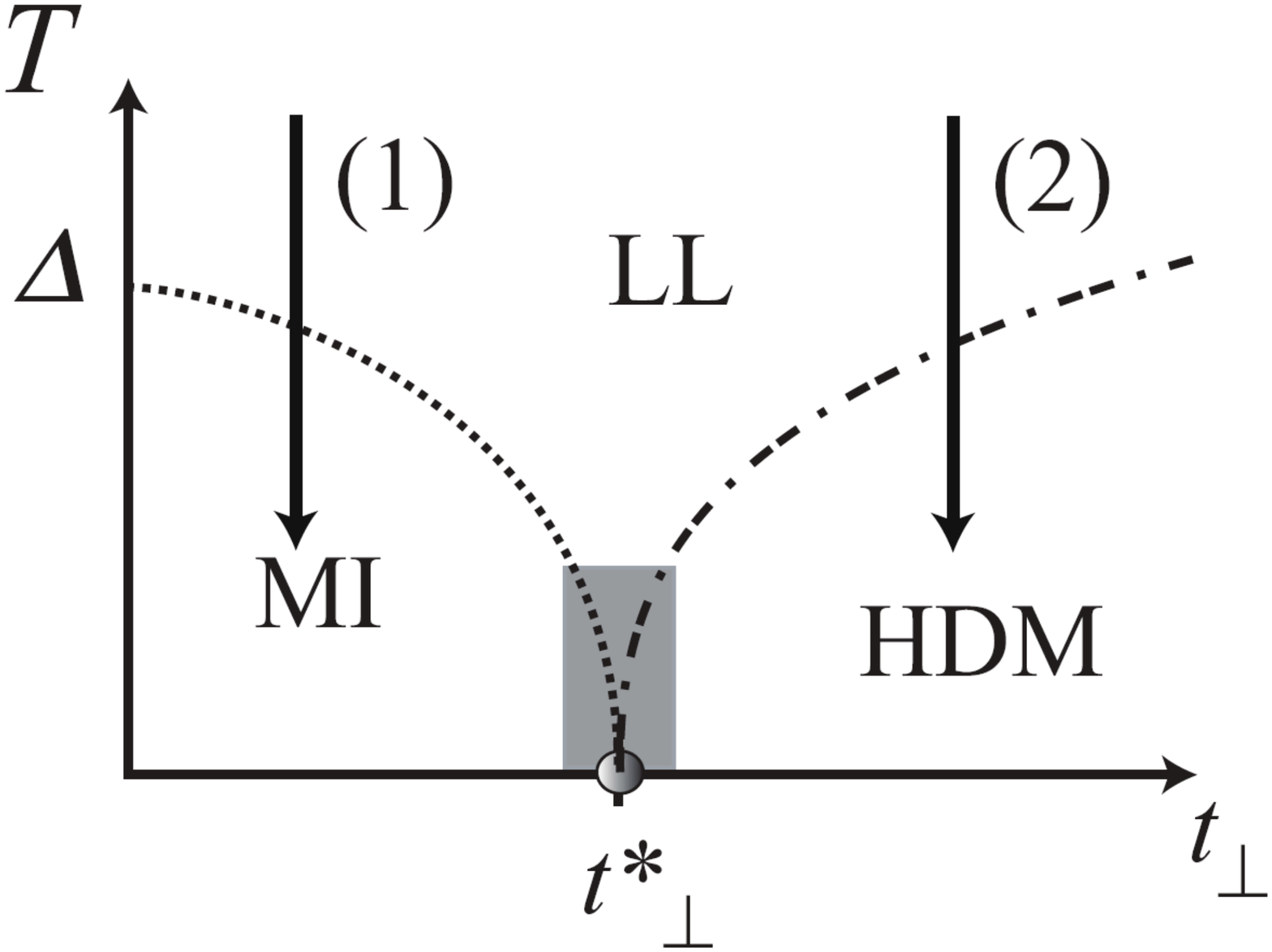}
\caption{\label{fig:deconfinement} Deconfinement transition:
quantum phase transition that takes place at $T=0$ as a function
of the interchain hopping $t_\perp$. At finite temperatures, one
can have a crossover between a Luttinger liquid (LL) and a Mott
insulator (MI) if $t_\perp<t^*_\perp$ by lowering the temperature
(arrow 1), or between a LL and a high metallic phase (HDM) if
$t_\perp>t^*_\perp$ (arrow 2). $\Delta$ is the Mott gap (From
Ref.~\cite{giamarchi_review_chemrev}).}
\end{center}
\end{figure}

Another property of quasi 1D systems that we shall mention in this
section is the {\it{deconfinement transition}}. It consists of a
quantum phase transition where the system goes from a
one-dimensional Mott insulator to a high dimensional metal with
increasing interchain hopping \cite{giamarchi_book_1d}. It appears
in commensurate systems which are generally, as we saw in the
previous section, Mott insulators with a gap $\Delta$ in the
charge excitations. A qualitative picture of the deconfinement
transition is given in Fig.~\ref{fig:deconfinement}. This
transition is more complex than the dimensional crossover, because
when it occurs electrons are able at the same time, to leave the
chains and to conduct. The critical value of $t_\perp$ at which
the deconfinement occurs is $t^*_{\perp}$.
One should solve the fulled coupled problem to obtain a critical
value for $t^*_{\perp}$. It is quite difficult to extract physical
properties in the deconfined phase. One known quantity is the
transverse conductivity that will be discussed in the next
chapter.

At this stage, we have seen the main properties of low-dimensional
systems related to transport physics. We know how to describe a
strongly correlated quasi 1D system, made of weakly coupled
Luttinger liquids. In addition, we can introduce the effects of
the lattice through the umklapp scattering review in this chapter.
All these will be implemented in Chapter~\ref{chap:Hall_quasi1d}.

\chapter{Transport in strongly correlated systems}\label{chap:transport}




In order to study the Hall effect in strongly correlated systems,
which is above all a transport measurement, we must first
investigate the various existing methods to treat transport in
systems with strong interactions. In this chapter we make a review
of two related methods and then use them to obtain the
conductivity in a one-dimensional system, with the purpose of
clarifying their domain of applicability.

A discussion about transport in correlated systems cannot start
without mentioning the most simple theory used to study transport
in nearly-free electron systems. This is the Drude theory of
metals \cite{ashcroft} (Drude constructed his theory on electrical
and thermal conduction in 1900). This theory is the precursor of
all modern transport theories and is still widely used. The Drude
model assumes electrons are classical particles that can move
freely and experience collisions with other electrons with a
probability per unit time $1/\tau$. The time $\tau$ is known as
the relaxation time. In accordance with {\it{Ohm's law}} there is
a relation between the potential drop $V$ and the current $I$
flowing along a wire, which is $V=IR$, where $R$ is the resistance
of the wire. This relation can be rewritten in terms of the
applied electric field $\vec{E}$ and the current density $\vec{j}$
as $\vec{E}=\rho\vec{j}$, where $\rho$ is the resistivity of the
wire. If $n$ electrons of charge $-|e|$ per unit volume move with
an average velocity $\vec{v}$, then the current density is
proportional to $\vec{v}$ and is given by $\vec{j}=-|e|n\vec{v}$.
After each collision, the momentum of the electrons is changed by
$\Delta\vec{p}=\vec{F}\Delta t$, where $\vec{F}=-|e|\vec{E}$ and
the average time $\Delta t$ between collisions is $\tau$. Then,
the average velocity of the electrons is given by
$\vec{v}_{\text{avg}}=-|e|\vec{E}\tau/m$, where $m$ is the
electron mass. Thus the current density can be rewritten as
\begin{equation}
\vec{j}=\frac{n|e|^2\tau}{m}\vec{E}.
\end{equation}
This formula is usually given in terms of the conductivity $\sigma=1/\rho$.
\begin{equation}\label{Drude_formula}
\vec{j}=\sigma\vec{E};\qquad \sigma=\frac{n|e|^2\tau}{m}.
\end{equation}
The formula on the right is the known as the Drude formula and
gives an estimate of the dc electrical conductivity in terms of
known quantities, except for the relaxation time $\tau$ which is
generally determined experimentally \cite{ashcroft}. 

In the case of an applied time-dependent Electric field,
$\vec{E}(t)=\text{Re}(\vec{E}(\omega)e^{-i\omega t})$, the current induced in
a metal $\vec{j}(\omega)$ is given by: $\vec{j}(\omega)=\sigma(\omega)\vec{E}(\omega)$,
where $\sigma(\omega)$ is known as the frequency dependent (or AC) conductivity.
In the Drude theory of Metal $\sigma(\omega)$ is equal to \cite{ashcroft} 
\begin{equation}\label{Drude_2}
\sigma(\omega)=\frac{\sigma_0}{1-i\omega\tau}, \qquad \sigma_0=\frac{n|e|^2\tau}{m}.
\end{equation}
This result reduces to the DC result of Eq.~(\ref{Drude_formula}) at $\omega=0$.
In the next sections we describe two quantum theories to compute the
conductivity $\sigma$. First, we discuss the linear response
theory used to compute ac conductivities ($\sigma(\omega)$), and
then we explain the memory function formalism that allows to
compute $\sigma(\omega)$ including the effect of interactions.
Finally we applied all these methods to the study of transport in
low-dimensional systems.

\section{Linear response theory and Kubo formulas}\label{sec:Kubo}

When an external field is applied to a physical system, its
response will depend on the magnitude of the perturbation. Many
experiments in condensed matter physics measure responses to
applied fields. For example, a sample is placed in an electric
field, a magnetic field or a temperature gradient to measure its
electric current, magnetization or thermal current, respectively.
In linear response theory, the applied field is assumed
sufficiently small, so that the system response increases linearly with
the intensity of the field, and non-linear terms can be neglected.

Although linear response theory applies to all possible fields
\cite{mahan_book} (Kubo first derived his equations for the
electrical conductivity in solids in 1957-1959), we will focus
here on the electrical conductivity tensor
$\sigma(\vec{r},\vec{r}',t-t')$ which relates the current response
$j_\alpha(\vec{r},t)$ to an electric field $E_\beta(\vec{r}',t')$
trough
\begin{equation}\label{linear_response}
j_\alpha(\vec{r},t)= \sum_{\beta}\int d\vec{r}'\int_{-\infty}^{t}dt'\,\sigma_{\alpha\beta}(\vec{r}-\vec{r}';t-t') E_\beta(\vec{r}',t'),
\end{equation}
where $E_\beta(\vec{r}',t')$ is the total electric field,
{\it{i.e}} the sum of the applied external field
$E_\beta^{\text{(ext)}}$ and the fields created by the charged
displaced in the solid. In Eq.~(\ref{linear_response}) we write
the conductivity as a function of the difference
($\vec{r}-\vec{r}'$) because we assume translational invariance in
the solid.

The main goal is to obtain an expression for the ac conductivity
$\sigma_{\alpha,\beta}(\vec{q},\omega)$, that is the Fourier
transform of $\sigma_{\alpha\beta}(\vec{r}-\vec{r}',t-t')$. Let us
consider a system described by a Hamiltonian $\mathcal{H}$, where
an electric field is applied. For this formalism to be well
defined, the system is considered unperturbed at $t=-\infty$ and
the perturbation, in this case the electric field, is turned on
slowly to be totally present at $t\sim0$. Then, at large times,
$t=+\infty$, the perturbation is turned off and the system returns
to its unperturbed state. $\vec{A}(\vec{r},t)$ is the time
dependent vector potential related to the electric field by
$\vec{E}(\vec{r},t)=-\partial\vec{A}(\vec{r},t)/\partial t.$ The
current is defined as the functional derivative of the Hamiltonian
with respect to the vector potential (unities are chosen in order
to have $c=1$, $\hbar=1$ and $k_B=1$)
\begin{equation}\label{def_current}
j_\alpha(\vec{r},t)=-\left.\frac{\delta\mathcal{H}}{\delta A_\alpha(\vec{r},t)}\right|_{\vec{A}^{\text{el}}=0},
\end{equation}
and it can be itself proportional to $\vec{A}$. We also define the
operator of the total current as $J_\alpha=\int d\vec{r}
j_\alpha(\vec{r})$. The vector potential also enters in the
Hamiltonian modifying the momentum $\Pi$ of the particles by
$\Pi-|e|\vec{A}$, with $|e|$ the elementary charge. Thus, to
capture all the linear response to the vector potential, one
should make a first order expansion in both the Hamiltonian and
the current. In the case of the current, the expansion of the
average value is given by
\begin{equation}\label{average_current}
\langle j_\alpha (\vec{r},t)\rangle = \sum_\beta \int d\vec{r}'dt' \frac{\delta \langle j_\alpha (\vec{r},t)\rangle}{\delta A_\beta(\vec{r}',t')} A_\beta (\vec{r}',t'),
\end{equation}
with $\langle\cdots\rangle$ the thermodynamic average taken with
respect to the Hamiltonian $\mathcal{H}$. Here we suppose that
$\mathcal{H}$ can be diagonalized in some basis for the average
$\langle\cdots\rangle$ to be computed analytically. In the same
manner, the total Hamiltonian is expanded to first order in
$\vec{A}$ giving
\begin{equation}
\mathcal{H} = \mathcal{H}[A=0] -  \int d\vec{r} \sum_\alpha j_\alpha^0(\vec{r},t) A_\alpha(\vec{r},t),
\end{equation}
where $j_\alpha^0(\vec{r},t)$ is the part of the current independent of $\vec{A}$.
Using the above result in Eq.~(\ref{average_current}) we obtain the following expression for the average current
\begin{equation}\label{average_current_2}
\langle j_\alpha (\vec{r},t)\rangle = \sum_\beta \int d\vec{r}'dt' \left[\left\langle\frac{\delta j_\alpha (\vec{r},t)}{\delta A_\beta(\vec{r}',t')}\right\rangle_{\mathcal{H}[A=0]} - \langle j_\alpha^0 (\vec{r},t);j_\beta^0 (\vec{r}',t')\rangle_{\text{ret}} \right]A_\beta (\vec{r}',t').
\end{equation}
$\langle j_\alpha^0 (\vec{r},t);j_\beta^0
(\vec{r}',t')\rangle_{\text{ret}}$ stands for the retarded
current-current correlation function \cite{mahan_book} defined as
\begin{equation}\label{retarded}
\langle j_\alpha (\vec{r},t);j_\beta
(\vec{r}',t')\rangle_{\text{ret}}=-i\theta(t-t')\left\langle
[j_\alpha (\vec{r},t),j_\beta (\vec{r}',t')] \right\rangle.
\end{equation}
In Eq.~(\ref{retarded}) we dropped the superscript $0$ to lighten
notation. Using the definition of the current given in
Eq.~(\ref{def_current}), we rewrite the first term in
Eq.~(\ref{average_current_2}) and we obtain for the average
current
\begin{eqnarray}\nonumber
\langle j_\alpha (\vec{r},t)\rangle &=& \sum_\beta \int d\vec{r}'dt' \left[-\left\langle\frac{\delta^2\mathcal{H}}{\delta A_\alpha(\vec{r},t)\delta A_\beta(\vec{r}',t')}\right\rangle_{A=0}\right.\\
&&\left.-\langle j_\alpha (\vec{r},t);j_\beta
(\vec{r}',t')\rangle_{\text{ret}} \right]A_\beta (\vec{r}',t').
\end{eqnarray}
As pointed out before, the Hamiltonian is a function of
$\Pi-|e|\vec{A}$ and thus the functional derivative with respect
to $\vec{A}$ can be written in terms of the momentum operator
$\Pi$,
\begin{equation}
\frac{\delta^2\mathcal{H}}{\delta A_\alpha(\vec{r},t)\delta A_\beta(\vec{r}',t')}\Big{|}_{A=0}=e^2\frac{\delta^2H}{\delta\Pi^2}\delta(\vec{r-r}')\delta(t-t')\delta_{\alpha\beta}.
\end{equation}
The average current is finally given by the following expression
\begin{eqnarray}\nonumber
\langle j_\alpha (\vec{r},t)\rangle &=& \sum_\beta \int d\vec{r}'dt' \left[-e^2\delta_{\alpha\beta}\left\langle\frac{\delta^2\mathcal{H}}{\delta \Pi^2}\right\rangle\delta(\vec{r-r}')\delta(t-t')\right.\\
&&\left.-\langle j_\alpha (\vec{r},t);j_\beta (\vec{r}',t')\rangle_{\text{ret}} \right]A_\beta (\vec{r}',t').
\end{eqnarray}
Making the Fourier transform
\begin{equation}
\langle j_\alpha (\vec{q},\omega)\rangle = \int d\vec{r} \int dt e^{-i\vec{q}\cdot\vec{r}}e^{i\omega t} \langle j_\alpha (\vec{r},t)\rangle,
\end{equation}
and using the relation between the vector potential and the
electric field in Fourier space:
$A_\alpha(\vec{q},\omega)=E_\alpha(\vec{q},\omega)/i\omega$, we
finally encounter an expression for the conductivity matrix
$\sigma_{\alpha\beta}$ describing the linear-response current
$j_\alpha(\vec{q},\omega)$ induced by an {\it{ac}} electric field
$E_\alpha(\vec{q},\omega)$
\begin{equation}
j_\alpha(\vec{q},\omega)=\sum_\beta \sigma_{\alpha\beta}(\vec{q},\omega)E_\beta(\vec{q},\omega),
\end{equation}
where $\sigma_{\alpha\beta}$ is
\begin{equation}\label{Kubo}
\sigma_{\alpha\beta}(\vec{q},\omega)=\frac{1}{i\omega}\left[\chi_{\alpha\beta}(0)\delta_{\alpha\beta}-\chi_{\alpha\beta}(\vec{q},\omega)\right].
\end{equation}
Eq.~(\ref{Kubo}) is known as the Kubo formula. The first term is called the diamagnetic term and is purely imaginary
\begin{equation}\label{diamagnetic}
\chi_{\alpha\alpha}(0)=-e^2\left\langle\frac{\partial^2\mathcal{H}}{\partial \Pi^2}\right\rangle = \left\langle\frac{\delta^2\mathcal{H}}{\delta A_\alpha^2}\right\rangle\Big{|}_{\vec{A}=0}.
\end{equation}
The second term is the retarded current-current correlation function in Fourier space
\begin{equation}\label{susceptibility}
\chi_{\alpha\beta}(\vec{q},\omega)=- i\int_{-\infty}^{\infty}dte^{i\omega(t-t')}\theta(t-t')\left\langle [j_\alpha (\vec{q},t),j_\beta (\vec{q},t')]\right\rangle.
\end{equation}
The real part of the conductivity is totally given by this term.
The {\it{dc}} conductivity is found by taking the limit
$\vec{q}\to0$ first and then the limit $\omega\to0$. If the order
of these limits is interchanged, one would obtain the
thermodynamic response of the system, where the limit $\omega\to0$
is taken first in order to have a static perturbation with
$\vec{q}\neq0$.

The retarded current-current correlation function is usually
computed in imaginary time $\tau$ using the Matsubara formalism
\cite{mahan_book}, because it is the standard way to perform the
calculations at nonzero temperatures. For this we first define the
current-current correlation function in imaginary time
\begin{equation}\label{matsubara_correlation}
\chi_{\alpha\beta}(\vec{q},\tau)= -\langle T_{\tau}j_{\alpha}^{\dagger}(\vec{q},\tau)j_{\beta}(\vec{q},0)\rangle,
\end{equation}
where $0<\tau<\beta$ and $\beta$ is the inverse temperature ($\beta=1/T$). $T_\tau$ is the
$\tau$-ordering operator, which arranges operators with earliest
$\tau$ to the right \cite{mahan_book}. Then, we perform the Fourier transform of
Eq.~(\ref{matsubara_correlation}):
\begin{equation}\label{X_omega}
\chi_{\alpha\beta}(\vec{q},i\omega_n)=\int_{0}^{\beta}d\tau e^{i\omega_n\tau}\chi_{\alpha\beta}(\vec{q},\tau).
\end{equation}
The frequencies $i\omega_n$ are called {\it{imaginary Matsubara
frequencies}} and have the following values: for bosons
$\omega_n=2n\pi/\beta$ and for fermions
$\omega_n=(2n+1)\pi/\beta$. To recover the retarded correlation
function from the Matsubara function we have to change the
imaginary frequencies to real frequencies making the analytical
continuation: $i\omega_n\to \omega+i\delta$,
\begin{equation}
\chi_{\alpha\beta}(\vec{q},i\omega_n)\xrightarrow[i\omega_n\to\omega+i\delta]{}
\chi_{\alpha\beta}(\vec{q},\omega),
\end{equation}
with $\delta=0^{+}$.
Finally by inserting the result in the Kubo formula (\ref{Kubo}) we obtain the conductivity.

Until now, we have seen how the conductivity can be expressed in
terms of a current-current correlation function plus a diamagnetic
term, via the Kubo formula (\ref{Kubo}). This formalism is
suitable for systems with a Hamiltonian that can be diagonalized,
{\it{i.e.}}, in which the averages $\langle\cdots\rangle$ can be
computed analytically. However, if we are interested in strongly
correlated systems, as is the case in this work, we must consider
interacting terms in the Hamiltonian that cannot be diagonalized.
Thus we must search for other approaches where the conductivity
can be calculated with interactions included. The Kubo formula is,
however, the most general expression to compute ac conductivities
and serves as a basis for the other approaches. The next section
is devoted to one of this approaches known as the memory matrix
formalism.

\section{Memory function formalism}\label{sec:memory_formalism}

To understand the essence of this formalism, we first need to
study some important properties of correlation functions. The
current-current correlation function defined in
Eqs.~(\ref{susceptibility})-(\ref{X_omega}), belongs to a larger
group of functions known as susceptibilities $\chi(\omega)$,
denoted as $\chi_{AB}(\omega)=\langle A; B\rangle$ at $\vec{q}=0$.
They are holomorphic functions for all complex frequencies
$\omega$. Integrating by parts Eq.~(\ref{susceptibility}) we can
extract the behavior at large frequencies (this is done in detail
in Sec.~\ref{high_frequency_RH})
\begin{equation}\label{asymptotic_X}
\chi_{AB}(\omega)=\frac{\langle[A, B]\rangle}{\omega}+\frac{\langle\left[[A,\mathcal{H}], B\right]\rangle}{\omega^2}\,\,\,
\text{for}\,\, \omega\to\infty,
\end{equation}
with $[A,B]$ the commutator between operators $A$ and $B$ at the
equal time, and $\mathcal{H}$ the total Hamiltonian of the system.
If $A$ and $B$ commute, it is the second term in
(\ref{asymptotic_X}) which gives the behavior at high frequency.
In the imaginary-time representation of the susceptibilities
$\chi_{AB}(\tau)=-\langle T_\tau A(\tau) B(0)\rangle$, the
operators obey the Heisenberg time evolution
$A(\tau)=e^{\mathcal{H}\tau}Ae^{-\mathcal{H}\tau}$, which implies
the following equation of motion $\partial_\tau A(\tau)
=[\mathcal{H},A(\tau)]$. With this, one can demonstrate that
susceptibilities obey the following equation of motion
\begin{equation}
\label{eq_motion}
 \omega\langle A; B\rangle=\langle[A,B]\rangle - \langle[\mathcal{H},A]; B\rangle =
\langle[A, B]\rangle + \langle A;[\mathcal{H},B]\rangle.
\end{equation}
In order to prove this, we have taken the time derivative of
$\chi_{AB}(\tau)$, applied the time homogeneity property $\langle
T_\tau A(\tau) B(0)\rangle=\langle T_\tau A(0) B(-\tau)\rangle$
and then Fourier transformed the whole equation. We recall that
the retarded correlation function $\chi(\omega)$ is obtained from
the Matsubara function $\chi(i\omega_n)$ by the analytical
continuation $i\omega_n\to\omega+i\delta$.

In general, susceptibilities can be represented as a spectral
integral (also called {\it{Lehmann representation }} (1954)) of
the form
\begin{equation}
\chi(i\omega_n)=
\int_{\infty}^{\infty}\frac{d\omega'}{2\pi}\frac{S(\omega')}{i\omega_n-\omega'},
\end{equation}
where the spectral function $S(\omega)$ is given by the imaginary
part of the retarded correlation function defined in
Eq.~(\ref{susceptibility})
\begin{equation}
S(\omega)=-2\text{Im}[\chi(\omega)]\qquad \text{and}\qquad
\chi(\omega\pm i\delta)=\text{Re}[\chi(\omega)]\pm
i\text{Im}[\chi(\omega)].
\end{equation}
One can also demonstrate that $\chi(\omega)$ verifies the
following symmetry properties
\begin{equation}\label{chi_properties}
\chi(\omega)= \chi(-\omega)\,\,\, \text{and}\,\,\,
\chi^*(\omega)=\chi(\omega^*).
\end{equation}
Thus, Re$[\chi(\omega)]$ and Im$[\chi(\omega)]$ are real and
satisfy Re$[\chi(\omega)]=$Re$[\chi(-\omega)]$ and
Im$[\chi(\omega)]=-$Im$[\chi(-\omega)]$. All these susceptibility
properties will help us define the memory function properly.

Now let us come back to the conductivity $\sigma$. We suppose a
system described by a Hamiltonian $\mathcal{H}=\mathcal{H}_0
+\mathcal{H}_{\text{int}}$, where $\mathcal{H}_{\text{int}}$ is an
interaction term that we do not know how to treat exactly. We will
start with the case where the conductivity goes only along one
direction and thus the Kubo formula (\ref{Kubo}) becomes
$\sigma_{\alpha\beta}(\omega)=\sigma(\omega)=\left[\chi(0)-\chi(\omega)\right]/i\omega$.
As pointed out before, the static conductivity, in a normal
conductor, is given by the diamagnetic term $\chi(0)$ ($\omega=0$
term in Eq.~(\ref{Kubo})) and for all $\omega$ we have:
\begin{equation}\label{inequality}
\chi(\omega)\ne \chi(0).
\end{equation}
The memory function formalism consist on representing the
conductivity by a relaxation or memory function $M(\omega)$
\cite{gotze_fonction_memoire}, as mentioned before. Let us
consider the function
\begin{equation}\label{memory1}
iM(\omega)=\frac{\omega \chi(\omega)}{\chi(0)-\chi(\omega)}.
\end{equation}
Due to inequality (\ref{inequality}) the memory function is also
an holomorphic function for all frequencies $\omega$. The behavior
of $\chi(\omega)$ at large frequencies shown in
Eq.~(\ref{asymptotic_X}) gives an asymptotically decrease of
$M(\omega)$ as $1/\omega^2$ when $\omega\to\infty$ (because the
commutator $[J,J]$ is zero). The fact that $M(\omega)$ vanishes at
infinite frequency will be determinant in the study of the
high-frequency Hall effect in Chapter~\ref{chap:Hall_effect}. The
susceptibility properties shown in Eq.~(\ref{chi_properties})
imply the following symmetry relations for the memory function:
$M^*(\omega)=M(\omega^*)$ and $M(\omega)=-M(-\omega)$.
Furthermore, it can also be represented by an spectral integral
\begin{equation}
M(i\omega_n)= \int
\frac{d\omega'}{2\pi}\frac{S(\omega')}{\omega'-\omega},
\end{equation}
where again
\begin{equation}
S(\omega')=-2\text{Im}[M(\omega)]\qquad \text{and}\qquad
M(\omega\pm i\delta)=\text{Re}[M(\omega)]\pm
i\text{Im}[M(\omega)].
\end{equation}
Thus, $\text{Re}[M(\omega)]$ and $\text{Im}[M(\omega)]$ are real
functions satisfying $\text{Re}[M(\omega)]=
-\text{Re}[M(-\omega)]$ and
$\text{Im}[M(\omega)]=\text{Im}[M(-\omega)]$. From
Eq.~(\ref{memory1}) we can rewrite the susceptibility in terms of
$M(\omega)$,
\begin{equation}\label{susceptibility_memory}
\chi(\omega)=\chi(0)\frac{iM(\omega)}{\omega+iM(\omega)}.
\end{equation}
Now that we have represented the current-current correlation
function in terms of the memory function, it can replaced in the
Kubo formula to derived an expression for the conductivity in
terms of $M(\omega)$,
\begin{equation}\label{sigma_Memory}
\sigma(\omega)=-i\frac{\chi(0)}{\omega+iM(\omega)}.
\end{equation}
This representation of the conductivity provides a correct way to
make perturbative expansions in small parameters (like a coupling
constant or density), that are not feasible on susceptibilities
due to their singular character at small frequencies. Comparing
Eq.~(\ref{sigma_Memory}) to the semiclassical formula
(\ref{Drude_2}), derived from the Drude model, it is obvious
that the memory function plays the role of the relaxation time
$\tau$ in the Drude conductivity. The advantage of the memory
function formalism is that $iM(\omega)$ gives a practical way to
obtain the conductivity which is expected to be non-singular when
$\omega\to0$. Eq.~(\ref{sigma_Memory}) will be frequently used in
this work to compute conductivity.

The next step is to see how the memory function is computed by
treating the interaction term $\mathcal{H}_{\text{int}}$
perturbatively. From the equation of motion (\ref{eq_motion}) one
can write
\begin{equation}\label{jj}
\omega\langle J;J \rangle= -\langle K;J \rangle
\end{equation}
where $K=[\mathcal{H},J]$ is an operator known as {\it{residual
force}}. It is given by the part of the Hamiltonian that does not
commute with the current, {\it{i.e.}}, the interaction term
$\mathcal{H}_{\text{int}}$ ($K=[\mathcal{H}_{\text{int}},J]$).
Furthermore, using the same equations of motion for operators $K$
and $J$, we have
\begin{equation}\label{Kj}
\omega\langle K;J \rangle= \langle[K,J]\rangle +\langle K;K \rangle.
\end{equation}
And from Eqs.~(\ref{jj}) and (\ref{Kj}) at $\omega=0$ we obtain
\begin{equation}\label{Kj_0}
\langle[K,J]\rangle = -\langle K;K \rangle_{\omega=0}.
\end{equation}
Using Eqs.~(\ref{jj})-(\ref{Kj_0}), and the current-current
correlator defined in (\ref{susceptibility}) at $q=0$ (where
operator $j$ becomes a total current operator $J$) we find
\begin{equation}
-\omega\chi(\omega)=\frac{\langle K;K \rangle-\langle K;K
\rangle_{\omega=0}}{\omega}.
\end{equation}
Until now we have worked with exact relations. However, we do not
know how to compute the correlator $\langle K;K\rangle$ with
interaction terms present in the Hamiltonian. Thus, we have to
make some approximations. First, we expand $iM(\omega)$ in
Eq.~(\ref{memory1}) at high enough frequencies where
$|\chi(\omega)/\chi(0)|$ is small
\begin{equation}
\chi(0) iM(\omega)\simeq\omega\chi(\omega).
\end{equation}
With this approximation we arrive to the following expression for the memory function
\begin{equation}\label{Memory_KK}
iM(\omega)\simeq-\frac{1}{\chi(0)}\frac{\langle K;K
\rangle_{\omega}-\langle K;K \rangle_{\omega=0}}{\omega}.
\end{equation}
Because the operator $K$ is proportional to the interacting term
$\mathcal{H}_{\text{int}}$, the thermodynamical average in
correlation $\langle K;K\rangle$ can be computed with the
Hamiltonian $\mathcal{H}_0$ (denoted $\langle \cdots\rangle^0$) to
get the lowest order in the interaction parameters,
\begin{equation}
iM(\omega)\simeq-\frac{1}{\chi(0)}\frac{\langle K;K \rangle^0_{\omega}-\langle K;K \rangle^0_{\omega=0}}{\omega}.
\end{equation}
Finally, calculating $iM(\omega)$ and replacing it on
Eq.~(\ref{sigma_Memory}), we obtain a result for the ac
conductivity at second order in the interaction parameters (each
$K$ contributes with one parameter), which is the lowest non-zero
order for the memory function. The application of these results
will be much more clear in Sec.~\ref{sec:transport_low_dim}, where
it will be used to obtain the conductivity in a one-dimensional
system.

In this entire discussion we have supposed a system with
longitudinal conductivity, where the memory matrix reduces to a
scalar function. However, if one is interested in the conductivity
in a plane (a more general case), the previous results must be
reobtained using relations involving matrices. In the next section
we will study the memory function formalism using the conductivity
tensor in order to obtain a matrix expression for $M(\omega)$.

\subsection{Matrix representation of the Memory function}\label{sec:memory_matrix}

In the study of the Hall effect, one is confronted with a problem
of conduction along different directions in the presence of an
applied magnetic field, as will be explained in
Chapter~\ref{chap:Hall_effect}. In this case, the previous
derivation for the memory function must be remade starting from
the conductivity tensor, in order to get a matrix representation
for $M(\omega)$. The conductivity tensor is given by
\begin{equation}
\bm{\sigma}=\left(\begin{matrix}\sigma_{xx}&\sigma_{xy}\\\sigma_{yx}&\sigma_{yy}\end{matrix}
\right).
\end{equation}
It is important to observe, for the following derivations, that
the non-diagonal terms of $\bm{\sigma}$  appear when a magnetic
field is applied perpendicular to the $x$-$y$ plane. Thus, for the
rest of this section we will suppose that a magnetic field is
indeed applied. Let us rewrite the conductivity tensor in terms of
the memory matrix $\bm{M}(\omega)$ as
\begin{equation}\label{sigma_M}
        i\bm{\sigma}(\omega)=\bm{\chi}(0)\left[\omega\mathbbm{1}+\bm{\Omega}
    +i\bm{M}(\omega)\right]^{-1}.
\end{equation}
Eq.~(\ref{sigma_M}) as well as the following definitions, are
enclosed in a general theory called the Mori theory, and details
can be found in Ref.~\cite{mori_theory_book}. As mentioned
previously, the advantage provided by the memory matrix formalism
is the possibility to make finite-order perturbation expansions
which are singular in the conductivities due to their resonance
structure \cite{gotze_fonction_memoire}. $\bm{\chi}(0)$ in
Eq.~(\ref{sigma_M}) is a diagonal matrix composed of the
diamagnetic susceptibilities in each direction,
\begin{equation}
\bm{\chi}(0)=\left(\begin{matrix}\chi_x(0)&0\\0&\chi_y(0)\end{matrix}
\right).
\end{equation}
The matrix $\mathbf{\Omega}$ in (\ref{sigma_M}) is called the
{\it{frequency}} matrix and is defined in terms of the equal-time
current-current correlator as \cite{mori_theory_book}
    \begin{equation}\label{Omega}
        \Omega_{\mu\nu}=\frac{1}{\chi_{\mu}(0)}
        \left\langle\big[J_{\mu},J_{\nu}\big]\right\rangle.
    \end{equation}
The frequency matrix gives the behavior of $\vec{\sigma}(\omega)$
at high frequencies because the memory matrix vanishes as
$1/\omega^2$ when $\omega\to\infty$, as mentioned in the previous
section. From Eq.~(\ref{sigma_M}) one can directly express the
memory matrix $\bm{M}(\omega)$ in terms of the conductivity
tensor,
\begin{equation}\label{Memory_matrix__definition}
i\bm{M}(\omega)=-i\bm{\sigma}^{-1}(\omega)\bm{\chi(0)}-\omega\mathbbm{1}-\bm{\Omega}.
\end{equation}
The diagonal terms $M_{xx}(\omega)$ and $M_{yy}(\omega)$, are
given by definition (\ref{memory1}) with $\chi(\omega)$ replaced
by $\chi_{xx}(\omega)$ and $\chi_{yy}(\omega)$, respectively (idem
for $\chi(0)$). For the off-diagonal terms we have
$M_{yx}(\omega)=-M_{xy}(\omega)$ (due to
$\sigma_{xy}=-\sigma_{yx}$), and we take $M_{xy}(\omega)$ because
it will be the term necessary in the description of the Hall
effect in Chapter~\ref{chap:Hall_effect}. $M_{xy}(\omega)$ written
in terms of the conductivities gives
    \begin{equation}\label{Memory_xy}
        iM_{xy}(\omega)=\frac{i\chi_y(0)\sigma_{xy}(\omega)}
        {\sigma_{xx}(\omega)\sigma_{yy}(\omega)+\sigma_{xy}^2(\omega)}
        -\Omega_{xy}.
   \end{equation}
Expressing the conductivities in terms of current susceptibilities
by means of the Kubo formula
$\sigma_{\mu\nu}=\frac{i}{\omega}[\chi_{\mu\nu}-\delta_{\mu\nu}\chi_{\mu}(0)]$,
the above expression  leads to
    \begin{equation}\label{Memory_susceptibilies}
        iM_{xy}(\omega)=\frac{\omega\chi_y(0)\chi_{xy}(\omega)}
        {\left[\chi_x(0)-\chi_{xx}(\omega)\right]
        \left[\chi_y(0)-\chi_{yy}(\omega)\right]}-\Omega_{xy}.
    \end{equation}
This representation, together with Eq.~(\ref{memory1}) for the
longitudinal terms, completely defines the Memory matrix. The
parity property in Eq.~(\ref{chi_properties}) is valid for the
longitudinal $\chi_{xx}(\omega)$ and $\chi_{yy}(\omega)$, giving
for the diagonal term $M_{xx}(\omega)=-M_{xx}(-\omega)$, and the
same for $M_{yy}(\omega)$. In the other hand, the transverse
susceptibility satisfies $\chi_{xy}(\omega)=-\chi_{xy}(-\omega)$
and from Eq.~(\ref{Memory_susceptibilies}) it is easy to prove
that $M_{xy}(\omega)=M_{xy}(-\omega)$.

Following the same procedure applied in the longitudinal case, we
rewrite Eq.~(\ref{Memory_susceptibilies}) at high enough
frequencies, such that $|\chi_{\mu\mu}(\omega)/\chi_{\mu}(0)|$ is
small. In this expansion we use the equation of motion for the
susceptibilities, as well as the relation
$[\mathcal{H}_0,J_{\mu}]=-\Omega_{\nu\mu}J_{\nu}$, with
$\mu,\nu=x,y$ and summation over repeated indices is implied. This
relation is the precondition necessary to obtain a regular
expansion of the memory function for all frequencies to leading
order in the interaction term \cite{mori_theory_book}. The
obtention of the following result will be presented in detail in
Sec.~\ref{sec:Hall_memory}. The off-diagonal term of the memory
matrix thus reads
    \begin{equation}\label{MofK}
        iM_{xy}(\omega)\simeq-\frac{1}{\chi_x(0)}\frac{\langle K_x;K_y\rangle_{\omega}}{\omega},
    \end{equation}
where $K_{\mu}$ are the \emph{residual forces} operators defined
in the previous section. In this case, they are given by the part
of the Hamiltonian which {\it{in the absence of magnetic field}}
does not commute with the currents, \textit{i.e.}
$K_{\mu}=[\mathcal{H}_{\text{int}},J_{\mu}]$. When a magnetic
field $\bm{B}$ is applied, the non-interacting Hamiltonian does
not commute anymore with the currents, that is why it must be
emphasized that $K$ operators are computed with $\bm{B}=0$. The
quantity $\langle K_x;K_y\rangle$ stands for the retarded
correlation function of the operators $K_{\mu}$. The obtention of
this correlator will be the key point in the study of the Hall
effect in a quasi 1D system in Chapter~\ref{chap:Hall_quasi1d}.

Expression (\ref{MofK}) does not contain the term at $\omega=0$
present in Eq.~(\ref{Memory_KK}), because the off-diagonal terms
of the memory matrix are even in frequency (therefore $\omega
iM_{xy}(\omega)$ is odd). Only the diagonal terms are odd in
$\omega$ and thus have the residual forces correlator evaluated at
zero frequency. The terms omitted in Eq.~(\ref{MofK}) are either
of second order in $|\chi_{\mu\mu}(\omega)/\chi_{\mu}(0)|$, or of
second order in the magnetic field. As will be seen in the next
chapter, Eq.~(\ref{MofK}) is the memory matrix element necessary
to compute the Hall resistivity.

Now that we have a technique to compute longitudinal and
transverse conductivities in strongly correlated systems, to
leading order in the interaction term, we will see in the next
sections one example of their application in low-dimensional
systems.

\section{Transport in low-dimensional systems}\label{sec:transport_low_dim}

After reviewing the main properties of low-dimensional systems in
Chapter~\ref{chap:strongly_correlated}, we will devote this
section to the study of "transport" properties in these systems.
For this we make use of the two formalisms discussed previously.
We will begin studying transport in purely one-dimensional systems
and then in quasi one-dimensional ones. We concentrate on systems
with commensurate fillings because in
Chapter~\ref{chap:Hall_quasi1d} we will investigate the Hall
effect on quasi 1D commensurate materials.

\subsection{1D systems without umklapp scattering}
Transport properties are commonly used as a probe for Luttinger
liquid (LL) behavior in low-dimensional systems. The measured
conductivity or resistivity helps determining the energy scales at
which the sample is in a Luttinger liquid regime. In
Sec.~\ref{sec:1D_case} we saw that the umklapp scattering is the
only interacting term producing dissipation in a Luttinger liquid.
Thus, if one removes this term for a moment and thus momentum is
conserved in all interaction processes, the system should behave
as a perfect conductor. In order to prove this we will compute the
conductivity in a LL without umklapp scattering, making use of the
Kubo formula presented in Sec.~\ref{sec:Kubo}. First, we calculate
the current using the charge part of the 1D Hamiltonian
(\ref{H_1D}) (because the transport properties do not affect the
spin degrees of freedom) in the form (again we put $\hbar=1$ and
$c=1$)
\begin{eqnarray}\label{H_1D_PI}
        \mathcal{H}_{\rho}&=&\int\frac{dx}{2\pi}\,\left\{u_{\rho}K_{\rho}
        \left[\pi\Pi_{\rho}(x)\right]^2+\frac{u_{\rho}}{K_{\rho}}
        \left[\nabla\phi_{\rho}(x)\right]^2\right\},
\end{eqnarray}
where $\Pi(x,t)=(1/\pi) \nabla\theta(x,t)$. The vector potential
enters only in the $\Pi$ part of the Hamiltonian via the
substitution $\Pi_{\rho}(x,t)\to\Pi_{\rho}(x,t)-eA(x,t)/\pi$.
Then, the current operator defined in Eq.~(\ref{def_current}) is
given by
\begin{equation}\label{1D_current}
 j(x,t)= e (\sqrt{2}u_\rho K_\rho)\Pi_{\rho}(x,t),
\end{equation}
where the factor $\sqrt{2}$ comes from the sum over spins
($\Pi_\rho=(\Pi_{\uparrow}+\Pi_{\downarrow})/\sqrt{2}$
\cite{giamarchi_book_1d}). In the same way we calculate the
diamagnetic term $\chi(0)$, defined in (\ref{diamagnetic}), which
for fermions with spins gives
\begin{equation}\label{diamagnetic_1d}
\chi(0)=-\frac{2e^2u_\rho K_\rho}{\pi}.
\end{equation}
As we saw in Sec.~\ref{sec:Kubo} the conductivity is equal to the
diamagnetic term plus the current-current correlation term. Using
(\ref{1D_current}) the current-current correlation function in
imaginary time (defined in Eq.~(\ref{matsubara_correlation})) thus
result
\begin{equation}\label{Pi_Pi_correlation}
\chi(x,x';\tau-\tau')=-\left(euK\right)^2\langle T_{\tau}\Pi(x,\tau)\Pi(x',\tau')\rangle.
\end{equation}
We remove the subscript $\rho$ in order to lighten the notation,
remembering that all these transport calculations affect only the
charge part of the 1D Hamiltonian. At this point we can write down
the Kubo formula (\ref{Kubo}) in bosonization language
\cite{giamarchi_umklapp_1d}:
\begin{equation}\label{sigma_chi}
\sigma(\omega)=\frac{i}{\omega}\left[\frac{e^22uK}{\pi}+\chi(\omega)\right],
\end{equation}
where $\chi(\omega)$ is the Fourier transform of
Eq.~(\ref{Pi_Pi_correlation}). Note that $8e^2uK$ plays the role of
the plasma frequency in the usual formulas for the conductivity
\cite{gotze_fonction_memoire}.

As we need to evaluate time-ordered correlation functions
$\chi(\omega)$ in order to obtain the conductivity, we give below
the necessary tools to compute correlation functions using
functional integral techniques. Some important results of
functional integration are presented here but we refer the reader
to the literature for a complete review \cite{negele_book}. The
partition function $Z=\text{Tr}\left(e^{-\beta
\mathcal{H}}\right)$ of Hamiltonian (\ref{H_1D_PI}) represented
via a functional integral is \cite{negele_book}
\begin{eqnarray}\label{partition}
Z&=&\int\mathcal{D}\phi(x,\tau)\mathcal{D}\Pi(x,\tau)e^{-S}\\ \nonumber
S&=&-\int_0^{\beta}d\tau\int dx[i\Pi(x,\tau)\partial_{\tau}\phi(x,\tau)-\mathcal{H}(\phi(x,\tau),\Pi(x,\tau))].
\end{eqnarray}
$S$ is the action in imaginary time associated with Hamiltonian
(\ref{H_1D_PI}). Time-ordered correlation functions of two
operators $\hat{A}$ and $\hat{B}$, functions of the operators
$\phi$ and $\Pi$, are defined in the functional integration
formalism by the formula
\begin{equation}\label{def_func_integral}
\langle T_\tau \hat{A}(x,\tau)\hat{B}(0,0) \rangle
=\frac{1}{Z}\int\mathcal{D}\phi(x,\tau)\mathcal{D}\Pi(x,\tau)A(\phi,\Pi)_{x,\tau}B(\phi,\Pi)_{0,0}\,e^{-S},
\end{equation}
where $Z$ and $S$ are given in Eq.~(\ref{partition}). From now on
we simply denote $\langle T_\tau ...\rangle$ as $\langle
...\rangle$. An important advantage of this technique, is that $A$
and $B$ on the right hand side of Eq.~(\ref{def_func_integral})
are the value of operators $\hat{A}$ and $\hat{B}$, respectively,
and thus have the properties of scalar fields, that are much
easier to deal with than operators \cite{negele_book}.
Another useful formula from functional integration (which we will
not prove here) is the one corresponding to the Fourier transform
of the correlation function $\langle u(r)u(r')\rangle$
\cite{negele_book}:
\begin{equation}\label{u_func_integral}
\langle u^*(q_1)u(q_2)\rangle =\frac{\int \mathcal{D}u[q]u^{*}(q_1)u(q_2)e^{-\frac{1}{2}\sum_{q}A(q)u^{*}(q)u(q)}}{\mathcal{D}u[q]e^{-\frac{1}{2}\sum_{q}A(q)u^{*}(q)u(q)}}=\frac{1}{A(q_1)\delta_{q_1,q_2}},
\end{equation}
where $A(q)$ is a diagonal matrix and, for real fields $u(r)$, we have $u^*(q)=u(-q)$.
Now we are ready to compute the correlation function in Eq.~(\ref{Pi_Pi_correlation}).
\begin{equation}
\langle \Pi(x,\tau)\Pi(x',\tau') \rangle = \frac{1}{Z}
\int\mathcal{D}\phi(x,\tau)\mathcal{D}\Pi(x,\tau)\Pi(x,\tau)\Pi(x',\tau')\,e^{-S}.
\end{equation}
To express the Fourier transforms we use the following notation:
$\vec{r}=(x,u\tau)$, $\vec{q}=(k,\omega_n/u)$, and
$e^{i\vec{q}\cdot\vec{r}}=e^{i(kx-\omega_n\tau)}$. Then
$\Pi(\vec{r})=\frac{1}{\beta\Omega}\sum_{\vec{q}}\Pi(\vec{q})e^{i\vec{q}\cdot\vec{r}}$,
with $\beta=1/T$ and $\Omega$ the volume of the system. Thus, the
above expression becomes
\begin{eqnarray}
\langle \Pi(\vec{r}_1)\Pi(\vec{r}_2) \rangle =
\frac{1}{(\beta\Omega)^2}\sum_{\vec{q}_1,\vec{q}_2} \langle
\Pi^*(\vec{q}_1)\Pi(\vec{q}_2) \rangle e^{-i\vec{q}_1\vec{r}_1}
e^{i\vec{q}_2\vec{r}_2}\\ \label{correlation_q} \langle
\Pi^*(\vec{q}_1)\Pi(\vec{q}_2)\rangle
=\frac{1}{Z}\int\mathcal{D}\phi(\vec{q})\mathcal{D}\Pi(\vec{q})\Pi^*(\vec{q}_1)\Pi(\vec{q}_2)
\,e^{-\mathcal{S}}.
\end{eqnarray}
The action given in Eq.~(\ref{partition}) must be also written in
Fourier space,
\begin{equation}
S=-\frac{1}{\beta\Omega}\sum_{\vec{q}}\left[\omega_n\phi(\vec{q})\Pi(-\vec{q})-\frac{uK\pi}{2}
\Pi(\vec{q})\Pi(-\vec{q})-\frac{u}{2\pi
K}k^2\phi(\vec{q})\phi(-\vec{q})\right].
\end{equation}
Next we complete the squares on the $\Pi$ part of the action to get
ride of the linear term. The action thus result
\begin{eqnarray}
S&=&\frac{1}{\beta\Omega}\sum_{\vec{q}}\frac{\omega_n^2}{2uK\pi}\phi(\vec{q})\phi(-\vec{q})
+\frac{1}{\beta\Omega}\sum_{\vec{q}}\frac{u}{2\pi K}k^2\phi(\vec{q})\phi(-\vec{q})\\ \nonumber &&+\frac{1}{\beta\Omega}\sum_{\vec{q}}\frac{uK\pi}{2}\left[\Pi(\vec{q})-\frac{\omega_n}{uK\pi}\phi(\vec{q})\right]\left[\Pi(-\vec{q})+\frac{\omega_n}{uK\pi}\phi(-\vec{q})\right].
\end{eqnarray}
Making the change of variable $\tilde{\Pi}(\vec{q})=\Pi(\vec{q})-\frac{\omega_n}{uK\pi}\phi(\vec{q})$, 
we obtain an action completely separable in $S_\phi$ and $S_{\tilde{\Pi}}$ part,
\begin{eqnarray}\nonumber
S_{\tilde{\Pi}}+S_\phi&=&\frac{1}{\beta\Omega}\left(\frac{1}{2}\right)\sum_{\vec{q}}\left[(uK\pi)\tilde{\Pi}(\vec{q})\tilde{\Pi}(-\vec{q}) +\left(\frac{\omega_n^2}{uK\pi}+\frac{u}{\pi K}k^2\right)\phi(\vec{q})\phi(-\vec{q})\right] \\ \label{S_phi_Pi}
&=& \frac{1}{\beta\Omega}\left(\frac{1}{2}\right)\sum_{\vec{q}}\left(\Pi^*_{\vec{q}},\,\phi^*_{\vec{q}} \right)\left(\begin{matrix} uK\pi&0\\0&\frac{\omega_n}{uK\pi}+\frac{uk^2}{K\pi}\end{matrix}
\right)\left(\Pi_{\vec{q}},\,\phi_{\vec{q}} \right)
\end{eqnarray}
The matrix representation defines the diagonal matrix $A(\vec{q})$
to be inserted in formula (\ref{u_func_integral}). Using the same
procedure, the partition function can also be separated in
$Z=Z_\phi Z_\Pi$. We rewrite the correlator (\ref{correlation_q})
in the new variables and simplify it using $(1/Z_\phi)\int
\mathcal{D}\phi\,e^{-S_\phi}=1$ (the same is valid for
$\tilde{\Pi}$),
\begin{eqnarray}\nonumber
\langle \Pi^*(\vec{q}_1)\Pi(\vec{q}_2)\rangle
&=&\frac{1}{Z_{\tilde{\Pi}}}\int\mathcal{D}\tilde{\Pi}(\vec{q})\tilde{\Pi}^*(\vec{q}_1)\tilde{\Pi}(\vec{q}_2)
e^{-S_{\tilde{\Pi}}}-\frac{\omega_n^2}{(uK\pi)^2}
\langle\phi^*(\vec{q}_1)\phi(\vec{q}_2)\rangle\\
&&+\frac{\omega_n}{uK\pi}\langle\tilde{\Pi}(\vec{q}_1)\phi^*(\vec{q}_2)\rangle
-\frac{\omega_n}{uK\pi}\langle\phi(\vec{q}_1)\tilde{\Pi}^*(\vec{q}_2)\rangle.
\end{eqnarray}
At this point, we can make use of formula (\ref{u_func_integral}).
It is evident that the crossed correlations
$\langle\tilde{\Pi}\phi^*\rangle$ and
$\langle\phi\tilde{\Pi}^*\rangle$ vanish because $A$ is a diagonal
matrix (see Eq.~(\ref{S_phi_Pi})). The $\langle\Pi^*\Pi\rangle$
correlation function thus gives
\begin{equation}\label{Pi_Pi_result}
\langle\Pi^*(\vec{q}_1)\Pi(\vec{q}_2)\rangle=-\frac{\omega_n^2}{(uK\pi)^2}
\langle\phi^*(\vec{q}_1)\phi(\vec{q}_2)\rangle +
\frac{1}{\pi uK}\delta_{\vec{q}_1,\vec{q}_2}.
\end{equation}
Moving to real space and multiplying the result
(\ref{Pi_Pi_result}) by $(euK)^2$  we obtain for the
current-current correlation function of
Eq.~(\ref{Pi_Pi_correlation})
\begin{equation}\label{Pi_Pi_tau}
(euK)^2\langle\Pi(x,\tau)\Pi(x',\tau')\rangle=-\frac{e^2}{\pi^2}\langle\partial_\tau\phi(x,\tau)
\partial_\tau\phi(x',\tau')\rangle +
\frac{e^2uK}{\pi}\delta(x-x')\delta(\tau-\tau').
\end{equation}
After summing over spins the above expression (a factor $\sqrt{2}$
appears for each $\Pi$ and $\phi$), the second term on the right
cancels exactly the diamagnetic term given in
(\ref{diamagnetic_1d}), when using Eq.~(\ref{sigma_chi}) to obtain
$\sigma(\omega)$. Now we Make the Fourier transform of
Eq.~(\ref{Pi_Pi_tau}),
\begin{equation}
\chi(k,i\omega_n)=\int dx e^{ikx}\int_0^\beta d\tau
e^{i\omega_n\tau}(euK)^2\langle\Pi(x,\tau)\Pi(0,0)\rangle.
\end{equation}
Thus the ac conductivity (at $k=0$) of a one-dimensional system
without umklapp scattering gives
\begin{eqnarray}
\sigma(\omega)&=&-\frac{2e^2}{\pi^2}i(\omega+i\delta)\langle \phi^*(k=0,\omega_n)\phi(k=0,\omega_n)\rangle_{i\omega_n\to\omega+i\delta}\\
&=&-\frac{e^2}{\pi^2}i(\omega+i\delta)\frac{\pi 2uK}{\omega_n^2}\big{|}_{i\omega_n\to\omega+i\delta}.
\end{eqnarray}
Here we have used again the functional integration formula
(\ref{u_func_integral}) to calculate the $\langle
\phi^*\phi\rangle$ correlation using the $\phi$ part of the action
given in (\ref{S_phi_Pi}) (and $Z_\phi$) at $k=0$ to get the long
wavelength ac conductivity. Finally, after analytical continuation
we obtain
\begin{equation}
\sigma(\omega)=\frac{e^2}{\pi}\frac{2iuK}{\omega+i\delta}=e^2(2uK)\left[\delta(\omega)+i\text{P}\frac{1}{\pi\omega} \right],
\end{equation}
where P denotes the Cauchy principal value. As expected, the
one-dimensional system is a perfect conductor with an infinite
static conductivity given by a delta function peak at $\omega=0$
(Drude peak). The weight of this peak is $2e^2uK$.

If we now include umklapp scattering in the system we would no
longer have a perfect conductor. Such scattering processes are
responsible for the appearance of a finite resistivity because
they produce momentum relaxation. The next section is dedicated to
the obtention of the conductivity in a 1D system in the presence
of umklapp scattering.

\subsection{1D systems with umklapp scattering}
As we saw in Sec.~\ref{sec:Mott_1D}, the umklapp processes arise
due to the presence of the lattice in systems with commensurate
fillings. Here we consider a $1/2$-filled system, but the
subsequent derivations can be done for any commensurate filling
\cite{giamarchi_book_1d}. In the following we make use of the
memory matrix formalism presented in
Sec.~\ref{sec:memory_formalism}, because the calculation of the
exact $\langle\phi\phi\rangle$ correlation function is not
feasible when umklapp scattering is present in the system. Let us
first write down the expression for the conductivity in the memory
function formalism, Eq.~(\ref{sigma_Memory}), in bosonization
language
\begin{equation}\label{sigma_umklapp}
\sigma(\omega)=\frac{e^2 2iu_\rho K_\rho}{\pi}\frac{1}{\omega+iM(\omega)}.
\end{equation}
In a one-dimensional system $M(\omega)$ is just a scalar function,
as the one discussed in Sec.~\ref{sec:memory_formalism}, given by
\footnote{This expression for $M(\omega)$ differs from the
expression in reference \cite{giamarchi_book_1d} by a factor $i$.
We make this choice of notation to be in accordance with the 2D
case where the memory matrix is defined as
$iM_{\alpha\beta}(\omega)$.}
\begin{equation}\label{Memory_1D}
iM(\omega)\simeq\frac{\left[\langle
K_u;K_u\rangle_\omega^0-\langle K_u;K_u\rangle_{\omega=0}^0
\right]/\omega}{-\chi(0)},
\end{equation}
where the force $K_u$ (the subscript $u$ is there to avoid
confusion with the LL parameter $K_\rho$) is
$K_u=[\mathcal{H}_u,J]$ and $\mathcal{H}_u$ is the umklapp
scattering term in a $1/2$-filled system,
\begin{equation}\label{umklapp_1}
\mathcal{H}_u=\frac{2\text{g}_{3}}{(2\pi a)^2}\int
dx\cos(\sqrt{8}\phi_\rho(x)).
\end{equation}
From the above expressions it is clear that if the current
commutes with the Hamiltonian, as in a 1D system without umklapp
scattering, the memory function is zero and one recovers a perfect
conductor.

The $K_u$ operators are easily computed calculating the commutator
between the umklapp operator (\ref{umklapp_1}) and the current
(\ref{1D_current}), using the commutation relations
(\ref{commutation_relations}) between bosonic fields $\phi$ and
$\theta$ (remember that $\Pi=\nabla\theta/\pi$),
\begin{equation}\label{Ku}
K_u=[\mathcal{H}_u,J]=\frac{8eg_3}{(2\pi a)^2}(u_\rho K_\rho)i\int
dx \sin(\sqrt{8}\phi_\rho(x,\tau)).
\end{equation}
The next step is to calculate the correlation $\langle
K_u;K_u\rangle$ and then insert the result in
Eq.~(\ref{Memory_1D}) to finally obtain the conductivity. We will
not compute here the full expression for the correlation $\langle
K_u;K_u\rangle$ which is obtained in detail in
Refs.~\cite{giamarchi_umklapp_1d,giamarchi_book_1d}. Instead, we
use scaling arguments to determine the exponents of the power-law
frequency (or temperature) dependence of the conductivity. Each
$K_u$ force depends linearly on the parameter $g_3$, thus the
memory function is of order $g_3^2$. The function
$\sin(\sqrt{8}\phi_\rho(x,\tau))$ in Eq.~(\ref{Ku}) behaves at
large distances as $(a\,\omega)^{2K_\rho}$
\cite{giamarchi_book_1d} (where the cutoff $a$ is of the order of
the lattice parameter), giving for the correlator
\begin{equation}
\langle K_u;K_u\rangle^0_{\omega}=\int dx\int_0^\beta d\tau
e^{i\omega_n\tau}\langle T_\tau
K_u(x,\tau)K_u(0,0)\rangle\big{|}_{i\omega_n\to\omega+i\delta}\simeq
g_3^2\omega^{4K_\rho-2}.
\end{equation}
The factor $-2$ in the exponent comes from the integrals in $\tau$
and $x$. Finally, we obtain the following frequency dependence for
the memory function
\begin{equation}
iM(\omega)\sim g_3^2\frac{1}{\omega} \omega^{4K_\rho-2}.
\end{equation}
At high frequency ($\omega\gg T$) the expression for the
conductivity (\ref{sigma_umklapp}) must be expanded in
$\omega^{-1}$: $\sigma(\omega)\sim \chi(0)\left[1/\omega +
iM(\omega)/\omega^2+\cdots\right]$. Therefore, $\sigma(\omega)$ is
given by a power-law dependence with an non-universal exponent
depending on interactions (typical behavior of a LL):
\begin{equation}\label{conductivity_1d}
\sigma(\omega)\sim \frac{1}{g_3^2} \omega^{4K_\rho-5}.
\end{equation}
\begin{figure}
\begin{center}
\includegraphics[width=8cm]{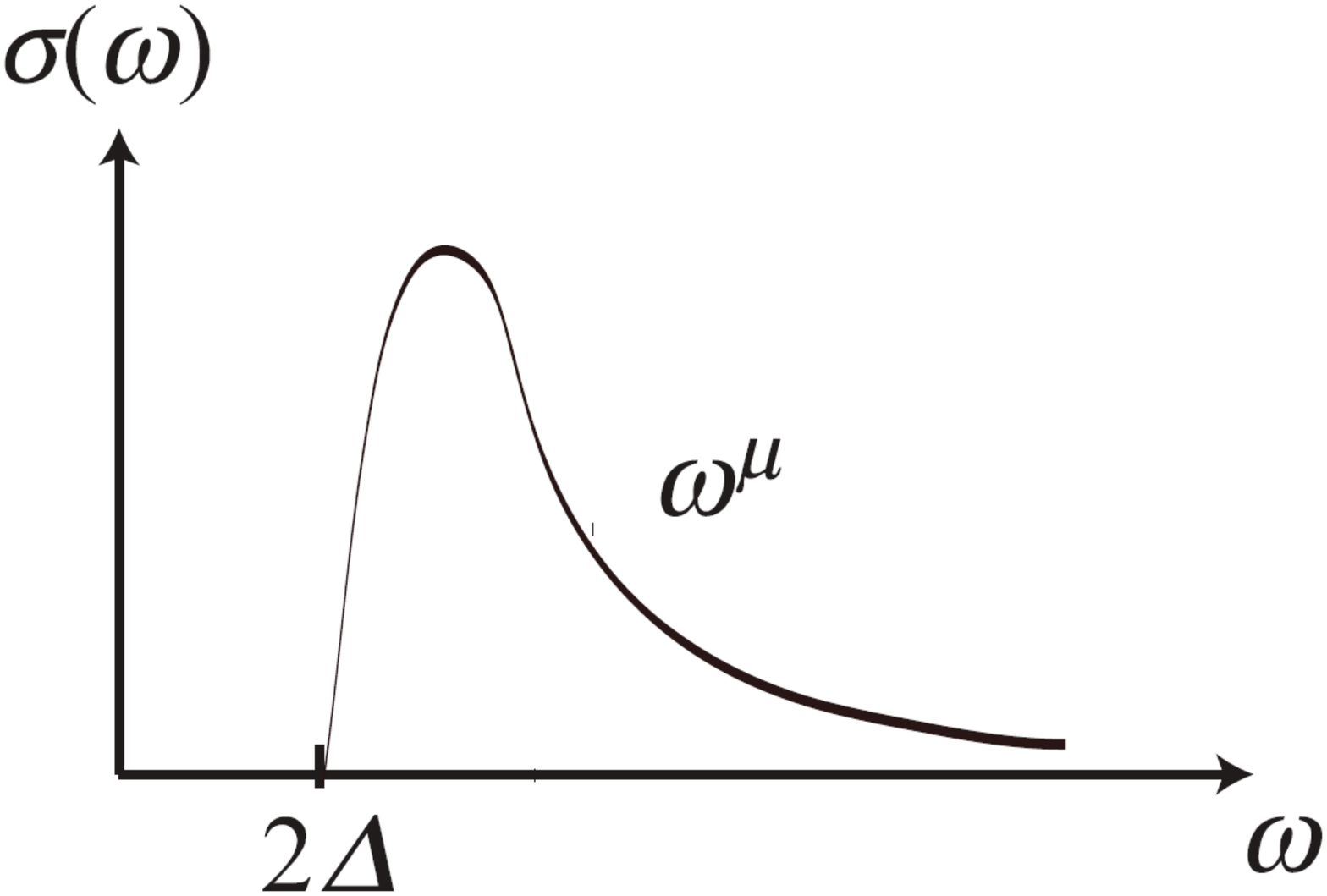}
\caption{\label{fig:cond_1d} ac conductivity in a one-dimensional
Mott insulator at commensurate fillings. At frequencies smaller
than the optical Mott gap $2\Delta$, the conductivity is zero and
at frequencies larger than $2\Delta$, it decays as a power-law
with an interaction-dependent exponent $\mu=3-4n^2K_\rho$, with
$n$ the commensurability order (From
Ref.~\cite{giamarchi_review_chemrev}).}
\end{center}
\end{figure}
Result (\ref{conductivity_1d}) is valid as long as a perturbative
expansion in $g_3$ is reasonable, {\it{i.e.}}, when the umklapp
operator is irrelevant (or marginal) \cite{giamarchi_book_1d}. In
this case the system is a perfect conductor with a regular part
given by the power-law dependence of Eq.~(\ref{conductivity_1d}):
$\text{Re}
[\sigma(\omega)]=D\delta(\omega)+\sigma_{\text{reg}}(\omega)$.
There is still a peak at $\omega=0$ (Drude peak). In the opposite
case, if the umklapp operator is relevant and leads to the opening
of a gap in the excitation spectrum, the above procedure will only
be valid at energy scales larger than the gap. For instance, in a
Mott insulator Eq.~(\ref{conductivity_1d}) is only valid for
frequencies larger than the optical Mott gap ($\omega\gg2\Delta$).
The conductivity is thus strongly affected by the Mott transition.
As mentioned before, we consider here only the case of the Mott-U
transition where the 1D Mott insulator becomes a Luttinger Liquid
(metallic phase) at a critical value of the interactions
$K_c=1/n^2$ (see Fig.~\ref{fig:mott_transition})
\cite{giamarchi_book_1d}. Fig.~\ref{fig:cond_1d} shows the zero
temperature ac conductivity in a Mott insulator. The ac
conductivity is zero for frequencies smaller than the optical Mott
gap $2\Delta$. At $\omega > 2\Delta$, the conductivity decays with
the power-law given in Eq.~(\ref{conductivity_1d}).

\begin{figure}
\begin{center}
\includegraphics[height=7cm,width=7.5cm]{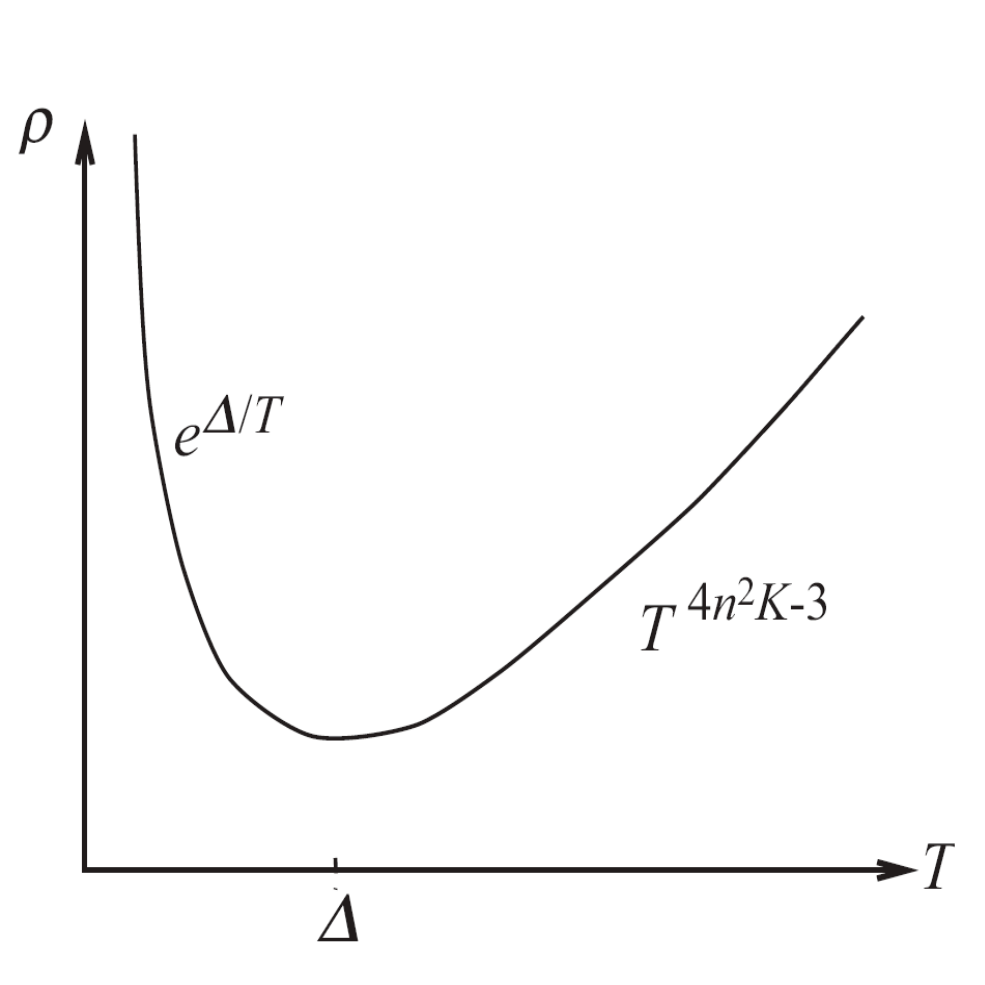}
\caption{\label{fig:resist_1d} Temperature dependence of the dc
resistivity in a one-dimensional Mott insulator with
commensurability $n$. At temperatures below the Mott gap $\Delta$,
the number of carriers is exponentially small giving an
exponential increase in resistivity. At $T>\Delta$ the dc
resistivity shows a power-law behavior typical of a Luttinger
Liquid (From Ref.~\cite{giamarchi_review_chemrev}).}
\end{center}
\end{figure}

The temperature dependence (at $\omega\ll T$) or dc conductivity
is obtained by the same scaling method, giving $\sigma(T)\sim
T^{3-4K_\rho}/g_3^2$. Fig.~\ref{fig:resist_1d} shows the
temperature dependence of the dc resistivity, which is related to
the conductivity as $\rho=1/\sigma$, in a 1D Mott insulator. At
temperatures smaller than the Mott gap $\Delta$, the resistivity
increases exponentially, and for temperatures larger than $\Delta$
there is again a power-law behavior
\begin{equation}
\rho(T)\sim T^{4K_\rho-3}.
\end{equation}
At the Mott transition, $K_\rho=K_c=1/n^2$, the ac conductivity
and dc resistivity lead back to universal exponents:
$\sigma(\omega)\sim 1/(\omega \ln(\omega)^2)$ and $\rho(T)\sim
T/\ln(1/T)^2$, respectively \cite{giamarchi_book_1d}.

Now that we have studied the transport properties of
one-dimensional systems, especially those of 1D Mott insulators,
we acquired the necessary tools to investigate the conductivity of
quasi 1D systems. These transport properties are essential for the
study of the Hall effect in weakly coupled Luttinger liquids in
Chapter~\ref{chap:Hall_quasi1d}.

\newpage

\subsection{Transport in quasi one-dimensional systems}\label{sec:transport_quasi1D}

In quasi one-dimensional systems, transport properties change from
the pure 1D case. In addition to the conductivity along the
chains, which remains essentially the same as studied before,
there is also a transverse conductivity $\sigma_{\perp}$. This
transverse conductivity is obtained at high temperatures or
frequencies by making a perturbative expansion in the interchain
coupling term, or by using a mean field approach
\cite{lopatin_q1d_magnetooptical,georges_organics_dinfiplusone}.
In this section we will use again a scaling analysis to obtain the
frequency (temperature) dependence of the transverse conductivity
$\sigma_{\perp}$ and we refer the reader to the literature
\cite{lopatin_q1d_magnetooptical,georges_organics_dinfiplusone}
for the complete expressions.

In order to obtain $\sigma_{\perp}$ we use again the Kubo formula
(\ref{Kubo}). Let us consider a system composed of weakly coupled
1D chains described by Hamiltonian of the form:
$\mathcal{H}=\mathcal{H}_{\text{1D}}+\mathcal{H}_\perp$ (see
Eqs.~(\ref{H_1d_continium}) and (\ref{hopping_term})),
\begin{eqnarray}\nonumber
         \mathcal{H}&=&\int dx\sum_{j\sigma}\left[v_{\text{F}}
        \psi_{j\sigma}^{\dagger}(x)\tau_3(-i\partial_x)
        \psi_{j\sigma}^{\phantom{\dagger}}(x)
        +g_2\,\psi_{j\sigma R}^{\dagger}(x)
        \psi_{j\sigma R}^{\phantom{\dagger}}(x)
        \psi_{j\sigma L}^{\dagger}(x)
        \psi_{j\sigma L}^{\phantom{\dagger}}(x) \right.\\ \label{Hamiltonian_quasi1D}
    &&\left. -t_{\perp}\left(\psi_{j\sigma}^{\dagger}(x)
        \psi_{j+1\sigma}^{\phantom{\dagger}}(x)+\text{h.c.}\right)\right],
\end{eqnarray}
where $j$ is the chain index. We suppose a system with spin
rotation symmetry $g_{1\perp}=g_{1\parallel}=0$ ($K_{\sigma}=1$).
We will keep the fermionic representation (to maintain the same
representation of the coupling term $\mathcal{H}_\perp$), knowing
that the terms corresponding to the 1D chains (first and second
terms in Eq.~(\ref{Hamiltonian_quasi1D})) are easily bosonized in
the form (~\ref{H_1D}), as explained in Sec.~\ref{sec:1D_case}.
With this Hamiltonian we can easily calculate the diamagnetic term
in the transverse direction using definition (\ref{diamagnetic}),
\begin{equation}\label{diamagnetic_quasi1D}
\chi_{\perp}(0)=-2e^2t_{\perp}a_y^2\!\int dx
        \left[\langle\psi_{0\uparrow}^{\dagger}(x)\psi_{1\uparrow}(x)\rangle+\text{h.c.}\right],
\end{equation}
with $a_y$ the lattice parameter in the transverse direction. The
thermodynamical average in Eq.~(\ref{diamagnetic_quasi1D}) must be
obtained to first order in the coupling term (first order in
$t_\perp<<1$), using standard perturbation theory
\cite{mahan_book}. For this we must expand the action in the
thermodynamical average to first order in $t_\perp$
\begin{equation}\label{t_perturbation}
\langle\psi_{0\uparrow}^{\dagger}(x)\psi_{1\uparrow}(x)\rangle=
-t_\perp\int_0^\beta d\tau_1\sum_j \delta_{j,0}
\langle\psi_{0\uparrow}^{\dagger}(x)\psi_{j\uparrow}(x,\tau_1)\rangle
\langle\psi_{j+1\uparrow}^{\dagger}(x,\tau_1)\psi_{1\uparrow}(x)\rangle
\end{equation}
The thermodynamical averages on the right hand side of
Eq.~(\ref{t_perturbation}) are taken with respect to the 1D part
of the Hamiltonian. The quantity
$\langle\psi_{\uparrow}^{\dagger}(x)\psi_{\uparrow}(x)\rangle$ is
just the equal-time Green's function of the 1D chains, who's
Fourier transform gives the occupation factor $n(k)\sim
\text{max}[\delta_k,T]^{\eta+1}$, with
$\eta=\frac{1}{4}(K_{\rho}+K_{\rho}^{-1})-\frac{1}{2}$, (see
Eq.~\ref{occupation_factor}). Using simple scaling arguments (we
must add a $T^{-2}$ factor to account for the $x$ and $\tau_1$
integrals), we find the temperature dependence of the transverse
diamagnetic term
\begin{equation}\label{chi_perp}
\chi_{\perp}(0)\sim t_\perp^2\, T^{-2} T^{2\eta+2}=t_\perp^2\, T^{-1+\frac{1}{2}(K_{\rho}+K_{\rho}^{-1})}.
\end{equation}
Note that the scaling analysis does not give the correct dimensions and
non-interacting limit for $\chi_\perp(0)$.
The detailed calculation gives $\sim t_\perp^2 \frac{T^{2\eta}}{\eta}$ \cite{lopatin_q1d_magnetooptical},
but we are interested here in the power-law dependence and for this the scaling analysis is sufficient. 
Result (\ref{chi_perp}) will be used in Chapter~\ref{chap:Hall_quasi1d} to
study the temperature dependence of the Hall coefficient.

To obtain the conductivity $\sigma_{\perp}$ we need the
current-current correlation function $\chi_{\perp}(\omega)$
defined in Sec.{\ref{sec:Kubo}}. The current along the transverse
direction is obtained using definition (\ref{X_omega}) and
Hamiltonian (\ref{Hamiltonian_quasi1D}),
\begin{equation}
J_\perp=-iet_{\perp}a_y\int dx\sum_{j\sigma}
        \left(\psi_{j\sigma}^{\dagger}(x)\psi_{j+1\sigma}^{\phantom{\dagger}}(x)-\text{h.c.}\right).
\end{equation}
Thus, the current-current correlation function in frequency is
given by (see definition (\ref{X_omega}))
\begin{equation}
\chi_{\perp}(\omega)=\left[-\int_{0}^{\beta}d\tau e^{i\omega_n\tau}\langle T_{\tau}J_{\perp}^{\dagger}(\tau)J_{\perp}(0)\rangle\right]_{i\omega_n\to\omega+i\delta}.
\end{equation}
Because each current operator contributes with a $t_\perp$, this
term is already of second order in this parameter ($t_\perp<<1$)
and thus the averages can be taken considering only the 1D part of
the Hamiltonian. Taking $\tau>0$, we have
\begin{eqnarray}\nonumber
\langle T_{\tau}J_{\perp}^{\dagger}(\tau)J_{\perp}(0)\rangle&=&-e^2t_{\perp}^2a_y^2\int dx\langle
\langle \psi_{j+1\sigma}^{\dagger}(x,\tau)\psi_{j+1\sigma}^{\phantom{\dagger}}(0,0)\rangle
\langle\psi_{j\sigma}^{\dagger}(0,0)\psi_{j\sigma}^{\phantom{\dagger}}(x,\tau)\rangle \\
&=& -e^2t_{\perp}^2a_y^2\int dx \sum_{r=R,L}
G_{r\sigma}(x,\tau)\left[1-G_{r\sigma}(x,\tau)\right],
\end{eqnarray}
with $G(x,\tau)_{r\sigma}\sim \text{max}[\omega,T]^{\eta+1}$, with
$\eta=\frac{1}{4}(K_{\rho}+K_{\rho}^{-1})-\frac{1}{2}$, the
Green's function of a one-dimensional system, with spin rotational
symmetry ($K_{\sigma}=1$). Now we can apply scaling arguments to
get the frequency (temperature) dependence of the current-current
correlation function
\begin{equation}
\chi_{\perp}(\omega,T)\sim \text{max}[\omega,T]^{2\eta}.
\end{equation}
Using the Kubo formula for the transverse conductivity $\sigma_{\perp}=[\chi_{\perp}(0)-\chi_{\perp}(\omega)]/i\omega$,
the result is, as for the longitudinal conductivity, a power law
\begin{eqnarray}\nonumber
\sigma_{\perp}(T\gg\omega,E^*)&\sim& T^{2\eta-1}\\
\sigma_{\perp}(\omega\gg T,E^*)&\sim& \omega^{2\eta-1}.
\end{eqnarray}
$E^{*}$ is the energy scale of the dimensional crossover: at
$T,\omega < E^*$ the system is in a high-dimensional metallic
state, whereas at $T,\omega > E^*$ the system is in a Luttinger
liquid state. A full expression of $\sigma_{\perp}$ can be found
in Ref.~\cite{lopatin_q1d_magnetooptical}. At this stage we have a
general knowledge of the transport properties in low-dimensional
system and, in the next chapter, we will take the challenge of
studying the Hall effect in these systems.

\chapter{The Hall effect}\label{chap:Hall_effect}




\section{The classical Hall effect}

\begin{figure}
\begin{center}
\includegraphics[width=8cm]{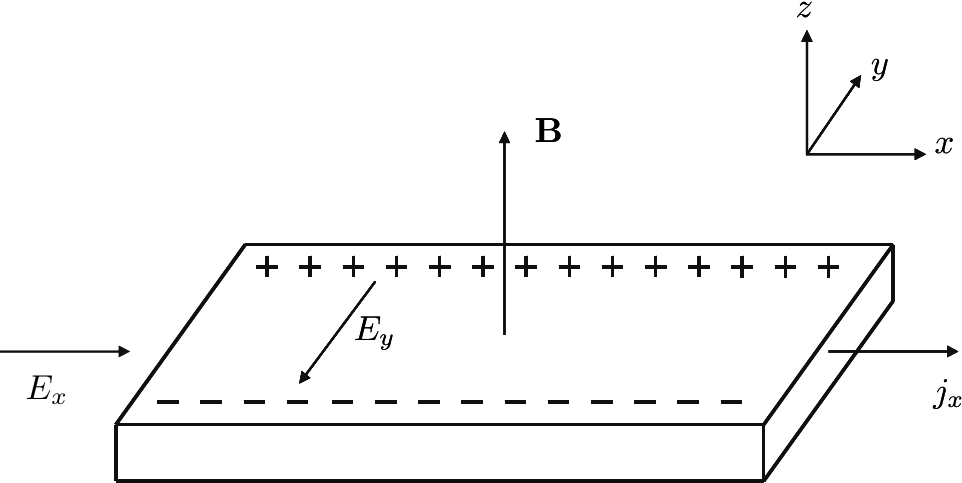}
\caption{\label{fig:Hall} Hall's experiment: an electric field
$E_x$ is applied in a metallic sample along the $x$-axis and a
magnetic field $\vec{B}$ is applied along the $z$-axis. The
electric field drives a current density $j_x$ along the
$x$-direction and the magnetic field deflects the electrons in the
negative $y$-direction with a Lorentz force equal to
$-(|e|/c)\vec{v}\mathbf{\times}\vec{B}$ ($\vec{v}$ is the charge's
velocity). The electrons accumulate on the sides creating an
electric field $E_y$ in the $y$-direction. At equilibrium, this
transverse electric field balances the Lorentz force.}
\end{center}
\end{figure}

The Hall effect was found by E. H. Hall in 1879 using an
experimental setup similar to the one shown in
Fig.~\ref{fig:Hall}. An electric field $E_x$ is applied along a
wire extending along the $x$-axis and a magnetic field $\vec{B}$
is applied along the $z$-axis. The electric field drives a current
density $j_x$ along the $x$-direction and the magnetic field
deflects the electrons in the negative $y$-direction with a
{\it{Lorentz}} force equal to
\begin{equation}
\vec{F}=-\frac{|e|}{c}\vec{v}\mathbf{\times}\vec{B}\, ,
\end{equation}
where $|e|$ is the electronic charge and $\vec{v}$ is the average
velocity of the charges. Because current cannot flow along the
$y$-axis, the charges accumulate on the sides of the wire. As they
accumulate, an electric field builds up in the $y$-direction that
opposes the motion of the charges and their further accumulation.
At equilibrium, this transverse field $E_y$ balances the Lorentz
force.
The Hall resistivity $\rho_{yx}$ is defined as the ratio between
the transverse electric field and the current along the wire
\cite{mahan_book}
\begin{equation}
\rho_{yx}=\frac{E_y}{j_x}.
\end{equation}
The Hall coefficient $\RH$ (or Hall constant) is just the Hall
resistivity divided by the magnetic field,
\begin{equation}
\RH=\frac{\rho_{yx}}{B}=\frac{E_y}{j_xB}.
\end{equation}
Assuming that the motion of the electrons is classical, their
equation of motion reads
\begin{equation}
m\vec{\dot{v}}=-|e|\left[\vec{E}+\frac{\vec{v}\mathbf{\times}\vec{B}}{c}
\right]-\frac{m\vec{v}}{\tau}
\end{equation} where $\tau$ is the
relaxation time for scattering, that was defined in the
description of the Drude model, at the beginning of the previous
chapter. In the steady state the time derivatives are zero and we
have
\begin{eqnarray}\label{Hall_eq_motion_1}
0&=&-|e|\left[E_x+\frac{v_yB}{c}\right]-\frac{mv_x}{\tau},\\ \label{Hall_eq_motion_2}
0&=&-|e|\left[E_y-\frac{v_xB}{c}\right]-\frac{mv_y}{\tau}.
\end{eqnarray}
The current density along the $x$-axis is given by $j_x=-|e|nv_x$.
Furthermore, the current density along the $y$-direction is
$v_y=0$. Then, multiplying Eqs.~(\ref{Hall_eq_motion_1}) and
(\ref{Hall_eq_motion_2}) by $n|e|\tau/m$ we obtain
\begin{equation}
\sigma E_x= j_x, \qquad E_y= -\frac{j_xB}{n|e|c}
\end{equation}
where $\sigma=n|e|^2\tau/m$ is the Drude model dc conductivity in
the absence of magnetic field (see Eq.~(\ref{Drude_formula})).
Then, the Hall coefficient results
\begin{equation}\label{Hall_constant}
\RH=-\frac{1}{n|e|c}.
\end{equation}
As can be seen in Eq.~(\ref{Hall_constant}) the Hall constant does
not depend on the parameters of the metal or the dimensions of the
sample, except for the density of carrier $n$. As a result, $\RH$
gives a direct measurement of the number and sign of the carriers
in the metal.

There are additional phenomena, related to the Hall effect, known
as the quantum Hall effect and the fractional quantum Hall effect
that occur in highly correlated electron systems where electrons
are allowed to move only in a plane and a strong magnetic field is
applied \cite{mahan_book}. In this work we study the regime where
the applied magnetic field $\vec{B}$ is supposed weak enough for
the linear response theory to be valid and the quantum effects do
not appear.

The Hall constant result obtained in the classical approximation
assumes a system where electrons move as free particles.
Therefore, it can only be used in metals where the free-electron
model works well. In order to understand the Hall effect in
correlated systems, we must take into account interactions, which
is generally a complicated task. The next sections are devoted to
the investigation of two different approaches for the problem of
the Hall effect in the presence of interactions. The first
approach consist on a high-frequency expansion where the Hall
coefficient is obtain in the limit of infinite-frequency. And the
second approach is the obtention of $\RH$ using the memory
function formalism studied in Chapter~\ref{chap:transport}.

\section{The infinite-frequency Hall constant}\label{high_frequency_RH}



The conductivity was expressed, in the previous chapter, by the
tensor $\sigma_{\mu\nu}$
\begin{equation}
\bm{\sigma}=\left(\begin{matrix}\sigma_{xx}&\sigma_{xy}\\\sigma_{yx}&\sigma_{yy}\end{matrix}\right).
\end{equation}
The Hall resistivity $\rho_{yx}=\sigma_{yx}^{-1}$, written in
terms of the conductivity tensor, is equal to
\begin{equation}\label{Hall_resistivity}
\rho_{yx}=\frac{\sigma_{xy}}{\sigma_{xx}\sigma_{yy}+\sigma_{xy}\sigma_{yx}}.
\end{equation}
To obtain a result for $\rho_{yx}$ by means of
Eq.~(\ref{Hall_resistivity}) when interactions are present in the
system is generally not feasible, thus some approximations must be
made. In the following, we develop a high-frequency expansion of
the conductivity, in order to obtain a simpler expression for the
ac Hall constant, as was proposed by Shastry {\it{et al.}} in
Ref.~\cite{Shastry_Hall}. Let us write the conductivities using
the Kubo formula Eq.~(\ref{Kubo}). The longitudinal conductivity
along the $x$-direction is
\begin{equation}
\sigma_{xx}(\omega)=\frac{1}{i\omega}\left[\chi_x(0)-\chi_{xx}(\omega)\right],
\end{equation}
where $\chi_{xx}(\omega)$ is given by Eq.~(\ref{susceptibility})
at $\vec{q}=0$. Without lost of generality we can take $t'=0$ in
Eq.~(\ref{susceptibility}). Integrating by parts, we have
\begin{equation}\label{int_by_parts}
\chi_{xx}(\omega)=-\frac{e^{i\omega t}}{\omega}\left\langle [J_x
(t),J_x (0)]\right\rangle\Big{|}_0^\infty +i\int_{0}^{\infty}dt
\frac{e^{i\omega t}}{\omega}\left\langle \left[[\mathcal{H},J_x
(t)],J_x (0)\right]\right\rangle.
\end{equation}
We have used the Heisenberg representation for the current
operator $J(t)=e^{i\mathcal{H}t}J(0)e^{-i\mathcal{H}t}$, with
$\mathcal{H}$ the total Hamiltonian of the system. The first term
in (\ref{int_by_parts}) vanishes because the currents commute at
$t=0$ and, as we said at the beginning of Sec.~\ref{Kubo}, the
perturbation (here the electric field) is assumed to vanish when
$t\to\pm\infty$, implying the vanishing of the current.
Integrating by parts again, the second term on the right side of
Eq.~(\ref{int_by_parts}) becomes
\begin{equation}
\chi_{xx}(\omega)=-\frac{e^{i\omega t}}{\omega^2}\left\langle
\left[[J_x (t),\mathcal{H}],J_x
(0)\right]\right\rangle\Big{|}_{0}^{\infty} +i\int_{0}^{\infty}dt
\frac{e^{i\omega t}}{\omega^2}\left\langle \left[[\mathcal{H},[J_x
(t),\mathcal{H}]],J_x (0)\right]\right\rangle.
\end{equation}
As before, the current vanishes at $t=\infty$ and therefore only
the $t=0$ contribution remains in the first term. Now, we can
write the high-frequency expansion of $\sigma_{xx}(\omega)$ as
\begin{equation}
\sigma_{xx}(\omega)=\frac{1}{i\omega}\left[\chi_x(0)+\frac{\langle [[J_x,\mathcal{H}],J_x]\rangle}{\omega^2}+\mathcal{O}\left(1/\omega^3\right)\right].
\end{equation}
The longitudinal conductivity in the $y$-direction,
$\sigma_{yy}(\omega)$, has the same high-frequency expansion with
$\chi_{x}(0)$ and $J_{x}$ replaced by $\chi_{y}(0)$ and $J_{y}$,
respectively. From the Kubo formula we have for the transverse
conductivity
\begin{equation}
\sigma_{xy}(\omega)=\frac{i}{\omega}\chi_{xy}(\omega).
\end{equation}
Integrating by parts the definition of $\chi_{xy}(\omega)$, as
before, the following high-frequency expansion results
\begin{equation}
\sigma_{xy}(\omega)=\frac{i}{\omega}\left[\frac{\langle[J_x,J_y]\rangle}{\omega}+\frac{\langle[[[J_x,\mathcal{H}],\mathcal{H}]J_y]\rangle}{\omega^3}+\mathcal{O}\left(1/\omega^5\right)\right].
\end{equation}
We can now insert the frequency expansion of the conductivities in
Eq.~(\ref{Hall_resistivity}). Since the conductivity $\sigma_{xy}$
is proportional to the magnetic field, the factor
$\sigma_{xy}\sigma_{yx}$ in the denominator of
Eq.~(\ref{Hall_resistivity}) is of second order in $\vec{B}$ and
thus can be neglected. Keeping only the linear term in $\vec{B}$
and zero-order term in $\omega$, we obtain for the Hall constant
\begin{equation}\label{inf_freq_RH}
\RH(\omega\to\infty)=-\frac{i}{\vec{B}}\frac{\langle[J_x,J_y]\rangle}{\chi_x(0)\chi_y(0)}.
\end{equation}
This expression for $\RH$ is known as the infinite-frequency Hall
constant \cite{Shastry_Hall}. We will use it in
Chapter~\ref{chap:Hall_quasi1d} to calculate the high-frequency
Hall coefficient on a triangular lattice. For this high-frequency
expansion to be valid in an experimental measurement, it requires
a probe frequency larger than any other energy scale in the
system, and it can be measured using the Faraday rotation
experiment.

Although Eq.~(\ref{inf_freq_RH}) seems to be a much simpler
expression for the Hall coefficient when comparing with
(\ref{Hall_resistivity}), we must remember that when interactions
are present in the system, they must be taking into account when
calculating the thermodynamical average $\langle\cdots\rangle$.
This is usually not a simple task. If interactions are small, they
can be treated using perturbation theory; if not, other formulas
to compute $\RH$ must be used. In the next section, we will use
the memory function formalism, studied in
Sec.~\ref{sec:memory_formalism}, to obtain an expression for the
Hall constant at leading order in the interaction term.


\section{The Hall constant in the memory function approach}\label{sec:Hall_memory}

As in the previous section, our main goal is to calculate the Hall
resistivity $\rho_{yx}$, and  consequently $\RH$, in terms of the
conductivity tensor $\sigma_{\mu\nu}$ using
Eq.~(\ref{Hall_resistivity}). However, in this section we ought to
obtain an expression for the Hall constant where interactions can
be included directly in the calculations, going beyond the
infinite-frequency limit. For this we use of the memory matrix
formalism. With the definitions given in
Eqs.~(\ref{sigma_M})-(\ref{Memory_matrix__definition}) we arrived
at an expression for the off-diagonal memory matrix element in
terms of the conductivities
\begin{equation}
        iM_{xy}(\omega)=\frac{i\chi_y(0)\sigma_{xy}(\omega)}
        {\sigma_{xx}(\omega)\sigma_{yy}(\omega)+\sigma_{xy}^2(\omega)}
        -\Omega_{xy}=i\chi_y(0)\rho_{xy}(\omega)-\Omega_{xy}.
   \end{equation}
From this expression it is straightforward to rewrite the Hall
coefficient, $\RH=\rho_{xy}/\vec{B}$, as
\begin{equation}\label{RH_memory}
\RH(\omega)=\frac{1}{i\chi_y(0)}\lim_{B\to0}
\frac{\Omega_{xy}+iM_{xy}(\omega)}{B}.
\end{equation}
Since the memory matrix vanishes as $\omega^{-2}$ at high
frequency, we see from Eq.~(\ref{Omega}) that the
infinite-frequency Hall constant, $\RH(\omega\to\infty)$, is
indeed given by Eq.~(\ref{inf_freq_RH}). The result
(\ref{RH_memory}) will be used to compute the Hall constant in a
strongly correlated quasi one-dimensional system in
Chapter~\ref{chap:Hall_quasi1d}.

Let us now return to the memory matrix element necessary to calculate $\RH$, given in Eq.~(\ref{Memory_susceptibilies}),
\begin{equation}
iM_{xy}(\omega)=\frac{\omega\chi_y(0)\chi_{xy}(\omega)}
        {\left[\chi_x(0)-\chi_{xx}(\omega)\right]
        \left[\chi_y(0)-\chi_{yy}(\omega)\right]}-\Omega_{xy}.
\end{equation}
This expression will be expanded at high frequencies to obtain the
result of $iM_{xy}(\omega)$ given in Eq.~(\ref{MofK}). At high
enough frequencies, where the quantity
$|\chi_{\mu\mu}(\omega)/\chi_{\mu}(0)|$ can be considered small,
the off-diagonal memory matrix term can be written as
\begin{equation}
iM_{xy}(\omega)\simeq\frac{\omega\chi_{xy}(\omega)}
        {\chi_x(0)}\left[1+\frac{\chi_{xx}(\omega)}{\chi_x(0)}+
\frac{\chi_{yy}(\omega)}{\chi_y(0)}\right]-\Omega_{xy}
\end{equation}
Owing to the equations of motion given in Eq.~(\ref{eq_motion}),
for the susceptibility $\chi_{xy}(\omega)$, the above expression
becomes
\begin{equation}
iM_{xy}(\omega)\simeq\left[\frac{\langle[J_x,J_y]\rangle}{\chi_x(0)}
-\frac{\langle[\mathcal{H},J_x];J_y\rangle}{\chi_x(0)}\right]
\left[1+\frac{\chi_{xx}(\omega)}{\chi_x(0)}+\frac{\chi_{yy}(\omega)}{\chi_y(0)}\right]
-\frac{\langle[J_x,J_y]\rangle}{\chi_x(0)}.
\end{equation}
The two terms involving $[J_{x},J_{y}]$ cancel. In the following
we will write the Hamiltonian as
$\mathcal{H}=\mathcal{H}_0+\mathcal{H}_{\text{int}}$,
\begin{eqnarray}\label{M_approx_1}
&&iM_{xy}(\omega)\simeq\Omega_{xy}\frac{\chi_{xx}(\omega)}{\chi_x(0)}+\Omega_{xy}\frac{\chi_{yy}(\omega)}{\chi_y(0)}\\
\nonumber
&&-\left[\frac{\langle[\mathcal{H}_0,J_x];J_y\rangle}{\chi_x(0)}
+\frac{\langle[\mathcal{H}_{\text{int}},J_x];J_y\rangle}{\chi_x(0)}\right]
\left[1+\frac{\chi_{xx}(\omega)}{\chi_x(0)}+\frac{\chi_{yy}(\omega)}{\chi_y(0)}\right].
\end{eqnarray}
Because there is always a magnetic field present in the Hall
experiment, the non-interacting part of the Hamiltonian
$\mathcal{H}_0$ does not commute with the currents. However,
following Mori's formalism \cite{mori_theory_book}, the commutator
$[\mathcal{H}_0,J_{\mu}]$ can be expressed as
\cite{lange_hall_constant,gotze_fonction_memoire}
\begin{equation}
[\mathcal{H}_0,J_{\mu}]=-J_{\nu}\Omega^0_{\nu\mu}
\end{equation}
with $\mu,\nu=x,y$ and summation over repeated indices is implied.
The superscript $0$ means that the average $\langle\cdots\rangle$
must be computed using the non-interacting Hamiltonian
$\mathcal{H}_0$. With this expression as well as the symmetry
property $\chi_x(0)\Omega_{xy}=-\chi_y(0)\Omega_{yx}$,
Eq.~(\ref{M_approx_1}) can be rewritten in the form
\begin{eqnarray}
&&iM_{xy}(\omega)\simeq\Omega_{xy}\frac{\chi_{xx}(\omega)}{\chi_x(0)}+\Omega_{xy}\frac{\chi_{yy}(\omega)}{\chi_y(0)}\\
\nonumber &&\left[-\Omega_{xy}\frac{\langle
J_y;J_y\rangle}{\chi_y(0)}
-\frac{\langle[\mathcal{H}_{\text{int}},J_x];J_y\rangle}{\chi_x(0)}\right]\left[1+\frac{\chi_{xx}(\omega)}{\chi_x(0)}+\frac{\chi_{yy}(\omega)}{\chi_y(0)}\right].
\end{eqnarray}
The second term in the right hand side is cancelled by the first
term in the square brackets. Now let us keep only the leading
terms in $|\chi_{\mu\mu}(\omega)/\chi_{\mu}(0)|$ (the correlator
$\langle[\mathcal{H}_{\text{int}},J_x];J_y\rangle$ is of the order
of $\chi_{\mu\mu}$),
\begin{equation}
iM_{xy}(\omega)\simeq-\frac{\langle[\mathcal{H}_{\text{int}},J_x];J_y\rangle} {\chi_x(0)}+\Omega_{xy}\frac{\chi_{xx}(\omega)}{\chi_x(0)}.
\end{equation}
As before, we use the equation of motion for the correlator
$\langle[\mathcal{H}_{\text{int}},J_x];J_y\rangle$ and for the
susceptibility $\chi_{xx}(\omega)$. Introducing the residual force
operator $K_x=[\mathcal{H}_{\text{int}},J_x]$, we have
\begin{equation}\nonumber
iM_{xy}(\omega)\simeq -\frac{\left(\langle[K_x,J_y]\rangle+\langle
K_x,[\mathcal{H},J_y]\rangle\right)}{\omega\chi_x(0)}
-\Omega_{xy}\frac{\langle
[\mathcal{H},J_x];J_x\rangle}{\omega\chi_x(0)}
\end{equation}
\begin{equation}\nonumber
=-\frac{\left(\langle[K_x,J_y]\rangle-\langle
K_x,J_x\rangle\Omega_{xy} +\langle
K_x,K_y\rangle\right)}{\omega\chi_x(0)}
-\Omega_{xy}\Omega_{yx}\frac{\langle
J_y;J_x\rangle}{\omega\chi_x(0)\chi_{x}(0)}
-\Omega_{xy}\frac{\langle K_x;J_x\rangle}{\omega\chi_x(0)}.
\end{equation}
The second term and last term in the right hand side cancel. Each
frequency matrix $\Omega_{\mu\nu}$ is of first order in the
magnetic field because the commutator $[J_x,J_y]$ is proportional
to $\vec{B}$. Thus, the fourth term above is of second order in
the magnetic field. Then, at high frequency, we end up with two
terms for the off-diagonal memory matrix element at first order in
$\vec{B}$
\begin{equation}\label{Memory_2_terms}
iM_{xy}(\omega)\simeq-\frac{\langle K_x,K_y\rangle}{\omega\chi_x(0)} -\frac{\langle[K_x,J_y]\rangle}{\omega\chi_x(0)}.
\end{equation}
The second term in Eq.~(\ref{Memory_2_terms}) is of first order in
the interactions parameters. However, it can be proved that the
leading terms of the memory matrix are of second order in the
interaction parameters, as obtained in
Sec.~\ref{sec:transport_low_dim} for a 1D system. In order to see
this, let us take a system where electrons interact via a Hubbard
interaction
$U\sum_{i}c^{\dagger}_{i\uparrow}c^{\phantom{\dagger}}_{i\downarrow}$.
The correlator $\langle[K_x,J_y]\rangle$ correspond to a
first-order expression of the frequency matrix
\cite{mori_theory_book}. The frequency matrix is easily traced
back to the number operator $n_{i\sigma}=\langle
c^{\dagger}_{i\sigma}c^{\phantom{\dagger}}_{i\sigma} \rangle$,
whose first-order contribution in $U$ vanishes. In consequence,
the only remaining term in Eq.~(\ref{Memory_2_terms}) is of
second-order in the interaction parameters,
\begin{equation}\label{Memory_final}
iM_{xy}(\omega)\simeq-\frac{\langle K_x,K_y\rangle}{\omega\chi_x(0)}.
\end{equation}
This is the expression that will be used in the following chapter
to compute the memory matrix contribution to the Hall constant
$\RH$, by means of Eq.~(\ref{RH_memory}). Because
Eq.~(\ref{Memory_final}) is already given at second order in the
scattering parameters, the thermodynamical average of $K$
operators can be computed with the free Hamiltonian, {\it{i.e.}},
putting interaction terms to zero. This facilitates enormously the
calculations, because these averages are usually straightforward
to compute without interaction. Even more in one-dimensional
systems, as explained in the preceding chapters.

At this point we have reviewed all the theoretical tools necessary
for our study of the Hall effect in two different models of
strongly correlated systems. In the next chapters, we will see how
all the previously studied formulas can be applied to an specific
theoretical model or to Hall measurements made in real
experimental compounds.

\chapter{Hall effect in strongly correlated quasi 1D
systems}\label{chap:Hall_quasi1d}




In the last decades, various experimental realizations of
low-dimensional systems have been achieved. Realizations as the
organic conductors \cite{jerome_review_chemrev}, carbon nanotubes
\cite{dresselhaus_book_fullerenes_nanotubes}, ultra cold atomic
gases \cite{stoferle_tonks_optical}, quantum wires
\cite{Bassler_quantum_wires}, quantum dots
\cite{Julien_quandum_dots,McEuen_quandum_dots} and others, have
largely stimulated the research in low-dimensional physics. Among
these, the organic conductors have become the model systems for
the study of quasi one-dimensional physics due to their highly
anisotropic molecular structures. They have been extensively
studied for more than twenty years now, since the discovery of a
superconducting state in their phase diagram
\cite{jerome_review_chemrev}. We will start this section with a
short review on the physical properties of these organic
conductors and some experimental facts that motivated the
theoretical study of the Hall effect in these strongly correlated
quasi 1D systems.

\section{Quasi one-dimensional organic conductors}
\begin{figure}
\begin{center}
\includegraphics[height=6cm,width=12cm]{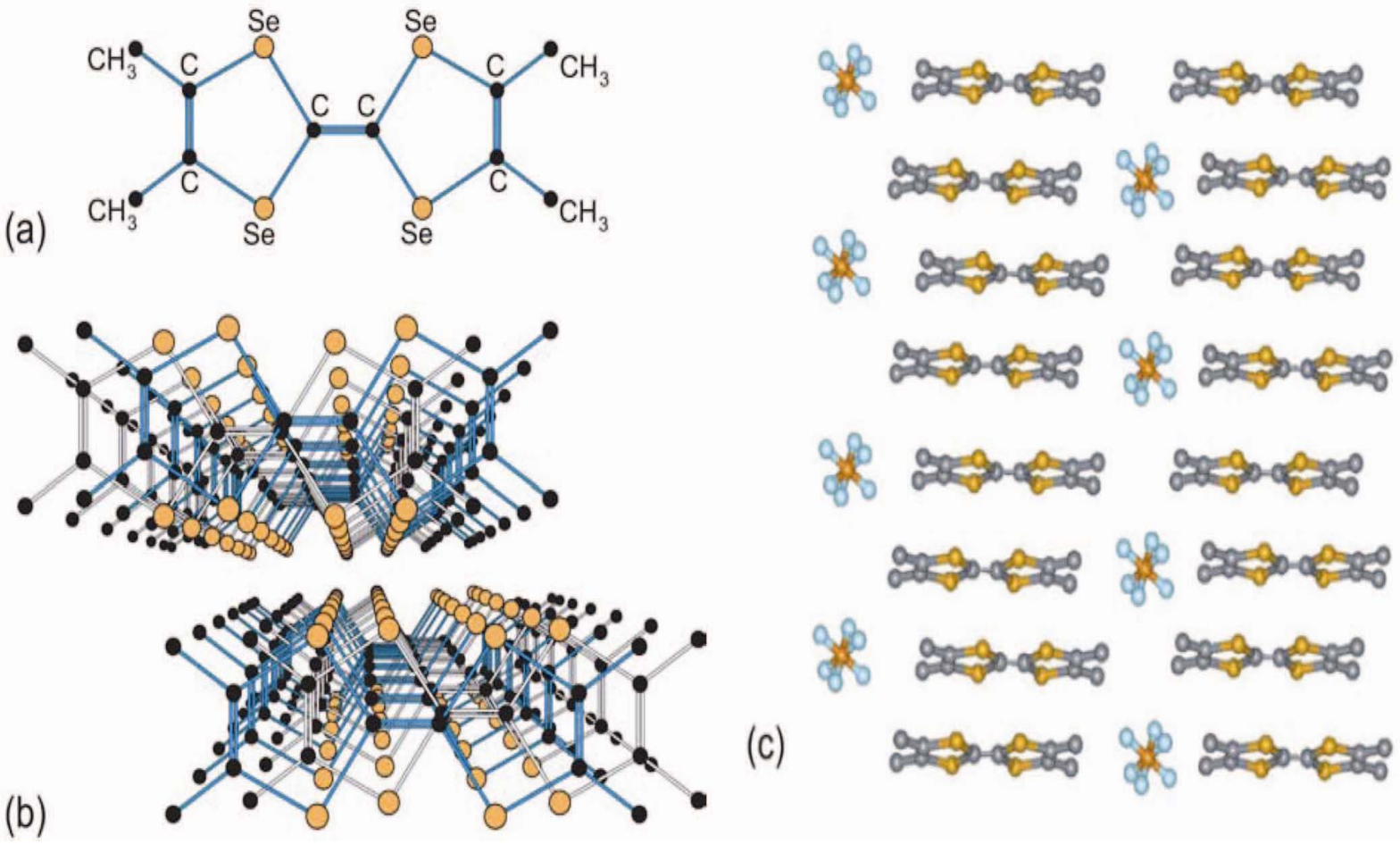}
\caption{\label{fig:Organic_structure} a) TMTSF molecule; b) view
along the stacks, $a$ direction; c) Along the $c$ direction the
stacks are separated by the X anions (PF$_6^{-}$ in the picture).
}
\end{center}
\end{figure}
\begin{figure}
\begin{center}
\includegraphics[width=8cm]{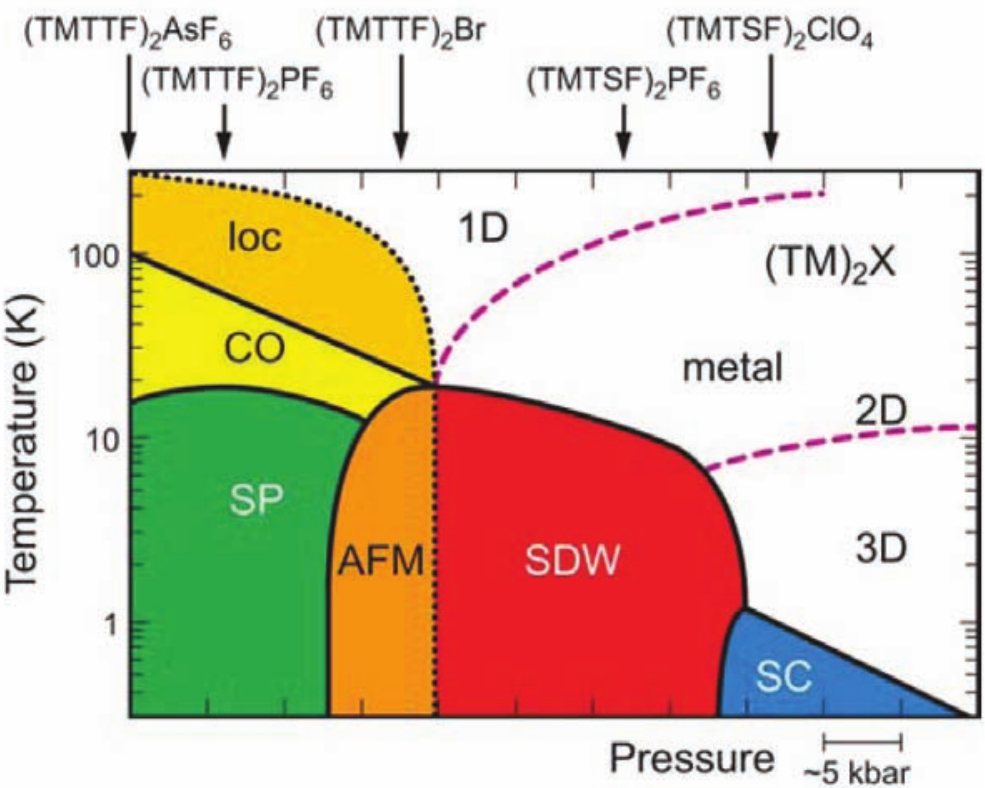}
\caption{\label{fig:Phase_organics} Phase diagram for the
(TM)$_2$X compounds. For the different compounds the
ambient-pressure position is indicated. Here loc stands for charge
localization, CO for charge ordering, SP for spin-Peierls, AFM for
antiferromagnet, SDW for spin density wave, and SC for
superconductor. The clear phase transitions are indicated with
solid lines and the dashed lines indicate crossovers. From
Ref.~\cite{Dressel_spin_charge}.}
\end{center}
\end{figure}

Within the various families of organic conductors
\cite{mori_review_chemrev} we will focus here in the properties of
Bechgaard (TMTSF-X) and Fabre (TMTTF-X) salts, which are very
suitable compounds for the study of quasi-1D physics, as will see
in the following. TMTTF-X stands for tetramethyltetrathiafulvalene
and TMTSF-X for tetramethyltetraselenafulvalene. In these
formulas, the X denotes an inorganic anion with various possible
symmetries: spherical (PF$_6$, AsF$_6$, SbF$_6$, TaF$_6$),
tetrahedral (BF$_4$, ClO$_4$, ReO$_4$) or triangular (NO$_3$).
TMTTF-X and TMTSF-X salts belong to a same family forming the
generic (TM)$_2$X phase diagram shown in
Fig.~\ref{fig:Phase_organics}. These salts have three main
conducting directions: the stacking direction of the molecules
called $a$, the transverse direction called $b$ and the direction
$c$, perpendicular to the $ab$ plane. The hopping integrals are of
the order of $t_a=3000$K, $t_b=300$K and $t_c=20$K. The molecular
structure of (TM)$_2$X conductors is depicted in
Fig~\ref{fig:Organic_structure}. With such anisotropy in their
molecular structure, where the overlap between electron clouds
along $a$ is $10$ times larger than the overlap between the stacks
in the transverse $b$ direction and $150$ times larger than the
one along $c$, the electronic structure can be seen as
one-dimensional with a slightly warped Fermi surface (see
Fig.~\ref{fig:Fermi_surface_1d}). This is what make these
compounds quasi 1D materials.

There exist a variety of ordered states in the phase diagram of
(TM)$_2$X compounds (see Fig.~\ref{fig:Phase_organics}):
spin-Peierls (SP), spin density wave (SDW), charge localized
(loc), charge ordered (CO), antiferromagnet (AFM) and
superconducting state (SC). In addition, there is a metallic state
which description changes from a one-dimensional Luttinger liquid
to a two- or three-dimensional Fermi liquid (dimensional crossover
studied in Sec.~\ref{sec:quasi_1d_systems}). Going from left to
right in the phase diagram, the materials get less one-dimensional
due to the increasing interaction in the second and third
directions. In order to study the Hall effect in these systems, we
will concentrate on the properties of the \textquotedblleft normal
phase\textquotedblright (metallic state) at high temperature,
where the one-dimensional behavior appears. We refer the reader to
the literature \cite{jerome_review_chemrev} for a review on the
different ordered states appearing in the phase diagram of Fig.~\ref{fig:Phase_organics}.

The insulating properties of the TMTTF compounds evolve towards
those of TMTSF, which are good conductors, trough an
insulator-metal transition when increasing pressure (or changing
the anions X), as can be seen in Fig.~\ref{fig:Phase_organics}.
The insulating behavior of the TMTTF family, as in
(TMTTF)$_2$PF$_6$ compound, is expected for a one-dimensional
Mott insulator. Such Mott insulator behavior could come from the
$1/4$-filled nature of the band or from the $1/2$-filled nature,
due to a small dimerization existing in these molecules, of the
order of $\Delta_d\sim100$K \cite{jerome_review_chemrev}. This
indicates that interactions have a large effect in the properties
of TMTTF family. In the case of the TMTSF family, due to their
metallic behavior at ambient pressure, the role of
interactions is more complicated to understand.

\begin{figure}
\begin{center}
\includegraphics[height=7cm,width=8cm]{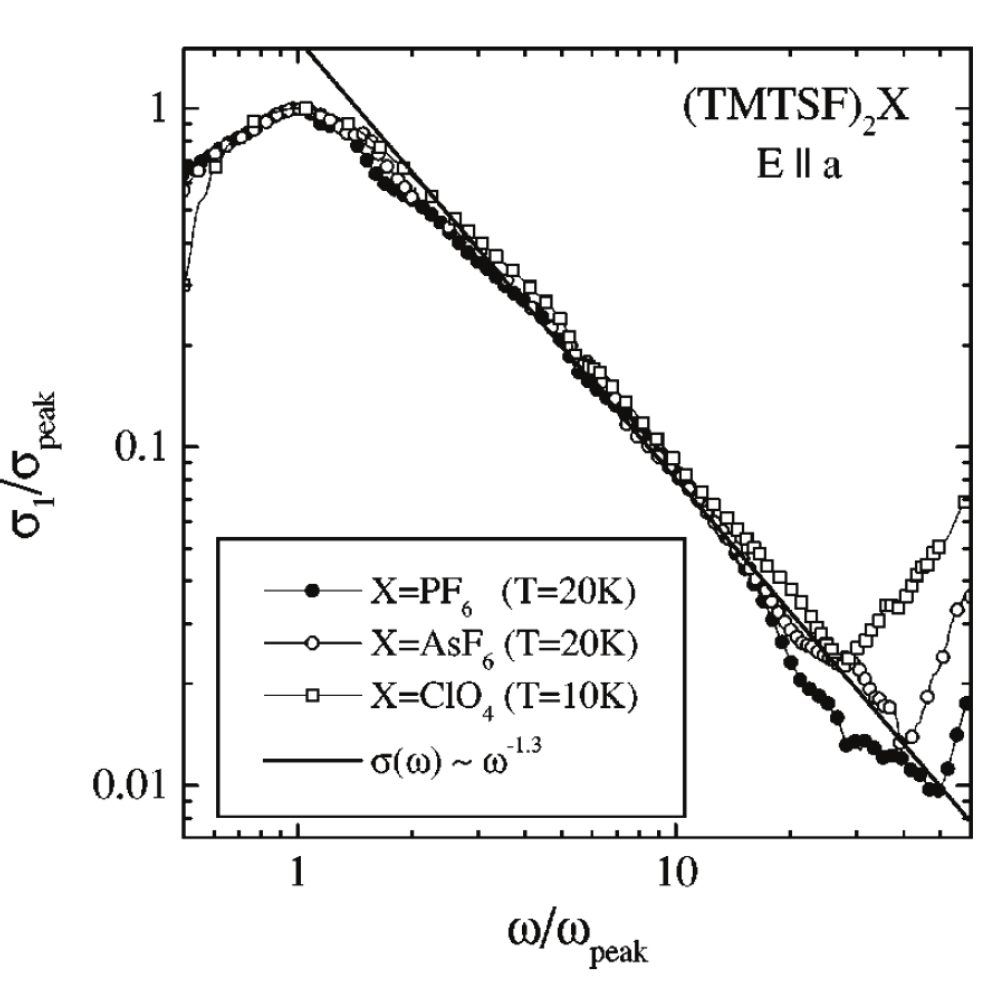}
\caption{\label{fig:cond_organics} Normalized conductivity along
the chain axis for different Bechgaard salts. The solid line shows
a fit of the form $\sigma(\omega)\sim\omega^{-\nu}$. From
Ref.~\cite{schwartz_electrodynamics}}
\end{center}
\end{figure}

Studies of the longitudinal transport ($a$ direction) have revealed signatures of LL properties 
\cite{schwartz_electrodynamics, giamarchi_mott_shortrev,jerome_review_chemrev}. Fig.~\ref{fig:cond_organics} shows the
optical conductivity along the chain axis ($a$-axis) in the TMTSF
family, where a power-law decay is found for $\sigma(\omega)$, in
agreement with the predicted conductivity in a Luttinger liquid (see
Eq.~(\ref{conductivity_1d})). The exponent of the power-law
behavior allowed an experimental determination of the
Luttinger parameter $K_{\rho}\simeq 0.23$
\cite{schwartz_electrodynamics}, consistent with an interpretation
of the insulating state as a quarter-filled Mott insulator in the
TMTSF family. Photoemission data has revealed a similar value for
$K_\rho$, but in a very large range of energies
\cite{vescoli_photoemission_tmtsf}, making these results more
difficult to interpret.
Transport transverse to the chains has given access to the
dimensional crossover between a pure 1D behavior and a more
conventional high-dimensional one \cite{moser_conductivite_1d,
henderson_transverse_optics_organics,giamarchi_review_chemrev,
jerome_review_chemrev}. Optical conductivity measurements give a
direct evidence of a deconfinement transition between a
one-dimensional insulator and a high-dimensional metallic regime,
when the observed gap is of the order of the interchain hopping
\cite{vescoli_photoemission_tmtsf}. Finally, measures of
resistivity along the chains and spin susceptibility, have shown
evidence of the spin-charge separation characteristic of a LL
\cite{Dressel_spin_charge}.

\begin{figure}
\begin{center}
\includegraphics[height=7cm,width=8cm]{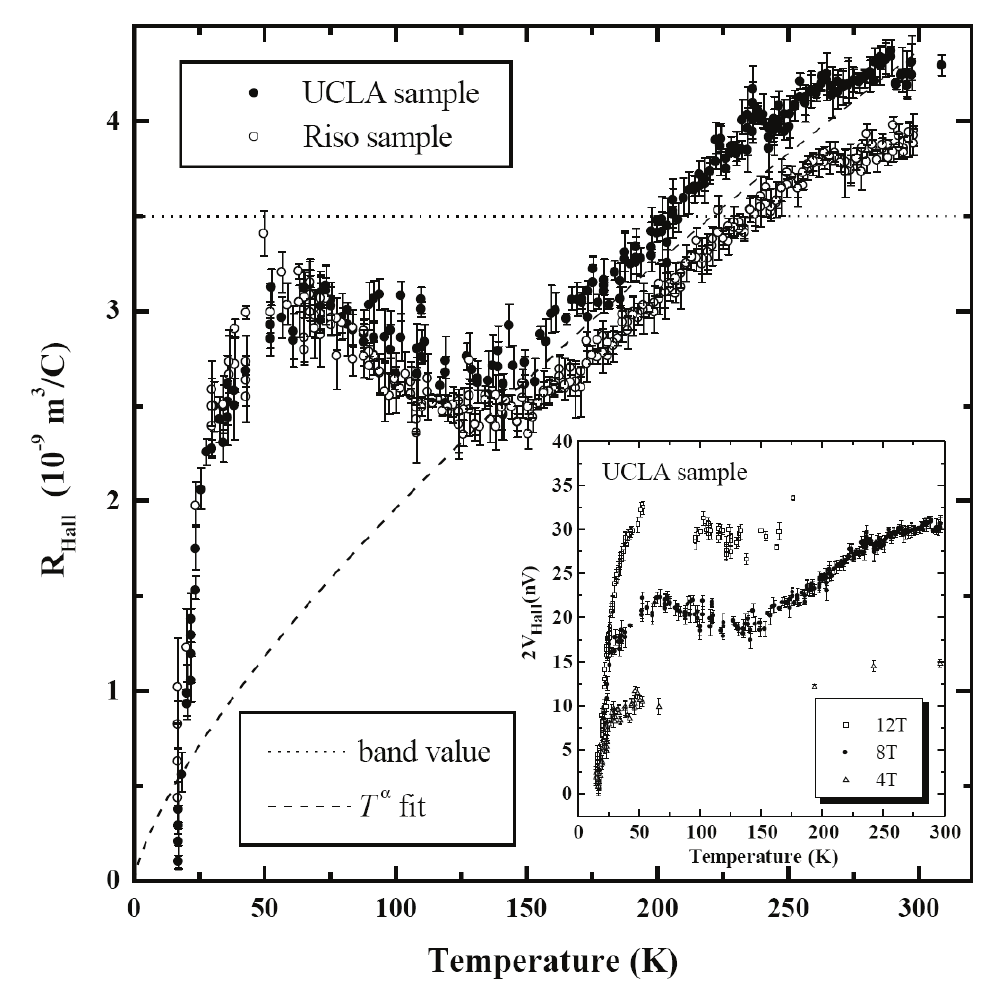}
\caption{\label{fig:Hall_Tdep} Temperature dependence of the Hall
constant in (TMTSF)$_2$PF$_6$ measured by Moser {\it{et al.}} in
Ref.~\cite{moser_hall_1d}. The current is applied along $a$ and
the magnetic field along $c$. The Hall voltage is measured along
$b$. The dashed line is a $T^\alpha$ power law fit with
$\alpha=0.73$.}
\end{center}
\end{figure}

\begin{figure}
\begin{center}
\includegraphics[width=8cm]{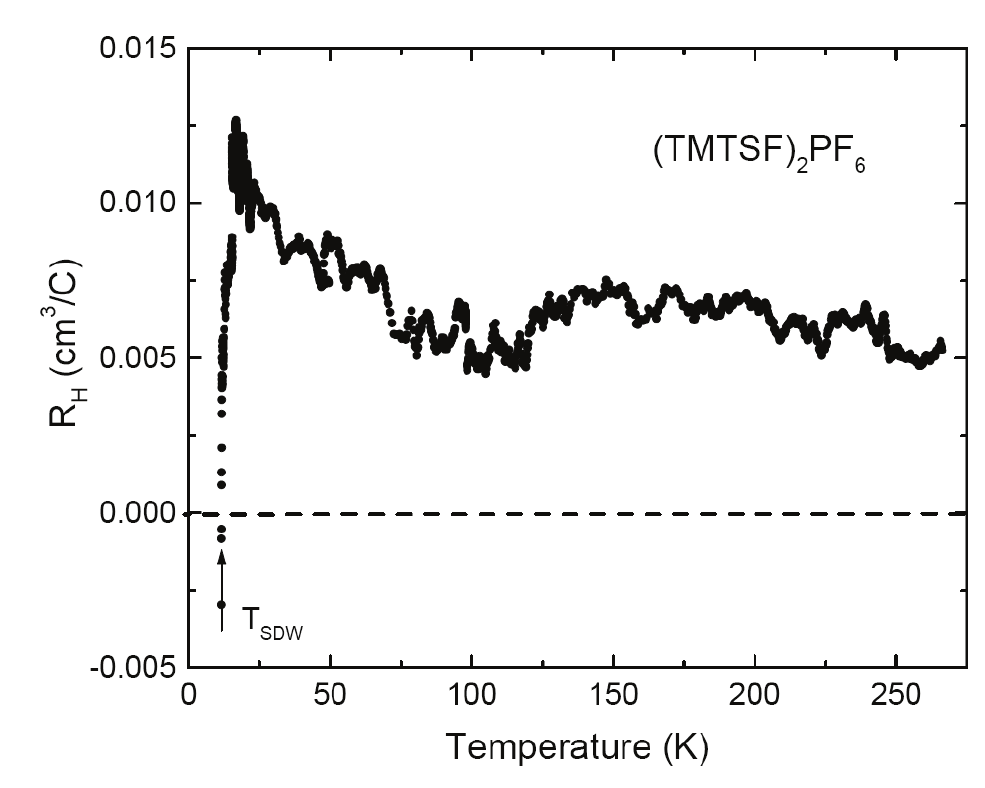}
\caption{\label{fig:Hall_Tindep} Temperature dependence of $\RH$
in the normal phase of (TMTSF)$_2$PF$_6$ measured by Mih\'{a}ly
{\it{et al.}} in Ref.~\cite{mihaly_hall_1d}. The current is
applied along $c$ and the magnetic field along $a$. The Hall
voltage is measured along $b$.}
\end{center}
\end{figure}

To probe further the consequences of correlations in these
compounds, several groups have undertaken the challenging
measurement of the Hall coefficient $\RH(T)$ \cite{moser_hall_1d,
mihaly_hall_1d, Korin-Hamzic_2003, Korin-Hamzic}. In particular,
two measurements of $\RH(T)$ were made in $2000$ by Moser {\it{et
al.}} \cite{moser_hall_1d} and Mih\'{a}ly {\it{et al.}}
\cite{mihaly_hall_1d}, in the organic conductor (TMTSF)$_2$PF$_6$.
The results are shown in Figs.~\ref{fig:Hall_Tdep} and
\ref{fig:Hall_Tindep}, respectively. In the former, the current is
applied along the $a$ axis and the magnetic field along the $c$
axis. Thus, the Hall voltage develops along the $b$ axis. They
investigated the temperature dependence of $\RH$ between $0$ and
$300$ K. In the normal state of (TMTSF)$_2$PF$_6$, at temperatures
$T>130$ K (a dimensional crossover to a 2D metal is expected at
$T\sim130K$), the Hall constant was found to be
temperature-dependent (see Fig.~\ref{fig:Hall_Tdep}), increasing
with $T$. The sign of $\RH$ was found to be positive (holelike).
In the second Hall experiment, shown in
Fig.~\ref{fig:Hall_Tindep}, the current was applied along the $c$
direction and the magnetic field parallel to the most conducting
direction ($a$ axis). The Hall voltage was measured along $b$.
They obtained a Hall constant independent of temperature for
$T>100$K and a positive sign for $\RH$. These results, different
depending on the direction of the applied magnetic field, proved
difficult to interpret due to a lack of theoretical understanding
of this problem. This prompted for a detailed theoretical analysis
of the Hall effect in quasi-1D systems. A first move in this
direction was reported in Ref.~\cite{lopatin_q1d_magnetooptical}
where the Hall coefficient of dissipationless weakly-coupled 1D
interacting chains was computed and found to be $T$-independent
and equal to the band value $\RH^0=1/nec$. This surprising result
shows that in this case $\RH$, unlike other transport properties,
is insensitive to interactions. However the assumption of
dissipationless chains is clearly too crude to be compared with
realistic systems for which a finite resistivity is induced by the
umklapp interactions \cite{giamarchi_umklapp_1d}.

This chapter is the object of publications \cite{leon_hall} and
\cite{proceedings_ISCOM06_Hall}. In the following we examine the
effect of umklapp scattering (see Sec.~\ref{sec:1D_case}) on the
temperature dependence of the Hall coefficient in quasi-1D
conductors, and we discuss the applications to the Hall
experiments mentioned above.

\section{The Hall effect in weakly coupled Luttinger
liquids}\label{Hall_quasi_1D}

We consider a model composed of weakly coupled $1/2$-filled 1D
chains. We take a $1/2$-filled band because the umklapp scattering
at $1/4$-filling is much more complicated to treat in the
calculations. But understanding the effect of the $1/2$-filled
umklapp scattering already gives an idea of the $1/4$-filled case.
With this model we compute $\RH(T)$ to leading order in the
umklapp scattering using the memory function approach explained
in Sec.~\ref{sec:memory_formalism}
 \cite{gotze_fonction_memoire,proceedings_ISCOM06_Hall}. We find that umklapp processes induce a
$T$-dependent correction to the free-fermion value $\RH^0$. This
correction decreases with increasing temperature as a power-law
with an exponent depending on interactions (Fig.~\ref{fig:graph}).
At the end, we discuss the implications for quasi-1D compounds.

\subsection{Model and methods}
\begin{figure}
\begin{center}
\includegraphics{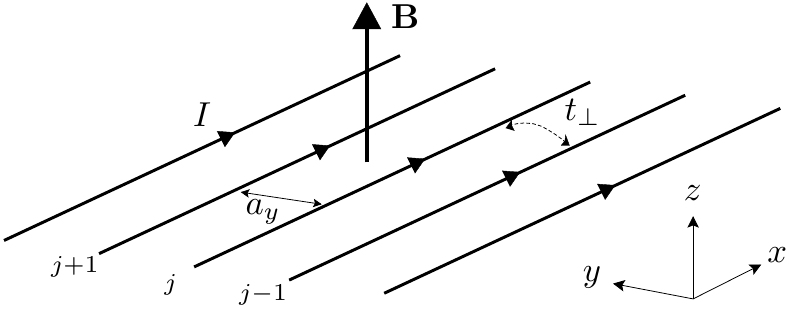}
\caption{\label{fig:model}
Schematics of the model. The chains and the current $I$ go along the $x$-axis,
the magnetic field $\mathbf{B}$ is applied along the $z$-axis, and the Hall voltage is
measured along the $y$-axis.}
\end{center}
\end{figure}
Our model is sketched in Fig.~\ref{fig:model}. We consider 1D chains coupled by
a hopping amplitude $t_{\perp}$ supposedly small compared to the in-chain
kinetic energy. As explained in Sec.~\ref{sec:1D_case},
the usual Luttinger liquid model of the 1D chains assumes that electrons
have a linear dispersion with a velocity $v_{\text{F}}$. For a strictly linear
band, however, the Hall coefficient vanishes identically owing to particle-hole
symmetry. A band curvature close to the Fermi momenta $\pm k_{\text{F}}$ is thus
necessary to get a finite $\RH$. We therefore take for the 1D chains of
Fig.~\ref{fig:model} the dispersion
    \begin{equation}\label{dispersion}
        \xi_{\pm}(k)=\pm v_{\text{F}}(k\mp k_{\text{F}})
        +\alpha (k\mp k_{\text{F}})^2.
    \end{equation}
The upper (lower) sign corresponds to right (left) moving electrons.
Eq.~(\ref{dispersion}) can be regarded as the minimal model which gives rise to
a Hall effect, while retaining most of the formal simplicity of the original LL
theory, and its wide domain of validity. In particular, this model is clearly
sufficient at low temperatures (compared to the electron bandwidth) since then
only electrons close to the Fermi points contribute to the conductivities.

Our purpose is to treat the umklapp term perturbatively. We express the
Hamiltonian as $\mathcal{H}_0+\mathcal{H}_u$ where $\mathcal{H}_u$ is the
umklapp scattering term and $\mathcal{H}_0$ reads
    \begin{multline}\label{Hamiltonian}
        \mathcal{H}_0=\int dx\sum_{j\sigma}\left[v_{\text{F}}
        \psi_{j\sigma}^{\dagger}\tau_3(-i\partial_x)
        \psi_{j\sigma}^{\phantom{\dagger}}
        -\alpha\,\psi_{j\sigma}^{\dagger}\partial_x^2
        \psi_{j\sigma}^{\phantom{\dagger}}\right.\\ \left.
        +g_2\,\psi_{j\sigma R}^{\dagger}
        \psi_{j\sigma R}^{\phantom{\dagger}}
        \psi_{j\sigma L}^{\dagger}
        \psi_{j\sigma L}^{\phantom{\dagger}}
        -t_{\perp}\left(\psi_{j\sigma}^{\dagger}
        \psi_{j+1,\sigma}^{\phantom{\dagger}}
        e^{-ieA_{j,j+1}}+\text{h.c.}\right)\right].
    \end{multline}
In Eq.~(\ref{Hamiltonian}) $j$ is the chain index, $\tau_3$ is a
Pauli matrix, and $A_{j,j'}=\int_j^{j'}\vec{A}\cdot d\vec{l}$. We
choose the Landau gauge $A_y=Bx$, such that $A_{j,j+1}=Bxa_y$ with
$a_y$ the interchain spacing.
$\psi^{\dagger}=(\psi^{\dagger}_R\;\psi^{\dagger}_L)$ is a
two-component vector composed of right- and left-moving electrons.
The second term in Eq.~(\ref{Hamiltonian}) is the band curvature,
the third term is the forward scattering and the last term
corresponds to the coupling between the chains
(Eq.~(\ref{hopping_term}) with a Peierls phase due to the presence
of the magnetic field). As mentioned before, we omit the
backscattering terms ($g_1$ processes) which are, for spin
rotationally invariant systems, marginally irrelevant
\cite{giamarchi_book_1d}. We therefore take
$g_{1\perp}=g_{1\parallel}=0$. At $1/2$ filling the umklapp term
reads (the bosonized version is given in Eq.~(\ref{umklapp_half}))
    \begin{equation}\label{Hu}
        \mathcal{H}_{u}=\frac{g_3}{2}\int dx\sum_{j\sigma}\left(
        \psi_{j\sigma R}^{\dagger}\psi_{j,-\sigma R}^{\dagger}
        \psi_{j\sigma L}^{\phantom{\dagger}}\psi_{j,-\sigma L}^{\phantom{\dagger}}
        +\text{h.c.}\right).
    \end{equation}
We will compute the ac Hall constant using the memory matrix formalism, where the Hall coefficient
$\RH=\rho_{yx}/B$ is given by (see Eq.~(\ref{RH_memory}))
    \begin{equation}\label{RH}
        \RH(\omega)=\frac{1}{i\chi_y(0)}\lim_{B\to0}
        \frac{\Omega_{xy}+iM_{xy}(\omega)}{B}.
    \end{equation}

From Hamiltonian (\ref{Hamiltonian}) and the definition for the
diamagnetic term, given in Sec.~\ref{sec:Kubo}, we obtain the
longitudinal and transverse diamagnetic terms
    \begin{subequations}\begin{eqnarray}
        \label{chix}
        \chi_x(0)&=&-\frac{2e^2 v_{\text{F}}}{\pi a_y},\\
        \label{chiy}
        \chi_y(0)&=&-2e^2t_{\perp}a_y^2\!\int dx
        \langle\psi_{0\uparrow}^{\dagger}(x)\psi_{1\uparrow}(x)
        e^{-ieBa_yx}+\text{h.c.}\rangle.
    \end{eqnarray}\end{subequations}
For the longitudinal diamagnetic term (\ref{chix}) we have used
the relation between the electron density $n$ and the Fermi
momentum: $na_y=k_{\text{F}}/\pi$ (since $na_y$ is the density per
one chain). It is also easily obtained from
Eq.~(\ref{diamagnetic_1d}) with $u_\rho K_\rho=v_\text{F}$. For
the evaluation of the frequency matrix in Eq.~(\ref{RH}), we write
down the current operators, making the functional derivatives of
the Hamiltonian with respect to the vector potential, as explained
in Sec.~\ref{sec:Kubo},
    \begin{subequations}\label{currents}\begin{eqnarray}
        J_x&=&e\int dx\sum_{j\sigma}\psi_{j\sigma}^{\dagger}(x)
        \left[v_{\text{F}}\tau_3+2\alpha(-i\partial_x)\tau_1\right]
        \psi_{j\sigma}^{\phantom{\dagger}}(x)\\ \label{J_y}
        J_y&=&-iet_{\perp}a_y\int dx\sum_{j\sigma}
        \left(\psi_{j\sigma}^{\dagger}\psi_{j+1,\sigma}^{\phantom{\dagger}}
        e^{-ieA_{j,j+1}}-\text{h.c.}\right)
    \end{eqnarray}\end{subequations}
To obtain the frequency matrix, we must calculate the commutator
between these current operators. Thus, using the standard
commutation relation for fermionic operators, the expression
resulting from Eq.~(\ref{Omega}) for $\Omega_{xy}$ is then
    \begin{equation}\label{Omegaxy}
        \Omega_{xy}=-i\frac{2\pi e\alpha t_{\perp}a_y^3B}{v_{\text{F}}}
        \!\int dx\,\langle\psi_{0\uparrow}^{\dagger}(x)
        \psi_{1\uparrow}^{\phantom{\dagger}}(x)
        e^{-ieBa_yx}+\text{h.c.}\rangle.
    \end{equation}
At this stage we can already evaluate the high-frequency limit of
$\RH$, because the memory matrix vanishes as $1/\omega^2$ if
$\omega\to\infty$. Thus, the effects of the umklapp disappear at
high frequency, and in this limit one recovers from
Eqs~(\ref{RH}--\ref{Omegaxy}) the result obtained for
dissipationless chains in Ref.~\cite{lopatin_q1d_magnetooptical},
namely that the Hall coefficient equals the band value $\RH^0$:
    \begin{equation}\label{RH0}
        \RH(\infty)=\RH^0=\frac{\pi\alpha a_y}{ev_{\text{F}}}.
    \end{equation}

\subsection{$\RH$ in the presence of umklapp and forward scattering}\label{sec:RH_umklapp}

As explained in Sec.~\ref{sec:Hall_memory} the memory matrix
element $M_{xy}(\omega)$, necessary to obtain $\RH(\omega)$, can
be reduced at high frequencies and linear order in $\mathbf{B}$ to
the calculation of the following average
    \begin{equation}\label{Memory_Kx_Ky}
        iM_{xy}(\omega)\approx-\frac{1}{\chi_x(0)}\frac{\langle K_x;K_y\rangle}{\omega}
    \end{equation}
with $K_{\mu}$ the \emph{residual forces} operators, which in this
case are $K_{\mu}=[\mathcal{H}_u,J_{\mu}]$, and $\langle
K_x;K_y\rangle$ stands for the retarded correlation function of
the operators $K_{\mu}$. Using the umklapp term (\ref{Hu}) and the
currents (\ref{currents}) \footnote{Here we use the following
properties for commutators: $[AB,CD]=A[B,CD]+[A,CD]B$, and
$[A,BC]=[A,B]_{+}C-B[A,C]_{+}$, where $[A,B]_{+}$ denotes the
anticommutator $AB+BA$.} we find
    \begin{subequations}\label{K}\begin{eqnarray}
        K_x&=&2ev_{\text{F}}g_3\int dx\sum_{j\sigma}
        \left(\psi_{j\sigma R}^{\dagger}\psi_{j,-\sigma R}^{\dagger}
        \psi_{j,-\sigma L}\psi_{j\sigma L}-\text{h.c.}\right)\\
        \nonumber
        K_y&=&iet_{\perp}g_3a_y\int dx\sum_{j\sigma}\sum_{b=L,R}
        \Big[e^{-ieA_{j,j+1}}
        \left(\psi_{j\sigma b}^{\dagger}\psi_{j,-\sigma b}^{\dagger}
        \psi_{j,-\sigma,-b}\psi_{j+1,\sigma,-b}\,\right.\\
        &&-\left.\psi_{j-1,\sigma b}^{\dagger}\psi_{j,-\sigma b}^{\dagger}
        \psi_{j,-\sigma,-b}\psi_{j\sigma,-b}\right)+\text{h.c.}\Big].\qquad
    \end{eqnarray}\end{subequations}
Note that each of the $K$'s is of first order in $g_3$, hence $M_{xy}$ is of
order $g_3^2$. The quantity $\langle K_x;K_y\rangle$ entering Eq.~(\ref{Memory_Kx_Ky})
is the real-frequency, long-wavelength limit of the correlator, which we
evaluate as
    \begin{equation}\label{KK}
        \langle K_x;K_y\rangle=-\int_0^{\beta} d\tau\,e^{i\Omega\tau}
        \langle T_{\tau}K_x(\tau)K_y(0)\rangle
        \Big|_{i\Omega\to\omega+i0^+},
    \end{equation}
where $i\Omega$ denotes the Matsubara frequency. The correlator
$\langle K_x;K_y\rangle$, at first order in $t_\perp$, vanishes
for $B=0$ or $\alpha=0$ as shown in Appendix~\ref{app:alpha0_B0}.
Thus, retaining only leading-order terms in $t_{\perp}$ and
$\alpha$, the first nonvanishing contribution in Eq.~(\ref{KK}) is
of order $\alpha t_{\perp}^2g_3^2B$, and involves three spatial
and three time integrations, which we were not able to perform in
full. Based on a scaling analysis, we can nevertheless extract the
temperature (or frequency) dependence of this contribution.

We evaluate the correlator $\langle K_x;K_y\rangle$ to first order
in $\alpha$ and $t_{\perp}$. Let's denote by
$\mathcal{H}_{\alpha}$ the curvature [second term in
Eq.~(\ref{Hamiltonian})], by $\mathcal{H}_{\perp}$ the inter-chain
hopping [fourth term in Eq.~(\ref{Hamiltonian})], and by
$\mathcal{H}_{0}$ the remaining parts of the Hamiltonian,
$\mathcal{H}_{1D}=\mathcal{H}_0-\mathcal{H}_{\alpha}-\mathcal{H}_{\perp}$.
Standard perturbation theory yields \cite{mahan_book}
    \begin{equation}\label{KKHcBC}
        \langle K_x;K_y\rangle=-\int d\tau\,e^{i\Omega\tau}
        \int d\tau_1d\tau_2
        \left\langle T_{\tau}K_x(\tau)K_y(0)\mathcal{H}_{\perp}(\tau_1)
        \mathcal{H}_{\alpha}(\tau_2)\right\rangle
    \end{equation}
where the average is taken with respect to $\mathcal{H}_{0}$. The
latter corresponds to a 1D chain and can be easily bosonized, as
shown in Sec.~\ref{sec:1D_case}. With the help of representation
(\ref{fermionic}) we bosonize each operator in Eq.~(\ref{KKHcBC}):
\begin{eqnarray}\label{Kx_Ky_bosonized}
    K_x&=&\frac{4iev_{\text{F}}g_3}{(2\pi a)^2}\int dx\sum_{j\sigma}
    \sin\left(\sqrt{8}\phi_{\rho}(x)\right)_j\\
    \nonumber
    K_y&=&\frac{iet_{\perp}g_3a_y}{(2\pi a)^2}
    \sum_{\langle j,j'\rangle}\sum_{\sigma b}\int dx\,\epsilon_{jj'}\Big(e^{-ieBa_yx}
    e^{\frac{i}{\sqrt{2}}\left\{3b\phi_{\rho}(x)-\epsilon_{jj'}\theta_{\rho}(x)
    -\sigma\left[b\phi_{\sigma}(x)+\epsilon_{jj'}\theta_{\sigma}(x)\right]\right\}_j}\\
    &&e^{\frac{i}{\sqrt{2}}\left\{b\phi_{\rho}(x)+\epsilon_{jj'}\theta_{\rho}(x)
+\sigma\left[b\phi_{\sigma}(x)+\epsilon_{jj'}\theta_{\sigma}(x)\right]\right\}_{j'}}+\text{h.c}\Big)
\end{eqnarray}
where $j$ and $j'$ are neighboring chains, $b=+1(-1)$ for
right(left) moving fermions, and $\epsilon_{jj'}=(-1)^{j'-j}$. For
the coupling term we have
\begin{eqnarray}
    \nonumber
    \mathcal{H}_{\perp}&=&-\frac{t_{\perp}}{2\pi a}\sum_{j\sigma b}\int dx\Big(e^{-ieBa_yx}
    e^{\frac{i}{\sqrt{2}}\left\{b\phi_{\rho}(x)-\theta_{\rho}(x)
    +\sigma\left[b\phi_{\sigma}(x)-\theta_{\sigma}(x)\right]\right\}_j}\\ \label{Hperp_bosonized}
    &&e^{-\frac{i}{\sqrt{2}}\left\{b\phi_{\rho}(x)-\theta_{\rho}(x)
    +\sigma\left[b\phi_{\sigma}(x)-\theta_{\sigma}(x)\right]\right\}_{j+1}}+\text{h.c.}\Big)
\end{eqnarray}
and for the band curvature term we take \cite{Haldane_JPC}
 \begin{equation}\label{band_curvature}
        \mathcal{H}_{\alpha}=\frac{\alpha}{2\pi a} \int dx\,
        \frac{\left(\nabla\phi_{\rho}\right)^3}{2}.
 \end{equation}
Next we will use a very helpful identity for the calculation of
correlators between functions of the fields $\phi$ and $\theta$.
The following identity can be proved using the functional integral
technique presented in Sec.~\ref{sec:transport_low_dim} and its
fully demonstrated in Ref.~\cite{giamarchi_book_1d}
\begin{eqnarray}\nonumber
    \langle\prod_ne^{i[A_n\phi(r_n)+B_n\theta(r_n)]}\rangle&=&
    \exp\Big\{-\frac{1}{2}{\sum_{n<m}}'-(A_nA_mK+B_nB_mK^{-1})F_1(r_n-r_m)\\ \label{condition_A_B}
    &+&(A_nB_m+B_nA_m)F_2(r_n-r_m)\Big\},
\end{eqnarray}
where $r\equiv(x,u\tau)$, the notation ${\sum}'$ means that the
sum is restricted to those terms for which
$\sum_nA_n=\sum_nB_n=0$, and $F_{1,2}$ are universal functions which for $(x,u\tau)\gg a$ are,
\begin{eqnarray}\label{F1_F2}
F_1(r)&=& \frac{1}{2}\log\left[\frac{\beta^2u^2}{\pi^2a^2}\left(\sinh^2(\frac{\pi x}{\beta u})+\sin^2(\frac{\pi\tau}{\beta})\right)\right]\\
F_2(r)&=&-i\text{Arg}\left[\tan(\frac{\pi y_a}{\beta u})+i\tanh(\frac{\pi x}{\beta u})\right],
\end{eqnarray}
where $y_{a}=u\tau+a\text{Sign}(\tau)$, and $a$ is a momentum cutoff.
The resulting expression for the correlator in Eq.~(\ref{KKHcBC})
is (see appendix~\ref{app:full_correlator})
 \begin{eqnarray}\label{scaling}
  \langle K_x;K_y\rangle&\sim& B\int d^2rd^2r_1d^2r_2\,
        e^{-3K_{\rho}F_1(r)}|r|\\ \nonumber
    &&e^{-K_{\rho}F_1(r-r_1)}
        e^{\frac{1}{2}(K_{\rho}-K_{\rho}^{-1}-2)F_1(r_1)}
        \frac{1}{|r_2|^3}.
 \end{eqnarray}
The factor $|r|$ results from the linearization in the magnetic
field $\mathbf{B}$. In Eq.~(\ref{scaling}) we have discarded all
factors involving the $F_2$ function, since they correspond to
angular integrals of the $r$ variables and therefore do not
contribute to the scaling dimension. At distances much larger than
the cutoff $a$ we have $e^{-AF_1(r)}\sim (a/|r|)^{A}$, and
therefore we find the high temperature, high frequency behavior as
    \begin{equation}
        \langle K_x;K_y\rangle\sim B
        \max(\omega,T)^{-3+4K_{\rho}-\frac{1}{2}(K_{\rho}-K_{\rho}^{-1})}.
    \end{equation}
As done in Sec.~\ref{sec:transport_quasi1D}, we follow the same
procedure for the diamagnetic term $\chi_y(0)$---however at zeroth
order in $\alpha$ and $B$---and find
    \begin{equation}
        \chi_y(0)\sim
        \max(\omega,T)^{-1+\frac{1}{2}(K_{\rho}+K_{\rho}^{-1})}.
    \end{equation}
Combining these expressions and collecting the relevant
prefactors we deduce
    \begin{equation}\label{KK-T}
        \frac{1}{i\chi_x(0)\chi_y(0)}\frac{\langle K_x;K_y\rangle}{\omega B}
        \sim \alpha\,g_3^2\,\max(\omega,T)^{3K_{\rho}-3},
    \end{equation}
where $K_{\rho}$ is the LL parameter in the charge sector (see Sec.~\ref{sec:1D_case}).
In the absence of interactions we have $K_{\rho}=1$, while $K_{\rho}<1$ ($K_{\rho}>1$) for
repulsive (attractive) interactions. If the interactions are strong and
repulsive ($K_\rho \ll 1$) the exponent in Eq.~(\ref{KK-T}) changes due to the
contraction \cite{giamarchi_book_1d} of the operators in $K_x$ and $K_y$, which
gives the relevant power-law in this case.
\begin{figure}
\begin{center}
\includegraphics{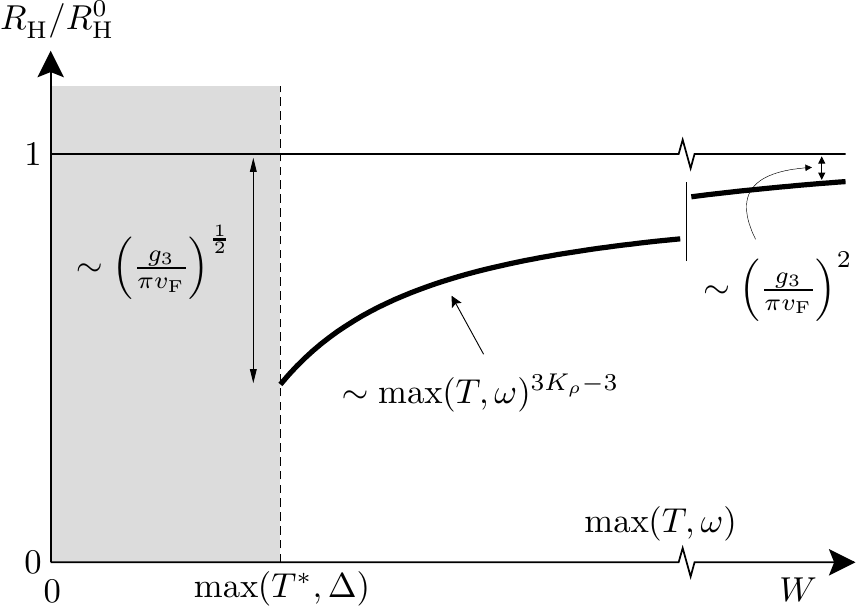}
\caption{\label{fig:graph}
Correction of the high-temperature/high-frequency Hall coefficient $\RH$ by the
umklapp scattering in weakly-coupled Luttinger liquids. $\RH^0$ is the value of
the Hall coefficient in the absence of umklapp scattering, Eq.~(\ref{RH0}), and
$W$ is the electron bandwidth. Our approach breaks down below some crossover
scale (dashed line, see text). In this figure we have assumed that $A$ in Eq.~(\ref{RH_result}) is negative.}
\end{center}
\end{figure}
Together with Eqs~(\ref{Memory_Kx_Ky}) and (\ref{RH}),
Eq.~(\ref{KK-T}) leads to our final expression for the Hall
coefficient:
    \begin{equation}\label{RH_result}
        \RH=\RH^0\left[1+A\left(\frac{g_3}{\pi v_{\text{F}}}\right)^2
        \left(\frac{T}{W}\right)^{3K_{\rho}-3}\right]
    \end{equation}
with $W$ the electron bandwidth.

Eq.~(\ref{RH_result}) shows that in
$1/2$-filled quasi-1D systems the umklapp scattering changes the absolute value
of the Hall coefficient with respect to the band value, which is only recovered
at high temperature or frequency. Note that Eq.~(\ref{RH_result}) also describes
the frequency dependence of $\RH$ provided $T$ is replaced by $\omega$. The
backscattering term $g_1$ (neglected here) could possibly give rise to
multiplicative logarithmic corrections to the power law in Eq.~(\ref{RH_result})
\cite{giamarchi_book_1d}. The sign of the dimensionless prefactor $A$ can only
be determined through a complete evaluation of $\langle K_x;K_y\rangle$ in
Eq.~(\ref{KK}), and is for the time being unknown. The available experimental
data are consistent with Eq.~(\ref{RH_result}) if one assumes that $A$ is
negative (see below), as illustrated in Fig.~\ref{fig:graph}.

Eq.~(\ref{RH_result}) would imply that in the non-interacting
limit $K_{\rho}\to1$ ($g_2\to0$) the correction to the Hall
coefficient behaves as $\log(T/W)$. In order to check this
prediction we have evaluated the correlator in Eq.~(\ref{KK}) for
$g_2=0$. This is done in the following section.


\subsection{$\RH$ in the presence of ukmlapp without forward scattering}

\begin{figure}
\begin{center}
\includegraphics{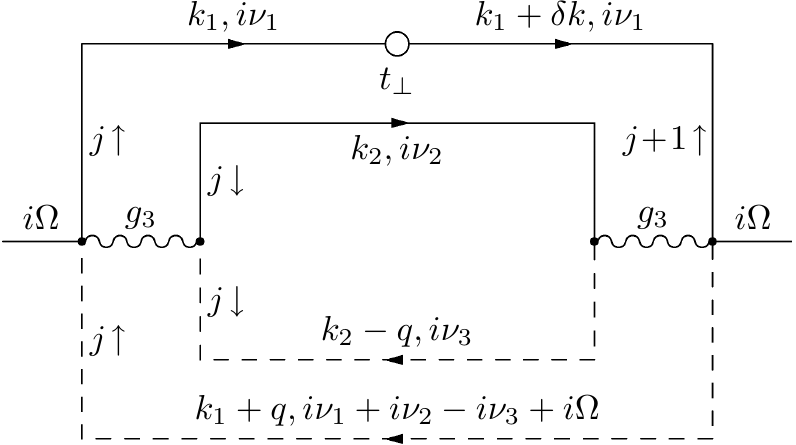}
\caption{\label{fig:diagram}
Example of a diagram appearing in Eq.~(\ref{KK}) at first order in $t_{\perp}$
and for $g_2=0$. The full (dashed) lines correspond to free right (left) moving
fermions, $j$ is the chain index, and the arrows represent up and down spins.
The magnetic field increases the momentum of the electron by $\delta k=eBa_y$.
}
\end{center}
\end{figure}

Here we provide the calculation of $\RH$ to leading order in $g_3$
but in the absence of forward scattering, $g_2=0$. Using
Eqs~(\ref{RH}) and (\ref{RH0}) we can express the zero-frequency
Hall coefficient in terms of $\RH^0$ and
$\text{Re}[M_{xy}(i0^+)]$. We then perform a Kramers-Kronig
transform, insert the free-fermion values of the diamagnetic
susceptibilities, $\chi_x(0)=-2e^2v_{\text{F}}/(\pi a_y)$ and
$\chi_y(0)=-4e^2t_{\perp}^2a_y/(\pi v_{\text{F}})$, and use
Eq.~(\ref{MofK}) to arrive at
\begin{equation}\label{free_1}
\RH(0)=\RH^0\left[1+\frac{v_{\text{F}}}{8e^3\alpha t_{\perp}^2a_y}
\frac{1}{B}\int\frac{d\omega}{\omega^2}\text{Im} \left(i\langle
K_x;K_y\rangle_0\Big{|}_{i\Omega\to\omega+i0^+}\right)\right]
\end{equation}
where $\langle K_x;K_y\rangle_0$ is to be evaluated to first order
in $B$. 
From Eq.~(\ref{K}) one sees that $\langle K_x;K_y\rangle_0$
involves 8 fermion fields and can be represented by diagrams like
the one displayed in Fig.~\ref{fig:diagram}. There are 32
different diagrams, but all of them can be expressed in terms of
only one function $A(i\Omega,B)$, whose expression is given by the
diagram in Fig.~\ref{fig:diagram}. This is done in
Appendix~\ref{app:Kx_Ky_g20}. We thus obtain,
    \begin{equation}
        \RH(0)=\RH^0\left\{1-\frac{4v_{\text{F}}^2g_3^2}{e\alpha}
        \int\frac{d\omega}{\omega^2}\phantom{\int}
        \text{Im}\left[A'(\omega+i0^+)-A(-\omega-i0^+)\right]\right\}
    \end{equation}
where $A'(i\Omega)=\partial A(i\Omega,B)/\partial B|_{B=0}$ and we
have pulled all prefactors from Eq.~(\ref{K}), as well as a factor
$t_{\perp}$ from the diagram, out of the definition of function
$A$. The explicit expression of $A'$ is (see
Appendix~\ref{app:Kx_Ky_g20})
    \begin{eqnarray}\nonumber
        A'(i\Omega)=\frac{e}{(2\pi)^3}\int dk_1dk_2dq\frac{d\,\xi_+(k_1)}{dk_1}
        \frac{1}{\beta^3}\sum_{\nu_1\nu_2\nu_3}\Big[\frac{1}{i\nu_1-\xi_+(k_1)}\Big]^3\\
	\label{eq:Aprime}
        \frac{1}{i\nu_2-\xi_+(k_2)}\frac{1}{i\nu_3-\xi_-(k_2-q)}
        \frac{1}{i\nu_1+i\nu_2-i\nu_3+i\Omega-\xi_-(k_1+q)}.
    \end{eqnarray}
The frequency summations in Eq.~(\ref{eq:Aprime}) are elementary,
and the various momentum integrals can also be evaluated
analytically to first order in $\alpha$, yielding (see
Appendix~\ref{app:Kx_Ky_g20})
    \begin{equation}
        \RH(0)=\RH^0\left[1-\frac{1}{16}\left(\frac{g_3}{\pi v_{\text{F}}}\right)^2
        \int\frac{d\omega}{\omega}
        \frac{(\beta\omega/4)^2-\sinh^2(\beta\omega/4)}
        {\tanh(\beta\omega/4)\sinh^2(\beta\omega/4)}\right].
    \end{equation}
The remaining energy integral is divergent and must be regularized. Cutting the
integral at the bandwidth $W$ and assuming $T\ll W$ we obtain the asymptotic
behavior
\begin{equation}\label{no-int}
        \RH=\RH^0\left[1+\frac{1}{8}
        \left(\frac{g_3}{\pi v_{\text{F}}}\right)^2
        \log\left(\frac{T}{W}\right)\right],
\end{equation}
consistent with Eq.~(\ref{RH_result}). For non-interacting
electrons, though, we see that the relative correction induced by
the $1/2$-filling umklapp is positive at $T<W$. Since all
properties are analytic in $K_{\rho}$, we can also deduce from
Eqs~(\ref{RH_result}) and (\ref{no-int}) that $A$ tends to
$[24(1-K_{\rho})]^{-1}$ in the limit $K_{\rho}\to1$. Note that
Eq.~(\ref{no-int}) would also apply to models in which $g_2 \sim
g_3$, such as the Hubbard model, while Eq.~(\ref{RH_result}) is
valid only when $g_3 \ll g_2$.

\subsection{Discussion and perspectives}

The result of Eq.~(\ref{RH_result}) shows that in $1/2$-filled
quasi 1D systems the umklapp processes induce a correction to the
free-fermion value (band value $\RH^0$) of the Hall coefficient
$\RH$, which depends on temperature as a power-law with an
exponent depending on interactions. At high temperatures or
frequencies, $\RH$ approaches the band value as shown in
Fig.~\ref{fig:graph}, implying that any fitting of experimental
data must be done with respect to the value of $\RH$ at high
temperature or frequency.

To study the range of validity of our result, one must consider
that at low temperature the quasi-1D systems generally enter
either in an insulating state characterized by a Mott gap
$\Delta$, or in a coherent two- or three-dimensional phase below a
temperature $T^*$ controlled by $t_{\perp}$, as explained in
Sec.~\ref{sec:quasi_1d_systems}. In either case our model of
weakly-coupled LL is no longer valid, as illustrated in
Fig.~\ref{fig:graph}. The variations of $\RH$ below
$\max(T^*,\Delta)$ can be very pronounced, and depend strongly on
the details of the materials. When the ground state is insulating,
for instance, $\RH(T)$ is expected to go through a minimum and
diverge like $e^{\Delta/T}$ as $T\to0$, reflecting the
exponentially small carrier density. Other behaviors, such as a
change of sign due to the formation of an ordered state or nesting
in the FL regime \cite{yakovenko_phenomenological_model_th}, can
also occur. The validity of Eq.~(\ref{RH_result}) is therefore
limited to the LL domain: $\max(T^*,\Delta)<\max(T,\omega)\ll W$.

For the case $\Delta>T^*$, we estimate the change of $\RH$ with
respect to $\RH^0$ at the crossover scale $\Delta$, for a system
with $g_3\ll U$, where $U$ is the Coulomb repulsion. The
umklapp-induced Mott gap in $1/2$-filled systems is given by
$\Delta/W\sim[g_3/(\pi v_{\text{F}})]^x$ with
$x=[2(1-K_{\rho})]^{-1}$ \cite{giamarchi_book_1d}. We thus find
that the largest correction is $\sim[g_3/(\pi
v_{\text{F}})]^{\frac{1}{2}}$ and has a universal exponent. On the
other hand, $\RH$ approaches the asymptotic value $\RH^0$ quite
slowly, and according to Eq.~(\ref{RH_result}) a correction of
$\sim[g_3/(\pi v_{\text{F}})]^2$ still exists at temperatures
comparable to the bandwidth.

The available Hall data in the TM family and in the geometry of
the present analysis \cite{moser_hall_1d,Korin-Hamzic_2003} show a
weak correction to the free fermion value which depends on
temperature. Some attempts to fit this behavior to a power law
have been reported (see Fig.~\ref{fig:Hall_Tdep})
\cite{moser_hall_1d}. However the analysis was performed by
fitting $\RH(T)$ to a power law starting at zero temperature. As
explained above, the proper way to analyze the Hall effect in such
quasi-1D systems is to fit the \emph{deviations} from the band
value starting from the high temperature limit. It would be
interesting to check whether a new analysis of the data would
provide good agreement with our results. However in these
compounds both $1/4$-filling and $1/2$-filling umklapp processes
are present. For the longitudinal transport, the $1/4$-filling
contribution dominates \cite{giamarchi_review_chemrev}. For the
Hall effect, the analysis in the presence of $1/4$-filling umklapp
is considerably more involved, but a crude evaluation of the
scaling properties of the corresponding memory matrix gives also a
weak power-law correction with an exponent
$2-16K_{\rho}+(K_{\rho}+K_{\rho}^{-1})/2$, and thus similar
effects, regardless of the dominant umklapp. The observed data is
thus consistent with the expected corrections coming from LL
behavior. However more work, both experimental and theoretical, is
needed for the TM family because of this additional complication.

Our result Eq.~(\ref{RH_result}) is however directly relevant for
$1/2$-filled organic conductors such as (TTM-TTP)I$_3$ and
(DMTSA)BF$_4$ \cite{mori_review_chemrev}. Hall measurements for
these compounds still remain to be performed. Comparison of the
Hall effect in these compounds with the one in $1/4$-filled
non-dimerized systems \cite{heuze_quarterfilled_refs,
jerome_review_chemrev} for which only $1/4$-filling umklapp is
present, could also help in understanding the dominant processes
for the TM family.

The other type of Hall measurements, shown in
Fig.~\ref{fig:Hall_Tindep}, were done in a different geometry from
the one used in the present theoretical work. In this case the
current flows along the least conducting direction $c$ and
the magnetic field is applied parallel to $a$, as explained
before. This implies a current flowing along a direction which is
incoherent at high temperature ($T>T^*$) and thus, the results for
$\RH$ are expected to differ from Eq~(\ref{RH_result}). To
describe the system in this geometry, the Hamiltonian must have
two coupling terms corresponding to the hopping along $b$ and $c$
directions and one must choose a gauge for the magnetic field that
determines the Peierl's phase appearing in the Hamiltonian, as
done for the model studied here. The longitudinal and transverse
currents will have the form of Eq.~(\ref{J_y}) (with a different
Peierl's phase), each one with the respective hopping amplitude
$t_b$ or $t_c$. The diamagnetic terms will be both similar to
Eq.~(\ref{chiy}), with $t_{\perp}$ replaced by $t_b$ or $t_c$ and
$a_y$ replaced by the respective lattice parameter. Since in this
geometry the magnetic field goes along the one-dimensional chains,
the question is whether or not signatures of Luttinger liquid
behavior will appear on $\RH$. With the information given until
now we can compute the first term of Eq.~(\ref{RH}) (the
high-frequency $\RH$ of Eq.~(\ref{inf_freq_RH})). For this we must
calculate the commutator between the currents, obtaining
\begin{equation}\label{RH0_second_geometry}
\RH(\omega\to\infty)=\frac{a_b a_c}{ec} \frac{\sum_{\alpha,\beta}
   \left\langle \psi^{\dagger}_{\alpha\beta} \psi^{\phantom{\dagger}}_{\alpha + 1\beta + 1}
   + \psi^{\dagger}_{\alpha\beta + 1} \psi^{\phantom{\dagger}}_{\alpha + 1\beta} + \text{h.c.} \right\rangle}
   {\sum_{\alpha, \beta}\left\langle
   \psi^{\dagger}_{\alpha\beta} \psi^{\phantom{\dagger}}_{\alpha + 1\beta} +\text{h.c.}
   \right\rangle  \left\langle \psi^{\dagger}_{\alpha
   \beta} \psi^{\phantom{\dagger}}_{\alpha\beta + 1} + \text{h.c.}
   \right\rangle},
\end{equation}
where $\alpha$ and $\beta$ are chain indexes, and $a_b$ and $a_c$
are the lattice parameters along $b$ and $c$ directions,
respectively. It is evident that the average in the numerator of
Eq.~(\ref{RH0_second_geometry}) must be expanded to first order in
$t_b$ and $t_c$ in order to obtain a nonzero result. Immediately,
many questions on the derivation of expression
(\ref{RH0_second_geometry}) appear: can we treat $t_c$
perturbatively even if the current goes along $c$?, is the 1D
Hamiltonian term necessary or just a Hubbard Hamiltonian will be
sufficient?, should one consider the 1D properties of the chains
to obtain the averages, even if they do not play a crucial role in
this geometry?, is the above expression temperature independent?, etc.
These and many other questions must be answered
in order to understand the Hall effect in the geometry consider by
Mih\'{a}ly {\it{et al.}} in Ref.~\cite{mihaly_hall_1d}. In this
work we did not solve this problem, but we hope to do it in the
near future.

\subsection{Conclusions}

In this chapter we have accomplished a theoretical study of the Hall
effect in a system made of weakly coupled $1/2$-filled chains, in
the presence of umklapp scattering. We obtained a Hall coefficient
$\RH$ given by the free-fermion value (band value $\RH^0$) plus a
correction term with a power-law dependence on temperature (or
frequency), due to the presence of umklapp scattering. This power-law is a
characteristic behavior of Luttinger liquids, where the exponent
depends on the interaction parameters. The Hall constant was also
computed for the system without forward scattering, resulting in a
logarithmic dependence on $T$ (or $\omega$), in agreement with the
zero interaction limit of the power-law. The results are not
directly applicable to the Hall data in the quasi 1D organic
conductors, reviewed at the beginning of this chapter, because
these are $1/2$- and $1/4$-filled compounds, but they allowed us
to reach the following conclusion: signatures of LL behavior
(power-law dependence) are expected to appear, at high
temperatures, in Hall measurements made in the geometry considered
here.







\chapter{Hall effect on the strongly correlated 2D triangular
lattice}\label{chap:Hall_triangular}




As we have seen in the preceding chapters, the interpretation of
the Hall effect in strongly correlated systems can result in a
very complicated task. Interactions can have a large effect in the
Hall resistivity, and this effect seams to increase as the
dimensionality of the system decreases. Furthermore, the
understanding of these effects is crucial for the investigation of
transport properties in strongly correlated systems, in particular
for the Hall effect.

We now know that there exist a variety of strongly correlated
systems with different geometries. Among these, the triangular
lattice exhibits a unique property: it has the smallest possible
closed loop with an odd number of steps (namely $3$), as explained
in Sec.~\ref{sec:Hubbard_triangular}. Anderson proposed that the
model could have a spin-liquid ground state at commensurate
fillings such as one electron per site
\cite{anderson_hgtc_hubbard}. These peculiarities make the
triangular lattice a very interesting system for the investigation
of the Hall effect. In particular, important differences between
the Hall effect in the square and triangular lattices were pointed
out by Shastry {\it{et al.}} in Ref.~\cite{Shastry_Hall}. Since
we have already studied the Hall effect in a quasi 1D system,
which can be also considered as a highly anisotropic square
lattice, the investigation of the same phenomena in the triangular
lattice can be a helpful way to understand the relation between
the Hall resistivity and the geometry of the underlying lattice in
two-dimensional systems.

There exist a variety of compounds with structures resembling the
triangular lattice. Among these the CoO$_2$ layered compounds
where the recent discovery of superconductivity in the hydrated
Na$_{x}$CoO$_2$ \cite{Takada_nature} have motivated a large number
of works. These materials, also known as cobaltates, are good
realizations of an isotropic 2D triangular lattice and have been
extensively investigated, both experimentally
\cite{Qian_cobaltates,laverock_cobaltates,Hasan_ARPES} and
theoretically \cite{Singh_band,Shastry_triangular_TJ_model,
Haerter_Peterson_Shastry}, in the past years. There are also
organic conductors of the BEDT (bis(ethylenedithio)) family
\cite{Seo_chemrev} where one finds various structures resembling
the anisotropic triangular system. This anisotropy, together with
the dimerization of the molecules present in some compounds of the
BEDT family, make these materials much more complicated to
describe from the theoretical point of view. That is why we will
concentrate here exclusively in the Na$_{x}$CoO$_2$ compound.

In the following we make a review of the most important properties
of Na$_{x}$CoO$_2$, which is the material considered in this work
for the application of our theoretical study of the Hall effect on
the two-dimensional triangular lattice.

\section{A triangular lattice compound: Na$_{x}$CoO$_2$}

\begin{figure}[tb]
\begin{center}
\includegraphics[width=5.5cm]{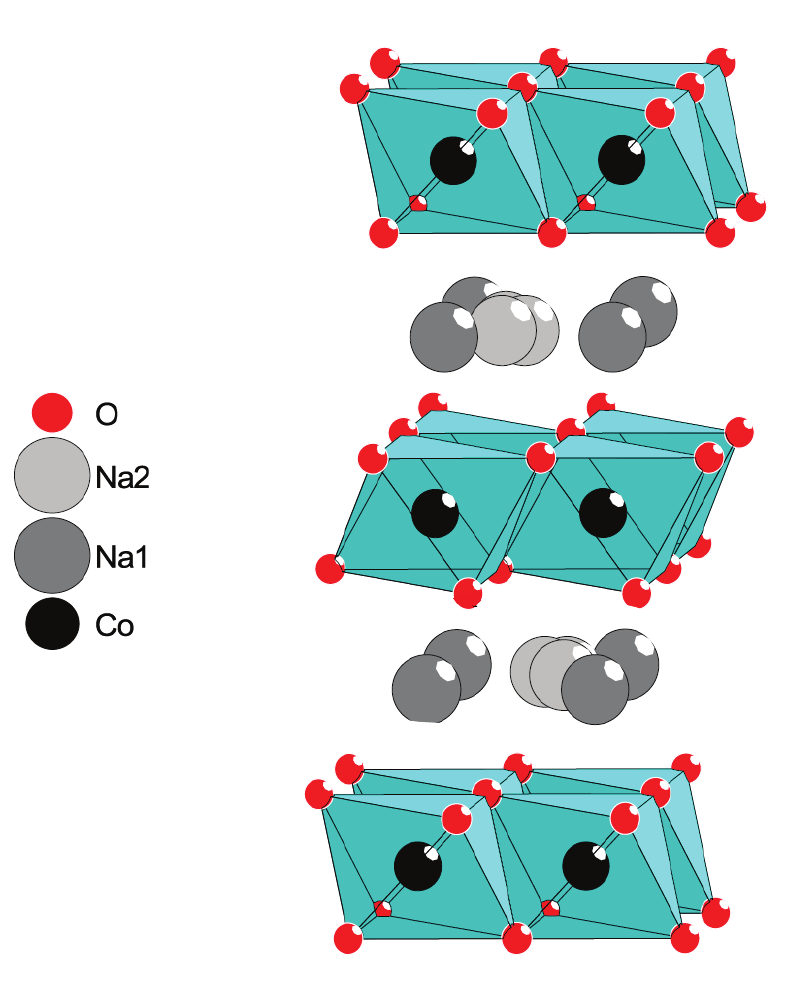}
\caption{\label{fig:cobaltate_structure} Molecular structure of
$\gamma$-Na$_{x}$CoO$_2$ phase with a space group symmetry of
$P6_3/mmc$ and lattice constants $a=2.84$ \AA$\,$ and $c=10.81$
\AA. The CoO$_2$ layers are intercalated with insulating layers of
Na$^+$ ions. Na$_1$ and Na$_2$ differ from their crystallographic
positions: Na$_1$ is situated on the vertical crossing Co atoms
above and below, and Na$_2$ is slightly displaced with respect to
the vertical crossing the center of the triangle formed by the Co
atoms. From Ref.~\cite{Vaulx_thesis}}
\end{center}
\end{figure}
The crystal structure of Na$_{x}$CoO$_2$ consist on
two-dimensional CoO$_2$ layers of edge-sharing tilted octahedra.
Each octahedra is composed by a Cobalt ion surrounded by six
Oxygen atoms at the vertices CoO$_6$. Within each CoO$_2$ layer,
the Co ions occupy the sites of a two-dimensional triangular
lattice. The CoO$_2$ layers are separated by insulating layers of
Na$^+$ ions. There exist four phases of Na$_{x}$CoO$_2$, with
slightly different structures, called $\alpha$, $\alpha'$, $\beta$
and $\gamma$. They differ by the stacking order of CoO$_2$ layers
and Na-O environments \cite{Son_NaxCoO2}. In this work we will
focus only on the $\gamma$ phase because it is the one used in the
Hall measurements. Fig.~\ref{fig:cobaltate_structure} shows the
structure of the $\gamma$-Na$_x$CoO$_2$, which has an hexagonal
structure with a space group symmetry of $P6_3/mmc$ and lattice
constants $a=2.84$ \AA$\,$ and $c=10.81$ \AA. Band-structure
calculations \cite{Singh_band} show that the O $2p$ orbital states
lie far below the Co $3d$ states and the chemical potential falls
within the band formed from $t_{2g}$ states in Co. Hence the
electrons donated by the Na ions are distributed among the Co
ions, a fraction ($\delta$) of which are in the Co$^{4+}$ state in
which $S=1/2$, while the rest $(1-\delta)$ are Co$^{3+}$ with $S=0$.
The elementary charge-transport process is the hopping of a hole
from Co$^{4+}$ to Co$^{3+}$. A large on-site repulsion excludes
double occupancy of a site by the holes.

\begin{figure}
\begin{center}
\includegraphics[width=8.5cm]{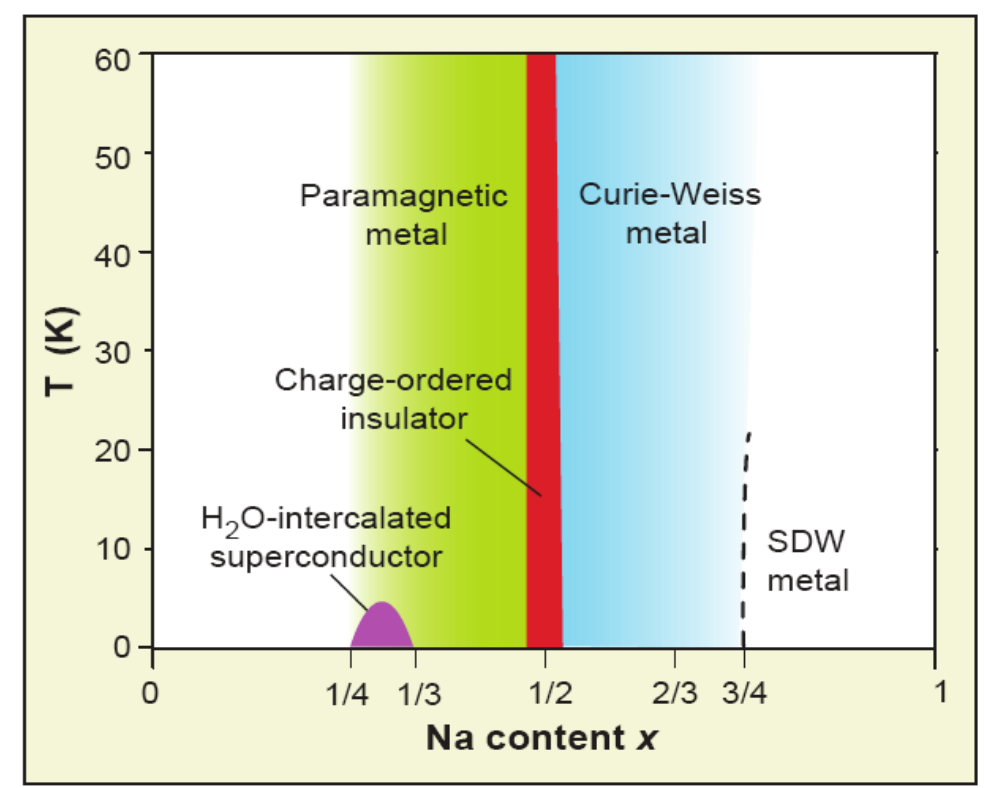}
\caption{\label{fig:cobaltates_phase} Phase Diagram of
Na$_{x}$CoO$_2$. The different order states are found changing the
doping of sodium atoms ($x$). From Ref.~\cite{Ong_Cava}}
\end{center}
\end{figure}

Different ordered states appear as a function of the doping $x$ of
Na ions. Fig.~\ref{fig:cobaltates_phase} shows the phase diagram
for Na$_{x}$CoO$_2$. A paramagnetic metal is found for
$x<\frac{1}{2}$ and for $x>\frac{1}{2}$ the system behaves as a
Curie-Weiss metal \cite{Foo_phase_diagram}. These two metallic states are separated by a
narrow charge-ordered insulating state at $x=\frac{1}{2}$. This
compound becomes superconducting below $5$K, when intercalated
with water forming Na$_{x}$CoO$_{2}\cdot y$H$_2$O, for
$\frac{1}{4}<x<\frac{1}{3}$. For values of the doping above
$\frac{3}{4}$, a Spin Density Wave metallic state appears
\cite{Foo_phase_diagram,Ong_Cava}. Na$_x$CoO2 presents also an
unusual enhanced thermopower at $x\sim 2/3$ which has been
recently related to the spin entropy carried by the holes in the
Curie-Weiss phase \cite{Wang_Hall}. Na$_{0.7}$CoO$_2$ is the host
compound from which Na$_x$ is varied to achieve superconductivity
in the hydrated compound.

The effective Hubbard interaction $U$ have been estimated in
band-structure calculations \cite{Singh_band} to be of the order
of $U\sim5$-$8$ eV and a Fermi surface with hexagonal character in
agreement with photoemission measurements
\cite{Hasan_ARPES,laverock_cobaltates}. The dispersion behavior
seen in photoemission data is consistent with a {\it{negative}}
sign of the single-particle hopping of the order of $t=10\pm2$ meV
\cite{Hasan_ARPES}. Therefore, the bandwidth (W) is estimated
between $70$ to $100$ meV making this system a real strongly
correlated one with $U\gg W$.
\begin{figure}
\begin{center}
\includegraphics[width=12cm]{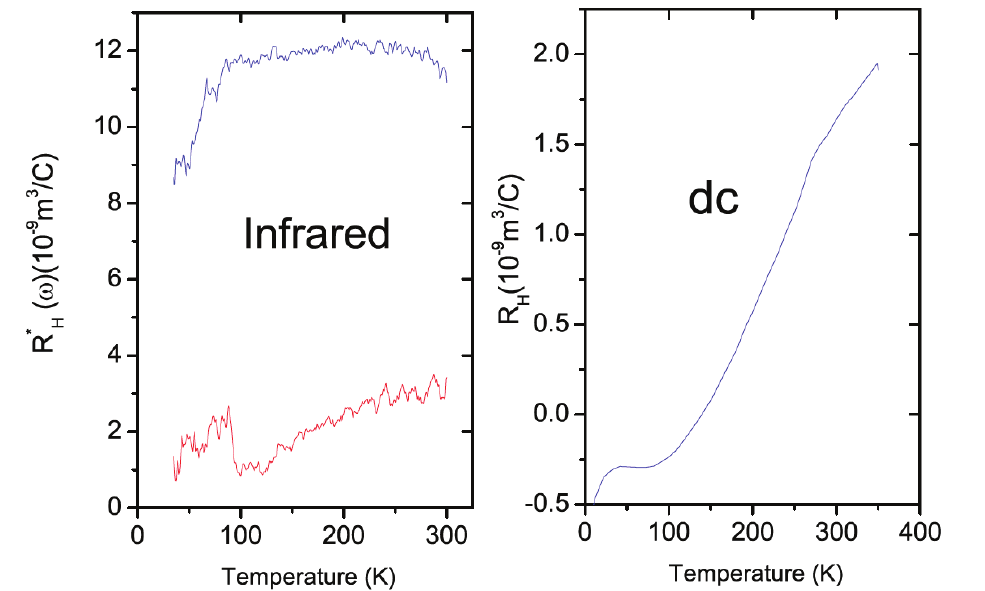}
\caption{\label{fig:Infrared_Hall} Temperature dependence of the
Hall constant measured in Na$_{0.7}$CoO$_2$ by Choi {\it{et. al.}}
\cite{Choi_Infrared}. (Left) Infrared Hall constant measured at
$\omega=1100$ cm$^{-1}$. The upper (lower) curve correspond to the
real (imaginary) part of $\RH(\omega)$ . (Right) dc Hall constant
($\omega=0$). The scale on the left differs from the one shown in
Ref.~\cite{Choi_Infrared} because it was corrected by the
authors.}
\end{center}
\end{figure}
Several Hall measurements have been undertaken in
Na$_{0.7}$CoO$_2$ \cite{Wang_Hall, Choi_Infrared}. The anomalous
linear increase of the dc Hall coefficient and a recent infrared
Hall measurement \cite{Choi_Infrared} have motivated recent
theoretical works \cite{Shastry_triangular_TJ_model,
Haerter_Peterson_Shastry, Shastry_Curie_Weiss, Koshibae_Hall} with
the aim of investigating the role of correlations in the Hall
effect, but many questions remain open.
Fig.~\ref{fig:Infrared_Hall} shows the Hall constant measured by
Choi {\it{et. al.}} \cite{Choi_Infrared} at an infrared frequency
and at $\omega=0$.

This chapter is the object of publication \cite{leon_triangular}.
In the following we make a theoretical study of the Hall effect in
a 2D triangular lattice where electrons interact via an onsite
Coulomb repulsion $U$. We calculate $\RH$ in the high frequency
limit studied in chapter~\ref{chap:Hall_effect}, and we cover the
whole range of interaction values using several approximation
schemes.


\section{Model and methods}

\begin{figure}[tb]
\begin{center}
\includegraphics[width=8cm]{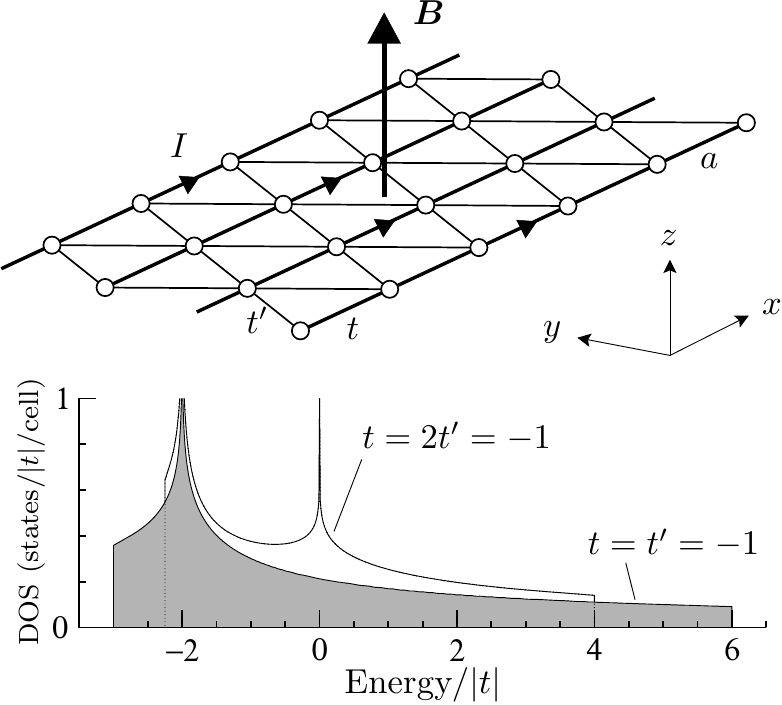}
\caption{\label{fig:model_triangular} Top: Two-dimensional
triangular lattice. $a$ is the lattice parameter, $t$ and $t'$ are
the hopping amplitudes for bonds along the $x$ direction and for
$\pm 60^\circ$ bonds, respectively. The unit-cell area is
$S=a^2\sqrt{3}/2$. The current $I$ flows along the $x$ axis, the
magnetic field $\vec{B}$ is applied along the $z$ axis, and the
Hall voltage is measured along the $y$ axis. Bottom:
Non-interacting density of states of the model in the cases
$t=t'=-1$ and $t=2t'=-1$, presented in
Sec.~\ref{sec:Hubbard_triangular}.}
\end{center}
\end{figure}

For this theoretical study, we consider an anisotropic triangular
lattice with nearest-neighbor hopping amplitudes $t$ and $t'$ and
an on-site Hubbard interaction $U$, with the structure sketched in
Fig.~\ref{fig:model_triangular}. The system is described by the Hubbard model
on the triangular lattice studied in Sec.~\ref{sec:Hubbard_triangular}:
\begin{equation}\label{H}
\mathcal{H}=- \sum_{\langle
ij\rangle\sigma}t_{ij}c^{\dagger}_{i\sigma}
c^{\phantom{\dagger}}_{j\sigma} + U\sum_i
n_{i\uparrow}n_{i\downarrow}
\end{equation}
where $c^{\dagger}_{\alpha}\,(c_\alpha)$ is the creation
(annihilation) fermion operator, $n_\alpha$ is the fermionic
number operator and $\langle ij\rangle$ are nearest-neighboring
sites. The dispersion relation for this model (see
Sec~\ref{sec:Hubbard_triangular}) is
\begin{equation}\label{dispertion}
\varepsilon_{\vec{k}} = -2t\cos(k_xa)
-4t'\cos(k_xa/2)\cos(k_ya\sqrt{3}/2).
\end{equation}

We assume that a current $I$ flows along the $x$ axis and a dc
magnetic field $\vec{B}$ is applied along $z$, hence a Hall
voltage develops along the $y$ axis (see
Fig.~\ref{fig:model_triangular}). We use the vector potential
$\vec{A}=\vec{A}^{\text{mag}}+\vec{A}^{\text{el}}$, where for the
magnetic part we choose, as in the previous chapter, the Landau
gauge $\vec{A}^{\text{mag}}=Bx\hat{\vec{y}}$, and
$\vec{A}^{\text{el}}$ describes the electric field. The coupling
between the lattice fermions and the electromagnetic field induces
a Peierls phase in the hopping amplitudes which change according
to $t_{ij}\to t_{ij}\exp(-ie\int_{i}^{j}\vec{A}\cdot\vec{dl})$.

We use Eqs.~(\ref{def_current}) and (\ref{diamagnetic}) for the
total current operator, $J_\mu=\int d\vec{r} j_\mu(\vec{r})$, and
the diamagnetic susceptibilities $\chi_\mu(0)$, respectively.
Performing the functional derivatives we find for the currents
\begin{subequations}\label{currents_triangular}
\begin{eqnarray}\nonumber
   J_x&=&ea\left[2t\sum_{\vec{k}\sigma}c^{\dagger}_{\vec{k}\sigma}c^{\phantom{\dagger}}_{\vec{k}\sigma}
    \sin(k_xa)+\right.\\ \label{Jx_triangular}
    &&+\left.t'\sum_{\vec{k}\sigma}\sin\left(\frac{k_xa}{2}+\frac{\eta a}{4}\right)
    \left(c^{\dagger}_{\vec{k}\sigma}c^{\phantom{\dagger}}_{\vec{k}+\vec{\eta}\sigma}
    e^{ik_y\sqrt{3}\frac{a}{2}}+\text{h.c.}\right)\right]\\ \label{Jy_triangular}
    J_y&=&-ea\sqrt{3}t'\sum_{\vec{k}\sigma}\cos\left(\frac{k_xa}{2}+
    \frac{\eta a}{4}\right)\left(ic^{\dagger}_{\vec{k}\sigma}c^{\phantom{\dagger}}_{\vec{k}
    +\vec{\eta}\sigma}e^{ik_y\sqrt{3}\frac{a}{2}}+\text{h.c.}\right),
\end{eqnarray}\end{subequations}
where we have defined the vector $\vec{\eta}=(\eta,0)$ with
$\eta=\sqrt{3}eBa/2$. The diamagnetic susceptibilities are:
\begin{subequations}\label{diamagnetic_terms}\begin{eqnarray}\nonumber
\chi_x(0)&=&-\frac{4e^2}{\sqrt{3}}\frac{1}{N}\sum_{\vec{k}}\Big[2t\cos(k_xa)
+t'\cos\left(\frac{k_xa}{2}\right)
\cos\left(k_y\sqrt{3}\frac{a}{2}\right)\Big]\langle n_{\vec{k}}\rangle\\
\label{Xx_triangular} \\
\chi_y(0)&=&-\frac{4\sqrt{3}e^2t'}{N}\sum_{\vec{k}}\cos\left(\frac{k_xa}{2}\right)
\cos\left(k_y\sqrt{3}\frac{a}{2}\right)\langle n_{\vec{k}}\rangle.
\end{eqnarray}\end{subequations}
As shown in Sec.~\ref{high_frequency_RH}, it is possible to
rewrite $\RH$ as a high-frequency series where the
infinite-frequency limit reads
\begin{equation}\label{RH_highw}
\RH(\omega\to\infty)=\lim_{B\rightarrow 0}\left(-\frac{i}{B
NS}\frac{\langle\left[J_x,J_y\right]\rangle}{\chi_x(0)\chi_y(0)}\right).
\end{equation}
The remaining contributions are expressed in terms of a memory
matrix (see Sec.~\ref{sec:Hall_memory}).
$\RH(\omega\to\infty)$ is expected to provide the dominant
contribution at any finite frequency. The memory matrix formalism
allows in principle to go beyond the infinite frequency
approximation and compute corrections at finite frequency
\cite{lange_hall_constant,leon_hall}. It leads, in particular, to
corrections due to interactions that vanish identically if $U=0$.
These corrections do not affect the sign of $\RH$ which should be
entirely determined by $\RH(\omega\to\infty)$. In the following we
shall consider only the infinite-frequency contribution to $\RH$,
Eq.~(\ref{RH_highw}), and adopt the notation
$\RH(\omega\to\infty)\equiv\RH$.

Strictly speaking, our results are valid provided the probing
frequency is larger than any other energy scale in the system,
$\omega > \max\lbrace U,t,T\rbrace$. The last two conditions,
$\omega > \max\lbrace t,T\rbrace$, are easily fulfilled
experimentally in known triangular compounds, while the condition
$\omega>U$ is more problematic. However, as we will discuss in
Sec. \ref{sec:discussion}, in certain limits our results coincide
with those obtained in Ref.~\cite{Shastry_triangular_TJ_model}
under the opposite assumption $\omega\ll U$, showing that this
condition is not stringent.

In order to evaluate Eq.~(\ref{RH_highw}), we calculate the
commutator $[J_x,J_y]$ from Eqs.~(\ref{currents_triangular}), and
we use the diamagnetic susceptibilities of
Eq.~(\ref{diamagnetic_terms}) to arrive at
\begin{eqnarray}\label{RH_AkBkCk}
\RH&=&\frac{S}{e}\frac{
\frac{1}{N}\sum_{\vec{k}}A_{\vec{k}}\langle n_{\vec{k}}\rangle}
{\frac{1}{N}\sum_{\vec{k}}B_{\vec{k}}\langle n_{\vec{k}}\rangle
\frac{1}{N}\sum_{\vec{k}}C_{\vec{k}}\langle n_{\vec{k}}\rangle},
\end{eqnarray}
with
\begin{eqnarray}\label{coeff}
\nonumber A_{\vec{k}}&=&\cos\left(\frac{k_xa}{2}\right)\cos(k_xa)
\cos\left(k_y\sqrt{3}\frac{a}{2}\right)\\ \nonumber
&&+\frac{1}{4}(t'/t)\left[\cos(k_xa)+\cos\left(k_y\sqrt{3}a\right)\right]\\
\nonumber
B_{\vec{k}}&=&2\cos(k_xa)+(t'/t)\cos\left(\frac{k_xa}{2}\right)
\cos\left(k_y\sqrt{3}\frac{a}{2}\right)\\
C_{\vec{k}}&=&\cos\left(\frac{k_xa}{2}\right)
\cos\left(k_y\sqrt{3}\frac{a}{2}\right).
\end{eqnarray}
As can be seen from Eq.~(\ref{RH_AkBkCk}) the high frequency Hall
coefficient depends only on the distribution function $\langle
n_{\vec{k}}\rangle$ as well as some geometrical factors. The
interaction term in Eq.~(\ref{H}) therefore only influences $\RH$
through its effect on $\langle n_{\vec{k}}\rangle$.
Another implication
of Eq.~(\ref{RH_AkBkCk}) is that at low temperature the behavior
of $\RH$ can be interpreted in terms of an effective carrier
concentration, as in the non-interacting case.

\section{Results}\label{sec:results}

In the following we evaluate $\RH$ in the whole domain of
interaction values $U$ with respect to the bandwidth $W=9|t|$ of
the system, by using four different approaches: exact calculation
at $U=0$, a perturbative expansion of the self-energy at
$U\lesssim W$, a local approximation to the self-energy, treated
with dynamical mean field theory (DMFT) at $U\gtrsim W$, and
finally the atomic limit of the self-energy at $U\gg W$.

\subsection{Non-interacting case}\label{sec:non-int}
In the non-interacting case there are various limits in which we
can obtain analytical results for $\RH^0$ ($\RH$ at $U=0$): at
zero temperature and band fillings near $n=0$ and $n=2$, and at
high temperature $T\gg W$. For intermediate fillings and
temperatures, we compute $\RH^0$ numerically by performing the sum
in Eq.~(\ref{RH_AkBkCk}) on a dense $2048\times2048$ discrete
$\vec{k}$-point mesh. Some details on the numerical sum over
momentum are given in Appendix~\ref{app_numerical}.

\subsubsection*{Zero temperature}

Here we restrict for simplicity to the isotropic case $t'= t$ and
we set the lattice parameter $a=1$. Close to the band edges we can
expand the various integrands of Eq.~(\ref{coeff}) and thus
perform the $\vec{k}$ integrals.

Near the bottom of the band the Fermi surface is made of two
nearly circular electron pockets around $(\frac{4\pi}{3},0)$ and
$(\frac{2\pi}{3},\frac{2\pi}{\sqrt{3}})$. In each pocket we have
$\xi_{\vec{k}}\equiv\varepsilon_{\vec{k}}-\mu\approx3t-\frac{3}{4}tk^2-\mu$,
where $k$ is the momentum measured from the pocket center, and
therefore $k_{\text{F}}^2=\frac{4}{3}(3-\mu/t)$. The corresponding
electron density is $n=k_{\text{F}}^2/\pi$. Writing similar
expansions of $A_{\vec{k}}$, $B_{\vec{k}}$, and $C_{\vec{k}}$
close to the pocket center and performing the Brillouin zone
integrations, we obtain the non-interacting Hall coefficient at
low electron density:
\begin{equation}\label{RH0_T0_1}
\RH^0(T=0)=\frac{1}{ne}\left[1-\frac{3\pi
n}{8}+\mathcal{O}(n^2)\right].
\end{equation}
At sufficiently low density we recover, in the above expression,
the classical result $\RH^0=1/ne$.

Near the top of the band the Fermi surface is a nearly circular
hole pocket centered at $\vec{k}=(0,0)$. Close to this point we
have $\xi_{\vec{k}}\approx-6t+\frac{3}{2}tk^2-\mu$, and therefore
$k_{\text{F}}^2=\frac{2}{3}(6+\mu/t)$. The corresponding density
is obtained by subtracting the contribution of the hole pocket
from the maximum density: $n_h=2-k_{\text{F}}^2/2\pi$. Similarly,
for the functions $A_{\vec{k}}$, $B_{\vec{k}}$, and $C_{\vec{k}}$
we have to subtract the contribution of the hole pocket from the
contribution of the whole Brillouin zone, which turns out to be
zero because
\begin{equation}\label{eq:zero}
\sum_{\vec{k}}A_{\vec{k}}=\sum_{\vec{k}}B_{\vec{k}}=\sum_{\vec{k}}C_{\vec{k}}=0.
\end{equation}
Thus, for low hole densities $n_h=2-n$ we find that the
non-interacting Hall coefficient is given by
\begin{equation}\label{RH0_T0_2}
\RH^0(T=0)=-\frac{1}{n_he}\left[1-\left(\frac{\pi
n_h}{4}\right)^2+\mathcal{O}(n_h^3)\right],
\end{equation}
and as $n_h\to0$ we have $\RH^0=-1/n_he$.

\begin{figure}
\begin{center}
\includegraphics[width=8cm]{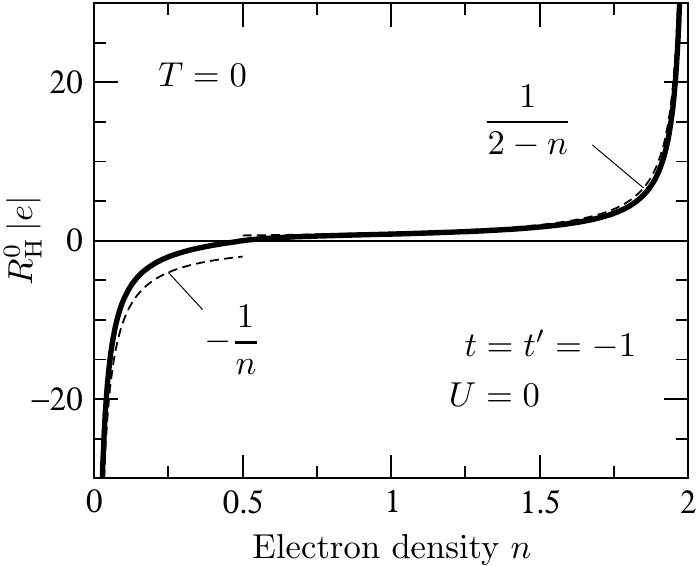}
\caption{\label{fig:RH0_T0} Non-interacting Hall coefficient
$\RH^0$ at zero temperature as a function of the electron density
$n$, for an isotropic triangular lattice with $t'=t=-1$. The
dashed lines indicates the classical behavior at low electron and
hole carrier densities.}
\end{center}
\end{figure}

The complete density dependence of $\RH^0$ calculated numerically
at zero temperature from Eq.~(\ref{RH_AkBkCk}) is displayed in
Fig.~\ref{fig:RH0_T0} and compared to the limiting cases
Eqs.~(\ref{RH0_T0_1}) and (\ref{RH0_T0_2}). It is clear from this
figure that the infinite-frequency $\RH$ follows the well-known
dependence of the dc Hall coefficient $\RH(\omega=0)$ on the
carrier charge density. This indicates a weak frequency dependence
of the non-interacting Hall coefficient at zero temperature, since
the dc result is recovered from the infinite frequency limit of
$\RH$. Furthermore this suggests, as we will discuss in more
details below, that the frequency dependence should not be too
crucial, even in the presence of interactions, for most band
fillings. At $U=0$ the sign of the Hall coefficient is entirely
given by the sign of the carriers, and it can be seen from
Fig.~\ref{fig:RH0_T0} how the sign changes at quarter filling when
the Fermi energy crosses the van Hove singularity of the DOS, and
the Fermi surface shape evolves from electron to hole like.

\subsubsection*{High temperature}\label{section:RH0_highT}

\begin{figure}
\begin{center}
\includegraphics[width=8cm]{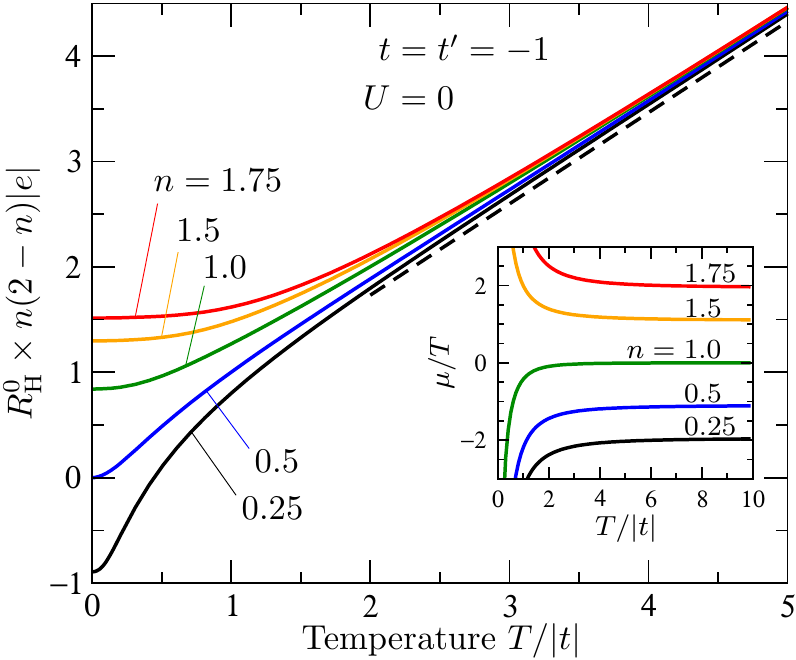}
\caption{\label{fig:RH_highT} Temperature and electron density
dependence of the non-interacting Hall coefficient $\RH^0$, for
the isotropic triangular lattice with $t=t'=-1$. The dashed line
shows the asymptotic behavior described by Eq.~(\ref{RH-T}).
Inset: Temperature and density dependence of the chemical
potential $\mu$, illustrating the relation $\mu\sim T$ at high
temperature. }
\end{center}
\end{figure}

If $T\gg t$ the distribution function $\langle
n_{\vec{k}}\rangle$, which reduces to the Fermi distribution at
$U=0$, can be expanded in power of $\beta=1/T$. This expansion
must be done at constant density $n$, which requires that
$\beta\mu$ remains finite as $\beta\to 0$, in other words $\mu\sim
T$ at high temperature. Taking this into account we can deduce the
relation between $\mu$ and $n$, $\exp(-\beta\mu)=2/n-1$, and write
the Fermi distribution as
\begin{equation}\label{expand}
\langle
n_{\vec{k}}\rangle=\frac{n}{2}-n(2-n)\varepsilon_{\vec{k}}\frac{\beta}{4}
 + \mathcal{O}(\beta^2).
\end{equation}
Due to Eq.~(\ref{eq:zero}) the $\vec{k}$-independent terms in
Eq.~(\ref{expand}) do not contribute to $\RH^0$, which in this
case takes the form:
\begin{equation}
\RH^0(T\gg t) =
-4T\frac{S}{e}\frac{1}{n(2-n)}\frac{\frac{1}{N}\sum_{\vec{k}}
A_{\vec{k}} \varepsilon_{\vec{k}}}{\frac{1}{N}\sum_{\vec{k}}
B_{\vec{k}}\varepsilon_{\vec{k}} \frac{1}{N}\sum_{\vec{k}}
C_{\vec{k}}\varepsilon_{\vec{k}}}.
\end{equation}
Performing the Brillouin zone integrations we obtain
\begin{equation}\label{RH-T}
\RH^0(T\gg
t)=\frac{T/t}{e}\frac{1}{n(2-n)}\frac{a^2\sqrt{3}}{2}\frac{3}{2+(t'/t)^2}.
\end{equation}
This result is plotted in Fig.~\ref{fig:RH_highT} together with
the numerically calculated full temperature and density
dependence. The most striking feature of Eq.~(\ref{RH-T}) is the
linear increase of $\RH^0$ with $T$. The same linear behavior was
obtained in Ref.~\cite{Motrunich_Lee} at $\omega=0$, indicating a
weak frequency dependence of $\RH^0$ at high temperature. Our
result shows that the $T$-linear dependence of $\RH$ is not due to
interactions but to the peculiar topology of the triangular
lattice. The sign of $\RH^0$ at high $T$ is determined by the sign
of $t$, irrespective of the density (see Fig.~\ref{fig:RH_highT}).
We attribute this property to the fact that at high enough
temperature the full band contributes to the Hall effect; hence
the sign of $\RH$ reflects the dominant nature, electron or
hole-like, of the band. As is clear from Fig.~\ref{fig:model_triangular}, for
$t<0$ the band is dominantly hole-like, while for $t>0$ it is
electron-like.

The relevance of result (\ref{RH-T}) is that even without
interactions, the Hall coefficient has a linear dependence at high
temperature due to the geometry of the system, emphasizing the
peculiarity of the triangular lattice. By contrast, on the square
lattice the same analysis yields a $T$-independent non-interacting
$\RH^0=\frac{2}{e}\left[\frac{1}{n}-\frac{1}{n(2-n)}\right]$ at
high temperature.

\subsection{Weakly interacting regime}

\begin{figure}
\begin{center}
\includegraphics[width=8cm]{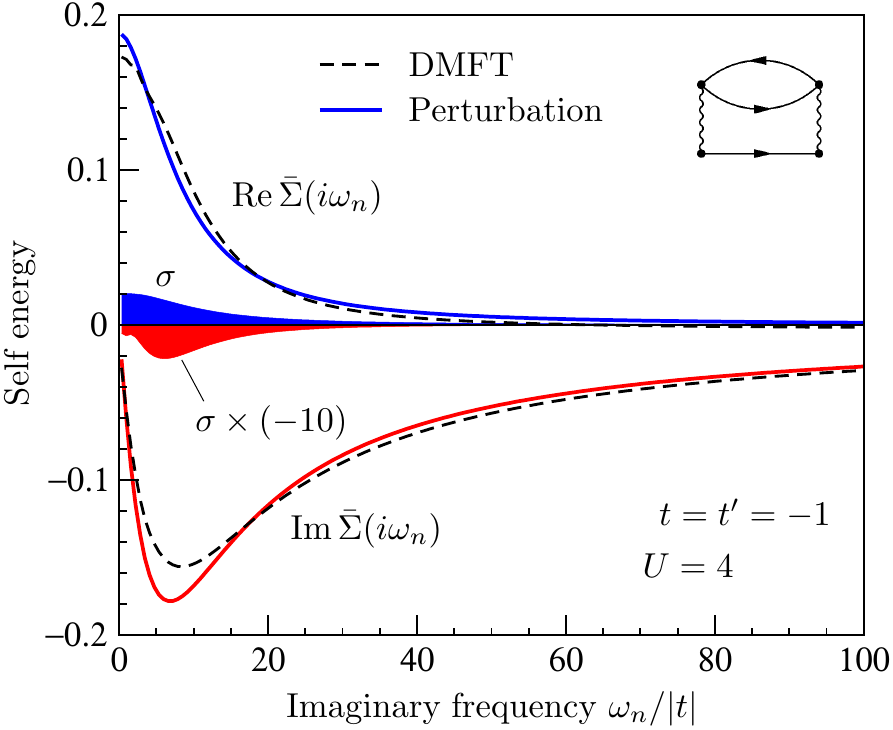}
\caption{\label{fig:Im_self} Brillouin zone average of the real
and imaginary parts of the self-energy Eq.~(\ref{self-E}) at low
temperature $T=0.1$, calculated using a $64\times64$
$\vec{k}$-point mesh (solid lines). The small standard deviations
$\sigma$ (shaded curves) illustrate the weak momentum dependence
of $\Sigma(\vec{k},i\omega_n)$. The dashed lines show the local
self-energy resulting from the DMFT calculation (see
Sec.~\ref{section:DMFT}). The density was set to $n=1.54$, which
is the value for Na$_{0.7}$CoO$_2$ (see
Sec.~\ref{sec:discussion}). Inset: Feynman diagram corresponding
to Eq.~(\ref{self-E}).}
\end{center}
\end{figure}

When interactions are present, the distribution function $\langle
n_{\vec{k}}\rangle$ can be expressed in terms of the one-electron
self-energy $\Sigma(\vec{k},i\omega_n)$ as:\cite{mahan_book}
\begin{equation}\label{n_k}
\langle
n_{\vec{k}}\rangle=\frac{1}{\beta}\sum_{\omega_n}\frac{e^{i\omega_n0^+}}{i\omega_n-\xi_{\vec{k}}-\Sigma(\vec{k},i\omega_n)},
\end{equation}
with $\omega_n=(2n+1)\pi T$ the odd Matsubara frequencies. In the
weak coupling regime $U\lesssim W$, we evaluate the self-energy
using conventional perturbation theory in $U$ and we keep only the
lowest order contributions of order $U^2$. For a local interaction
like the Hubbard term in Eq.~(\ref{H}) there is only one diagram
which is drawn in Fig.~\ref{fig:Im_self}. The standard
diagrammatic rules yield the following expression for the
self-energy:
\begin{eqnarray}\label{self-E}
&&\Sigma(\vec{k},i\omega_n)=-\frac{U^2}{N^2}\sum_{\vec{k}_1\vec{k}_2}\\\nonumber
&&\frac{f(\xi_{\vec{k}_2})\left[f(\xi_{\vec{k}_1})-f(\xi_{\vec{k}+\vec{k}_1-\vec{k}_2})\right]-f(\xi_{\vec{k}_1})
f(-\xi_{\vec{k}+\vec{k}_1-\vec{k}_2})}{i\omega_{n}+\xi_{\vec{k}_1}-\xi_{\vec{k}_2}-\xi_{\vec{k}+\vec{k}_1-\vec{k}_2}}
\end{eqnarray}
where $f(\xi_{\vec{k}})$ is the Fermi distribution function.

The numerical evaluation of Eq.~(\ref{self-E}) is demanding due to
the double momentum integration. This is particularly time
consuming because our calculations are done at fixed density, and
thus require to calculate $\Sigma(\vec{k},i\omega_n)$ many times
in order to determine the chemical potential. However it turns out
that the momentum dependence of $\Sigma(\vec{k},i\omega_n)$ in
Eq.~(\ref{self-E}) is weak. This is illustrated in
Fig.~\ref{fig:Im_self} where we plot the Brillouin zone average of
the self-energy, $\bar{\Sigma}(i\omega_n)$, as well as its
standard deviation. The weak momentum dependence allows us to
compute $\Sigma(\vec{k},i\omega_n)$ on a coarse (typically
$16\times16$) $\vec{k}$-point mesh, and then to interpolate using
splines (see Appendix~\ref{app_numerical}) onto a dense mesh for
the evaluation of $\langle n_{\vec{k}}\rangle$ and eventually
$\RH$. The Matsubara sum in Eq.~(\ref{n_k}) also requires special
attention: the formal regularization of the divergence through the
exponential factor is not suitable for a numerical evaluation of
the sum. We therefore rewrite Eq.~(\ref{n_k}) as
\begin{equation}\label{n_k_2}
\langle
n_{\vec{k}}\rangle=\frac{1}{2}+\frac{1}{\beta}\sum_{\omega_n}\left(\frac{1}{i\omega_n-\xi_{\vec{k}}-\Sigma(\vec{k},i\omega_n)}-\frac{1}{i\omega_n}\right).
\end{equation}
The $\omega_n$ sum is now convergent and can be efficiently
calculated via the truncation at some large frequency and the
analytical evaluation of the remaining terms using an asymptotic
expansion of the self-energy.

The $\RH$ resulting from perturbation theory are valid in the
regime $ U < W\ll\omega$, with $W=9|t|$ the bandwidth of the
system. As already anticipated the effect of a small $U$ on the
distribution $\langle n_{\vec{k}}\rangle$ is a subtle broadening,
and as a result the dependence of $\RH$ on $U$ is very weak at low
$U$. Fig.~\ref{fig:RH_vs_U} provides an illustration of this weak
dependence. As a consequence the non-interacting results of
Sec.~\ref{sec:non-int} are expected to give a fairly good account
of the Hall effect for an interaction strength smaller than the
bandwidth $W$.

An important observation which we can make from our perturbative
calculations is that the momentum dependence of the self-energy is
very small, \textit{i.e.} the self-energy is almost local in real
space. This suggests to approach the strong-coupling regime
$U\gtrsim W$ by assuming that the self-energy is \emph{exactly}
local. In the following section we study such local approximations
to the self-energy, and we compare them to the result of the
perturbation theory.

\subsection{Strongly interacting regime}
Assuming that the self-energy is local in first approximation, we
investigate here two models for $\Sigma(i\omega_n)$ and their
implications for the Hall coefficient $\RH$. The first approach is
based on the dynamical mean field theory (DMFT)
\cite{georges_dmft} and requires to solve a difficult
self-consistent quantum impurity problem. Due to numerical
difficulties this method cannot be pushed to very high
interactions and/or very low temperature. Our second approach is
based on a simple analytical form for $\Sigma(i\omega_n)$, which
is expected to be valid at $U\gg W$, and allows us to express
$\langle n_{\vec{k}}\rangle$ analytically in this limit.

\subsubsection*{DMFT}\label{section:DMFT}

The DMFT approximation provides the exact solution of the problem
under the assumption that the self-energy is local. In this
framework the self-energy is expressed as:
\begin{equation}\label{sigma}
\Sigma(i\omega_n) =
\mathcal{G}_0^{-1}(i\omega_n)-\mathcal{G}^{-1}(i\omega_n)
\end{equation}
where $\mathcal{G}_0$ is an effective propagator describing the
time evolution of the fermions in the absence of interaction, and
$\mathcal{G}$ is the full propagator, which takes into account the
local Hubbard interaction. The calculation of $\mathcal{G}$ from a
given $\mathcal{G}_0$ amounts to solve the problem of a quantum
impurity embedded in a bath. We do it by means of the quantum
Monte Carlo Hirsh-Fye algorithm \cite{hirsch_fye_qmc} as described
in Ref.~\cite{georges_dmft} (see Appendix~\ref{app_DMFT}). From
the requirement that $\mathcal{G}$ coincides with the local
Green's function of the lattice, \textit{i.e.}
\begin{equation}
\mathcal{G}(i\omega_n)=\frac{1}{N}\sum_{\vec{k}}\frac{1}{i\omega_n-\xi_{\vec{k}}-\Sigma(i\omega_n)},
\end{equation}
one can deduce the self-consistency condition
\begin{equation}\label{eq:self-consistency}
\mathcal{G}_0^{-1}(i\omega_n)=1/\tilde{D}\left[i\omega_n-\Sigma(i\omega_n)\right]+\Sigma(i\omega_n),
\end{equation}
where $\tilde{D}(z)\equiv \int d\xi\,D(\xi)/(z-\xi)$ is the
Hilbert transform of the DOS $D(\xi)$ corresponding to the
triangular lattice and shown in Fig.~\ref{fig:model_triangular}. Once the
self-consistent $\mathcal{G}_0(i\omega_n)$ is obtained, the
corresponding self-energy $\Sigma(i\omega_n)$ is injected in
Eq.~(\ref{n_k_2}) to compute $\RH$.

\begin{figure}
\begin{center}
\includegraphics[width=8cm]{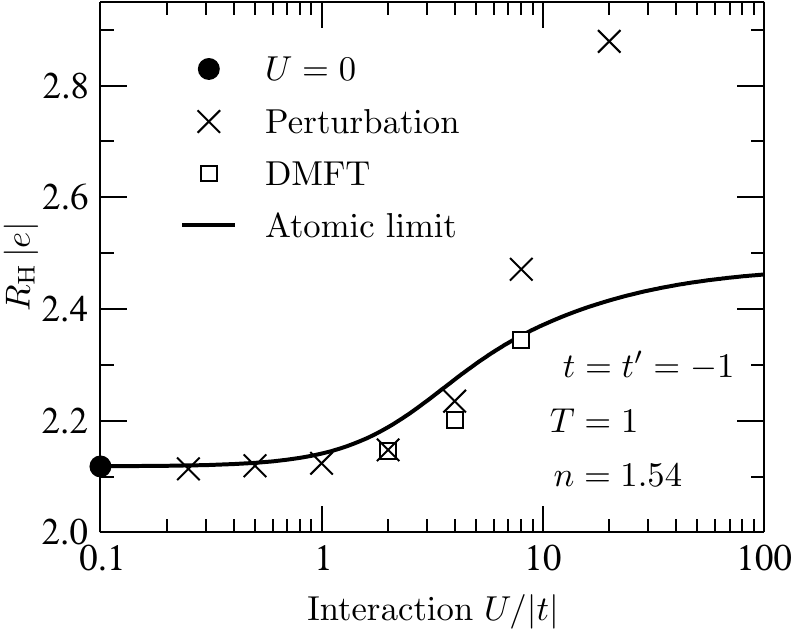}
\caption{\label{fig:RH_vs_U} Evolution of the high-frequency Hall
coefficient with $U$ calculated using different approximations at
$T=|t|$ and $n=1.54$, for an isotropic triangular lattice with
$t=t'=-1$.}
\end{center}
\end{figure}

In Fig.~\ref{fig:Im_self} we compare the DMFT self-energy with the
Brillouin zone average of the perturbative expression
Eq.~(\ref{self-E}), both calculated at $U=4$. It can be seen that
the frequency dependence and the order of magnitude of the two
quantities are very similar, suggesting that the self-energy is
dominated by the $U^2$ term and therefore the domain of validity
of the perturbation theory is not limited to very small $U$. On
the other hand it shows that the DMFT, although it is not a
perturbative approach, provides a smooth transition from the weak
to the strong-coupling regimes. This is further illustrated in
Fig.~\ref{fig:RH_vs_U} where we see that the values of $\RH$
calculated by perturbation theory and DMFT coincide up to
$U\approx4|t|$. At not too low temperature the DMFT calculation is
reliable up to interaction strengths comparable to the bandwidth
$W$. We have performed DMFT calculations at $U>W$, but since these
results could be affected by systematic statistical errors in the
Monte-Carlo summation, they are not shown in
Fig.~\ref{fig:RH_vs_U} (Appendix~\ref{app_DMFT}). At $U\gg W$ it
is expected that the DMFT result approaches the atomic limit in
which accurate calculations can be performed, as discussed in the
next paragraph.

\subsubsection*{Atomic limit}\label{atomic_limit}

In the limit of very strong interactions $U\gg W$ we assume that
the self-energy approaches its atomic limit given by the
expression: (see Appendix \ref{app_atomic}):
\begin{equation}\label{atomic-self}
\Sigma_{\text{at}}(i\omega_n)=\frac{nU}{2}+\frac{n/2(1-n/2)U^2}{i\omega_n+\mu_{\text{at}}-(1-n/2)U}
\end{equation}
with $\mu_{\text{at}}$ the chemical potential in the atomic limit,
not to be confused with the lattice chemical potential $\mu$.
Using this expression in Eq.~(\ref{n_k}) it is possible to
evaluate analytically the sum over Matsubara frequencies and thus
to obtain a closed expression for $\langle n_{\vec{k}}\rangle$
(Appendix \ref{app_atomic}). In Fig.~\ref{fig:RH_vs_U} we show the
Hall coefficient calculated with the atomic limit of the
self-energy in the whole range of interaction values. $\RH$
obviously converges to the non-interacting limit at low $U$ since
the atomic self-energy vanishes at $U=0$, and provides a good
interpolation between the weak and the strong-coupling regimes. At
intermediate values $U\sim W$ the atomic limit is not reliable,
although it gives the correct order of magnitude for $\RH$.
Fig.~\ref{fig:RH_vs_U} also shows that $\RH$ saturates at
sufficiently large $U$.

\begin{figure}
\begin{center}
\includegraphics[width=8cm]{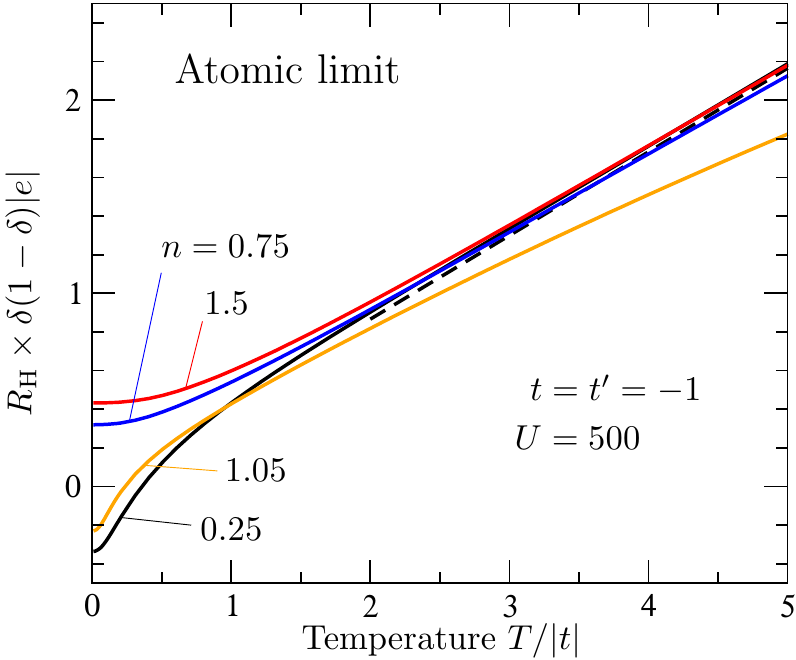}
\caption{\label{fig:atomic_limit} Temperature and density
dependence of the high frequency Hall coefficient calculated in
the atomic limit at $U=500|t|$. The dashed line shows the
asymptotic behavior, Eq.~(\ref{RHinf-T}), and $\delta =|n-1|$. For
densities close to half-filling ($n=1.05$), $\RH(T)$ deviates from
the asymptotic behavior (see text).}
\end{center}
\end{figure}

In Fig.~\ref{fig:atomic_limit} we display the temperature and
density dependence of $\RH$ at $U=500|t|$, which is a
typical value for the cobaltate compounds as discussed in the next
section. We have selected four densities corresponding to the
bottom and top of the lower and upper Hubbard bands (see also
Fig.~\ref{fig:hubbard_bands} below). Like for $U=0$ we find a
$T$-linear increase of $\RH$ at $T\gtrsim W$. Due to the Mott gap,
however, the density dependence of the slope is not the same as
for $U=0$. The slope can be obtained explicitly by sending $U$ to
$+\infty$ and performing the high-temperature expansion as in
Sec.~\ref{section:RH0_highT} 
The result is
\begin{equation}\label{RHinf-T}
        \RH^{U=\infty}(T\gg t)=\frac{T/t}{e}\frac{1}{\delta(1-\delta)}
        \frac{a^2\sqrt{3}}{4}\frac{3}{2+(t'/t)^2},
    \end{equation}
very similar to Eq.~(\ref{RH-T}) except that the slope
$\propto[4\delta(1-\delta)]^{-1}$ replaces $[2n(2-n)]^{-1}$, where
$\delta=|n-1|$ measures the departure from half-filling. The
$U=\infty$ result of Eq.~(\ref{RHinf-T}) is displayed in
Fig.~\ref{fig:atomic_limit}, and correctly describes our
high-temperature results at $U=500|t|$. The differences observed
at $n=1.05$ in Fig.~\ref{fig:atomic_limit} reflect the fact that
close to half-filling the slope of the high temperature $\RH$
depends strongly on the interaction and is not saturated even at
$U=500|t|$ (see also Fig.~7). Away from half-filling the $U$
dependence of the slope is weaker, and Eq.~(\ref{RHinf-T}) is valid for lower
interaction values.

\section{Discussion and perspectives}\label{sec:discussion}

The various approximations presented above allow us to calculate
the Hall coefficient on the triangular lattice for all
interactions strengths $U$ and all temperatures $T$. The main
limitation of our approach, in view of a comparison with
experimental systems, is that our results are in principle valid
in the limit $W,U \ll \omega$, because they are based on a
high-frequency expansion. The first criterion, $W \ll \omega$, is
not too difficult to satisfy for realistic compounds if the
measurement of the Hall effect is performed at optical
frequencies. The second criterion, $U\ll\omega$, seems more
problematic since interaction strengths can be as large as several
electron volts, at the upper edge of the mid-ultraviolet frequency
domain. However, we have seen (Fig.~\ref{fig:RH0_T0}) that at
$U=0$ and $T=0$ the Hall coefficient calculated at $\omega=\infty$
coincides with the $\omega=0$ dc value, and at $U=0$ and $T\gg t$,
we obtained the $\omega=0$ results of Ref.~\cite{Motrunich_Lee}.
All this suggests that the frequency dependence of $\RH$ is weak
in the non-interacting case.

\begin{figure}
\begin{center}
\includegraphics[width=8cm]{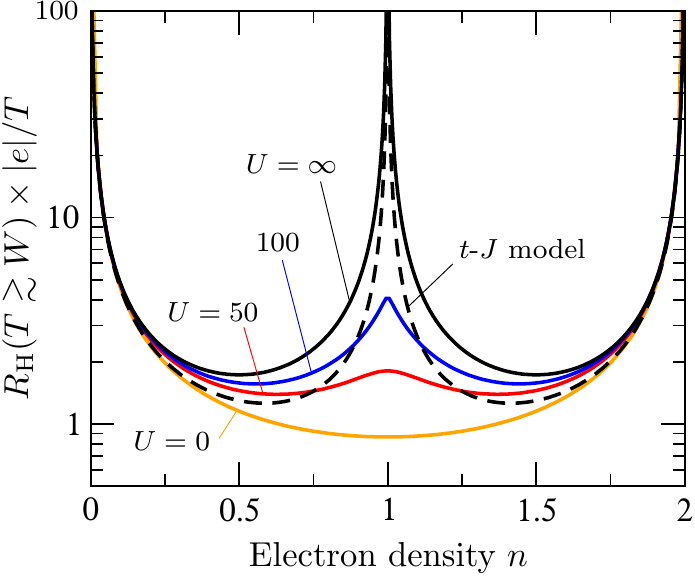}
\caption{\label{fig:comparison} Density dependence of the Hall
coefficient within the $T$-linear regime at $U=0$ and in the
atomic limit at $U\gg|t|$ (solid lines), compared to the result of
the $t$-$J$ model (Ref.~\cite{Shastry_triangular_TJ_model}, dashed
line). }
\end{center}
\end{figure}

At the other extreme of the parameter space, $U=\infty$ and $T\gg
W$, we can compare the result of the atomic limit approximation
with the result of the $t$-$J$ model
\cite{Shastry_triangular_TJ_model}. In the latter model $U$ is
considered infinite from the outset, so that the high-frequency
and high temperature expansion of
Ref.~\cite{Shastry_triangular_TJ_model} is in fact valid at
frequencies $\omega<U$. We plot in Fig.~\ref{fig:comparison} the
density dependence of $\RH$ obtained in both models at $U=\infty$
and  $T\gtrsim W$. The small quantitative difference between the
atomic limit at $U=\infty$ and the $t$-$J$ model shows that these
two ways of treating the $U=\infty$ limit are not equivalent: they
differ, in particular, in the renormalization of the kinetic
energy by the interaction. However the two models give away from 
half filling very similar behaviors. This reinforces the idea
that the frequency dependence of $\RH$ is weak. Exact
diagonalization on small clusters also indicate such a weak
frequency dependence \cite{Haerter_Peterson_Shastry}. This
strongly suggests that our results could also be valid at
$\omega<U$, and therefore be relevant to interpret experiments
performed in this regime.
The atomic-limit approach has the advantage to give access to the
full temperature dependence (Fig.~\ref{fig:atomic_limit}) as well
as the $U$-dependence as shown in Fig.~\ref{fig:comparison}, while
the calculation of Ref.~\cite{Shastry_triangular_TJ_model} is
valid at $U=\infty$ and $T\gg W$.

The evolution of $\RH$ with temperature is of particular interest
since it is most easily probed experimentally. A linear increase
of $\RH$ with temperature, without saturation at high $T$, was
reported in Ref.~\cite{Shastry_triangular_TJ_model} for the
$t$-$J$ model. Our results show that the Coulomb interaction is
not responsible for this effect which is also present at $U=0$
(Fig.~\ref{fig:RH_highT}) and is therefore a consequence of the
peculiar geometry of the triangular lattice. However the
interaction controls the density dependence of the slope which
changes smoothly from $[2n(2-n)]^{-1}$ at $U=0$ to
$[4\delta(1-\delta)]^{-1}$ at $U=\infty$. This is further
corroborated in Fig.~\ref{fig:comparison}.

\begin{figure}
\begin{center}
\includegraphics[width=8cm]{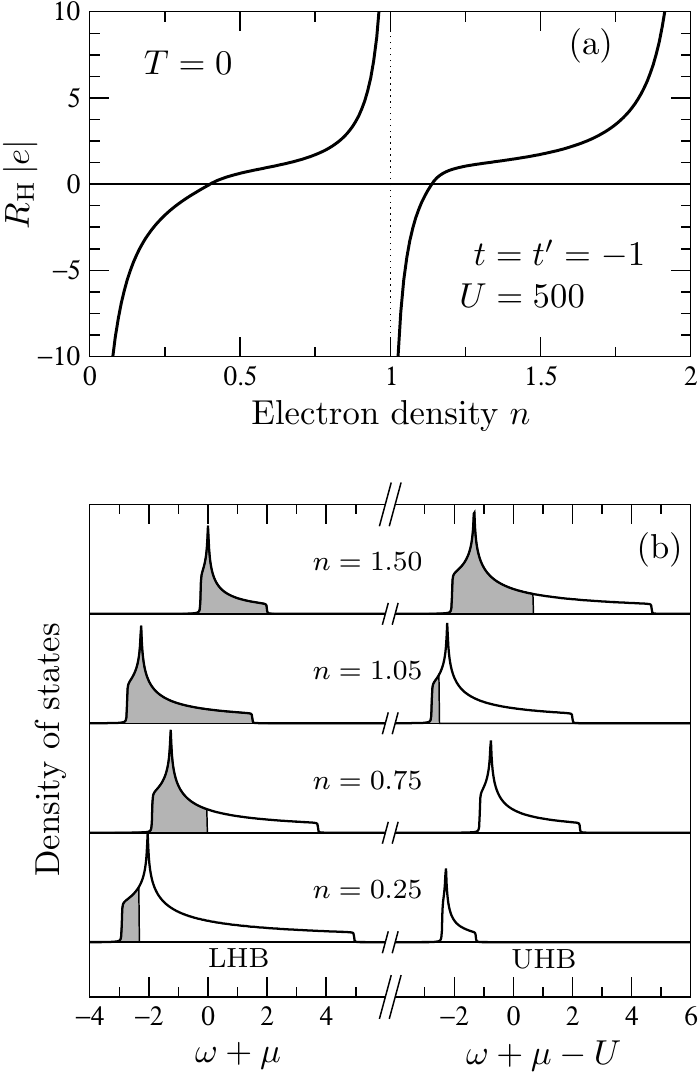}
\caption{\label{fig:hubbard_bands} (a) Density dependence of the
Hall coefficient at $T=0$ and $U=500|t|$ calculated in the atomic
limit approximation. (b) Density of states for various electron
densities, showing the lower (LHB) and upper (UHB) Hubbard bands.
The shaded regions indicate the occupied states and the position
of the chemical potential. }
\end{center}
\end{figure}

The sign of $\RH$ turns out to be independent of $n$ and $U$ at
high temperature, unlike in the square lattice where $\RH$ changes
sign at $n=1$. The situation is different at $T=0$. In the
non-interacting case $\RH$ changes sign at quarter filling and can
be simply interpreted in terms of the carrier density
(Fig.~\ref{fig:RH0_T0}). We have also investigated the $T=0$
density dependence of $\RH$ at large $U$ as shown in
Fig.~\ref{fig:hubbard_bands}. The interpretation in terms of the
carrier density remains qualitatively valid, provided one takes
into account the splitting of the DOS into the lower and upper
Hubbard bands. These two bands are displayed in
Fig.~\ref{fig:hubbard_bands}b, where it can also be seen that the
DOS keeps qualitatively the same shape as for $U=0$, but the width
of each band varies strongly with the density $n$. Due to this
band renormalization the sign change of $\RH$ at $n<1$ does not
occur at quarter filling, but a little below. Comparing
Fig.~\ref{fig:hubbard_bands}a with Fig.~\ref{fig:comparison} one
easily understands why the temperature dependence of $\RH$ is more
pronounced slightly above $n=0$ and $n=1$, where $\RH$ changes
from negative at $T=0$ to positive at high $T$, than slightly
below $n=1$ and $n=2$ where its stays positive (see also
Fig.~\ref{fig:atomic_limit}).

\begin{figure}
\begin{center}
\includegraphics[width=8cm]{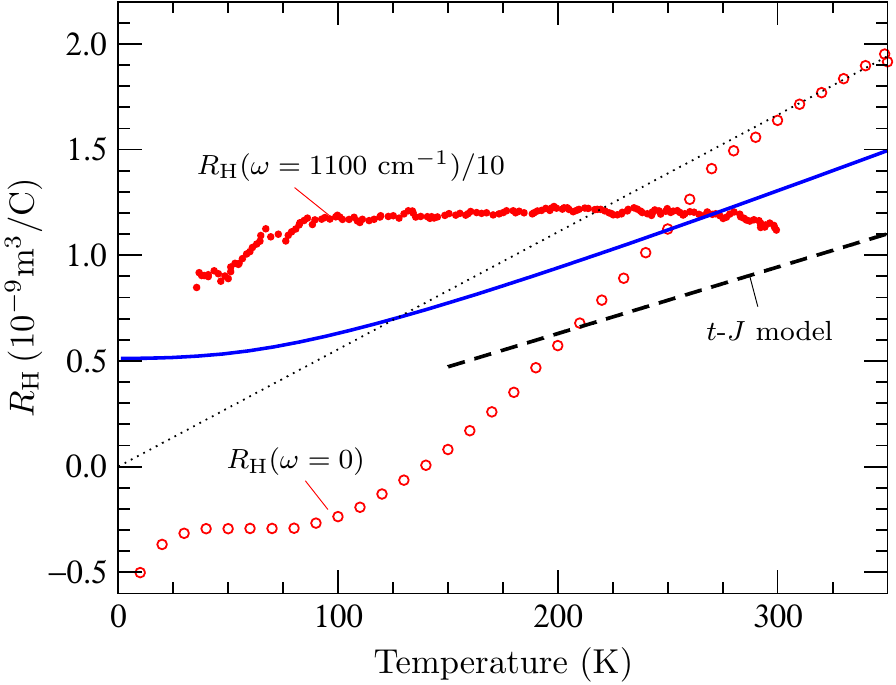}
\caption{\label{fig:cobaltates} Comparison of the Hall coefficient
in Na$_{0.7}$CoO$_2$ as measured by Choi \textit{et al.}
\cite{Choi_Infrared} at $\omega=0$ (empty circles) and
$\omega=1100$~cm$^{-1}$ (full circles) with available theoretical
models. The blue solid line is the high-frequency result in the
atomic limit and the dashed line shows the high-temperature result
in the $t$-$J$ model \cite{Shastry_triangular_TJ_model}.
Theoretical curves are calculated at $t=t'=-10$~meV, $U=5$~eV, and
$n=1.54$. The dotted line shows the best linear fit of the
experimental data at high temperature (see text).}
\end{center}
\end{figure}

Let us now discuss the application of our theoretical results to
the Hall measurements performed at finite frequency by Choi
\textit{et al.} \cite{Choi_Infrared} on the cobaltate
Na$_{0.7}$CoO$_2$ (see Fig.~\ref{fig:Infrared_Hall}). As mentioned
before, ARPES measurements \cite{Hasan_ARPES} indicate that the
triangular lattice is isotropic with an estimated hopping
amplitude of $t=-10$~meV and an effective Hubbard energy $U\sim5$
eV. From the radius of the Fermi-surface hole pocket observed in
ARPES, $k_{\text{F}}=0.65\pm0.1$~\AA$^{-1}$, we deduce an electron
density $n=1.54$. Choi \textit{et al.} measured the temperature
dependence of both the dc and ac Hall coefficients up to room
temperature (see Fig.~\ref{fig:Infrared_Hall}). The ac measurement
was performed at $\omega=1100$~cm$^{-1}\approx12|t|$. The
experimental conditions thus satisfy $T, W<\omega\ll U$. In
Fig.~\ref{fig:cobaltates} we plotted together the experimental and
theoretical curves. As can be seen in the figure, there is a
factor of $10$ between the measured $\RH$ at finite and zero
frequency. The order of magnitude of our theoretical curve,
obtained with the parameters measured with ARPES, is in agreement
with the dc Hall data and not with the ac Hall data, as expected.
We do not have, at the moment, any clear explanation for this
discrepancy. One would have to extend the theoretical approach in
order to cover the domain of intermediate frequencies and more
infrared measurements are needed in order to assure that the
experimental data is correct. In any case, we compare our results
with the dc Hall data at high temperatures and the results are
discussed below.

The behavior of the dc $\RH$ above $T=250$~K is consistent with
the linear increase predicted by the various theoretical models.
By adjusting these models on the dc experimental data at high
temperature (dotted line in Fig.~\ref{fig:cobaltates}) we obtain
independent determinations of the hopping amplitude $t$, namely
$t=-7.4$~meV using the atomic limit model Eq.~(\ref{RHinf-T}) and
$t=-5.7$~meV using the $t$-$J$ model. This values are in good
agreement with the ARPES results. We note, however, that there are
discrepancies between different sets of experimental data
\cite{Choi_Infrared, Wang_Hall}.

The organic conductors of the BEDT family present several
compounds with an anisotropic triangular structure. Unfortunately
we are not aware of any measurements of the ac Hall effect which
we could compare to our calculations, although measurements have
been done at zero frequency in these
materials.\cite{Sushko_Hall_organics, Katayama_Hall_organics}

\section{Conclusions}

The theoretical high-frequency Hall coefficient in the
two-dimensional triangular lattice exhibits two different
characteristic behaviors at low and high temperatures: near $T=0$,
$\RH$ resembles the classical dc Hall coefficient $1/qn^*$ where
$q$ and $n^*$ are the carrier charge and density, respectively; at
temperatures higher than the bandwidth, $\RH$ shows a remarkable
$T$-linear behavior with a density- and interaction-dependent
slope. These conclusions apply provided the probing frequency is
larger that the other energy scales of the problem, and that the
electron self-energy remains essentially local for strong
interactions.

Although we argued that the frequency dependence of $\RH$ is
probably weak, it is clear that for understanding the anomalously
large $\RH(\omega)$ measured experimentally in Na$_{0.7}$CoO$_2$
in the mid-infrared range, one would have to extend the approach
in order to cover the domain of intermediate frequencies.
Concerning the self-energy, we do not exclude that this quantity
could present a non-negligible momentum dependence for strong
interactions. This could affect the Hall coefficient, especially
at low temperature. Such a momentum dependence is indeed expected
for ordered ground states at and close to half-filling due to
antiferromagnetism. However, the influence of a momentum-dependent
self-energy would be reduced on the Hall effect, because the
expression Eq.~(\ref{RH_AkBkCk}) for the Hall coefficient averages
the distribution function over the Brillouin zone. In addition we
do not expect a strong momentum dependence far from half-filling,
as in the case of the $n=1.54$ cobaltate considered here.

Another possibility is that the simple one-band model considered
in this study would not suffice to capture the detailed properties
of the materials \cite{Weber_triangular}. Experiments conducted as
a function of $\omega$, as well as measurements of other materials
with a triangular structure, would be very helpful to elucidate
the peculiarities of the Hall effect in triangular compounds.

\chapter{Conclusions}

This work was devoted to the investigation of the Hall effect in
two different theoretical models of strongly correlated systems: a
system made of weakly coupled Luttinger liquids and the
two-dimensional triangular lattice. In order to have the necessary
theoretical tools to tackle these problems, we started with a
review on the main properties of systems with strong interactions,
focusing on two-dimensional and low-dimensional systems (1D and
quasi 1D). Because the Hall effect is first of all a transport
phenomenon, we reviewed the formalisms existing in the literature
to treat transport properties in strongly interacting systems, and
we dedicated a whole chapter to the specific treatment of the Hall
effect. The conclusions corresponding to each of the models
studied are presented in what follows.
\newline

The study of the Hall effect in a quasi 1D system was in part
motivated by various Hall measurements made on quasi 1D organic
conductors. This work consisted on the application of the memory
matrix formalism to obtain the Hall constant ($\RH$) in weakly
coupled $1/2$-filled chains, in the presence of umklapp
scattering. The geometry of the model was chosen in order to have
the current flowing along the 1D chains. We computed the
temperature and frequency dependence of $\RH$ taking into account
particle-particle interactions and particle-lattice interactions
(umklapp processes). We obtained a Hall coefficient $\RH$ given by
the free-fermion value (band value $\RH^0$) plus a correction term
with a power-law dependence on temperature (or frequency), due to
the presence of umklapp scattering. These power-law dependencies
are signatures of Luttinger liquid behavior, where the exponents
depend on the interaction parameters. The Hall coefficient was
also computed for the system without forward scattering
(particle-particle interaction), resulting in a logarithmic
dependence on $T$ (or $\omega$), in agreement with the zero
interaction limit of the power-law. Because the organic conductors
are $1/2$-filled and $1/4$- filled systems at the same time, our
theoretical results are not directly applicable to explain the
available Hall data, but they allowed us to make valuable
conclusions. Firstly, the proper way to analyze the Hall data,
made in such quasi-1D systems (in the same geometry of our model),
is to fit the \emph{deviations} from the band value starting from
the high temperature limit. And secondly, signatures of LL
behavior (power-law dependence) are expected to appear, at high
temperatures, in Hall measurements made in the geometry considered
here. A second geometry, where the magnetic field is applied along
the 1D chains, can be considered for this system and the question
of the appearance of LL signatures in this case remains open.
\newline

The second model studied in this work was the two-dimensional
triangular lattice with an onsite interaction $U$. This
theoretical study was partially motivated by Hall experiments made
at finite and zero frequency on the sodium cobalt oxide. We
computed the infinite-frequency Hall constant, where the probing
frequency must be larger that the other energy scales of the
problem, for all interactions strengths $U$ and all temperatures
$T$. We obtained an $\RH(\omega\to\infty)$ with two different
characteristic behaviors at low and high temperatures: near $T=0$,
$\RH$ resembles the classical dc Hall coefficient $1/qn^*$ where
$q$ and $n^*$ are the carrier charge and density, respectively; at
temperatures higher than the bandwidth, $\RH$ presented a
$T$-linear behavior with a density- and interaction-dependent
slope. These results were applied to the Hall data measured on
cobaltates at finite (infrared) and zero frequency. The finite
frequency data showed an order of magnitude of difference with our
theoretical results. To understand this discrepancy, one would
have to extend the approach in order to cover the domain of
intermediate frequencies. The behavior of the dc $\RH$ at high
temperature was consistent with the linear increase predicted
theoretically. By adjusting the model on the dc experimental data
at high temperature we obtain independent determinations of the
hopping amplitude in accordance with the experimental values. An
extension to the work done here can be the calculation of the
memory matrix term for the Hall constant to leading order in $U$.
This would give the behavior of the Hall constant at lower
frequencies.
\newline

After summarizing the conclusions corresponding to each one of the
models studied, we must point out that the study of the Hall
effect in strongly correlated system is a very complicated problem
from the theoretical point of view. Interactions can have a large
effect on the transport properties. In particular, for the Hall
resistivity these effects seem to increase when the dimensionality
of the system decreases. However, with this work we have learned
that there are feasible ways to treat this problem and that each
geometry must be treated in a different way. In any case, much
more needs to be done in order to understand form a theoretical
point of view the large number of experiments and the real meaning
of the Hall resistivity in strongly correlated systems.

\appendix

\chapter{Appendix for the study of the Hall effect in quasi 1D systems}




\small

\section{Correlator $\langle K_x;K_y\rangle$ at zero order in $\alpha$ and $B$}\label{app:alpha0_B0}

The operators $K_x, K_y$ and $\mathcal{H}_{\perp}$ are given in
boson representation in Eqs.~(\ref{Kx_Ky_bosonized}) and
(\ref{Hperp_bosonized}). Here we replace $b=+1(-1)$ by $r=+1(-1)$
for right(left) moving fermions in the expression for $K_y$. To
obtain the correlator at zero order in the band curvature term and
magnetic field, we take Eq.~(\ref{KKHcBC}) with
$\mathcal{H}_\alpha=0$ and we put $B=0$ in expressions
(\ref{Kx_Ky_bosonized}) and (\ref{Hperp_bosonized}),
\begin{equation}\label{average_expand_t}
 \langle K_x;K_y\rangle=-\int d\tau\,e^{i\Omega\tau}
 \int d\tau_1\left\langle T_{\tau}K_x(\tau)K_y(0)^{(0)}
\mathcal{H}_{\perp}^{(0)}(\tau_1)\right\rangle.
\end{equation}
We denote by $\left\langle \ldots \right\rangle$ the time-ordered
correlation function and the superindice $(0)$ means zero order in
$B$. All the terms are multiplied  by the factor: $\left(2e
v_{\text{F}}g_3/(2\pi a)^2\right)$ $\left(ie t_{\perp} g_3a_y/(2
\pi a)^2\right)$ $\left(t_{\perp}/2 \pi a \right)$. We will use
the letter $C_1$ to denote it in the following calculations. In
bosonization language, the average (\ref{average_expand_t}) gives
\begin{eqnarray}
&&= C_1\sum_{r, \sigma} \left\langle \left[ e^{i \sqrt{8} \phi_{\rho} (r_1)} -
   e^{- i \sqrt{8} \phi_{\rho} (r_1)} \right]_{{\color{red}  j}} \times \right.\\ \nonumber
&&\left[ e^{\frac{i}{\sqrt{2}} \left[ 3  r \phi_{\rho} ( r_2) - \theta_{\rho} ( r_2) +
   \sigma \left( -  r \phi_{\sigma} ( r_2) - \theta_{\sigma} ( r_2) \right)
   \right]_{{\color{red} j}}} e^{- \frac{i}{\sqrt{2}} \left[ -  r
   \phi_{\rho} ( r_2) - \theta_{\rho} ( r_2) + \sigma \left( -  r \phi_{\sigma}
   ( r_2) - \theta_{\sigma} ( r_2) \right) \right]_{{\color{red} j + 1}}} +
   h.c. \right] \\ \nonumber
&&\times\sum_{ r', \sigma'} \left[ e^{\frac{i}{\sqrt{2}} \left[  r'
   \phi_{\rho} ( r_3) - \theta_{\rho} ( r_3) + \sigma' \left(r'\phi_{\sigma}
   ( r_3) - \theta_{\sigma} ( r_3) \right) \right]_{{\color{red}  j}}} e^{-
   \frac{i}{\sqrt{2}} \left[r' \phi_{\rho} ( r_3) - \theta_{\rho} ( r_3) +
   \sigma' \left(  r' \phi_{\sigma} ( r_3) - \theta_{\sigma} ( r_3) \right)
   \right]_{{\color{red} j + 1}}} \right. \\ \nonumber
&&+\left.\left.e^{- \frac{i}{\sqrt{2}}\left[  r'\phi_{\rho} ( r_3) - \theta_{\rho} (r_3) + \sigma' \left(r' \phi_{\sigma}
   ( r_3) - \theta_{\sigma} ( r_3) \right) \right]_{{\color{red}  j}}}
   e^{\frac{i}{\sqrt{2}} \left[ r' \phi_{\rho} ( r_3) - \theta_{\rho} ( r_3) +
   \sigma' \left(  r' \phi_{\sigma} ( r_3) - \theta_{\sigma} ( r_3) \right)
   \right]_{{\color{red}j + 1}}} \right] \right\rangle
\end{eqnarray}
with $r_i=(x_i,u\tau_i)$ Due to result (\ref{condition_A_B}), the
surviving terms are those satisfying $\sum_i A_i =\sum_i B_i = 0$,
where $A_i$ and $B_i$ are defined in Eq.~(\ref{condition_A_B}).
Separating spin and charge parts, we have (the common prefactor
$C_1$ is omitted in other to lighten notation)
\begin{eqnarray}\nonumber
=\sum_{\sigma} \left\langle \left[ e^{i \sqrt{8} \phi_{\rho, {\color{red}
   j}} (r_1)} \left( e^{\frac{i}{\sqrt{2}} \left[ - 3 \phi_{\rho} (r_2) -
   \theta_{\rho} (r_2) + \sigma \left( \phi_{\sigma} (r_2) - \theta_{\sigma}
   (r_2) \right) \right]_{{\color{red} j}}} e^{- \frac{i}{\sqrt{2}}
   \left[ \phi_{\rho} (r_2) - \theta_{\rho} (r_2) + \sigma \left(
   \phi_{\sigma} (r_2) - \theta_{\sigma} (r_2) \right) \right]_{{\color{red}
   j + 1}}} \right.\right.\right.\\ \nonumber
+\left. e^{- \frac{i}{\sqrt{2}} \left[ 3 \phi_{\rho} (r_2) -
   \theta_{\rho} (r_2) + \sigma \left( - \phi_{\sigma} (r_2) - \theta_{\sigma}
   (r_2) \right) \right]_{{\color{red} j}}} e^{\frac{i}{\sqrt{2}} \left[
   - \phi_{\rho} (r_2) - \theta_{\rho} (r_2) + \sigma \left( - \phi_{\sigma}
   (r_2) - \theta_{\sigma} (r_2) \right) \right]_{{\color{red} j + 1}}}
   \right)\\ \nonumber
- e^{- i \sqrt{8} \phi_{\rho, {\color{red} j}} (r_1)} \left(
   e^{\frac{i}{\sqrt{2}} \left[ 3 \phi_{\rho} (r_2) - \theta_{\rho} (r_2) +
   \sigma \left( - \phi_{\sigma} (r_2) - \theta_{\sigma} (r_2) \right)
   \right]_{{\color{red} j}}} e^{- \frac{i}{\sqrt{2}} \left[ -
   \phi_{\rho} (r_2) - \theta_{\rho} (r_2) + \sigma \left( - \phi_{\sigma}
   (r_2) - \theta_{\sigma} (r_2) \right) \right]_{{\color{red} j + 1}}} \right.\\ \nonumber
+\left.\left.e^{- \frac{i}{\sqrt{2}} \left[ - 3 \phi_{\rho} (r_2) - \theta_{\rho} (r_2)
   + \sigma \left( \phi_{\sigma} (r_2) - \theta_{\sigma} (r_2) \right)
   \right]_{{\color{red} j}}} e^{\frac{i}{\sqrt{2}} \left[ \phi_{\rho}
   (r_2) - \theta_{\rho} (r_2) + \sigma \left( \phi_{\sigma} (r_2) -
   \theta_{\sigma} (r_2) \right) \right]_{{\color{red} j + 1}}} \right)
   \right]\times \\\nonumber
 \sum_{r', \sigma'} \left[ e^{\frac{i}{\sqrt{2}} \left[ r'
   \phi_{\rho} (r_3) - \theta_{\rho} (r_3) + \sigma' \left( r' \phi_{\sigma}
   (r_3) - \theta_{\sigma} (r_3) \right) \right]_{{\color{red} j}}} e^{-
   \frac{i}{\sqrt{2}} \left[ r' \phi_{\rho} (r_3) - \theta_{\rho} (r_3) +
   \sigma' \left( r' \phi_{\sigma} (r_3) - \theta_{\sigma} (r_3) \right)
   \right]_{{\color{red} j + 1}}} \right.\\ \nonumber
+ \left.\left. e^{- \frac{i}{\sqrt{2}} \left[ r'
   \phi_{\rho} (r_3) - \theta_{\rho} (r_3) + \sigma' \left( r' \phi_{\sigma}
   (r_3) - \theta_{\sigma} (r_3) \right) \right]_{{\color{red} j}}}
   e^{\frac{i}{\sqrt{2}} \left[ r' \phi_{\rho} (r_3) - \theta_{\rho} (r_3) +
   \sigma' \left( r' \phi_{\sigma} (r_3) - \theta_{\sigma} (r_3) \right)
   \right]_{{\color{red} j + 1}}} \right] \right\rangle
\end{eqnarray}
Now, making explicitly all the products in the above expression we
obtain four different terms,
\begin{eqnarray*}
= \left\langle \sum_{\sigma} e^{i \sqrt{8} \phi_{\rho, {\color{red}
   j}} (r_1)} \left( e^{\frac{i}{\sqrt{2}} \left[ - 3 \phi_{\rho} (r_2) -
   \theta_{\rho} (r_2) + \sigma \left( \phi_{\sigma} (r_2) - \theta_{\sigma}
   (r_2) \right) \right]_{{\color{red} j}}} e^{- \frac{i}{\sqrt{2}}
   \left[ \phi_{\rho} (r_2) - \theta_{\rho} (r_2) + \sigma \left(
   \phi_{\sigma} (r_2) - \theta_{\sigma} (r_2) \right) \right]_{{\color{red}
   j + 1}}} \right) \right.\\
\left. \sum_{\sigma'} \left( e^{- \frac{i}{\sqrt{2}} \left[ \phi_{\rho}
   (r_3) - \theta_{\rho} (r_3) + \sigma' \left( \phi_{\sigma} (r_3) -
   \theta_{\sigma} (r_3) \right) \right]_{{\color{red} j}}}
   e^{\frac{i}{\sqrt{2}} \left[ \phi_{\rho} (r_3) - \theta_{\rho} (r_3) +
   \sigma' \left( \phi_{\sigma} (r_3) - \theta_{\sigma} (r_3) \right)
   \right]_{{\color{red} j + 1}}} \right) \right\rangle \\
+ \left\langle
   \sum_{\sigma} e^{i \sqrt{8} \phi_{\rho, {\color{red} j}} (r_1)} \left(
   e^{- \frac{i}{\sqrt{2}} \left[ 3 \phi_{\rho} (r_2) - \theta_{\rho} (r_2) +
   \sigma \left( - \phi_{\sigma} (r_2) - \theta_{\sigma} (r_2) \right)
   \right]_{{\color{red} j}}} e^{\frac{i}{\sqrt{2}} \left[ - \phi_{\rho}
   (r_2) - \theta_{\rho} (r_2) + \sigma \left( - \phi_{\sigma} (r_2) -
   \theta_{\sigma} (r_2) \right) \right]_{{\color{red} j + 1}}} \right)
   \right.\\
   \left. \sum_{\sigma'} \left( e^{\frac{i}{\sqrt{2}} \left[ - \phi_{\rho}
   (r_3) - \theta_{\rho} (r_3) + \sigma' \left( - \phi_{\sigma} (r_3) -
   \theta_{\sigma} (r_3) \right) \right]_{{\color{red} j}}} e^{-
   \frac{i}{\sqrt{2}} \left[ - \phi_{\rho} (r_3) - \theta_{\rho} (r_3) +
   \sigma' \left( - \phi_{\sigma} (r_3) - \theta_{\sigma} (r_3) \right)
   \right]_{{\color{red} j + 1}}} \right) \right\rangle \\
- \left\langle
   \sum_{\sigma} e^{- i \sqrt{8} \phi_{\rho, {\color{red} j}} (r_1)}
   \left( e^{\frac{i}{\sqrt{2}} \left[ 3 \phi_{\rho} (r_2) - \theta_{\rho}
   (r_2) + \sigma \left( - \phi_{\sigma} (r_2) - \theta_{\sigma} (r_2) \right)
   \right]_{{\color{red} j}}} e^{- \frac{i}{\sqrt{2}} \left[ -
   \phi_{\rho} (r_2) - \theta_{\rho} (r_2) + \sigma \left( - \phi_{\sigma}
   (r_2) - \theta_{\sigma} (r_2) \right) \right]_{{\color{red} j + 1}}}
   \right) \right.\\
   \left. \sum_{\sigma'} \left( e^{- \frac{i}{\sqrt{2}} \left[ - \phi_{\rho}
   (r_3) - \theta_{\rho} (r_3) + \sigma' \left( - \phi_{\sigma} (r_3) -
   \theta_{\sigma} (r_3) \right) \right]_{{\color{red} j}}}
   e^{\frac{i}{\sqrt{2}} \left[ - \phi_{\rho} (r_3) - \theta_{\rho} (r_3) +
   \sigma' \left( - \phi_{\sigma} (r_3) - \theta_{\sigma} (r_3) \right)
   \right]_{{\color{red} j + 1}}} \right) \right\rangle \\
- \left\langle
   \sum_{\sigma} e^{- i \sqrt{8} \phi_{\rho, {\color{red} j}} (r_1)}
   \left( e^{- \frac{i}{\sqrt{2}} \left[ - 3 \phi_{\rho} (r_2) - \theta_{\rho}
   (r_2) + \sigma \left( \phi_{\sigma} (r_2) - \theta_{\sigma} (r_2) \right)
   \right]_{{\color{red} j}}} e^{\frac{i}{\sqrt{2}} \left[ \phi_{\rho}
   (r_2) - \theta_{\rho} (r_2) + \sigma \left( \phi_{\sigma} (r_2) -
   \theta_{\sigma} (r_2) \right) \right]_{{\color{red} j + 1}}} \right)
   \right.\\
   \left. \sum_{\sigma'} \left( e^{\frac{i}{\sqrt{2}} \left[ \phi_{\rho} (r_3)
   - \theta_{\rho} (r_3) + \sigma' \left( \phi_{\sigma} (r_3) -
   \theta_{\sigma} (r_3) \right) \right]_{{\color{red} j}}} e^{-
   \frac{i}{\sqrt{2}} \left[ \phi_{\rho} (r_3) - \theta_{\rho} (r_3) + \sigma'
   \left( \phi_{\sigma} (r_3) - \theta_{\sigma} (r_3) \right)
   \right]_{{\color{red} j + 1}}} \right) \right\rangle
\end{eqnarray*}
Summing over the spins $\sigma$ and $\sigma'$, and using the result shown in Eq.~(\ref{condition_A_B}), we finally obtain
that at zero order in the band curvature ($\alpha$) and magnetic field ($B$), the correlator $\langle K_x;K_y\rangle$ vanishes,
\begin{eqnarray*}
&=& + 2e^{- \frac{1}{2} \left[ 6 K_{\rho} F_1 (r_1 - r_2) + 2 K_{\rho}
      F_1 (r_1 - r_3) - \left( K_{\rho} - K_{\rho}^{- 1} \right) F_1 (r_2 -
      r_3) \right]}\times \\
 &&\, e^{-\frac{1}{2} \left[ - 2 F_2 (r_1 - r_2) + 2 F_2 (r_1 -
      r_3) \right]}e^{- \left[ \left( \frac{1}{2} K_{\sigma} + \frac{1}{2} K_{\sigma}^{- 1}
      \right) F_1 (r_2 - r_3) + F_2 (r_2 - r_3) \right]} \\
&+& 2e^{-\frac{1}{2}\left[ 6 K_{\rho} F_1 (r_1 - r_2) + 2 K_{\rho} F_1 (r_1 - r_3) - \left(
      K_{\rho} - K_{\rho}^{- 1} \right) F_1 (r_2 - r_3) \right]} \times\\
&&\, e^{-\frac{1}{2} \left[ 2 F_2 (r_1 - r_2) - 2 F_2 (r_1 - r_3) \right]}
    e^{- \left[ \left( \frac{1}{2} K_{\sigma} + \frac{1}{2} K_{\sigma}^{- 1}
      \right) F_1 (r_2 - r_3) - F_2 (r_2 - r_3) \right]}  \\
&-&e^{-\frac{1}{2}\left[ 6 K_{\rho} F_1 (r_1 - r_2) + 2 K_{\rho} F_1 (r_1 - r_3) - \left(
      K_{\rho} - K_{\rho}^{- 1} \right) F_1 (r_2 - r_3) \right]}\times \\
&&\, e^{-\frac{1}{2} \left[ 2 F_2 (r_1 - r_2) - 2 F_2 (r_1 - r_3) \right]}
      e^{- \left[ \left( \frac{1}{2} K_{\sigma} + \frac{1}{2} K_{\sigma}^{- 1}
      \right) F_1 (r_2 - r_3) - F_2 (r_2 - r_3) \right]} \\
&-& e^{-\frac{1}{2}\left[ 6 K_{\rho} F_1 (r_1 - r_2) + 2 K_{\rho} F_1 (r_1 - r_3) - \left(
      K_{\rho} - K_{\rho}^{- 1} \right) F_1 (r_2 - r_3) \right]}\times \\
&&\, e^{-\frac{1}{2} \left[ - 2 F_2 (r_1 - r_2) + 2 F_2 (r_1 - r_3) \right]}
      e^{- \left[ \left( \frac{1}{2} K_{\sigma} + \frac{1}{2} K_{\sigma}^{- 1}
      \right) F_1 (r_2 - r_3) + F_2 (r_2 - r_3) \right]} = 0
\end{eqnarray*}
This can also be verified applying spatial inversion and
particle-hole symmetry in correlator $\langle K_x; K_y\rangle.$

\section{Full expression for the correlator $\langle K_x; K_y\rangle$ at first order in $\alpha,B$ and $t_{\perp}$}\label{app:full_correlator}

Here we will repeat the same type of calculations done in
Appendix~\ref{app:alpha0_B0}, but in this case we consider the
band curvature term in the correlator $\langle K_x; K_y\rangle$
and we take to first order in $B$ the terms depending on the
magnetic field ($K_y$ and $\mathcal{H}_{\perp}$). Thus the
correlator is given in Eq.~(\ref{KKHcBC}). Again we take the
bosonized form for each operator (see Sec.~\ref{sec:RH_umklapp})
and we put all the prefactors together in a constant named $C_2$.
The resulting expression is
\begin{eqnarray}\nonumber
&&= \sum_{r, \sigma} \left( \frac{i e B (x_2+x_3) a_y}{c} \right) C_2\left\langle
   \left[ e^{i \sqrt{8} \phi_{\rho} (r_1)} - e^{- i \sqrt{8} \phi_{\rho}
   (r_1)} \right]_{{\color{red} j}} \right. \\ \nonumber
&&\left[ e^{\frac{i}{\sqrt{2}} \left[ 3
   r \phi_{\rho} (r_2) - \theta_{\rho} (r_2) + \sigma \left( - r \phi_{\sigma}
   (r_2) - \theta_{\sigma} (r_2) \right) \right]_{{\color{red} j}}} e^{-
   \frac{i}{\sqrt{2}} \left[ - r \phi_{\rho} (r_2) - \theta_{\rho} (r_2) +
   \sigma \left( - r \phi_{\sigma} (r_2) - \theta_{\sigma} (r_2) \right)
   \right]_{{\color{red} j + 1}}} - h.c. \right]\\ \nonumber
&&\sum_{r', \sigma'}
   \left[ e^{\frac{i}{\sqrt{2}} \left[ r' \phi_{\rho} (r_3) - \theta_{\rho}
   (r_3) + \sigma' \left( r' \phi_{\sigma} (r_3) - \theta_{\sigma} (r_3)
   \right) \right]_{{\color{red} j}}} e^{- \frac{i}{\sqrt{2}} \left[ r'
   \phi_{\rho} (r_3) - \theta_{\rho} (r_3) + \sigma' \left( r' \phi_{\sigma}
   (r_3) - \theta_{\sigma} (r_3) \right) \right]_{{\color{red} j + 1}}}\right.\\ \nonumber
&&+\left.e^{- \frac{i}{\sqrt{2}} \left[ r' \phi_{\rho} (r_3) - \theta_{\rho} (r_3) +
   \sigma' \left( r' \phi_{\sigma} (r_3) - \theta_{\sigma} (r_3) \right)
   \right]_{{\color{red} j}}} e^{\frac{i}{\sqrt{2}} \left[ r' \phi_{\rho}
   (r_3) - \theta_{\rho} (r_3) + \sigma' \left( r' \phi_{\sigma} (r_3) -
   \theta_{\sigma} (r_3) \right) \right]_{{\color{red} j + 1}}}\right]\\
&&\left. \times(\nabla\phi_{\rho}(r_4))^3\right\rangle
\end{eqnarray}

There are again four surviving terms, all multiplied by $\left(i e B (x_2+x_3) a_y/c\right)C_2$
\begin{eqnarray*}
 &&= \left\langle \left[\sum_{\sigma} e^{i \sqrt{8} \phi_{\rho, {\color{red}
   \alpha}} (r_1)} \left( e^{\frac{i}{\sqrt{2}} \left[ - 3 \phi_{\rho} (r_2) -
   \theta_{\rho} (r_2) + \sigma \left( \phi_{\sigma} (r_2) - \theta_{\sigma}
   (r_2) \right) \right]_{{\color{red} \alpha}}} e^{- \frac{i}{\sqrt{2}}
   \left[ \phi_{\rho} (r_2) - \theta_{\rho} (r_2) + \sigma \left(
   \phi_{\sigma} (r_2) - \theta_{\sigma} (r_2) \right) \right]_{{\color{red}
   \alpha + 1}}} \right) \right.\right.\\
 &&\left. \sum_{\sigma'} \left( e^{- \frac{i}{\sqrt{2}} \left[ \phi_{\rho}
   (r_3) - \theta_{\rho} (r_3) + \sigma' \left( \phi_{\sigma} (r_3) -
   \theta_{\sigma} (r_3) \right) \right]_{{\color{red} \alpha}}}
   e^{\frac{i}{\sqrt{2}} \left[ \phi_{\rho} (r_3) - \theta_{\rho} (r_3) +
   \sigma' \left( \phi_{\sigma} (r_3) - \theta_{\sigma} (r_3) \right)
   \right]_{{\color{red} \alpha + 1}}} \right) \right\rangle\\
 &&- \left\langle
   \sum_{\sigma} e^{i \sqrt{8} \phi_{\rho, {\color{red} \alpha}} (r_1)} \left(
   e^{- \frac{i}{\sqrt{2}} \left[ 3 \phi_{\rho} (r_2) - \theta_{\rho} (r_2) +
   \sigma \left( - \phi_{\sigma} (r_2) - \theta_{\sigma} (r_2) \right)
   \right]_{{\color{red} \alpha}}} e^{\frac{i}{\sqrt{2}} \left[ - \phi_{\rho}
   (r_2) - \theta_{\rho} (r_2) + \sigma \left( - \phi_{\sigma} (r_2) -
   \theta_{\sigma} (r_2) \right) \right]_{{\color{red} \alpha + 1}}} \right)
   \right. \\
 &&\left. \sum_{\sigma'} \left( e^{\frac{i}{\sqrt{2}} \left[ - \phi_{\rho}
   (r_3) - \theta_{\rho} (r_3) + \sigma' \left( - \phi_{\sigma} (r_3) -
   \theta_{\sigma} (r_3) \right) \right]_{{\color{red} \alpha}}} e^{-
   \frac{i}{\sqrt{2}} \left[ - \phi_{\rho} (r_3) - \theta_{\rho} (r_3) +
   \sigma' \left( - \phi_{\sigma} (r_3) - \theta_{\sigma} (r_3) \right)
   \right]_{{\color{red} \alpha + 1}}} \right) \right\rangle \\
&&- \left\langle
   \sum_{\sigma} e^{- i \sqrt{8} \phi_{\rho, {\color{red} \alpha}} (r_1)}
   \left( e^{\frac{i}{\sqrt{2}} \left[ 3 \phi_{\rho} (r_2) - \theta_{\rho}
   (r_2) + \sigma \left( - \phi_{\sigma} (r_2) - \theta_{\sigma} (r_2) \right)
   \right]_{{\color{red} \alpha}}} e^{- \frac{i}{\sqrt{2}} \left[ -
   \phi_{\rho} (r_2) - \theta_{\rho} (r_2) + \sigma \left( - \phi_{\sigma}
   (r_2) - \theta_{\sigma} (r_2) \right) \right]_{{\color{red} \alpha + 1}}}
   \right) \right.\\
&& \left. \sum_{\sigma'} \left( e^{- \frac{i}{\sqrt{2}} \left[ - \phi_{\rho}
   (r_3) - \theta_{\rho} (r_3) + \sigma' \left( - \phi_{\sigma} (r_3) -
   \theta_{\sigma} (r_3) \right) \right]_{{\color{red} \alpha}}}
   e^{\frac{i}{\sqrt{2}} \left[ - \phi_{\rho} (r_3) - \theta_{\rho} (r_3) +
   \sigma' \left( - \phi_{\sigma} (r_3) - \theta_{\sigma} (r_3) \right)
   \right]_{{\color{red} \alpha + 1}}} \right)\right\rangle \\
&&+ \left\langle
   \sum_{\sigma} e^{- i \sqrt{8} \phi_{\rho, {\color{red} \alpha}} (r_1)}
   \left( e^{- \frac{i}{\sqrt{2}} \left[ - 3 \phi_{\rho} (r_2) - \theta_{\rho}
   (r_2) + \sigma \left( \phi_{\sigma} (r_2) - \theta_{\sigma} (r_2) \right)
   \right]_{{\color{red} \alpha}}} e^{\frac{i}{\sqrt{2}} \left[ \phi_{\rho}
   (r_2) - \theta_{\rho} (r_2) + \sigma \left( \phi_{\sigma} (r_2) -
   \theta_{\sigma} (r_2) \right) \right]_{{\color{red} \alpha + 1}}} \right)\right. \\
 &&\left.\left. \sum_{\sigma'} \left( e^{\frac{i}{\sqrt{2}} \left[ \phi_{\rho} (r_3)
   - \theta_{\rho} (r_3) + \sigma' \left( \phi_{\sigma} (r_3) -
   \theta_{\sigma} (r_3) \right) \right]_{{\color{red} \alpha}}} e^{-
   \frac{i}{\sqrt{2}} \left[ \phi_{\rho} (r_3) - \theta_{\rho} (r_3) + \sigma'
   \left( \phi_{\sigma} (r_3) - \theta_{\sigma} (r_3) \right)
   \right]_{{\color{red} \alpha + 1}}} \right)\right](\nabla\phi_{\rho}(r_4))^3 \right\rangle
\end{eqnarray*}
Summing over the spins $\sigma$ and $\sigma'$, and using result (\ref{condition_A_B}) we write the
above expressions in terms of functions $F_1$ and $F_2$
\begin{eqnarray}\nonumber
&&= \left( \frac{i e}{c} H (x_2+x_3) a_y \right) \left[ 4 e^{- \frac{1}{2} \left[ 6
   K_{\rho} F_1 (r_1 - r_2) + 2 K_{\rho} F_1 (r_1 - r_3) - \left( K_{\rho} -
   K_{\rho}^{- 1} \right) F_1 (r_2 - r_3) \right]}\right.\\ \nonumber
 &&\times e^{- \frac{1}{2} \left[-2F_2 (r_1 - r_2) + 2 F_2 (r_1 - r_3) \right]}
   e^{- \left[ \left( \frac{1}{2} K_{\sigma} + \frac{1}{2} K_{\sigma}^{- 1}
   \right) F_1 (r_2 - r_3) + F_2 (r_2 - r_3) \right]} \\ \nonumber
 &&-4 e^{- \frac{1}{2}\left[ 6 K_{\rho} F_1 (r_1 - r_2) + 2 K_{\rho} F_1 (r_1 - r_3) - \left(
   K_{\rho} - K_{\rho}^{- 1} \right) F_1 (r_2 - r_3) \right]} e^{- \frac{1}{2}
   \left[ 2 F_2 (r_1 - r_2) - 2 F_2 (r_1 - r_3) \right]}\\
 &&\left. e^{- \left[ \left( \frac{1}{2} K_{\sigma} + \frac{1}{2}
   K_{\sigma}^{- 1} \right) F_1 (r_2 - r_3) - F_2 (r_2 - r_3)\right]}\right]
   \left(\frac{a}{|r_4|}\right)^3
\end{eqnarray}
The functions $F_1(r)$ and $F_2(r)$ are given in
Eq.~(\ref{F1_F2}). The factor $\left(a/|r_4|\right)^3$ is the
behavior of $\nabla\phi_{\rho}$ at large distances $(x,u\tau)\gg
a$ \cite{giamarchi_book_1d}. Because the model studied in
Sec.~\ref{sec:RH_umklapp} is spin rotational invariant, we have
$K_\sigma=1$. The function $F_2(r)$ gives the angular part in the
above expression and thus, does not play any role in the scaling
analysis, where only power-law dependencies are involved. Keeping
only the terms with $F_1$ functions and making the following
change of variables: $r_1=r$, $r_2=0$, $r_3=r_1$ and $r_4=r_2$; we
obtain result (\ref{scaling}). We take only the magnitude for
$x_3$ in the prefactor $\frac{i e}{c} B (x_3) a_y$ because, as we
said before, angular parts can be neglected in the scaling
analysis.

\section{Full expression for $\langle K_x;K_y\rangle$ with $g_2=0$}\label{app:Kx_Ky_g20}

For the analytical obtention of correlator $\langle K_x;K_y\rangle$ with $g_2=0$, we will need
in several instances the average value
$\langle\psi_{j\sigma R}^{\dagger}(x,\tau)\psi_{j+1,\sigma R}(y,0)\rangle_0$
to first order in $t_{\perp}$. We note that in the absence of magnetic field this
quantity equals the free Green's function of the 2D lattice:
    \begin{eqnarray*}
        &&\langle\psi_{j\sigma R}^{\dagger}(x,\tau)
        \psi_{j+1,\sigma R}(y,0)\rangle_0=
        G_{\sigma R}(y-x,a_y,-\tau)=\\
        &&\frac{1}{\beta\mathcal{S}}\sum_{\omega_n}\sum_{kk_{\perp}}
        G_{\sigma R}(k,k_{\perp},i\omega_n)
        e^{ik(y-x)}e^{ik_{\perp}a_y}e^{i\omega_n\tau}\\
        &=&\frac{1}{\beta\mathcal{S}}\sum_{\omega_n}\sum_{kk_{\perp}}
        \frac{e^{ik(y-x)}e^{ik_{\perp}a_y}e^{i\omega_n\tau}}
        {i\omega_n-\xi_R(k)+2t_{\perp}\cos(k_{\perp}a_y)}\\
        &=&\frac{1}{\beta\mathcal{S}}\sum_{\omega_n}\sum_{kk_{\perp}}\left[
        \frac{1}{i\omega_n-\xi_R(k)}-2t_{\perp}\cos(k_{\perp}a_y)
        \left(\frac{1}{i\omega_n-\xi_R(k)}\right)^2+\mathcal{O}(t_{\perp}^2)\right]
        e^{ik(y-x)}e^{ik_{\perp}a_y}e^{i\omega_n\tau}\\
        &=&-t_{\perp}\frac{1}{\beta L_x}\sum_{\omega_n}\sum_k
        \left(\frac{1}{i\omega_n-\xi_R(k)}\right)^2e^{ik(y-x)}e^{i\omega_n\tau}\\
        &=&-t_{\perp}\int dx_1d\tau_1\,\mathcal{G}_{j\sigma R}(x_1,\tau_1)
        \mathcal{G}_{j\sigma R}(y-x-x_1,-\tau-\tau_1).
    \end{eqnarray*}
With a finite magnetic field we can use another method: in
general, in the presence of a time-independent perturbation, the
Green's function obeys Dyson's equation:
$G(\bm{r},\bm{r}',\omega)=G_0(\bm{r},\bm{r}',\omega)+\int
d\bm{r}_1d\bm{r}_2G_0(\bm{r},\bm{r}_1,\omega)\Sigma(\bm{r}_1,\bm{r}_2,\omega)G(\bm{r}_2,\bm{r}',\omega)$.
In our case $G_0(\bm{r},\bm{r}',\omega)$ is the Green's function
in the absence of interaction and for $t_{\perp}=0$, \textit{i.e.}
$G_0(\bm{r},\bm{r}',\omega)=\delta_{jj'}\mathcal{G}(x-x',\omega)$
with $j$ the chain index. On the other hand, the self-energy is
simply
$\Sigma(\bm{r}_1,\bm{r}_2,\omega)=\Sigma_{jj'}(x_1,x_2,\omega)=-t_{\perp}\delta(x_1-x_2)(\delta_{j',j+1}e^{-i\delta
kx_1}+\delta_{j',j-1}e^{i\delta kx_1})$. The magnetic field $B$
enters in the shift of momentum $\delta k=eB/c$. At first order in
$t_{\perp}$, we have:
    \begin{eqnarray}\label{eq:G0jjp}
        \nonumber
        G_{jj'}(x,x',\omega)&=&\delta_{jj'}\mathcal{G}(x-x',\omega)
        +\sum_{kl}\int dx_1dx_2\delta_{jk}\mathcal{G}(x-x_1,\omega)\times\\
        \nonumber
        &&(-t_{\perp})\delta(x_1-x_2)\left(\delta_{l,k+1}e^{-i\delta kx_1}
        +\delta_{l,k-1}e^{i\delta kx_1}\right)\delta_{lj'}\mathcal{G}(x_2-x',\omega)\\
        \nonumber
        &=&\delta_{jj'}\mathcal{G}(x-x',\omega)-t_{\perp}\delta_{j',j+1}
        \int dx_1\,\mathcal{G}(x-x_1,\omega)e^{-i\delta kx_1}
        \mathcal{G}(x_1-x',\omega)\\ &&-t_{\perp}\delta_{j',j-1}
        \int dx_1\,\mathcal{G}(x-x_1,\omega)e^{i\delta kx_1}
        \mathcal{G}(x_1-x',\omega).
    \end{eqnarray}
Thus, we obtain an expression for the Green's function to first
order in $t_{\perp}$ and in presence of a magnetic field $B$,
    \begin{eqnarray}
        \nonumber
        &&\langle\psi_{j\sigma R}^{\dagger}(x,\tau)
        \psi_{j+1,\sigma R}(y,0)\rangle_0=G_{j+1,j}(y,x,-\tau)=\\ \nonumber
        &&-t_{\perp}\frac{1}{\beta}\sum_{\omega_n}
        \int dx_1\,\mathcal{G}_{j\sigma R}(y-x_1,i\omega_n)e^{i\delta kx_1}
        \mathcal{G}_{j\sigma R}(x_1-x,i\omega_n)e^{-i\omega_n(-\tau)}\\ \label{eq:interchain}
        &=&-t_{\perp}\int dx_1d\tau_1\,\mathcal{G}_{j\sigma R}(y-x_1,\tau_1)
        e^{i\delta kx_1}\mathcal{G}_{j\sigma R}(x_1-x,-\tau-\tau_1).
    \end{eqnarray}
This last expression reduces to our previous result if $B=0$. The
result (\ref{eq:interchain}) will be used in the following
computation of the correlator $\langle K_x;K_y\rangle$. Starting
from Eq.~(\ref{free_1}), we write the complete expression for
correlator $\langle K_x;K_y\rangle$
\begin{eqnarray}\label{eq:KxKy0}
    &&i\langle K_x;K_y\rangle_0=2e^2 v_{\text{F}} t_{\perp}g_3^2a_y\int dx\sum_{j\sigma}
    \int dy\sum_{j'\sigma'}\sum_{b=L,R}\int_0^{\beta}d\tau\,e^{i\Omega\tau}\\ \nonumber
    &&\left\langle T_{\tau}\left(\psi_{j\sigma R}^{\dagger}\psi_{j,-\sigma R}^{\dagger}
    \psi_{j,-\sigma L}\psi_{j\sigma L}-\text{h.c.}\right)
    \left[e^{-i\delta ky}\left(
    \psi_{j'\sigma'b}^{\dagger}\psi_{j',-\sigma'b}^{\dagger}
    \psi_{j',-\sigma',-b}\psi_{j'+1,\sigma',-b}\right.\right.\right.\\ \nonumber
    &&-\left.\left.\left.\psi_{j'\sigma'b}^{\dagger}\psi_{j'+1,-\sigma'b}^{\dagger}
    \psi_{j'+1,-\sigma',-b}\psi_{j'+1,\sigma',-b}\right)
    +\text{h.c.}\right]\right\rangle_0.
\end{eqnarray}
In the first parenthesis the operators are evaluated at position $x$ and time
$\tau$, while in the second parenthesis they are at position $y$ and time $0$.
We consider the first term:
    \begin{equation}\label{eq:term1}
        e^{-i\delta ky}\langle\underbrace{\psi_{j\sigma R}^{\dagger}
        \psi_{j,-\sigma R}^{\dagger}\psi_{j,-\sigma L}\psi_{j\sigma L}}_{(x,\tau)}
        \underbrace{\psi_{j'\sigma'b}^{\dagger}\psi_{j',-\sigma'b}^{\dagger}
        \psi_{j',-\sigma',-b}\psi_{j'+1,\sigma',-b}}_{(y,0)}\rangle_0.
    \end{equation}
We can decouple this term in only two different ways. The first is
    \begin{equation}
        e^{-i\delta ky}
        \langle\psi_{j\sigma R}^{\dagger}\psi_{j'+1,\sigma',-b}\rangle_0
        \langle\psi_{j,-\sigma R}^{\dagger}\psi_{j',-\sigma',-b}\rangle_0
        \langle\psi_{j,-\sigma L}\psi_{j',-\sigma'b}^{\dagger}\rangle_0
        \langle\psi_{j\sigma L}\psi_{j'\sigma'b}^{\dagger}\rangle_0
    \end{equation}
and it clearly implies that $\sigma'=\sigma$ and $b=L$. The chain indices
$j$ and $j'$ must be such that the fermions are on the same or nearest-neighbor
chains, since we work at first order in $t_{\perp}$. The only solution is
$j=j'$, since the other possibility, $j'=j-1$, gives a term of order $t_{\perp}^3$.
Hence the term reads
    \begin{equation}
        \delta_{\sigma\sigma'}\delta_{bL}\delta_{jj'}e^{-i\delta ky}
        \langle\psi_{j\sigma R}^{\dagger}(x,\tau)\psi_{j+1,\sigma R}(y,0)\rangle_0
        \mathcal{G}_{j,-\sigma R}(y-x,-\tau)
        \mathcal{G}_{j,-\sigma L}(x-y,\tau)\mathcal{G}_{j\sigma L}(x-y,\tau)
    \end{equation}
with $\mathcal{G}$ the free propagator. Using Eq.~(\ref{eq:interchain}), we find
that the first decoupling yields the
following contribution to $i\langle K_x;K_y\rangle_0$ [we temporarily omit the
prefactor in Eq.~(\ref{eq:KxKy0})]:
    \begin{eqnarray}
        \nonumber
        &&-t_{\perp}\int dxdydx_1d\tau d\tau_1\sum_{j\sigma}\sum_{j'\sigma'}
        \sum_{b=L,R}e^{i\Omega\tau}\delta_{\sigma\sigma'}\delta_{bL}
        \delta_{jj'}e^{-i\delta k(y-x_1)}\times\\ \nonumber
        &&\mathcal{G}_{j\sigma R}(y-x_1,\tau_1)
        \mathcal{G}_{j\sigma R}(x_1-x,-\tau-\tau_1)
        \mathcal{G}_{j,-\sigma R}(y-x,-\tau)
        \mathcal{G}_{j,-\sigma L}(x-y,\tau)
        \mathcal{G}_{j\sigma L}(x-y,\tau)\\
        \nonumber
        &=&-t_{\perp}\frac{1}{\beta^3}\sum_{i\nu_1,i\nu_2,i\nu_3}\int dxdydx_1
        \sum_{j\sigma}e^{-i\delta k(y-x_1)}\times\\ \nonumber
        &&\mathcal{G}_{j\sigma R}(y-x_1,i\nu_1)
        \mathcal{G}_{j\sigma R}(x_1-x,i\nu_1)
        \mathcal{G}_{j,-\sigma R}(y-x,i\nu_2)\\ \nonumber
        &&\times\mathcal{G}_{j,-\sigma L}(x-y,i\nu_3)
        \mathcal{G}_{j\sigma L}(x-y,i\nu_1+i\nu_2-i\nu_3+i\Omega)\\
        \nonumber
        &=&-2t_{\perp}\frac{N_y}{L_x^2}
        \sum_{k_1k_2q}\frac{1}{\beta^3}\sum_{i\nu_1,i\nu_2,i\nu_3}
        \mathcal{G}_{\uparrow R}(k_1,i\nu_1)
        \mathcal{G}_{\uparrow R}(k_1+\delta k,i\nu_1)
        \mathcal{G}_{\downarrow R}(k_2,i\nu_2)\\ \nonumber
        &&\times\mathcal{G}_{\downarrow L}(k_2-q,i\nu_3)
        \mathcal{G}_{\uparrow L}(k_1+q,i\nu_1+i\nu_2-i\nu_3+i\Omega)\\
        &\equiv&-2t_{\perp}A(i\Omega,B).
    \end{eqnarray}
This corresponds to the diagram shown in Fig.~\ref{fig:diagram}.
The 2nd decoupling of Eq.~(\ref{eq:term1}) is
    \begin{equation}
        e^{-i\delta ky}
        \langle\psi_{j\sigma R}^{\dagger}\psi_{j',-\sigma',-b}\rangle_0
        \langle\psi_{j,-\sigma R}^{\dagger}\psi_{j'+1,\sigma',-b}\rangle_0
        \langle\psi_{j,-\sigma L}\psi_{j'\sigma'b}^{\dagger}\rangle_0
        \langle\psi_{j\sigma L}\psi_{j',-\sigma'b}^{\dagger}\rangle_0
    \end{equation}
and implies $\sigma'=-\sigma$, $b=L$, and $j=j'$:
    \begin{equation}
        \delta_{\sigma,-\sigma'}\delta_{bL}\delta_{jj'}e^{-i\delta ky}
        \mathcal{G}_{j\sigma R}(y-x,-\tau)
        \langle\psi_{j,-\sigma R}^{\dagger}(x,\tau)\psi_{j+1,-\sigma R}(y,0)\rangle_0
        \mathcal{G}_{j,-\sigma L}(x-y,\tau)\mathcal{G}_{j\sigma L}(x-y,\tau).
    \end{equation}
This is the same expression as for the first decoupling provided we change $\sigma$
into $-\sigma$. It will therefore give the same contribution $-2t_{\perp}A(i\Omega,B)$.

We now move to the second term:
    \begin{equation}
        -e^{-i\delta ky}\langle\underbrace{
        \psi_{j,-\sigma L}^{\dagger}
        \psi_{j\sigma L}^{\dagger}\psi_{j\sigma R}\psi_{j,-\sigma R}}_{(x,\tau)}
        \underbrace{\psi_{j'\sigma'b}^{\dagger}\psi_{j',-\sigma'b}^{\dagger}
        \psi_{j',-\sigma',-b}\psi_{j'+1,\sigma',-b}}_{(y,0)}\rangle_0
    \end{equation}
Again there are only two decoupling:
    \begin{eqnarray*}
        \nonumber
        &&-\delta_{\sigma,-\sigma'}\delta_{bR}\delta_{jj'}e^{-i\delta ky}
        \langle\psi_{j,-\sigma L}^{\dagger}\psi_{j'+1,\sigma',-b}\rangle_0
        \langle\psi_{j\sigma L}^{\dagger}\psi_{j',-\sigma',-b}\rangle_0
        \langle\psi_{j\sigma R}\psi_{j',-\sigma'b}^{\dagger}\rangle_0
        \langle\psi_{j,-\sigma R}\psi_{j'\sigma'b}^{\dagger}\rangle_0\\
        &&=-\delta_{\sigma,-\sigma'}\delta_{bR}\delta_{jj'}e^{-i\delta ky}
        \langle\psi_{j,-\sigma L}^{\dagger}(x,\tau)\psi_{j+1,-\sigma L}(y,0)\rangle_0
        \mathcal{G}_{j\sigma L}(y-x,-\tau)
        \mathcal{G}_{j\sigma R}(x-y,\tau)
        \mathcal{G}_{j,-\sigma R}(x-y,\tau)\qquad\\
        \nonumber
        &&-\delta_{\sigma\sigma'}\delta_{bR}\delta_{jj'}e^{-i\delta ky}
        \langle\psi_{j,-\sigma L}^{\dagger}\psi_{j',-\sigma',-b}\rangle_0
        \langle\psi_{j\sigma L}^{\dagger}\psi_{j'+1,\sigma',-b}\rangle_0
        \langle\psi_{j,\sigma R}\psi_{j'\sigma'b}^{\dagger}\rangle_0
        \langle\psi_{j,-\sigma R}\psi_{j',-\sigma'b}^{\dagger}\rangle_0\\
        &&=-\delta_{\sigma\sigma'}\delta_{bR}\delta_{jj'}e^{-i\delta ky}
        \mathcal{G}_{j,-\sigma L}(y-x,-\tau)
        \langle\psi_{j\sigma L}^{\dagger}(x,\tau)\psi_{j+1,\sigma L}(y,0)\rangle_0
        \mathcal{G}_{j\sigma R}(x-y,\tau)
        \mathcal{G}_{j,-\sigma R}(x-y,\tau).
    \end{eqnarray*}
The calculation is very similar to the previous one: just exchange
$R$ and $L$ and add a minus sign. Hence these two terms contribute
    \begin{eqnarray}
        \nonumber
        &&=4t_{\perp}\sum_{k_1k_2q}\frac{1}{\beta^3}\sum_{i\nu_1,i\nu_2,i\nu_3}
        \mathcal{G}_{\uparrow L}(k_1,i\nu_1)
        \mathcal{G}_{\uparrow L}(k_1+\delta k,i\nu_1)
        \mathcal{G}_{\downarrow L}(k_2,i\nu_2)\\ \nonumber
        &&\mathcal{G}_{\downarrow R}(k_2-q,i\nu_3)
        \mathcal{G}_{\uparrow R}(k_1+q,i\nu_1+i\nu_2-i\nu_3+i\Omega)\\
        \nonumber
        &=&4t_{\perp}\sum_{k_1k_2q}\frac{1}{\beta^3}\sum_{i\nu_1,i\nu_2,i\nu_3}
        \mathcal{G}_{\uparrow R}(k_1,i\nu_1)
        \mathcal{G}_{\uparrow R}(k_1-\delta k,i\nu_1)
        \mathcal{G}_{\downarrow R}(k_2,i\nu_2)\\ \nonumber
        &&\mathcal{G}_{\downarrow L}(k_2-q,i\nu_3)
        \mathcal{G}_{\uparrow L}(k_1+q,i\nu_1+i\nu_2-i\nu_3+i\Omega)\\
        &=&4t_{\perp}A(i\Omega,-B),
    \end{eqnarray}
where we have used the fact that $\mathcal{G}_{\sigma L}(k,i\omega)=\mathcal{G}_{\sigma R}(-k,i\omega)$.
The third term is
    \begin{equation}
        -e^{-i\delta ky}\langle\underbrace{\psi_{j\sigma R}^{\dagger}
        \psi_{j,-\sigma R}^{\dagger}\psi_{j,-\sigma L}\psi_{j\sigma L}}_{(x,\tau)}
        \underbrace{\psi_{j'\sigma'b}^{\dagger}\psi_{j'+1,-\sigma'b}^{\dagger}
        \psi_{j'+1,-\sigma',-b}\psi_{j'+1,\sigma',-b}}_{(y,0)}\rangle_0
    \end{equation}
and can be decoupled as follows
    \begin{eqnarray}
        \nonumber
        &&-\delta_{\sigma\sigma'}\delta_{bL}\delta_{j',j-1}e^{-i\delta ky}
        \langle\psi_{j\sigma R}^{\dagger}\psi_{j'+1,\sigma',-b}\rangle_0
        \langle\psi_{j,-\sigma R}^{\dagger}\psi_{j'+1,-\sigma',-b}\rangle_0
        \langle\psi_{j,-\sigma L}\psi_{j'+1,-\sigma'b}^{\dagger}\rangle_0
        \langle\psi_{j\sigma L}\psi_{j'\sigma'b}^{\dagger}\rangle_0\\ \nonumber
        &&=-\delta_{\sigma\sigma'}\delta_{bL}\delta_{j',j-1}e^{-i\delta ky}
        \mathcal{G}_{j\sigma R}(y-x,-\tau)
        \mathcal{G}_{j,-\sigma R}(y-x,-\tau)
        \mathcal{G}_{j,-\sigma L}(x-y,\tau)
        \langle\psi_{j\sigma L}(x,\tau)\psi_{j-1,\sigma L}^{\dagger}(y,0)\rangle_0\\
        \nonumber
        &&-\delta_{\sigma,-\sigma'}\delta_{bL}\delta_{j',j-1}e^{-i\delta ky}
        \langle\psi_{j\sigma R}^{\dagger}\psi_{j'+1,-\sigma',-b}\rangle_0
        \langle\psi_{j,-\sigma R}^{\dagger}\psi_{j'+1,\sigma',-b}\rangle_0
        \langle\psi_{j,-\sigma L}\psi_{j'\sigma'b}^{\dagger}\rangle_0
        \langle\psi_{j\sigma L}\psi_{j'+1,-\sigma'b}^{\dagger}\rangle_0\\ \nonumber
        &&=-\delta_{\sigma,-\sigma'}\delta_{bL}\delta_{j',j-1}e^{-i\delta ky}
        \mathcal{G}_{j\sigma R}(y-x,-\tau)
        \mathcal{G}_{j,-\sigma R}(y-x,-\tau)
        \langle\psi_{j,-\sigma L}(x,\tau)\psi_{j-1,-\sigma L}^{\dagger}(y,0)\rangle_0
        \mathcal{G}_{j\sigma L}(x-y,\tau)
    \end{eqnarray}
Both terms are equivalent except for the sign of $\sigma$.
From Eq.~(\ref{eq:G0jjp}) we read that
    \begin{equation}
        \langle\psi_{j\sigma L}(x,\tau)\psi_{j-1,\sigma L}^{\dagger}(y,0)\rangle_0
        =G_{j,j-1}(x,y,\tau)
        =-t_{\perp}\int dx_1d\tau_1\,\mathcal{G}_{j\sigma L}(x-x_1,\tau_1)
        e^{i\delta kx_1}\mathcal{G}_{j\sigma L}(x_1-y,\tau-\tau_1)
    \end{equation}
and we then find that these two terms contribute $4t_{\perp}A(-i\Omega,B)$. We
continue the game with the fourth term:
    \begin{equation}
        e^{-i\delta ky}\langle\underbrace{\psi_{j,-\sigma L}^{\dagger}
        \psi_{j\sigma L}^{\dagger}\psi_{j\sigma R}\psi_{j,-\sigma R}}_{(x,\tau)}
        \underbrace{\psi_{j'\sigma'b}^{\dagger}\psi_{j'+1,-\sigma'b}^{\dagger}
        \psi_{j'+1,-\sigma',-b}\psi_{j'+1,\sigma',-b}}_{(y,0)}\rangle_0
    \end{equation}
which has the decoupling
    \begin{eqnarray}
        \nonumber
        &&\delta_{\sigma,-\sigma'}\delta_{bR}\delta_{j',j-1}e^{-i\delta ky}
        \langle\psi_{j,-\sigma L}^{\dagger}\psi_{j'+1,\sigma',-b}\rangle_0
        \langle\psi_{j\sigma L}^{\dagger}\psi_{j'+1,-\sigma',-b}\rangle_0
        \langle\psi_{j\sigma R}\psi_{j'+1,-\sigma'b}^{\dagger}\rangle_0
        \langle\psi_{j,-\sigma R}\psi_{j'\sigma'b}^{\dagger}\rangle_0\\ \nonumber
        &&=\delta_{\sigma,-\sigma'}\delta_{bR}\delta_{j',j-1}e^{-i\delta ky}
        \mathcal{G}_{j,-\sigma L}(y-x,-\tau)
        \mathcal{G}_{j\sigma L}(y-x,-\tau)
        \mathcal{G}_{j\sigma R}(x-y,\tau)
        \langle\psi_{j,-\sigma R}(x,\tau)\psi_{j-1,-\sigma R}^{\dagger}(y,0)\rangle_0\\
        \nonumber
        &&\delta_{\sigma\sigma'}\delta_{bR}\delta_{j',j-1}e^{-i\delta ky}
        \langle\psi_{j,-\sigma L}^{\dagger}\psi_{j'+1,-\sigma',-b}\rangle_0
        \langle\psi_{j\sigma L}^{\dagger}\psi_{j'+1,\sigma',-b}\rangle_0
        \langle\psi_{j\sigma R}\psi_{j'\sigma'b}^{\dagger}\rangle_0
        \langle\psi_{j,-\sigma R}\psi_{j'+1,-\sigma'b}^{\dagger}\rangle_0\\ \nonumber
        &&=\delta_{\sigma\sigma'}\delta_{bR}\delta_{j',j-1}e^{-i\delta ky}
        \mathcal{G}_{j,-\sigma L}(y-x,-\tau)
        \mathcal{G}_{j\sigma L}(y-x,-\tau)
        \langle\psi_{j\sigma R}(x,\tau)\psi_{j-1,\sigma R}^{\dagger}(y,0)\rangle_0
        \mathcal{G}_{j,-\sigma R}(x-y,\tau)
    \end{eqnarray}
and contributes $-4t_{\perp}A(-i\Omega,-B)$. There are four
additional terms to consider, for which we look only at the first
of the two equivalent decoupling:
    \begin{eqnarray}\nonumber
        &&e^{i\delta ky}\langle\underbrace{\psi_{j\sigma R}^{\dagger}
        \psi_{j,-\sigma R}^{\dagger}\psi_{j,-\sigma L}\psi_{j\sigma L}}_{(x,\tau)}
        \underbrace{\psi_{j',-\sigma',-b}^{\dagger}\psi_{j'+1,\sigma',-b}^{\dagger}
        \psi_{j'\sigma'b}\psi_{j',-\sigma'b}}_{(y,0)}\rangle_0\\ \nonumber
        &=&2\delta_{\sigma,-\sigma'}\delta_{bR}\delta_{jj'}e^{i\delta ky}
        \langle\psi_{j\sigma R}^{\dagger}\psi_{j',-\sigma'b}\rangle_0
        \langle\psi_{j,-\sigma R}^{\dagger}\psi_{j'\sigma'b}\rangle_0
        \langle\psi_{j,-\sigma L}\psi_{j'+1,\sigma',-b}^{\dagger}\rangle_0
        \langle\psi_{j\sigma L}\psi_{j',-\sigma',-b}^{\dagger}\rangle_0\\ \nonumber
        &=&2\delta_{\sigma,-\sigma'}\delta_{bR}\delta_{jj'}e^{i\delta ky}
        \mathcal{G}_{j\sigma R}(y-x,-\tau)
        \mathcal{G}_{j,-\sigma R}(y-x,-\tau)
        \langle\psi_{j,-\sigma L}(x,\tau)\psi_{j+1,-\sigma L}^{\dagger}(y,0)\rangle_0
        \mathcal{G}_{j\sigma L}(x-y,\tau)\\
        &\hookrightarrow& -4t_{\perp}A(-i\Omega,-B),
    \end{eqnarray}
    \begin{eqnarray}\nonumber
        &&-e^{i\delta ky}\langle\underbrace{\psi_{j,-\sigma L}^{\dagger}
        \psi_{j\sigma L}^{\dagger}\psi_{j\sigma R}\psi_{j,-\sigma R}}_{(x,\tau)}
        \underbrace{\psi_{j',-\sigma',-b}^{\dagger}\psi_{j'+1,\sigma',-b}^{\dagger}
        \psi_{j'\sigma'b}\psi_{j',-\sigma'b}}_{(y,0)}\rangle_0\\ \nonumber
        &=&-2\delta_{\sigma\sigma'}\delta_{bL}\delta_{jj'}e^{i\delta ky}
        \langle\psi_{j,-\sigma L}^{\dagger}\psi_{j',-\sigma'b}\rangle_0
        \langle\psi_{j\sigma L}^{\dagger}\psi_{j'\sigma'b}\rangle_0
        \langle\psi_{j\sigma R}\psi_{j'+1,\sigma',-b}^{\dagger}\rangle_0
        \langle\psi_{j,-\sigma R}\psi_{j',-\sigma',-b}^{\dagger}\rangle_0\\ \nonumber
        &=&-2\delta_{\sigma\sigma'}\delta_{bL}\delta_{jj'}e^{i\delta ky}
        \mathcal{G}_{j,-\sigma L}(y-x,-\tau)
        \mathcal{G}_{j\sigma L}(y-x,-\tau)
        \langle\psi_{j\sigma R}(x,\tau)\psi_{j+1,\sigma R}^{\dagger}(y,0)\rangle_0
        \mathcal{G}_{j,-\sigma R}(x-y,\tau)\\
        &\hookrightarrow& 4t_{\perp}A(-i\Omega,B)
    \end{eqnarray}
    \begin{eqnarray}\nonumber
        &&-e^{i\delta ky}\langle\underbrace{\psi_{j\sigma R}^{\dagger}
        \psi_{j,-\sigma R}^{\dagger}\psi_{j,-\sigma L}\psi_{j\sigma L}}_{(x,\tau)}
        \underbrace{\psi_{j'+1,-\sigma',-b}^{\dagger}\psi_{j'+1,\sigma',-b}^{\dagger}
        \psi_{j'\sigma'b}\psi_{j'+1,-\sigma'b}}_{(y,0)}\rangle_0\\ \nonumber
        &=&-2\delta_{\sigma,-\sigma'}\delta_{bR}\delta_{j',j-1}e^{i\delta ky}
        \langle\psi_{j\sigma R}^{\dagger}\psi_{j'+1,-\sigma'b}\rangle_0
        \langle\psi_{j,-\sigma R}^{\dagger}\psi_{j'\sigma'b}\rangle_0
        \langle\psi_{j,-\sigma L}\psi_{j'+1,\sigma',-b}^{\dagger}\rangle_0
        \langle\psi_{j\sigma L}\psi_{j'+1,-\sigma',-b}^{\dagger}\rangle_0\\ \nonumber
        &=&-2\delta_{\sigma,-\sigma'}\delta_{bR}\delta_{j',j-1}e^{i\delta ky}
        \mathcal{G}_{j\sigma R}(y-x,-\tau)
        \langle\psi_{j,-\sigma R}(x,\tau)\psi_{j-1,-\sigma R}^{\dagger}(y,0)\rangle_0
        \mathcal{G}_{j,-\sigma L}(x-y,\tau)
        \mathcal{G}_{j\sigma L}(x-y,\tau)\\
        &\hookrightarrow& 4t_{\perp}A(i\Omega,-B),
    \end{eqnarray}
    \begin{eqnarray}\nonumber
        &&e^{i\delta ky}\langle\underbrace{\psi_{j,-\sigma L}^{\dagger}
        \psi_{j\sigma L}^{\dagger}\psi_{j\sigma R}\psi_{j,-\sigma R}}_{(x,\tau)}
        \underbrace{\psi_{j'+1,-\sigma',-b}^{\dagger}\psi_{j'+1,\sigma',-b}^{\dagger}
        \psi_{j'\sigma'b}\psi_{j'+1,-\sigma'b}}_{(y,0)}\rangle_0\\ \nonumber
        &=&2\delta_{\sigma\sigma'}\delta_{bL}\delta_{j',j-1}e^{i\delta ky}
        \langle\psi_{j,-\sigma L}^{\dagger}\psi_{j'+1,-\sigma'b}\rangle_0
        \langle\psi_{j\sigma L}^{\dagger}\psi_{j'\sigma'b}\rangle_0
        \langle\psi_{j\sigma R}\psi_{j'+1,\sigma',-b}^{\dagger}\rangle_0
        \langle\psi_{j,-\sigma R}\psi_{j'+1,-\sigma',-b}^{\dagger}\rangle_0\\ \nonumber
        &=&2\delta_{\sigma\sigma'}\delta_{bL}\delta_{j',j-1}e^{i\delta ky}
        \mathcal{G}_{j,-\sigma L}(y-x,-\tau)
        \langle\psi_{j\sigma L}(x,\tau)\psi_{j-1,\sigma L}^{\dagger}(y,0)\rangle_0
        \mathcal{G}_{j\sigma R}(x-y,\tau)
        \mathcal{G}_{j,-\sigma R}(x-y,\tau)\\
        &\hookrightarrow& -4t_{\perp}A(i\Omega,B).
    \end{eqnarray}
At this point we have calculate all the terms and we have found that
    \begin{equation}
        i\langle K_x;K_y\rangle_0=2e^2 v_{\text{F}} t_{\perp}g_3^2a_y
        (-8t_{\perp})\left[A(i\Omega,B)-A(i\Omega,-B)
        -A(-i\Omega,B)+A(-i\Omega,-B)\right].
    \end{equation}
It is clear enough that $i\langle K_x;K_y\rangle_0$ vanishes at $B=0$. The first-order
term is
    \begin{equation}
        i\frac{\langle K_x;K_y\rangle_0}{\mathcal{S}}=-32e^2 v_{\text{F}} t_{\perp}^2g_3^2a_yB
        \left[A'(i\Omega)-A'(-i\Omega)\right]+\mathcal{O}(B^2).
    \end{equation}
with
$A'(i\Omega)=\frac{1}{\mathcal{S}}dA(i\Omega,B)/dB|_{B=0}=\frac{ea_y}{\mathcal{S}}dA(i\Omega,B)/d\delta
k|_{\delta k=0}$. For $\RH$ this implies
    \begin{equation}
        \RH(0)=\RH^0\left\{1-\frac{4\pi v_{\text{F}}^2 g_3^2}{e\alpha}
        \int\frac{d\omega}{\omega^2}\,\frac{1}{\pi}\text{Im}
        \left[A'(\omega+i0^+)-A'(-\omega-i0^+)\right]\right\}.
    \end{equation}
The last step is to evaluate $A'(\omega)$:
    \begin{eqnarray*}
        A'(i\Omega)&=&\frac{ea_y}{\mathcal{S}}\frac{N_y}{L_x^2}
        \sum_{k_1k_2q}\frac{1}{\beta^3}\sum_{i\nu_1,i\nu_2,i\nu_3}
        \mathcal{G}_{\uparrow R}(k_1,i\nu_1)
        \left[\frac{d}{d\delta k}
        \mathcal{G}_{\uparrow R}(k_1+\delta k,i\nu_1)\right]_{\delta k=0}\times\\
        &&\qquad\mathcal{G}_{\downarrow R}(k_2,i\nu_2)
        \mathcal{G}_{\downarrow L}(k_2-q,i\nu_3)
        \mathcal{G}_{\uparrow L}(k_1+q,i\nu_1+i\nu_2-i\nu_3+i\Omega)\\
        &=&\frac{ea_yN_yL_x}{\mathcal{S}}\int\frac{dk_1dk_2dq}{(2\pi)^3}
        \frac{d\xi^+_{k_1}}{dk_1}\frac{1}{\beta^3}\sum_{i\nu_1,i\nu_2,i\nu_3}
        \mathcal{G}^3_{\uparrow R}(k_1,i\nu_1)\times\\
        &&\qquad\mathcal{G}_{\downarrow R}(k_2,i\nu_2)
        \mathcal{G}_{\downarrow L}(k_2-q,i\nu_3)
        \mathcal{G}_{\uparrow L}(k_1+q,i\nu_1+i\nu_2-i\nu_3+i\Omega).
    \end{eqnarray*}
The Matsubara sums are elementary (once third-order poles are under control, and yield
\begin{eqnarray*}
    A'(i\Omega)&=&\frac{e}{(2\pi)^3}\int dk_1dk_2dq\,\frac{d\xi^+_{k_1}}{dk_1}
    F_0\left\{\frac{F_1}{i\Omega-a}-\frac{F_2}{(i\Omega-a)^2}+\frac{F_3}{(i\Omega-a)^3}\right\}\\
    F_0&=&f(\xi^+_{k_2})\left[f(\xi^-_{k_1+q})+f(\xi^-_{k_2-q})-1\right]
        -f(\xi^-_{k_1+q})f(\xi^-_{k_2-q})\\
    F_1&=&\frac{1}{2}f''(\xi^+_{k_1}),\qquad F_2=f'(\xi^+_{k_1}),\qquad
    F_3=f(\xi^+_{k_1})-f(\xi^-_{k_1+q}+\xi^-_{k_2-q}-\xi^+_{k_2})\\
    a&=&\xi^-_{k_1+q}+\xi^-_{k_2-q}-\xi^+_{k_1}-\xi^+_{k_2}
\end{eqnarray*}
where $f$ is the Fermi function $f(\xi)=(e^{\beta\xi}-1)^{-1}$.
At this stage we shift $k_1$ and $k_2$ by $k_{\text{F}}$, and $q$ by $-2k_{\text{F}}$.
Then the energy of the right movers $\xi^+(k_{1,2})= v_{\text{F}}(k_{1,2}-k_{\text{F}})+\alpha(k_{1,2}-k_{\text{F}})^2$,
which depend only on one momentum, becomes $E^+_{k_{1,2}}= v_{\text{F}} k_{1,2}+\alpha k_{1,2}^2$.
At the same time $\xi^-_{k_1+q}=- v_{\text{F}}(k_1+q+k_{\text{F}})+\alpha(k_1+q+k_{\text{F}})^2$ becomes
$E^-(k_1+q)=- v_{\text{F}}(k_1+q)+\alpha(k_1+q)^2$, and finally
$\xi^-_{k_2-q}$ becomes $\xi^-_{k_2-q+4k_{\text{F}}}=E^-(k_2-q)$. This last equality
holds because we are at half-filling so that $4k_{\text{F}}\equiv0$. We also move to
adimensional variables $\tilde{k}=\beta v_{\text{F}} k$:
    \begin{equation}
    A'(i\Omega)=\frac{e}{(2\pi)^3}\frac{1}{(\beta v_{\text{F}})^3}
    \int d\tilde{k}_1d\tilde{k}_2d\tilde{q}\, v_{\text{F}}(1+2\tilde{\alpha}\tilde{k}_1)
    \tilde{F}_0\left\{\frac{\tilde{F}_1}{i\Omega-\tilde{a}}
    -\frac{\tilde{F}_2}{(i\Omega-\tilde{a})^2}+\frac{\tilde{F}_3}{(i\Omega-\tilde{a})^3}\right\}
    \end{equation}
where $\tilde{\alpha}=\frac{\alpha}{\beta v_{\text{F}}^2}$ and the
$\tilde{F}$'s and $\tilde{a}$ are the $F$'s and $a$ with momenta
$k$ replaced by $\tilde{k}/(\beta v_{\text{F}})$ and $\xi$
replaced by $E$. For example, $\xi^+_k$ must be replaced by
$E^+_{\tilde{k}/(\beta
v_{\text{F}})}=(\tilde{k}+\tilde{\alpha}\tilde{k}^2)/\beta$. We
see that the temperature disappears completely from the momentum
integrals. We then take away the tildes and rewrite:
\def\eps{\varepsilon}
    \begin{eqnarray}\nonumber
        A'(i\Omega)&=&\frac{e}{(2\pi)^3}\frac{ v_{\text{F}}\beta^3}{(\beta v_{\text{F}})^3}
        \int dk_1dk_2dq\,(1+2\tilde{\alpha}k_1)F_0\times\\ \label{eq:Ap}
        &&\left\{\frac{F_1}{i\Omega\beta-a}
        -\frac{F_2}{(i\Omega\beta-a)^2}+\frac{F_3}{(i\Omega\beta-a)^3}\right\}\\
        \nonumber
        F_0&=&g(\eps^+_{k_2})\left[g(\eps^-_{k_1+q})+g(\eps^-_{k_2-q})-1\right]
            -g(\eps^-_{k_1+q})g(\eps^-_{k_2-q})\\
        \nonumber
        F_1&=&\frac{1}{2}g''(\eps^+_{k_1}),\qquad
        F_2=g'(\eps^+_{k_1}),\qquad
        F_3=g(\eps^+_{k_1})-g(\eps^-_{k_1+q}+\eps^-_{k_2-q}-\eps^+_{k_2})\\
        \nonumber
        g(x)&=&f(x/\beta)=(e^x-1)^{-1},\qquad\eps^{\pm}_k=\pm k+\tilde{\alpha}k^2\\
        \nonumber
        a&=&\eps^-_{k_1+q}+\eps^-_{k_2-q}-\eps^+_{k_1}-\eps^+_{k_2}.
    \end{eqnarray}
We treat this expression to first order in $\alpha$. For this we
introduce a new quantity: $A''(i\Omega)=\frac{(2\pi)^3
v_{\text{F}}^2}{e}dA'(i\Omega)/d\tilde{\alpha}|_{\tilde{\alpha}=0}$,
in such a way that $A'(i\Omega)=\frac{e}{(2\pi)^3
v_{\text{F}}^2}\frac{\alpha}{\beta
v_{\text{F}}^2}A''(i\Omega)+\mathcal{O}(\alpha^2)$. Taking into
account the various factors we arrive at
    \begin{equation}\label{eq:RHn}
    \RH(0)=\RH^0\left\{1-\frac{1}{2}\left(\frac{g_3}{\pi v_{\text{F}}}\right)^2\frac{1}{\beta}
        \int\frac{d\omega}{\omega^2}\,\frac{1}{\pi}
        \text{Im}\left[A''(\omega+i0^+)-A''(-\omega-i0^+)\right]\right\}.
    \end{equation}
Taking the imaginary part in Eqs~(\ref{eq:RHn}) and (\ref{eq:Ap}) leads to delta functions
which allows one to do the $k_1$ integration directly using the identity:
    \begin{eqnarray}\nonumber
        &&\frac{1}{\pi}\text{Im}\left[\frac{1}{(\beta\omega-a+i0^+)^n}
        -\frac{1}{(-\beta\omega-a-i0^+)^n}\right]=\\
        &&\frac{1}{(n-1)!}\left[(-1)^n\delta^{(n-1)}(\beta\omega-a)-
        \delta^{(n-1)}(\beta\omega+a)\right].
    \end{eqnarray}
The remaining $k_2$ and $q$ integrals can be done exactly yielding
    \begin{equation}\label{eq:RHnp1}
        \RH(0)=\RH^0\left\{1-\frac{1}{16}\left(\frac{g_3}{\pi v_{\text{F}}}\right)^2
        \int\frac{d\omega}{\omega}\,\frac{(\beta\omega/4)^2-\sinh^2(\beta\omega/4)}
        {\tanh(\beta\omega/4)\sinh^2(\beta\omega/4)}\right\}.
    \end{equation}
We see that the integrand becomes $-\omega^{-1}$ at high frequency, and therefore
the $\omega$ integral diverges as $-\log\,\omega_{\text{max}}$ where $\omega_{\text{max}}$ is the
cutoff. We can extract the divergent term by rewriting the integral as
    \begin{eqnarray}\nonumber
        \int\frac{d\omega}{\omega}\,\frac{(\beta\omega/4)^2-\sinh^2(\beta\omega/4)}
        {\tanh(\beta\omega/4)\sinh^2(\beta\omega/4)}
        &\sim&2\int_0^{\infty} dx\left[\frac{1}{x}\frac{x^2-\sinh^2x}{\tanh x\sinh^2 x}
        +\frac{1}{3+x}\right]-2\int_0^{\beta W/2}\frac{dx}{3+x}\\
        \nonumber
        &=&-2.0622-2\log\left(1+\frac{\beta W}{6}\right)\\
        \nonumber
        &\sim&-2\log\left(\frac{W}{T}\right).
    \end{eqnarray}
Our final result is thus
    \begin{equation}\label{eq:RHfree}
        \RH(0)\sim\RH^0\left[1+\frac{1}{8}\left(\frac{g_3}{\pi v_{\text{F}}}\right)^2
        \log\left(\frac{W}{T}\right)\right].
    \end{equation}

\chapter{Appendix for the study of the Hall effect on the triangular lattice}





\section{Some numerical details}\label{app_numerical}

To obtain the full temperature and density dependence of $\RH$
using Eq~(\ref{RH_AkBkCk}) we computed the sum over momentum
numerically. For this we took a reduced zone of the reciprocal
space of the 2D triangular lattice, defined by vectors
$\vec{b}_1=\frac{2\pi}{a}(1,-\frac{1}{\sqrt{3}})$ and
$\vec{b}_2=\frac{2\pi}{a}(0,\frac{2}{\sqrt{3}})$ (see
Sec.~\ref{sec:Hubbard_triangular}). We then performed an $N\times
N$ grid discretization of this zone (discrete mesh), with $N^2$ the total
number of sites. Due to the $\vec{k}$-periodicity of coefficients
$A_{\vec{k}}$, $B_{\vec{k}}$ and $C_{\vec{k}}$, and that of
dispersion relation $\varepsilon_{\vec{k}}$, which are the
$\vec{k}$-dependent quantities in Eq~(\ref{RH_AkBkCk}), it can be
proved that for the description of the whole reciprocal space, is
enough to make the sum over a triangle defined by points
P$_1=(0,0)$; P$_2=\frac{2\pi}{a}(1,-\frac{1}{\sqrt{3}})$ and
P$_3=\frac{2\pi}{a}(1,0)$. Each $\vec{k}$-point in the triangle
must be weighted according to the number of times it appears when
the full reduced zone is recovered, by making reflections of the
triangle. In the following, some short routines are presented
(written in fortran 95). They allow to obtain the diamagnetic
terms and therefore $\RH$ (in unities of $\RH\times |e|c$ with
c=1) at $U=0$. In this routines, "Nt" denotes the number of sites (the total number given by Nt$^2$)
and "dist" is the Fermi distribution function (where the
temperature and density enter),
\newline
\newline
\fbox{\parbox[t]{13cm}{
\ttfamily
  \textbf{function p(Nt)} ({\it{gives the proper weigh to each point in the
  triangle}})\newline 
    Implicit None \newline
    integer, intent(in) :: Nt \newline
    integer, dimension((Nt/2+2)*Nt/2) :: p \newline
    integer :: i,m  \newline
    p=4;  p(1:Nt-1)=2 \newline
    do i=1,Nt/2;  m=i*(Nt-(i-2));  p(m)=2;  p(m-1)=2 \newline
    enddo \newline
    p(Nt)=1 ; p((2+Nt/2)*Nt/2)=1 \newline
 \textbf{end function p}
}}\newline
\newline
\newline
This funtion $p(Nt)$ is used in the two following routines.
\newline
\fbox{\parbox[t]{13.1cm}{
\ttfamily
\textbf{function X(dist)}   ({\it{ returns the diamagnetic terms $\chi_x(0)$ and $\chi_y(0)$}})
\newline Implicit None \newline
  real,dimension(:),intent(in) :: dist  \newline
  real,  dimension(2):: X  \newline
  integer, dimension((Nt/2+2)*Nt/2) :: wei  \newline
  integer :: k,ki,kj            \newline
  real:: kx,ky              \newline
  wei=p(Nt) ;   X = zero      \newline
  do k=1,size(dist)     \newline
     kj=floor((Nt+two)/2-sqrt((Nt+two)**2/4-k)+1.d-10)\newline ki=k-kj*(Nt-kj) \newline
     kx=mod(ki,Nt)*two*pi/(a*Nt)\newline ky=(-mod(ki,Nt)+2*kj)*two*pi/(a*Nt*sqrt(3.d0))  \newline
     X(1)=X(1)+wei(k)*(two*cos(kx*a)+
\newline(tper/tpar)*cos(kx*a/two)*cos(ky*sqrt(3.d0)*a/two))*dist(k)
     \newline
     X(2)=X(2)+wei(k)*cos(kx*a/two)*cos(ky*sqrt(3.d0)*a/two)*dist(k)\newline
  enddo \newline
  X(1)= X(1)/Nt**2 ; X(2)= X(2)/Nt**2 \newline \textbf{end function X}}} 
\newline

Once we have the diamagnetic terms, we are ready to compute the Hall coefficient $\RH$ by means of the following routine.
\newline
\newline
\newline
\fbox{\parbox[t]{13.1cm}{
\ttfamily
\textbf{function RH(dist)} ({\it{returns the Hall constant
$\RH$}})\newline Implicit None \newline
  real(dp),dimension(:),intent(in) :: dist  \newline
  real(dp), dimension(2):: diamag   \newline
  real(dp) :: RH,term1,term2,kx,ky  \newline
  integer, dimension((Nt/2+2)*Nt/2) :: wei \newline
  integer:: k,ki,kj \newline
  diamag=X(dist);  RH= zero;  term1=zero;  term2=zero;  wei=p(Nt) \newline
  do k=1,size(dist)\newline
     kj=floor((Nt+two)/2-sqrt((Nt+two)**2/4-k)+1.d-10)\newline ki=k-kj*(Nt-kj)\newline
     kx=mod(ki,Nt)*two*pi/(a*Nt) \newline
     ky=(-mod(ki,Nt)+2*kj)*two*pi/(a*Nt*sqrt(3.d0))  \newline
     term1=term1+\newline wei(k)*cos(kx*a)*cos(kx*a/two)*cos(ky*sqrt(3.d0)*a/two)*dist(k)    \newline
     term2=term2+\newline wei(k)*(cos(kx*a)+cos(ky*sqrt(3.d0)*a))*dist(k)/4.d0   \newline
  enddo \newline
  RH = -((a**2)*sqrt(3.d0)/2.d0)*(term1+\newline
(tper/tpar)*term2)/(diamag(1)*diamag(2)*Nt**2) \newline
\textbf{end function RH}}}
\newline

In the  $U\ne0$ case, the above routines must be adapted to obtain
the self-energy and consequently $\langle n_{\vec{k}}\rangle$,
with the different methods explained in Sec.~\ref{sec:results}.
Due to the weak momentum of $\Sigma(\vec{k},i\omega_n)$, we can
compute it in a $16\times16$ mesh and then, using the function
shown below, interpolate it into an $N\times N$ mesh as large as
possible (in our work $1024\times1024$):
\fbox{\parbox[t]{13cm}{
\ttfamily
\textbf{function spline (v1,v2,A)} ({\it{returns a linear interpolation of the array A to the vector v) (v1 must
        be in the range [0,size(A,1)] and v2 in the range [0,size(A,2)].}})
\newline
         Implicit None \newline
         real(dp),                      intent(in) :: v1, v2 \newline
         complex(dp), dimension(0:,0:), intent(in) :: A      \newline
         complex(dp)                               :: spline  \newline
         integer :: N, M, i, j; real(dp) x, y\newline
	 complex(dp) :: A1, A2, A3,
         A4\newline
         N=size(A,1); M=size(A,2) \newline
         if(v1<zero.or.v1>real(N,dp)) then \newline
            stop 'Stopped => Out-of-range vector v1 in lib:spline'
            \newline
         else if(v2<zero.or.v2>real(M,dp)) then \newline
            stop 'Stopped => Out-of-range vector v2 in lib:spline'
            \newline
         endif \newline
         i=int(v1); j=int(v2); x=v1-i; y=v2-j; i=mod(i,N);
         j=mod(j,M)\newline
         A1=A(i,j); A2=A(mod(i+1,N),j) \newline
         A3=A(mod(i+1,N),mod(j+1,M)); A4=A(i,mod(j+1,M)) \newline
         spline=A1+(A2-A1)*x+(A4-A1)*y+(A1-A2+A3-A4)*x*y \newline
       \textbf{end function spline}}}
\newline
\newline

Then, for each element $(i,j)$ of the $N_0\times N_0$ original complex matrix $\Sigma(\vec{k},i\omega_n)$ we find by interpolation the element $(iN_0/N,jN_0/N)$ with $N\times N$ the size of the final matrix we want to obtain. In the above routine, v$_1$, v$_2$ and A correspond to $iN_0/N$, $jN_0/N$ and $\Sigma$ respectively.

\section{Calculation of the DMFT self-energy}\label{app_DMFT}

We evaluate the local self-energy in the DMFT framework using the
Hirsh-Fye algorithm \cite{hirsch_fye_qmc} as described in
Ref.~\cite{georges_dmft}. In this method the imaginary-time
axis $[0,\beta[$ is cut into $L$ slices, and the Trotter formula
is used in each time slice in order to single out the Hubbard
interaction. In a second step the interaction is decoupled via the
introduction of an Ising variable in every time slice. The Green's
function $\mathcal{G}(\tau)$ is finally calculated by averaging
over the ensemble of configurations of the Ising variables using a
Monte-Carlo sampling and local updates. In our calculations at
$\beta=1$ and $U\leqslant20$ we take $L=128$ and we keep $10^6$
out of the $\sim 10^8$ configurations visited. The numerical
accuracy of the calculated $\mathcal{G}(\tau)$ is estimated to be
$\sim10^{-3}$ at the highest $U$ values, and closer to $10^{-4}$
at $U\lesssim8$. We believe that the accuracy at $U>8$ is not
sufficient to get a reliable self-energy; hence we did not
evaluate $\RH$ at $U>8$ in Fig.~\ref{fig:RH_vs_U}.

\begin{figure}[tb]
\includegraphics[width=8cm]{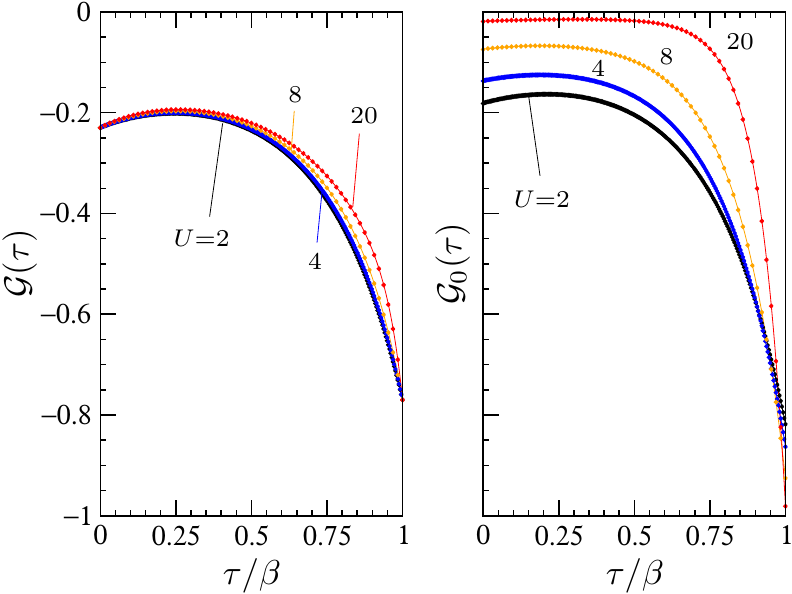}
\caption{\label{fig:DMFT} DMFT imaginary-time propagators
calculated at $\beta=1$ and $n=1.54$ for various interaction
strengths. The symbols show the QMC results on the discrete-time
mesh. The solid lines are the fit, Eq.~(\ref{eq:fit_poles}), used
to evaluate the self-energy shown in Fig.~\ref{fig:Im_self}. }
\end{figure}

In order to calculate the self-energy and solve the DMFT
self-consistency condition, Eq.~(\ref{eq:self-consistency}), we
need to Fourier transform $\mathcal{G}(\tau)$ from imaginary time
to imaginary Matsubara frequencies $i\omega_n$. In traditional
implementations of the algorithm this step is performed through a
cubic spline interpolation of $\mathcal{G}(\tau)$. Because cubic
splines are non-analytic, however, the resulting Fourier series
are unreliable at frequencies above $\sim L/\beta$. Instead of an
interpolation, we have performed a fit of $\mathcal{G}(\tau)$. Our
fitting function is a discrete form of the spectral
representation, $\mathcal{G}(\tau)=-\int
d\varepsilon\,A(\varepsilon) e^{-\varepsilon\tau}f(-\varepsilon)$,
which we express as
    \begin{equation}\label{eq:fit_poles}
        \mathcal{G}(\tau)=-\sum_{j=1}^{M}A_je^{-\varepsilon_j\tau}f(-\varepsilon_j)
    \end{equation}
with $A_j\geqslant0$ and $\sum_{j=1}^M A_j=1$. The number $M$ of
poles $\varepsilon_j$ and their weight $A_j$ are determined by
adding more and more terms in Eq.~(\ref{eq:fit_poles}), until the
fitted function matches all QMC data points within a numerical
tolerance, which we take as the estimated accuracy of
$\mathcal{G}(\tau)$. The Fourier transform is then simply
    \begin{equation}\label{eq:continuation}
        \mathcal{G}(i\omega_n)=\sum_{j=1}^{M}\frac{A_j}{i\omega_n-\varepsilon_j}.
    \end{equation}
The calculated self-consistent propagators $\mathcal{G}(\tau)$ and
$\mathcal{G}_0(\tau)$ are displayed in Fig.~\ref{fig:DMFT},
together with the fits to Eq.~(\ref{eq:fit_poles}).

Solving Eq.~(\ref{eq:self-consistency}) at fixed electron density
requires to determine the chemical potential $\mu$
self-consistently. In our calculations we perform the search for
both the self-consistent $\mathcal{G}_0$ and $\mu$ in one shot
using a global minimization procedure. As a result the
self-consistent solution can be reached in typically less that 20
iterations.

\section{Self-energy and distribution function in the atomic limit}
\label{app_atomic}

In the case $U\gg t$ one can treat the Hamiltonian Eq.~(\ref{H}) using a
perturbative expansion in $t_{ij}/U$. The atomic limit is the zeroth-order term
of this development, and it corresponds to a collection of disconnected sites
with four possible states on each site. This limit is not very useful since
there is no hopping and thus no possible transition below the Hubbard energy
$U$. In order to retain the low-energy dynamics of the problem we adopt an
hybrid approach, where the free dispersion is used in the lattice Green's
function together with the self-energy evaluated in the atomic limit. The atomic
self-energy is obtained by diagonalizing the Hamiltonian Eq.~(\ref{H}) with
$t_{ij}=0$, which leads to the atomic Green's function
    \begin{equation}\label{Gatomic}
        G_{\text{at}}(i\omega_n)=\frac{1-n/2}{i\omega_n+\mu_{\text{at}}}+
        \frac{n/2}{i\omega_n+\mu_{\text{at}}-U},
    \end{equation}
while the non-interacting $G_{0,\text{at}}=1/(i\omega_n+\mu_{\text{at}})$
results by putting $U=0$. From Dyson's equation, $\Sigma=G_0^{-1}-G^{-1}$, we
deduce the atomic self-energy displayed in Eq.~(\ref{atomic-self}). Here
$\mu_{\text{at}}$ is the chemical potential in the \emph{true} atomic
limit---\textit{i.e.} the limit where the lattice Green's function takes the
form (\ref{Gatomic}), and therefore the electron density is given by
$n=(2-n)f(-\mu_{\text{at}})+nf(U-\mu_{\text{at}})$ with $f$ the Fermi function.
We can invert this relation and express $\mu_{\text{at}}$ explicitly in terms of
the electron density as
    \begin{equation*}
        \mu_{\text{at}}=-\frac{1}{\beta}\log\left[{\textstyle\frac{1}{n}-1+
        \sqrt{\left(\frac{1}{n}-1\right)^2+
        e^{-\beta U}\left(\frac{2}{n}-1\right)}}\right].
    \end{equation*}
Using the atomic self-energy Eq.~(\ref{atomic-self}) as an approximation to the
exact self-energy in Eq.~(\ref{n_k}), we evaluate analytically the lattice
distribution function $\langle n_{\vec{k}}\rangle$. Let's first remark that
    \begin{equation*}
        \frac{1}{i\omega_n-\xi_{\vec{k}}-\Sigma_{\text{at}}(i\omega_n)}=
        \frac{Z_{\vec{k}}}{i\omega_n-E^+_{\vec{k}}}+
        \frac{1-Z_{\vec{k}}}{i\omega_n-E^-_{\vec{k}}}
    \end{equation*}
with
    \begin{eqnarray*}
        E^{\pm}_{\vec{k}}&=&(\xi_{\vec{k}}\pm\Delta_{\vec{k}}+U-\mu_{\text{at}})/2\\
        Z_{\vec{k}}&=&\frac{\xi_{\vec{k}}+\Delta_{\vec{k}}+U-\mu_{\text{at}}}
            {4\Delta_{\vec{k}}}\times\\ \nonumber
            &&\frac{(\xi_{\vec{k}}+\Delta_{\vec{k}}-U+\mu_{\text{at}})
            (\mu_{\text{at}}-U+nU/2)+n\mu_{\text{at}}U}{\xi_{\vec{k}}
            (\mu_{\text{at}}-U+nU/2)+n\mu_{\text{at}}U/2}\\
        \Delta_{\vec{k}}&=&\sqrt{(\xi_{\vec{k}}+U+\mu_{\text{at}})^2+2(n-2)
            (\xi_{\vec{k}}+\mu_{\text{at}})U}.
    \end{eqnarray*}
As a result the Matsubara sum in Eq.~(\ref{n_k}) is easily performed to yield
    \begin{equation}
        \langle n_{\vec{k}}\rangle_{\text{at}}=Z_{\vec{k}}f(E^+_{\vec{k}})+
        (1-Z_{\vec{k}})f(E^-_{\vec{k}}).
    \end{equation}

Within this approximation it is also straightforward to perform the infinite $U$
limit. Taking into account that both $\mu$ and $\mu_{\text{at}}$ are either of
order $t$ (if $n<1$) or of order $U$ (if $n>1$) we find that $Z_{\vec{k}}$
approaches $n/2$ as $U$ increases toward $+\infty$. Likewise, if $n<1$ we have
$E^+_{\vec{k}}\sim U$ and $E^-_{\vec{k}}\sim t$ while if $n>1$ we have
$E^+_{\vec{k}}\sim t$ and $E^-_{\vec{k}}\sim-U$. Hence we find
     \begin{equation*}
        \langle n_{\vec{k}}\rangle_{\text{at}}^{U=\infty}=
        \begin{cases}
        \left(1-\frac{n}{2}\right)f\big[\left(1-\frac{n}{2}\right)\xi_{\vec{k}}-
        \frac{n}{2}\mu_{\text{at}}\big]&n<1\\[1em]
        \frac{n}{2}f\big[\frac{n}{2}\tilde{\xi}_{\vec{k}}-
        \left(1-\frac{n}{2}\right)\tilde{\mu}_{\text{at}}\big]+1-\frac{n}{2}&n>1
        \end{cases}
     \end{equation*}
where we have introduced
$\tilde{\mu}_{\text{at}}\equiv\mu_{\text{at}}-U=-\frac{1}{\beta}\log[(1-n/2)/(n-
1)]$ and $\tilde{\xi}_{\vec{k}}\equiv\varepsilon_{\vec{k}}-\tilde{\mu}$ with
$\tilde{\mu}=\mu-U$. For the purpose of evaluating the high-temperature behavior
of the Hall coefficient at $U=\infty$, we finally expand the distribution
function in powers of $\beta$ following the procedure described in
Sec.~\ref{section:RH0_highT}:
    \begin{equation*}
        \langle n_{\vec{k}}\rangle_{\text{at}}^{U=\infty}=
        \begin{cases}
        \frac{n}{2}-n(1-n)\varepsilon_{\vec{k}}\frac{\beta}{2}
            +\mathcal{O}(\beta^2)&n<1\\[1em]
        \frac{n}{2}-(n-2)(1-n)\varepsilon_{\vec{k}}\frac{\beta}{2}
            +\mathcal{O}(\beta^2)&n>1
        \end{cases}
    \end{equation*}
Comparing with Eq.~(\ref{expand}), which is valid at $U=0$, we see that the only
difference between the high-temperature behaviors of $\RH$ at $U=0$ and
$U=\infty$ is the $n$-dependent slope, and we easily deduce that
    \begin{equation*}
        \RH^{U=\infty}(T\gg t)=
        \begin{cases}
            \frac{T/t}{e}\frac{1}{n(1-n)}\frac{a^2\sqrt{3}}{4}\frac{3}{2+(t'/t)^2}
            &n<1\\[1em]
            \frac{T/t}{e}\frac{1}{(n-2)(1-n)}\frac{a^2\sqrt{3}}{4}\frac{3}{2+(t'/t)^2}
            &n>1
        \end{cases}
    \end{equation*}
By introducing $\delta=|n-1|$ which measures the doping with respect to half-filling,
these two cases can be recast in one single expression shown in Eq.~(\ref{RHinf-T}).






\begin{center}
\vspace*{1.5cm} {\Large \bfseries Acknowledgements}
\end{center}
\vspace{0.5cm}

This PhD thesis would not have been possible without the help and
support of the people around me, to only some of whom it is
possible to give particular mention here.

Above all, I would like to thank my principal supervisor, Prof.
Thierry Giamarchi, who gave me the opportunity to come to
Switzerland for the accomplishment of this research. His hard
work, support, enthusiasm and scholarship have set an example I
hope to match someday. The patience, support and dedication of my
second advisor, Dr. Christophe Berthod, has been invaluable for
the achievement of this thesis, for which I am extremely grateful.

I am also grateful to Prof. Andrew Millis who visited us for six
month and collaborated intensively in the second part of this
work. I would like to thank Prof. B. S. Shastry, Prof. H. D. Drew
and Prof. S. Brown for valuable discussions.

I would like to acknowledge the financial support given by the
Swiss National Science Foundation through Division II and MaNEP,
and the finantial and academic support of the Universit\'{e} de
Gen\`{e}ve, in particular the Condensed Matter Physics Department
and its staff.

Specials thank to my colleagues in Giamarchi's group: Anibal,
Alejandro, Corinna, Sebastian, Alejandro, Pierre, Vivien,
Elisabeth and Mikhail; and those at the Ecole de Physique: Dook,
Heidi, Anna-Sabina, Alexander, Giorgio, Pablo, Yanina, Silvia,
No\'{e}; for their kind support and friendship along these four
years in Geneva.

I would like to thank also the VTT group of the Ecole de Physique:
Charly, Robert, G\'{e}raldine, Loren, Enrico, Tania, Alexander;
with whom I discovered the countryside of Geneva and Switzerland.

Special thanks to Aisha, Yamila, Aur\'{e}lie and others that I
have already mentioned, for making life outside work much more
fun. And so many thanks to the group of Torino, where I spent so
unforgettable weekends during my Ph.D. I would like to thank also
my friends from Caracas, you will always be in my heart.

Finally, I have no words to express the gratitude that I feel to
my parents, Chichi and Rina, who have taught me everything I know
about life; to my brother, Edel, who gives me so much energy to
continue every day and to my lovely Jose Enrique, who makes my
life so joyful. This thesis is dedicated to you, my family.







\bibliographystyle{Latex/Classes/nature}      
\renewcommand{\bibname}{References}






\end{document}